\documentclass[onecolumn]{aastex631}

\shorttitle{Molecular deuterations in SCCs}
\shortauthors{Yang et al.}

\begin{document}

\title{Molecular deuterations in massive Starless Clump Candidates}

\author{Kai Yang}\thanks{E-mail: kyang@nju.edu.cn}
\affiliation{School of Astronomy and Space Science, Nanjing University, 163 Xianlin Avenue, Nanjing 210023, P.R.China}
\affiliation{Key Laboratory of Modern Astronomy and Astrophysics (Nanjing University), Ministry of Education, Nanjing 210023, P.R.China}
\affiliation{I. Physikalisches Institut, Universit\"at zu K\"oln, Z\"ulpicher Str. 77, 50937 K\"oln, Germany}

\author{Junzhi Wang}
\affiliation{School of Physical Science and Technology, Guangxi University, Nanning 530004, P.R.China}

\author{Keping Qiu}\thanks{E-mail: kpqiu@nju.edu.cn}
\affiliation{School of Astronomy and Space Science, Nanjing University, 163 Xianlin Avenue, Nanjing 210023, P.R.China}
\affiliation{Key Laboratory of Modern Astronomy and Astrophysics (Nanjing University), Ministry of Education, Nanjing 210023, P.R.China}

\author{Tianwei Zhang}
\affiliation{I. Physikalisches Institut, Universit\"at zu K\"oln, Z\"ulpicher Str. 77, 50937 K\"oln, Germany}
\affiliation{Research Center for Intelligent Computing Platforms, Zhejiang Laboratory, Hangzhou 311100, P.R.China}

\begin{abstract}

Deuterated molecules are valuable probes for investigating the evolution and the kinematics in the earliest stages of star formation.
In this study, we conduct a comprehensive investigation by performing a single point survey of 101 starless clump candidates, and carrying out on-the-fly (OTF) observations of 11 selected sources, focusing on deuterated molecular lines using the Institut de Radioastronomie Millim\'{e}trique (IRAM) 30-m telescope.
In the single-point observation, we make 46 detections for DCO$^{+}$ \textit{J}=1$-$0, 12 for DCN \textit{J}=1$-$0, 51 for DNC \textit{J}=1$-$0, 7 for N$_{2}$D$^{+}$ \textit{J}=1$-$0, 20 for DCO$^{+}$ \textit{J}=2$-$1, and 10 for DCN \textit{J}=2$-$1. The starless clump candidates (SCCs) with deuterated molecule detections exhibit lower median kinetic temperatures and narrower H$_{2}$CO (1$_{(0,1)}$$-$0$_{(0,0)}$) median full width at half maximum (FWHM) compared to those without such detections, while simultaneously displaying similar median values of 1.1mm intensity, mass, and distance. Furthermore, our OTF observations reveal that deuterated molecules predominantly have peaks near the 1.1mm continuum peaks, with the DCO$^{+}$ \textit{J}=1$-$0 emission demonstrating higher intensity in the deuterated peak region compared to the DCN and DNC \textit{J}=1$-$0 emissions. Additionally, the majority of emissions from deuterated molecules and $^{13}$C$-$isotopologues exhibit peak positions close to those of the 1.1mm continuum peaks. By analyzing the 20$^{\prime\prime}$$\times$20$^{\prime\prime}$ regions with strongest deuterated emissions in the OTF observations, we estimated deuterated abundances of 0.004$-$0.045, 0.011$-$0.040, and 0.004$-$0.038 for D$_{\rm frac}$(HCN), D$_{\rm frac}$(HCO$^{+}$), and D$_{\rm frac}$(HNC), respectively. The differential detection of deuterated molecular lines in our OTF observations could be attributed to variations in critical densities and formation pathways.

\end{abstract}

\keywords{Massive stars (732);  Interstellar abundances (832); Interstellar molecules (849); Stars formation (1569)}


\section{Introduction}

High-mass stars play a significant role in dominating the energetics of the interstellar medium (ISM) through their feedback mechanisms (e.g., \citealt{2013Natur.499..450B}; \citealt{2015EAS....75..227S}). Several theories have been developed to investigate the formation of high-mass stars \citep{2003ApJ...585..850M,2006A&A...454L..51B,2009MNRAS.400.1775S,2018ARA&A..56...41M,2020ApJ...900...82P}. A pivotal difference among these models lies in their treatment of the early stage of this process. However, our comprehension of the initial phases of high-mass star formation remains incomplete when derived from observation. Therefore, revealing initial conditions of massive stars is paramount to gaining insights into their formation process.

During early stages of high-mass star forming regions, they are deeply embedded in their parental molecular clouds. Consequently, recent observational studies focus on studying the cold gas (e.g., \citealt{2021A&A...649A..21W}; \citealt{2022MNRAS.515.5219G}; \citealt{2022ApJ...925..144S}; \citealt{2022A&A...667A.136V}; \citealt{2023A&A...676A..78G}) and dust (e.g., \citealt{2016ApJ...822...59S}; \citealt{2019ApJ...886..102S}; \citealt{2022ApJ...936..169R}; \citealt{2023ApJ...949..109L}; \citealt{2023ApJ...950..148M}) in high-mass star forming regions at (sub-)millimeter wavelengths.

Deuterated molecules are useful diagnostic tools for studying cold ($\sim$10 K) and dense environments (n$_{\rm H_{2}}$ $>$ 10$^{5}$ cm$^{-3}$) (e.g., \citealt{2011A&A...529L...7F}; \citealt{2011A&A...530A.118P}). Due to the small zero-point energy differences which ensure that deuterium is preferentially bonded into molecules compared to hydrogen, deuterated molecules can occur effectively at a previous cold phase in the molecular evolution of the gas with very low temperatures and D atoms in molecules can be much enhanced over the cosmic D/H abundance ratio, which is $\sim$1.5 $\times$ 10$^{-5}$ \citep{2000A&A...361..388R}. 
At temperatures below 20 K and assuming low ortho-to-para H$_{2}$ ratio, deuterium fractionation increases mainly due to the effective reaction \citep{2000A&A...361..388R},
\begin{equation}\label{Eq_deut}
\rm H_{3}^{+} + HD \rightleftharpoons H_{2}D^{+} + H_{2} +\Delta E\\
\end{equation}
This reaction is exothermic, resulting in a significant increase in H${_2}$D$^{+}$ abundance. Furthermore, under cold ($\textless$10 K) and dense ($>$10$^{5}$ cm$^{-3}$) conditions, deuterium fractionation is further enhanced due to the depletion interaction between CO and H$_{2}$D$^{+}$ (e.g., \citealt{1999ApJ...523L.165C}).
This explains why the processes that form the high abundances of several deuterated molecular lines occur in massive starless clumps and they are promising candidates for us to study the physical and kinematic conditions where star forms. 

Various deuterated molecules have been studied towards star-forming regions at an early evolutionary stage, such as deuterated NH$_{3}$ (e.g., \citealt{2010A&A...517L...6B}; \citealt{2011A&A...530A.118P}; \citealt{2022MNRAS.512.4934L}), deuterated H$_{2}$CO (e.g., \citealt{2007A&A...464..245R}; \citealt{2011A&A...527A..39B}; \citealt{2021A&A...653A..45Z}), deuterated CH$_{3}$OH (e.g., \citealt{2014A&A...569A..27B}; \citealt{2015A&A...575A..87F}; \citealt{2022A&A...667A.136V}), DCN (e.g., \citealt{2001ApJS..136..579T}; \citealt{2015A&A...579A..80G}), DNC (e.g., \citealt{2003ApJ...594..859H}; \citealt{2012ApJ...747..140S}; \citealt{2014MNRAS.440..448F}), DCO$^{+}$ (e.g., \citealt{2002ApJ...565..331C}; \citealt{2011A&A...534A.134M}) and N$_{2}$D$^{+}$ (e.g., \citealt{2003A&A...403L..37C}; \citealt{2005ApJ...619..379C}; \citealt{2013ApJS..207...27A}). Among them, DCO$^{+}$ is thought to form in the gas phase and is expected to be abundant below $\sim$ 30 K \citep{1989ApJ...340..906M}. The pathway of DCN is much more complicated and can form in both cold and warm regions \citep{2007ApJ...660..441W,2013JPCA..117.9959R}. In addition to this, due to their relatively high abundances and accessible rotational transitions, DCO$^{+}$ and DCN are among the primary tracers for studying the physical conditions in the early stage of massive star formation. N$_{2}$D$^{+}$ is frequently found in the early phases of protostellar cores (e.g., \citealt{2011A&A...529L...7F}; \citealt{2019A&A...621L...7G}; \citealt{2022ApJ...939..102L}).

As powerful tracers of cold pre-stellar phase in star-forming regions, many deuterated molecules had been found in Galactic low-mass and high-mass star-forming regions in the past few decades (e.g., \citealt{2003A&A...403L..37C}; \citealt{2005ApJ...619..379C}; \citealt{2007A&A...470..221C}; \citealt{2010ApJ...718..666F}), as well as in regions thought to be precursors of massive stars and stellar clusters (e.g., \citealt{2006A&A...460..709F}; \citealt{2009A&A...499..233F}; \citealt{2011A&A...529L...7F}; \citealt{2007A&A...467..207P}; \citealt{2012ApJ...751..135P}). Both DCN and DCO$^{+}$ have been observed toward proto-planetary disks \citep{2017ApJ...835..231H}. Observations of DCN have been done toward OMC-1 \citep{1992A&A...256..595S} and Orion Bar \citep{2006A&A...454L..47L}, and detections of DCO$^{+}$ have been confirmed toward dark clouds \citep{2000A&A...356.1039T,2001ApJS..136..579T}. However, there have been few surveys of deuterated molecules toward a sample of star-forming regions and the deuterium chemistry in high-mass star formation needs more investigations.

The deuterium fraction $D_{\rm frac}$ is a ratio between the abundance of a deuterated molecule and that of its hydrogenated counterpart. Many deuterated molecules (e.g., DCN and DCO$^{+}$) have been used to study their deuterium fractions. In an ALMA observation toward seven proto-planetary disks \citep{2017ApJ...835..231H}, DCO$^{+}$/HCO$^{+}$ and DCN/HCN abundance ratios range from 0.02$-$0.06 and 0.005$-$0.08, respectively. DCO$^{+}$/HCO$^{+}$ of 2 sources in \citet{2000A&A...356.1039T} are 0.18 and 0.02. Observations of the abundance of interstellar deuterated molecules offer a valuable probe into the situation of star formation. 

In this paper, we present the results of a single-point survey of deuterated molecular lines toward 101 massive starless clump candidates and OTF observations towards 11 selected sources. The outline of this paper is as follows: we introduce the observations and data reduction in Section \ref{Obs_Data} and describe the main results in Section \ref{Res}. In Section \ref{Dis}, we present the analysis and discussion of the results, and a summary is given in Section \ref{Sum}.


\section{Sample and Observations} \label{Obs_Data}

Based on observational diagnostics related to star formation activity: compact 70 $\mu$m sources, mid-IR color-selected young stellar objects (YSOs), H$_{2}$O and CH$_{3}$OH masers, and UCH {\scriptsize II} regions, \citet{2016ApJ...822...59S} identified 2223 starless clump candidates devoid of indicators of star formation activities from the 1.1mm Bolocam Galactic Plane Survey. Futhermore, \citet{2018ApJ...862...63C} blindly selected 101 targets from this SCC catalog, specifically those with NH$_{3}$(1,1) detections, in order to investigate potential inflow signatures. In this study \citep{2018ApJ...862...63C}, HCO$^{+}$ \textit{J}=1$-$0 was detected in all sources, except for two sources with non-detections at the NH$_{3}$ velocity. In addition, six clumps exhibited blue asymmetric, self-absorbed line profiles in the presence of inflow.

The 101 massive starless clump candidates, listed in Table \ref{Tab_clumps}, are also chosen as our source sample. These targets are situated in the first quadrant of the Galactic plane, with a median distance of $\sim$4.1 kpc. A comparison of the mass and mass surface density between the selected sample and the 2223 complete SCCs in \citet{2018ApJ...862...63C} reveals that the selected sample adequately represents the mass range of the complete catalog. However, the selected sample exhibits higher peak mass surface densities, approximately twice as much as those found in the complete SCC sample. For the identification of line detection, the $v_{LSR}$ of NH$_{3}$ reported in \citet{2016ApJ...822...59S} is regarded as the systemic velocity for each source.

Our observation was performed with the IRAM 30-m telescope at Pico Veleta, Spain, in May 2019. We checked the pointing about every two hours with nearby strong millimeter emitting quasi-stellar objects, and the focus after starting of the observation, sunrise, and sunset with Mars and Jupiter. We started from single-point line survey of deuterated molecular lines, such as DCN, DCO$^{+}$, N$_{2}$D$^{+}$ \textit{J}=1$-$0, and DCN, DCO$^{+}$ \textit{J}=2$-$1 toward SCCs. A position-switching mode was used and each source was observed in two on-off cycles, with each cycle lasting four minutes. We set the off position (1$_{\cdot}^{\circ}$0, 0$_{\cdot}^{\circ}$0) in R.A. and DEC. from the on position. The Eight Mixer Receiver (EMIR) with dual-polarization and the Fourier Transform Spectrometers (FTS) backend with 8 GHz frequency coverage and 195 kHz spectral resolution were used to record signals. The 3 mm (E0) band and 2 mm (E1) band of the EMIR receiver were used to cover two frequency ranges, 71.3-79 GHz and 138.3-146 GHz, respectively. The typical system temperatures are 134 K and 161 K at 72 GHz and 144 GHz, respectively. The beam sizes of the IRAM 30-m telescope are roughly 34$^{\prime\prime}$ and 17$^{\prime\prime}$ at 72 GHz and 144 GHz, respectively. The root-mean-square (rms) noise levels are 61.4 mK at 72 GHz and 66.9 mK at 144 GHz. The velocity resolution and aperture efficiency at different frequencies of deuterated molecular lines are shown in Table \ref{obs_para}.

After single-point observation, we chose 11 sources with strong flux densities of DCO$^{+}$ and DNC \textit{J}=1$-$0 or the most detection of the deuterated molecular lines and then carried out the on-the-fly (OTF) mode mapping observation toward them. We used the OTF PSW observing mode with EMIR to cover a field of view of 2$^{\prime}$ $\times$ 2$^{\prime}$. The scan spacing is 6$^{\prime\prime}$. The FTS backend with 8 GHz frequency coverage and 195 kHz spectral resolution was also used in our OTF mode mapping. On the one hand, we set the frequency ranges of EMIR to be the same as in single-point observation. We obtained a sensitivity of 44.3 mK (T$_{a}^{\ast}$) in 1.5 hours for each source and a system temperature of 183 K at 72 GHz. On the other hand, to cover H$^{13}$CN, H$^{13}$CO$^{+}$ \textit{J}=1$-$0 and N$_{2}$D$^{+}$ \textit{J}=2$-$1, we also mapped each source for 0.5 hours with two frequency ranges, 86-94 GHz and 148-156 GHz, set respectively for the 3 mm (E0) band and 2 mm (E1) band. The sensitivity was 60.2 mK and the system temperature was around 137 K at 87 GHz. Table \ref{obs_para} also lists the other detailed observational information on different molecular lines.
 
The CLASS and GREG packages of GILDAS\footnote{http://www.iram.fr/IRAMFR/GILDAS} are used to reduce data from the single-point and OTF observation. We fit and subtract first-order polynomial baselines of the spectrum. The rest frequencies of the molecular lines from the Cologne Database for Molecular Spectroscopy \citep[CDMS;][]{2001A&A...370L..49M,2005JMoSt.742..215M,2016JMoSp.327...95E}. The velocity-integrated intensities of these molecular lines are derived from a single-component Gaussian fit to the spectra. We convert the antenna temperature (T$_{a}^{\ast}$) to the main beam brightness temperature (T$_{\rm mb}$) with T$_{\rm mb}$ = T$_{a}^{\ast}$ $\cdot$ F$_{\rm eff}$/B$_{\rm eff}$, where the forward efficiency (F$_{\rm eff}$) and the beam efficiency (B$_{\rm eff}$) from the IRAM 30-m homepage\footnote{http://www.iram.es/IRAMES/mainWiki/Iram30mEfficiencies} are listed in Table \ref{obs_para}. The final data cube is produced by combining all data and re-gridding the image to a pixel size of 10$^{\prime\prime}$ at 3 mm band and 6$^{\prime\prime}$ at 2 mm band, approximately one third of the beam sizes.


\section{Results} \label{Res}

We made single-point observations towards the 101 starless clump candidates with the IRAM 30-m telescope and further conducted OTF mode observation towards 11 selected sources to obtain the spatial distribution of deuterated molecules and their $^{13}$C$-$isotopologues. DCO$^{+}$ and DNC \textit{J}=1$-$0 have higher detection rates than DCN and N$_{2}$D$^{+}$ \textit{J}=1$-$0. 1$-$0 rotational lines of deuterated molecules and $^{13}$C$-$isotopologues have peaks near the 1.1~mm continuum peaks. The results of the single-point observations are described in Section \ref{res_sp} and the OTF mode observations are shown in Section \ref{res_OTF}.

\subsection{Single-point observations}\label{res_sp}

\subsubsection{Deuterated molecules}

\begin{figure*}
\centering
\includegraphics[width=165mm]{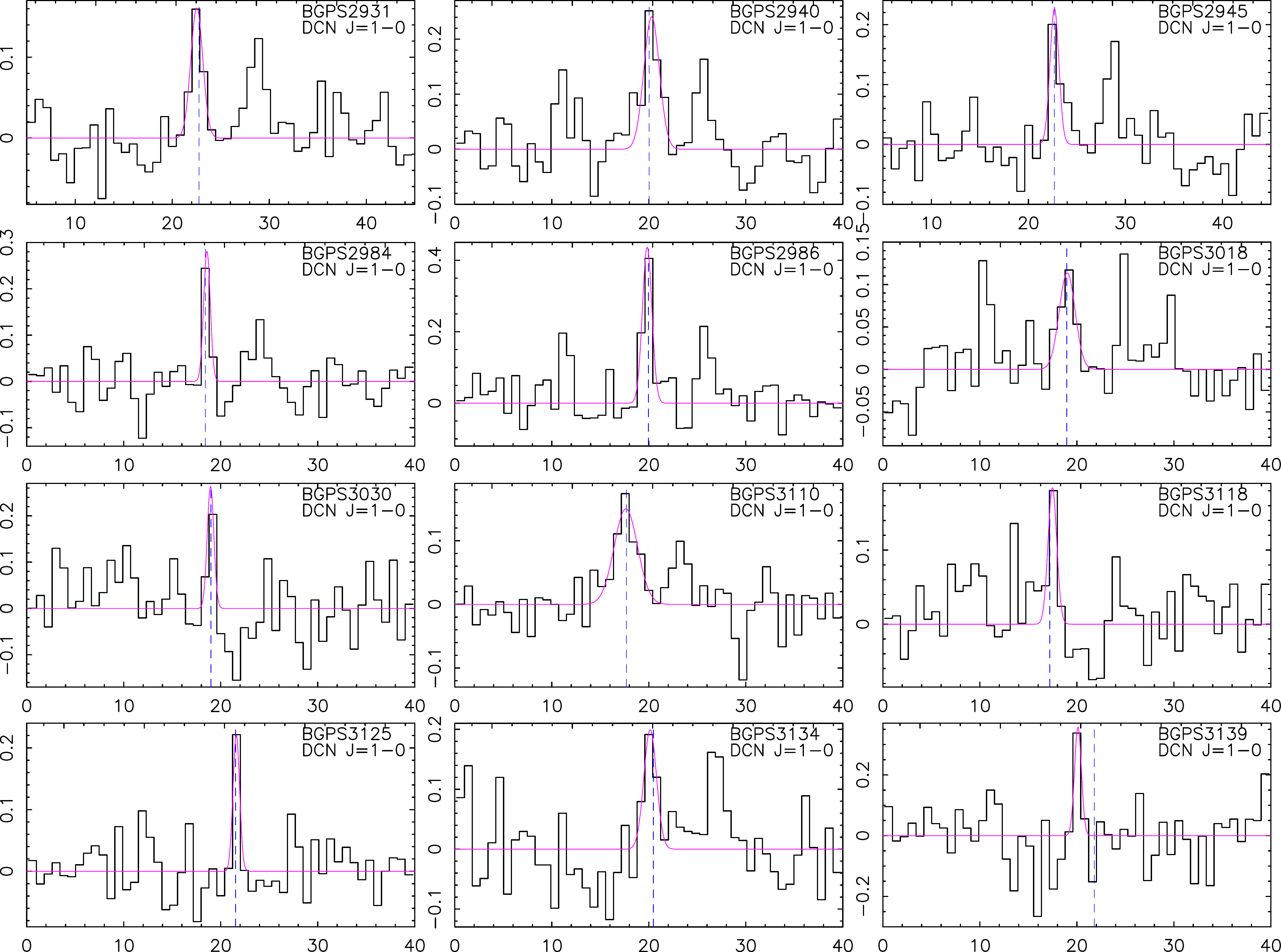}  
\caption{Line profiles of the 12 DCN \textit{J}=1$-$0 detected clumps. The Gaussian fitting result for each spectrum is shown in magenta. The blue dashed vertical line represents the LSR velocity of NH$_{3}$ for each SCC derived from \citet{2016ApJ...822...59S}. The x-axis is velocity in km s$^{-1}$, and the y-axis is T$_{\rm mb}$ in kelvin.}
\label{DCN1-0}
\end{figure*}

\begin{figure*}
\centering
\includegraphics[width=165mm]{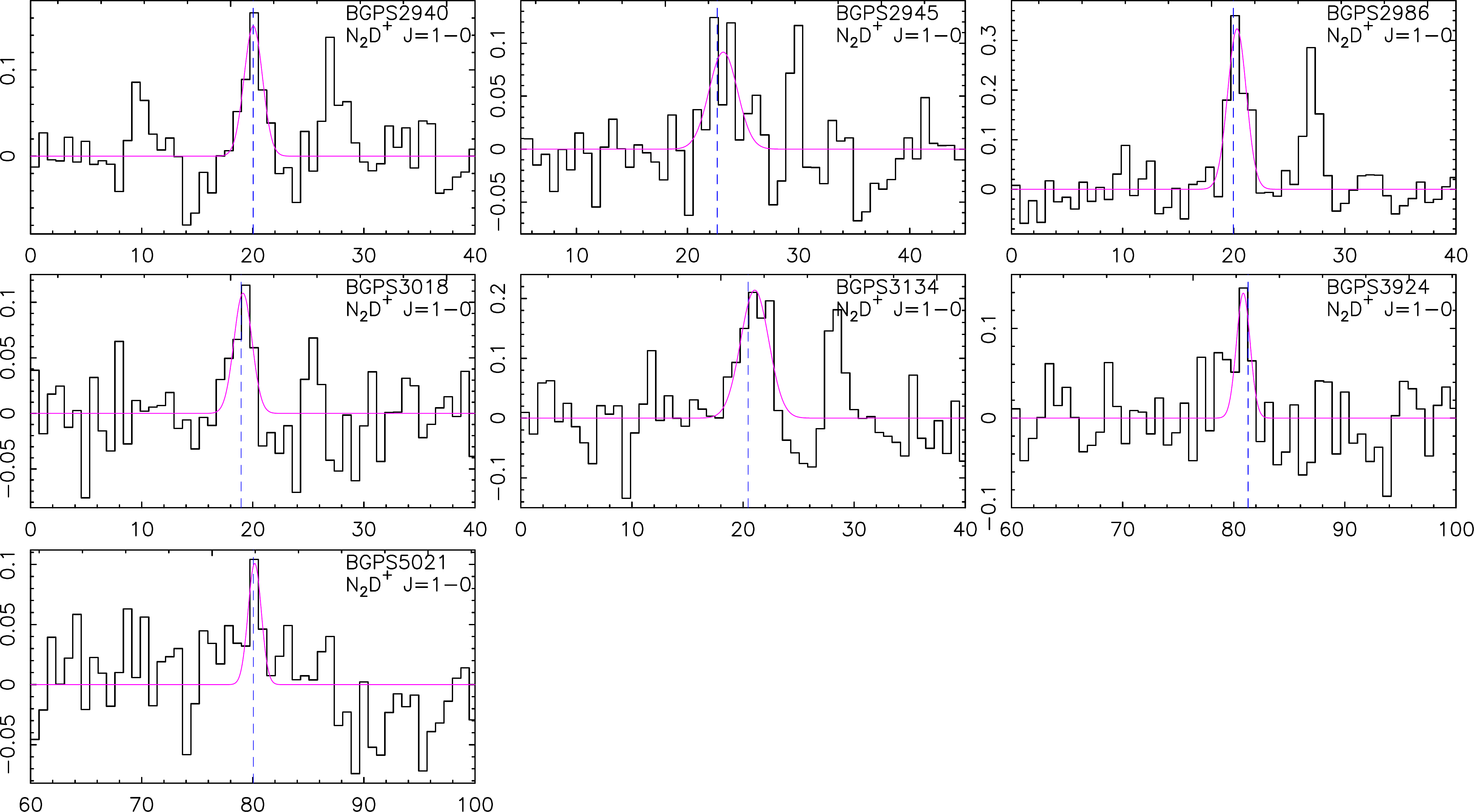}  
\caption{Line profiles of the 7 N$_{2}$D$^{+}$ \textit{J}=1$-$0 detected clumps. The Gaussian fitting result for each spectrum is shown in magenta. The blue dashed vertical line represents the LSR velocity of NH$_{3}$ for each SCC derived from \citet{2016ApJ...822...59S}. The x-axis is velocity in km s$^{-1}$, and the y-axis is T$_{\rm mb}$ in kelvin.}
\label{N2D+1-0}
\end{figure*}

\begin{figure*}
 \begin{center}
\includegraphics[width=16.5cm]{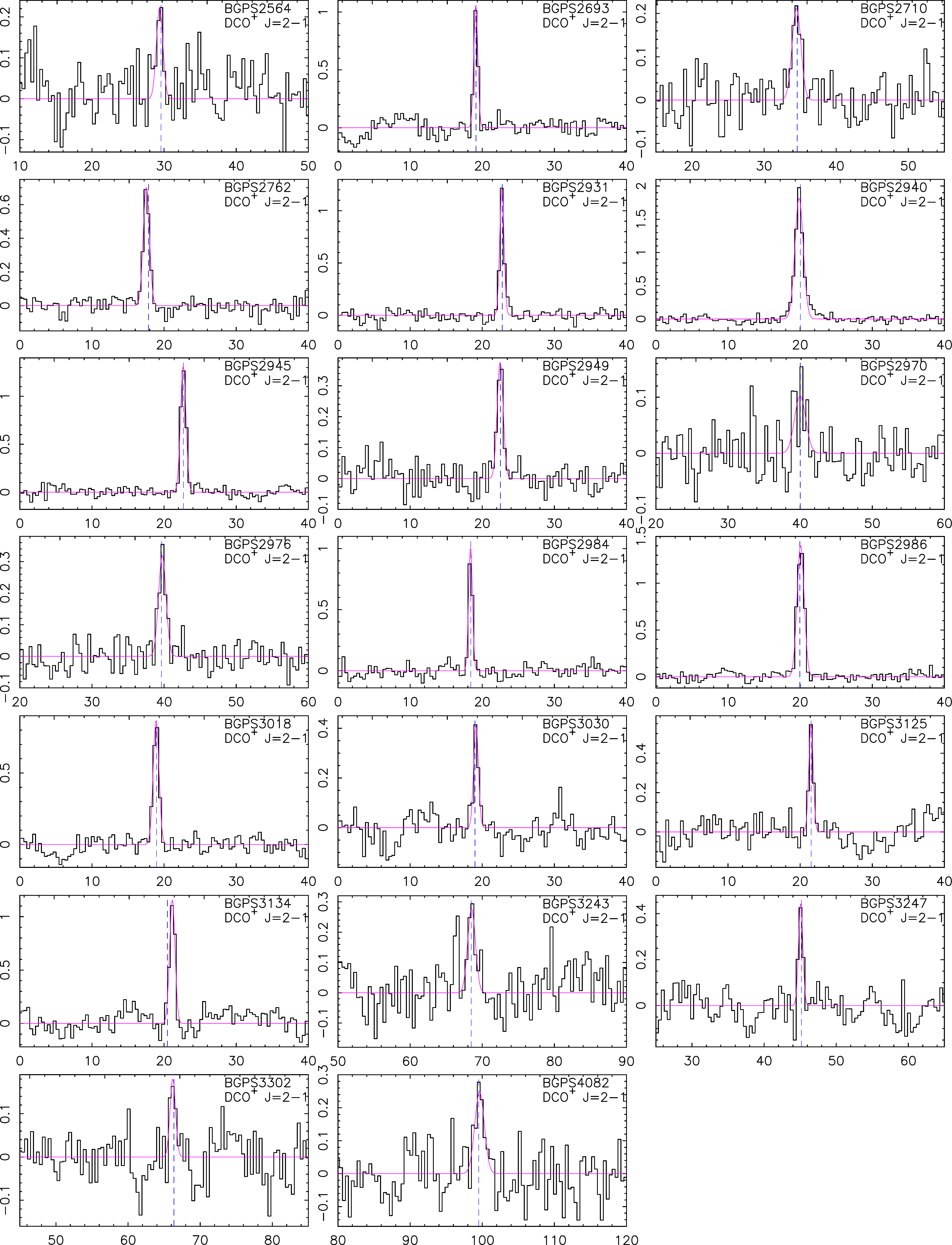}
\caption{Line profiles of the 20 DCO$^{+}$ \textit{J}=2$-$1 detected clumps. The Gaussian fitting result for each spectrum is shown in magenta. The blue dashed vertical line represents the LSR velocity of NH$_{3}$ for each SCC derived from \citet{2016ApJ...822...59S}. The x-axis is velocity in km s$^{-1}$, and the y-axis is T$_{\rm mb}$ in kelvin.}
\label{DCO+2-1}
\end{center}
\end{figure*}

\begin{figure*}
\centering
\includegraphics[width=165mm]{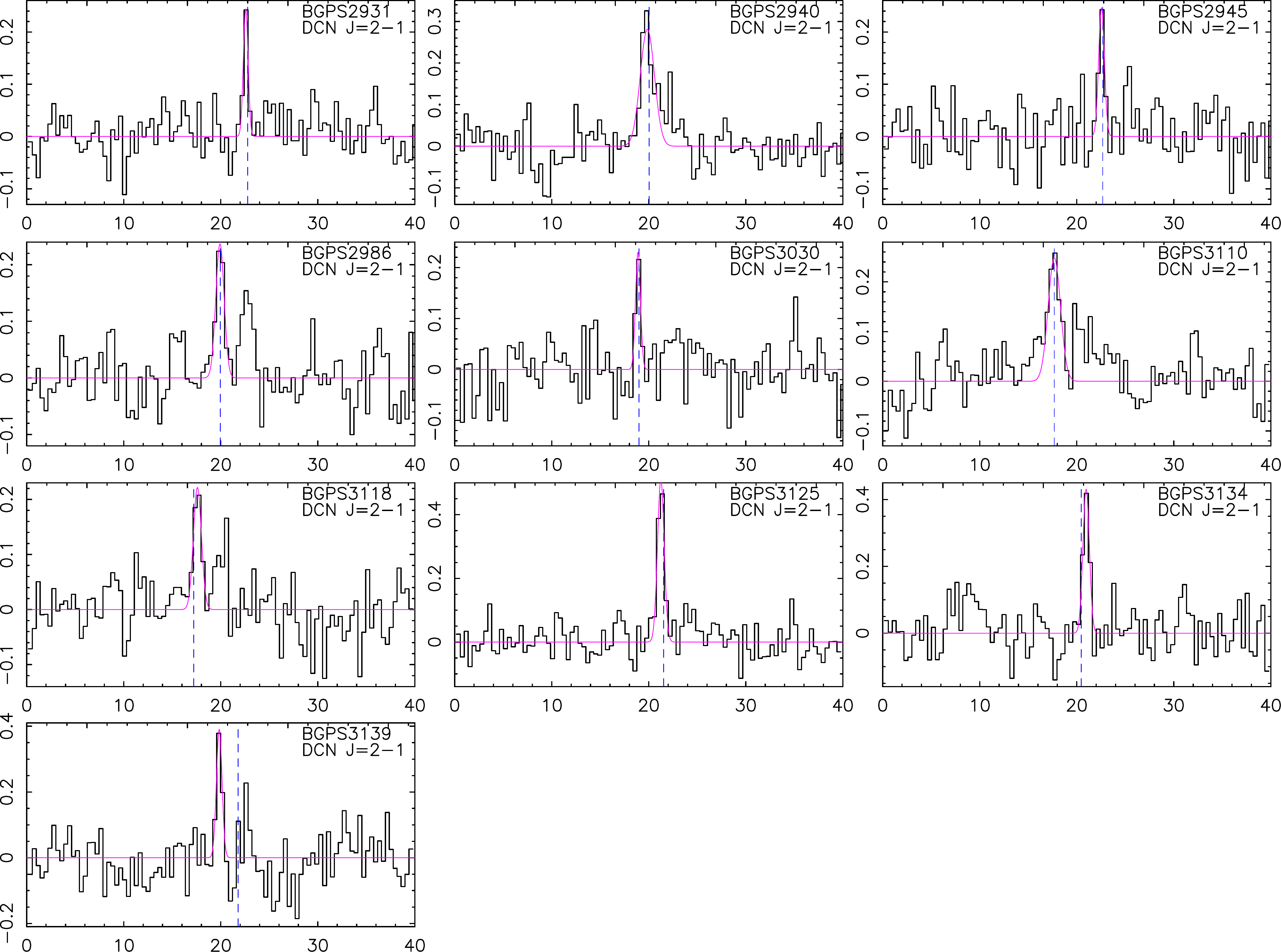}  
\caption{Line profiles of the 10 DCN \textit{J}=2$-$1 detected clumps. The Gaussian fitting result for each spectrum is shown in magenta. The blue dashed vertical line represents the LSR velocity of NH$_{3}$ for each SCC derived from \citet{2016ApJ...822...59S}. The x-axis is velocity in km s$^{-1}$, and the y-axis is T$_{\rm mb}$ in kelvin.}
\label{DCN2-1}
\end{figure*}

In total, the DCO$^{+}$, DCN, DNC, N$_{2}$D$^{+}$ \textit{J}=1$-$0, and DCO$^{+}$, DCN \textit{J}=2$-$1 emissions are detected in 46, 12, 51, 7, 20, and 10 out of the 101 sources with velocity integrated intensities above 3$\sigma_{\rm area}$, respectively. The rms of the velocity integrated intensity $\sigma_{\rm area}$ is derived from the single component Gaussian fitting. Table \ref{Tab_clumps} presents the detection status of these deuterated molecular lines and Figure \ref{DCO+1-0}$-$\ref{DCN2-1} displays the detected spectra. Gaussian fitting parameters of DCO$^{+}$, DCN, and DNC \textit{J}=1$-$0 detections are listed in Table \ref{Res_Single1}, and the results of N$_{2}$D$^{+}$ \textit{J}=1$-$0, and DCO$^{+}$, DCN \textit{J}=2$-$1 detections are shown in Table \ref{Res_Single2}. 
Owing to the limited signal-to-noise level and velocity resolution, we have not applied deconvolution to the data. As a result, we present the observed velocity width directly in the tables.

\begin{figure*}
\centering
\includegraphics[width=80mm]{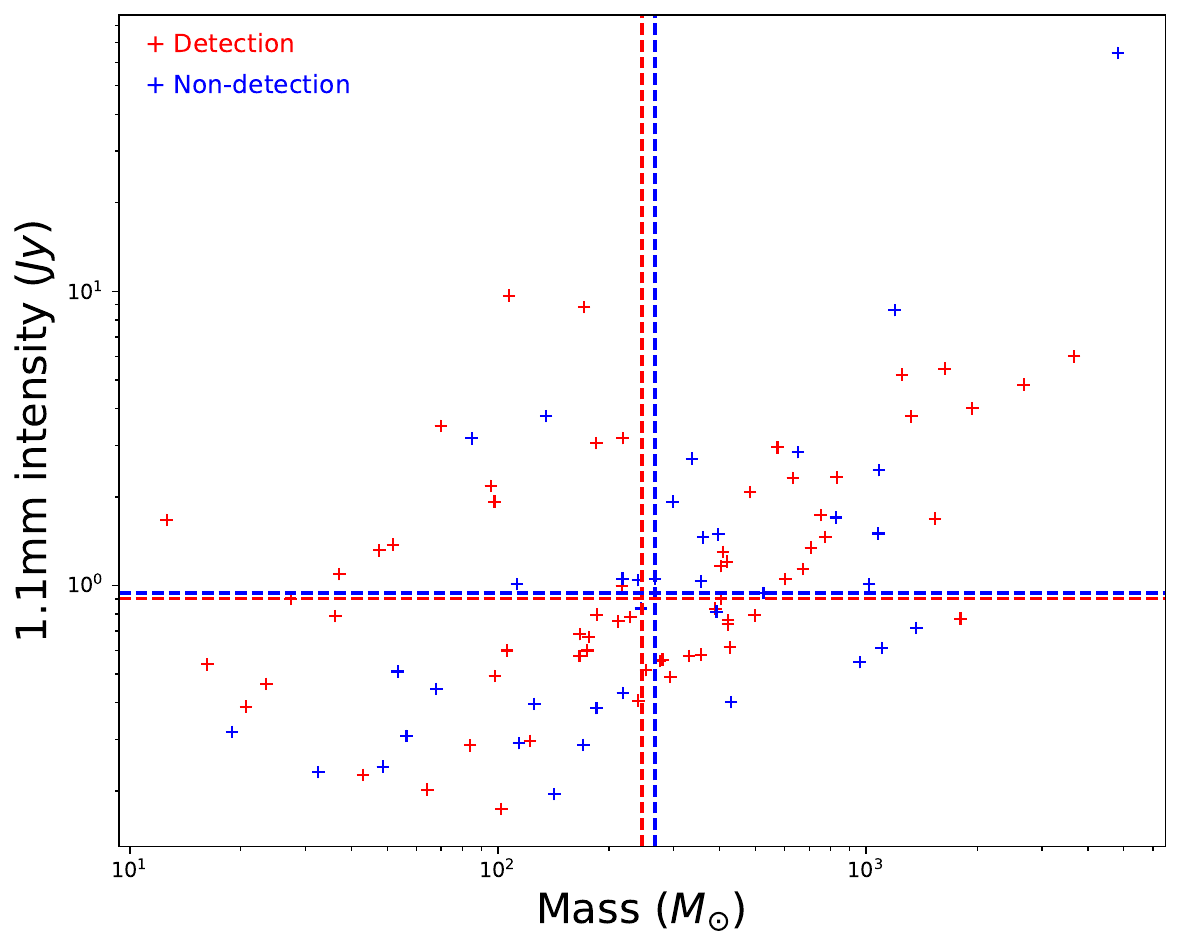}  
\includegraphics[width=80mm]{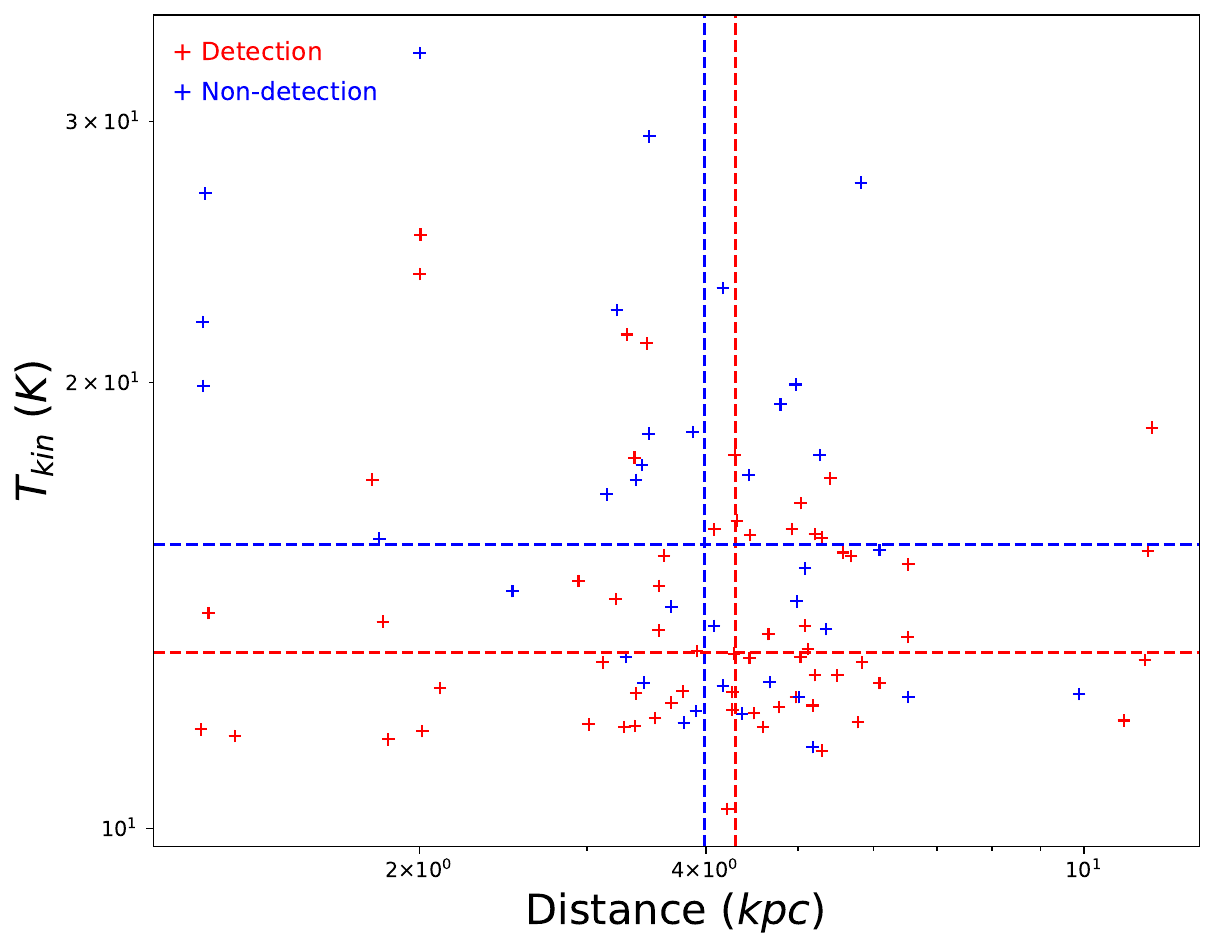} 
\caption{1.1mm intensities versus masses (left panel) and kinetic temperatures obtained from NH$_{3}$ observations versus distances (right panel) for SCCs with and without deuterated molecule detections. The dashed lines represent the median values.}
\label{fig_clumps}
\end{figure*}

Among 101 SCCs observed, 64 of them exhibit detections of deuterated molecules (DCO$^{+}$, DCN, DNC, N$_{2}$D$^{+}$), while 37 do not. We compare their physical properties obtained from \citet{2016ApJ...822...59S} and visualize the results in Figure \ref{fig_clumps}. The SCCs with and without deuterated molecule detections share similar median values for three parameters, including 1.1 mm intensity (0.90 and 0.94 Jy), mass (246.2 and 266.9 $M_{\odot}$), and distance (4.3 and 4.1 kpc). However, the SCCs with deuterated molecule detections display a lower median kinetic temperature (13.2 K) obtained from NH$_{3}$ observations compared to those without such detections (15.7 K).

For the 1$-$0 rotational lines of DCO$^{+}$, DCN, DNC, and N$_{2}$D$^{+}$, five sources, BGPS2940, BGPS2945, BGPS2986, BGPS3018, and BGPS3134, are detected with all four lines. Among the 101 sources, 5, 28, and 26 sources are detected with three, two, and one lines in the four 1$-$0 rotational lines, respectively. We do not detect deuterated line emissions above 3$\sigma_{\rm area}$ towards the rest 37 sources. 

Sources detected with emissions of DCO$^{+}$ and DCN \textit{J}=2$-$1 are also detected with their 1$-$0 rotational lines. For the 46 DCO$^{+}$ \textit{J}=1$-$0 sources, the peak flux densities of 20 sources which are also detected with DCO$^{+}$ \textit{J}=2$-$1 emissions have a range of 0.094 to 1.684 K, the remaining 26 sources without DCO$^{+}$ \textit{J}=2$-$1 detection have peak flux densities ranging from 0.100 to 0.689 K. Among 12 DCN \textit{J}=1$-$0 sources, there are 10 sources also detected with DCN \textit{J}=2$-$1 emissions.

There are 39 sources detected with both DCO$^{+}$ and DNC \textit{J}=1$-$0 emissions, and the peak flux density ratios of DCO$^{+}$/DNC range from 0.39 to 2.86 with a median value of 1.45, which indicates that DCO$^{+}$ \textit{J}=1$-$0 mostly has a stronger emission than DNC \textit{J}=1$-$0. DNC \textit{J}=1$-$0 has a higher FWHM median value (1.6 km s$^{-1}$) than that of DCO$^{+}$ \textit{J}=1$-$0 (1.2 km s$^{-1}$) towards these sources.

\begin{figure*}
\centering
\includegraphics[width=59mm]{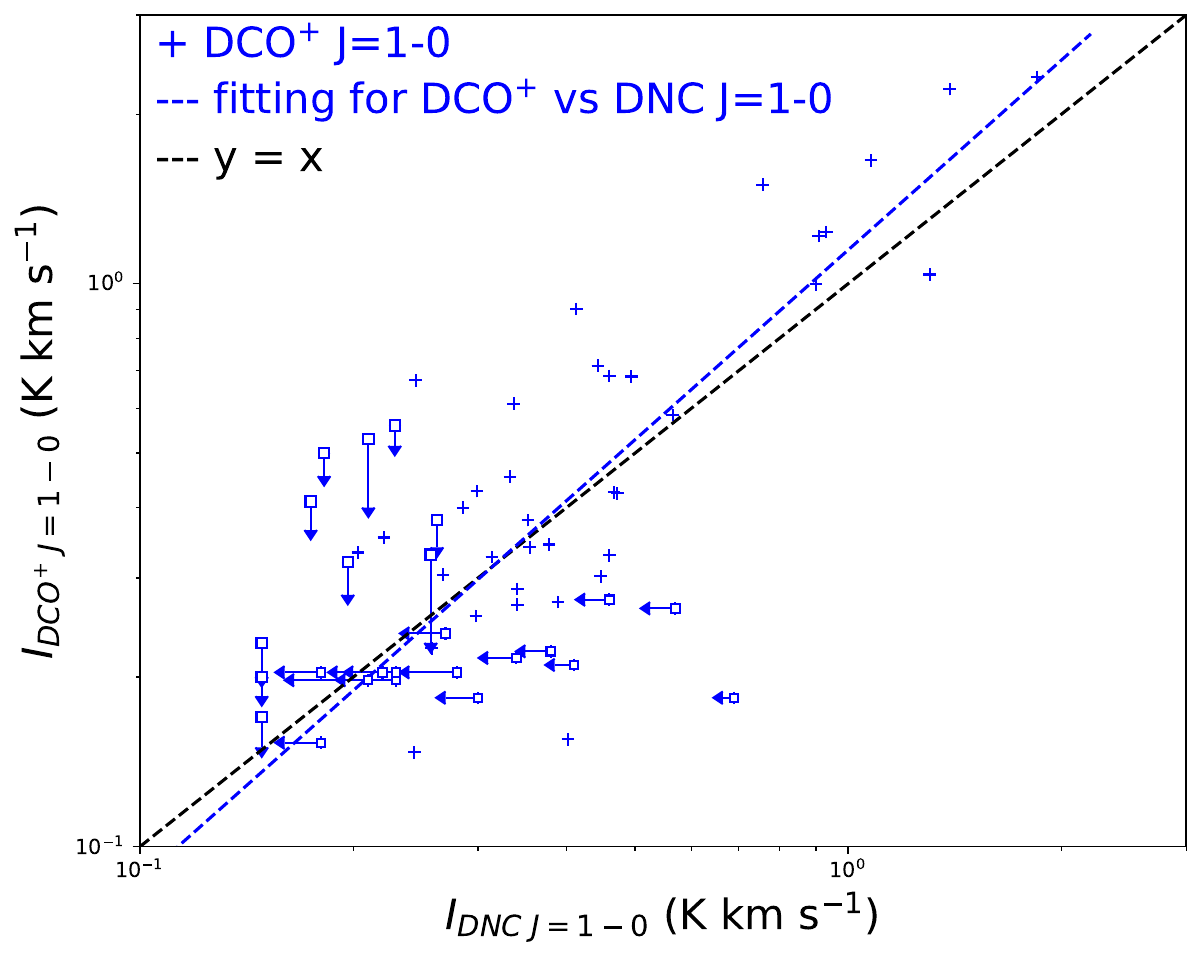}  
\includegraphics[width=59mm]{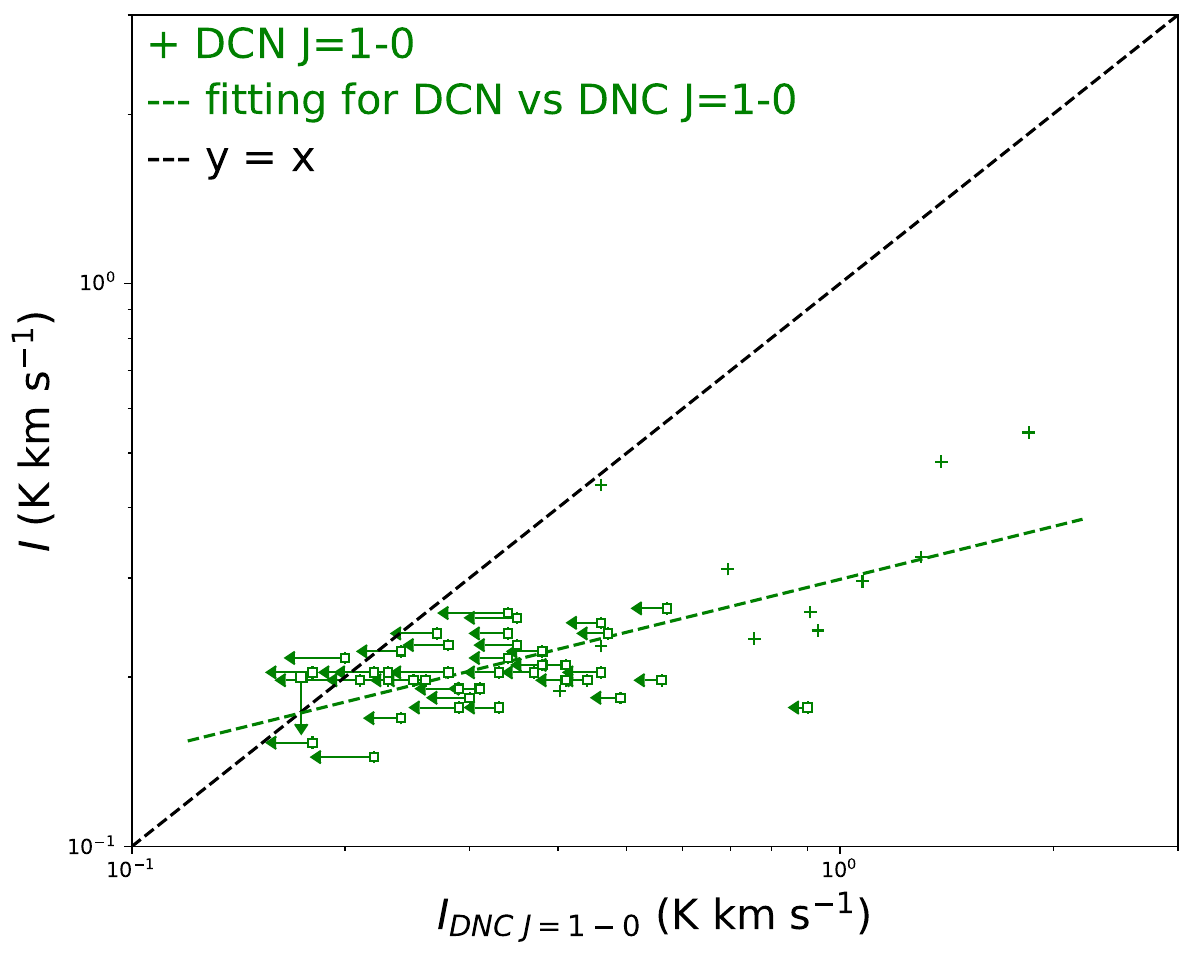}  
\includegraphics[width=59mm]{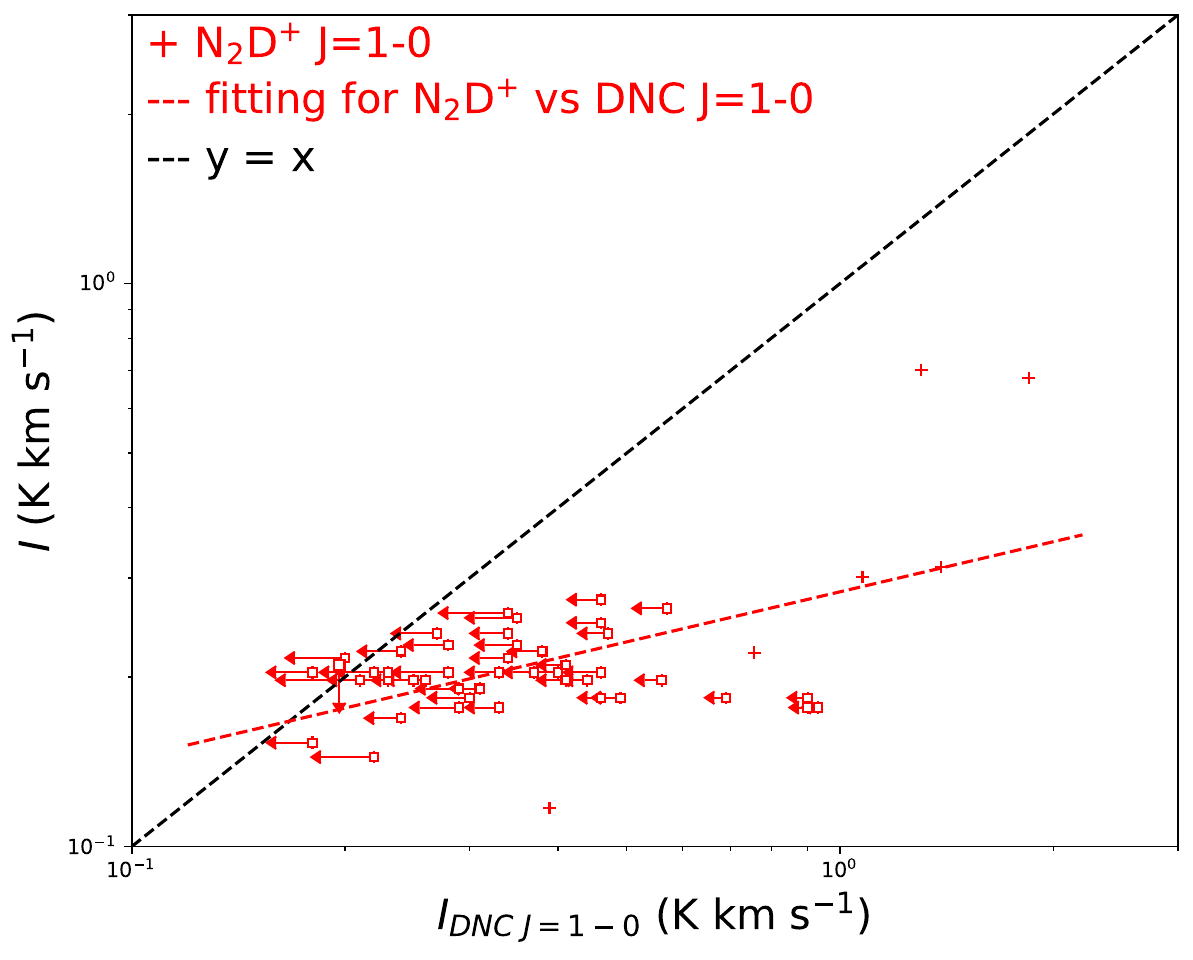}  
\caption{The integrated intensities of DCO$^{+}$ (left), DCN (middle), and N$_{2}$D$^{+}$ (right) \textit{J}=1$-$0 in relation to DNC \textit{J}=1$-$0. The colorful dashed lines represent the fitting results. The black dashed lines represent y = x.}
\label{fig_intensity_relation}
\end{figure*}

We present the integrated intensities of DCO$^{+}$, DCN, and N$_{2}$D$^{+}$ \textit{J}=1$-$0 in relation to DNC \textit{J}=1$-$0 in Figure \ref{fig_intensity_relation}. The upper limit values for the non-detected lines are estimated by 3$\sigma_{\rm area}$. These relations could be fitted by the following equations: log($I_{DCO^{+}}$) = 1.12 $\times$ log($I_{DNC}$) + 0.06, log($I_{DCN}$) = 0.31 $\times$ log($I_{DNC}$) $-$ 0.53, and log($I_{N_{2}D^{+}}$) = 0.30 $\times$ log($I_{DNC}$) $-$ 0.55. The positive slopes in these equations indicate that higher DNC \textit{J}=1$-$0 integrated intensity is associated with a higher integrated intensity for the other three deuterated molecular lines. In addition, DCO$^{+}$ and DNC \textit{J}=1$-$0 exhibit comparable integrated intensities, whereas the intensities of DCN, and N$_{2}$D$^{+}$ \textit{J}=1$-$0 are comparatively weaker, which indicates that the molecular lines with higher detection rates have stronger intensities than those with lower detection rates.

\subsubsection{H$_{2}$CO}

Information from simultaneously obtained H$_{2}$CO (1$_{(0,1)}$$-$0$_{(0,0)}$) at 72.837951 GHz can be used to probe dense gas properties in these sources. H$_{2}$CO (1$_{(0,1)}$$-$0$_{(0,0)}$) emissions are detected in 91 sources with velocity integrated intensities above 3$\sigma_{\rm area}$, most of which can be fitted with a single-component Gaussian profile. Their parameters and spectra are shown in Table \ref{Res_Single3} and Figure \ref{H2CO}. Sixty among these H$_{2}$CO (1$_{(0,1)}$$-$0$_{(0,0)}$) sources are detected with deuterated line emissions, their FWHMs range from 0.8 to 4.6 km s$^{-1}$ with a median value of 2.1 km s$^{-1}$. For the remaining 31 sources, their FWHMs range from 0.8 to 4.8 km s$^{-1}$ with a higher median value of 2.8 km s$^{-1}$. 

In order to get a better understanding of the spatial distribution of these deuterated molecular lines in these regions at the early stage of star-formation, after taking the sources with strong emissions or most detection of the deuterated molecular lines into consideration, we finally choose 11 sources as our targets to conduct our continuous OTF mode observation.

\subsection{OTF observations toward selected sources}\label{res_OTF}

\begin{figure*}
 \begin{center}
\includegraphics[width=18cm]{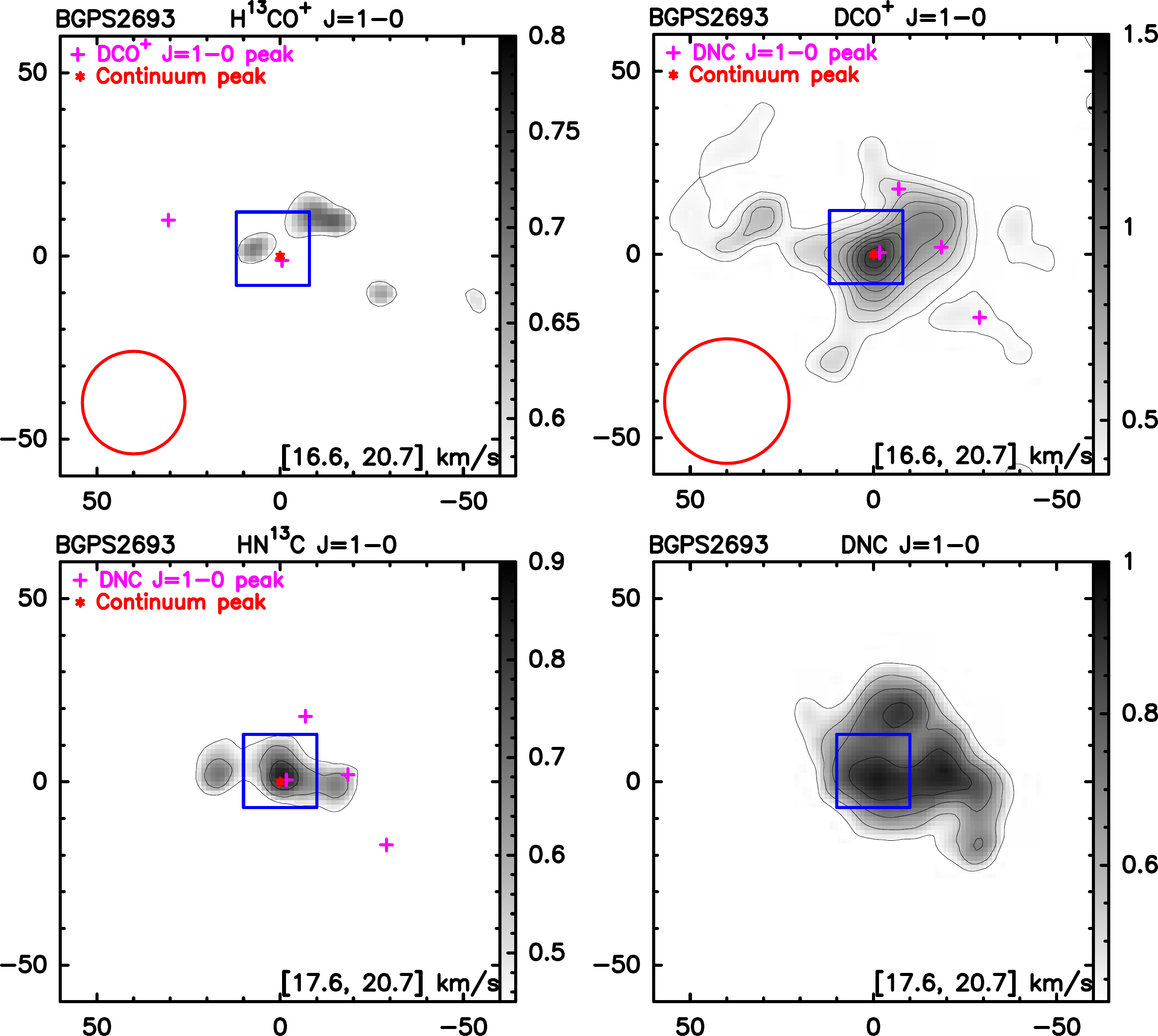}
\caption{The velocity integrated maps of deuterated and $^{13}$C-isotopic molecular lines for BGPS2693. The integrated velocity range, shown at the right-bottom corner, is derived after combining the Gaussian fitting line widths for each pair of deuterated molecular line and $^{13}$C$-$isotopologue.
For H$^{13}$CO$^{+}$, DCO$^{+}$, HN$^{13}$C, and DNC \textit{J}=1$-$0, the contours start from 3$\sigma$ in steps of 1$\sigma$, with $\sigma$ = 0.19, 0.12, 0.15, and 0.14 K km s$^{-1}$, respectively.
The crosses represent peaks in the velocity integrated maps of different molecular lines and the red star represents the 1.1mm continuum peak obtained from \citet{2016ApJ...822...59S}. The beam sizes for H$^{13}$CO$^{+}$ and DCO$^{+}$ \textit{J}=1$-$0 of $\sim$28$^{\prime\prime}$ and $\sim$34$^{\prime\prime}$ are shown as red circles at the bottom-left corners in their integrated intensity maps. The averaged spectra for column density estimation are derived in the regions enclosed by blue squares.}
\label{fig_BGPS2693}
\end{center}
\end{figure*}

\begin{figure*}
 \begin{center}
\includegraphics[width=12cm]{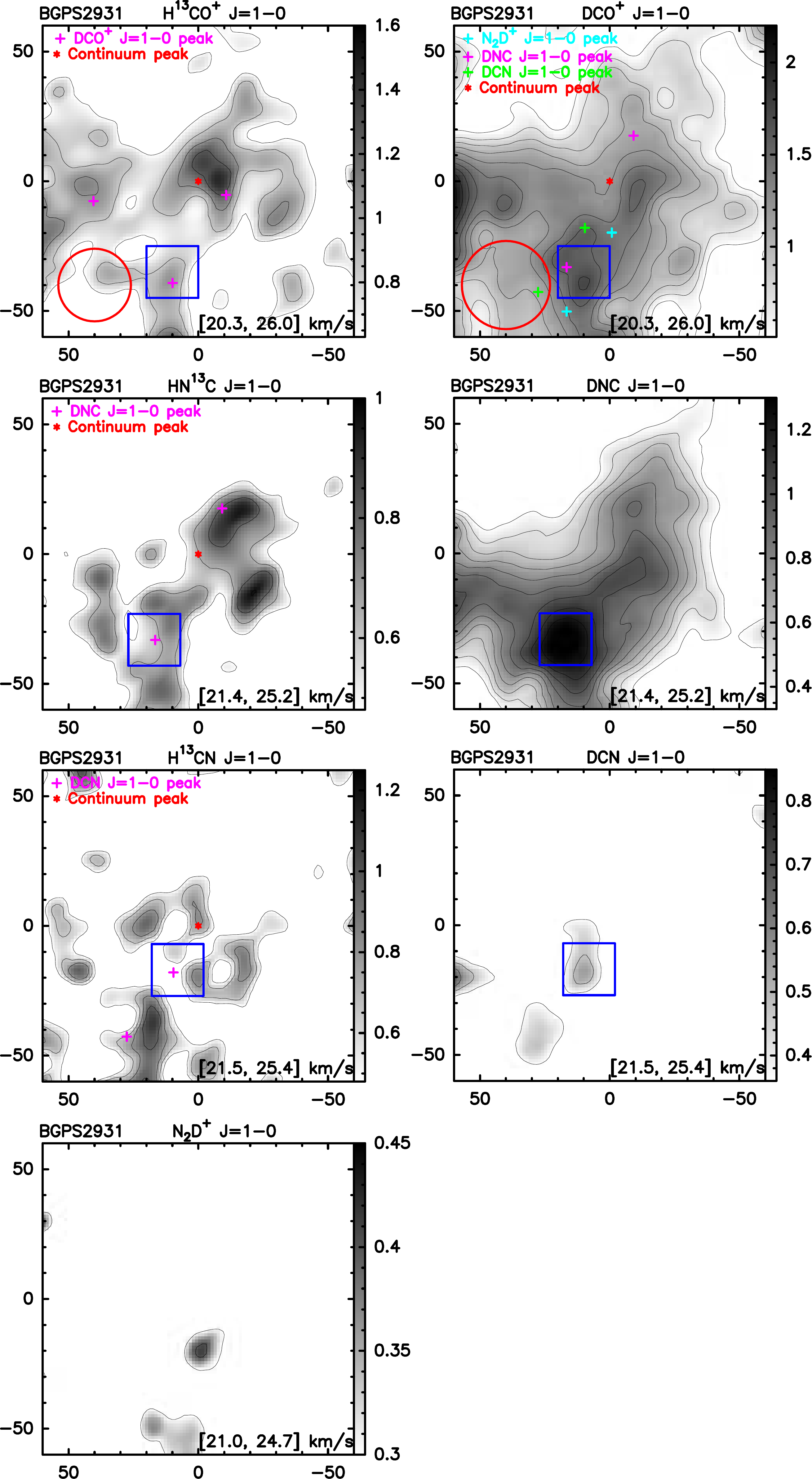}
\caption{The velocity integrated maps of deuterated and $^{13}$C-isotopic molecular lines for BGPS2931. The integrated velocity range, shown at the right-bottom corner, is derived after combining the Gaussian fitting line widths for each pair of deuterated molecular line and $^{13}$C$-$isotopologue.
For H$^{13}$CO$^{+}$, DCO$^{+}$, HN$^{13}$C, DNC, H$^{13}$CN, DCN, and N$_{2}$D$^{+}$ \textit{J}=1$-$0, the contours start from 3$\sigma$ in steps of 1$\sigma$, with $\sigma$ = 0.21, 0.18, 0.16, 0.11, 0.16, 0.12, and 0.10 K km s$^{-1}$, respectively.
The crosses represent peaks in the velocity integrated maps of different molecular lines and the red star represents the 1.1mm continuum peak obtained from \citet{2016ApJ...822...59S}. The beam sizes for H$^{13}$CO$^{+}$ and DCO$^{+}$ \textit{J}=1$-$0 of $\sim$28$^{\prime\prime}$ and $\sim$34$^{\prime\prime}$ are shown as red circles at the bottom-left corners in their integrated intensity maps. The averaged spectra for column density estimation are derived in the regions enclosed by blue squares.}
\label{fig_BGPS2931}
\end{center}
\end{figure*}

\begin{figure*}
 \begin{center}
\includegraphics[width=12cm]{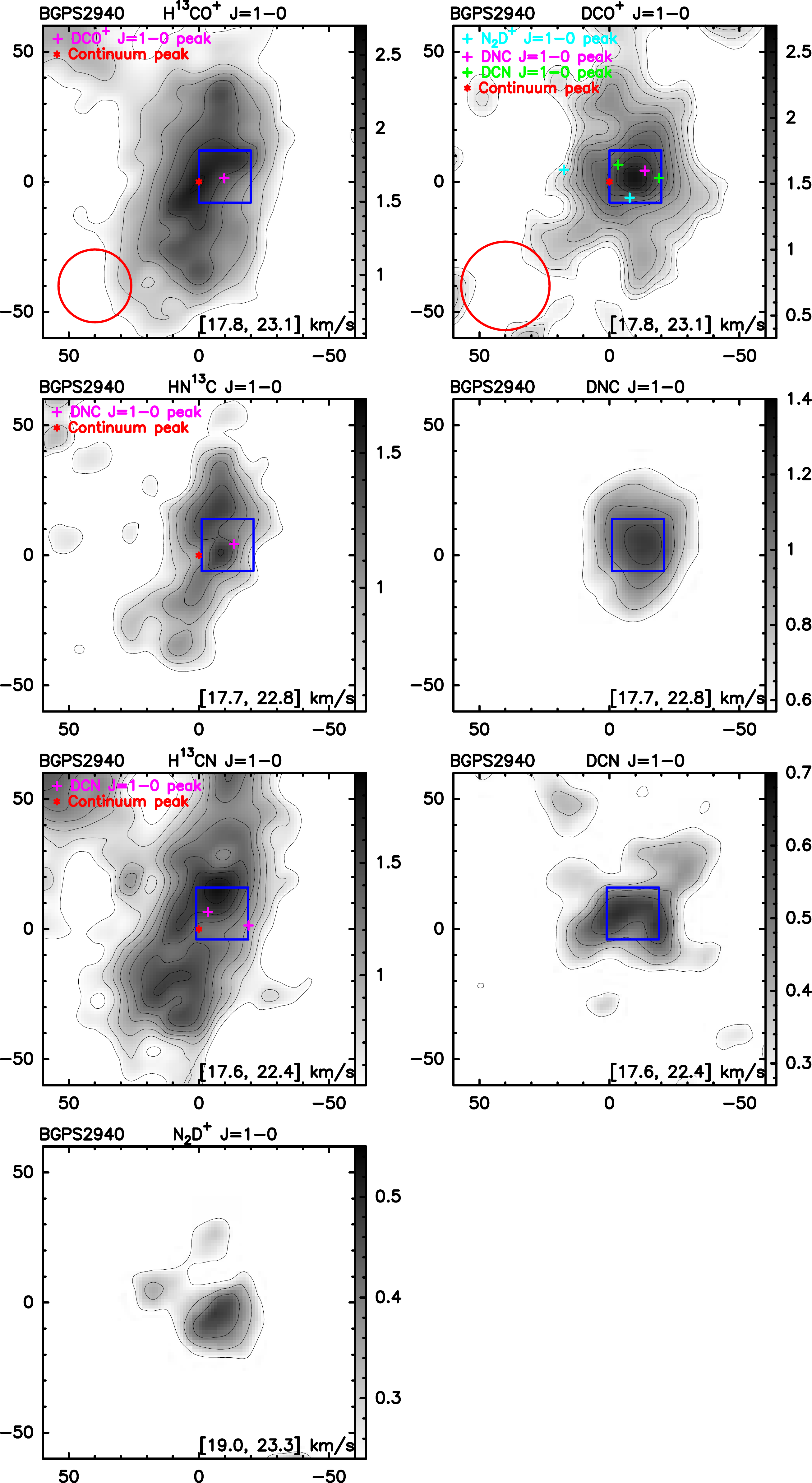}
\caption{The velocity integrated maps of deuterated and $^{13}$C-isotopic molecular lines for BGPS2940. The integrated velocity range, shown at the right-bottom corner, is derived after combining the Gaussian fitting line widths for each pair of deuterated molecular line and $^{13}$C$-$isotopologue.
For H$^{13}$CO$^{+}$ and DCO$^{+}$ \textit{J}=1$-$0, the contours start from 3$\sigma$ in steps of 2$\sigma$, with $\sigma$ = 0.19 and 0.11 K km s$^{-1}$, respectively.
For HN$^{13}$C, DNC, H$^{13}$CN, DCN, and N$_{2}$D$^{+}$ \textit{J}=1$-$0, the contours start from 3$\sigma$ in steps of 1$\sigma$, with $\sigma$ = 0.18, 0.19, 0.17, 0.09, and 0.08 K km s$^{-1}$, respectively.
The crosses represent peaks in the velocity integrated maps of different molecular lines and the red star represents the 1.1mm continuum peak obtained from \citet{2016ApJ...822...59S}. The beam sizes for H$^{13}$CO$^{+}$ and DCO$^{+}$ \textit{J}=1$-$0 of $\sim$28$^{\prime\prime}$ and $\sim$34$^{\prime\prime}$ are shown as red circles at the bottom-left corners in their integrated intensity maps. The averaged spectra for column density estimation are derived in the regions enclosed by blue squares.}
\label{fig_BGPS2940}
\end{center}
\end{figure*}

\begin{figure*}
 \begin{center}
\includegraphics[width=18cm]{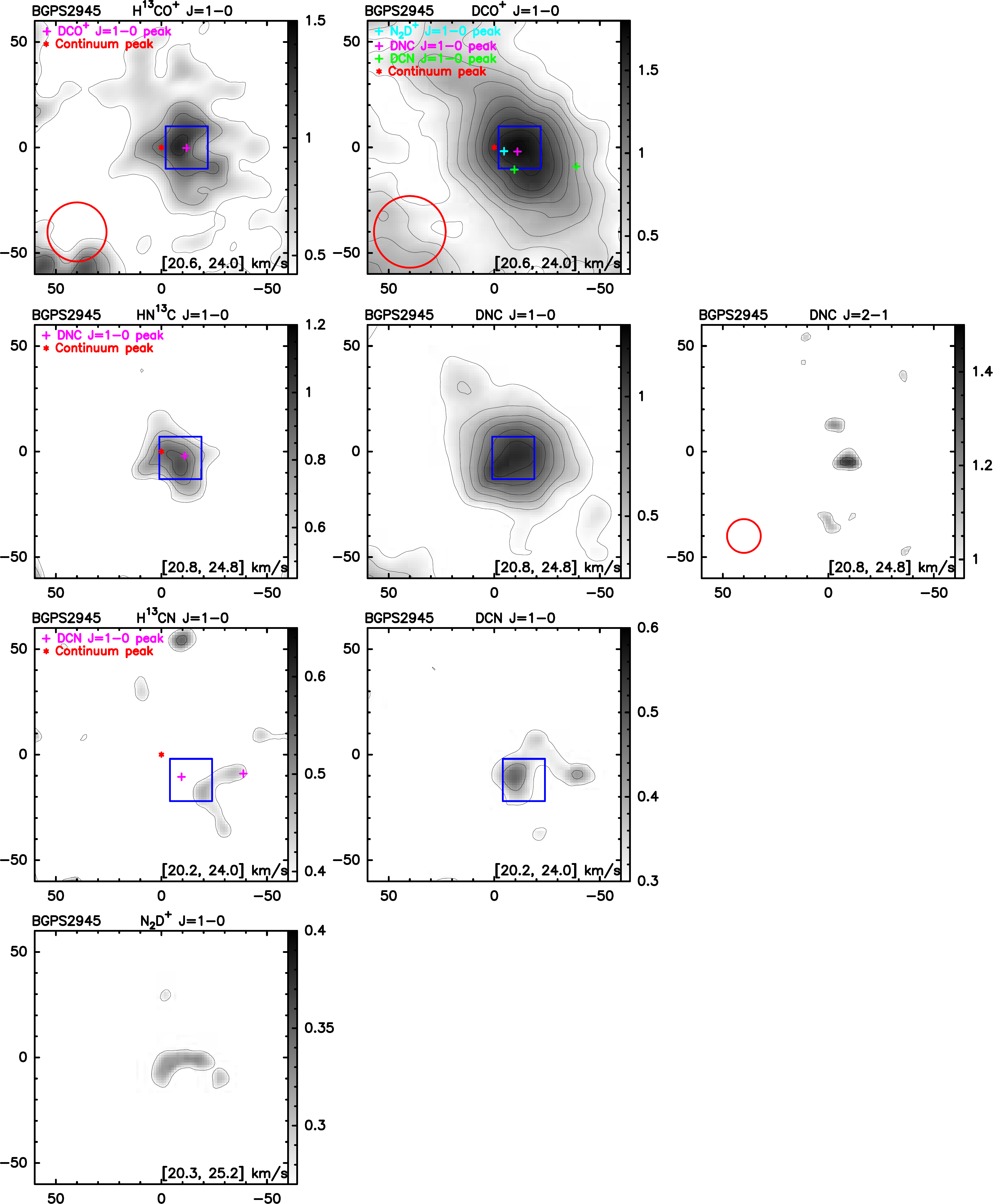}
\caption{The velocity integrated maps of deuterated and $^{13}$C-isotopic molecular lines for BGPS2945. The integrated velocity range, shown at the right-bottom corner, is derived after combining the Gaussian fitting line widths for each pair of deuterated molecular line and $^{13}$C$-$isotopologue.
For H$^{13}$CO$^{+}$, DCO$^{+}$, and DNC \textit{J}=1$-$0, the contours start from 3$\sigma$ in steps of 2$\sigma$, with $\sigma$ = 0.14, 0.09, and 0.08 K km s$^{-1}$, respectively.
For HN$^{13}$C, H$^{13}$CN, DCN, N$_{2}$D$^{+}$ \textit{J}=1$-$0, and DNC \textit{J}=2$-$1, the contours start from 3$\sigma$ in steps of 1$\sigma$, with $\sigma$ = 0.15, 0.13, 0.10, 0.09, and 0.32 K km s$^{-1}$, respectively.
The crosses represent peaks in the velocity integrated maps of different molecular lines and the red star represents the 1.1mm continuum peak obtained from \citet{2016ApJ...822...59S}. The beam sizes for H$^{13}$CO$^{+}$, DCO$^{+}$ \textit{J}=1$-$0 and DNC \textit{J}=2$-$1 of $\sim$28$^{\prime\prime}$, $\sim$34$^{\prime\prime}$, and $\sim$16$^{\prime\prime}$ are shown as red circles at the bottom-left corners in their integrated intensity maps. The averaged spectra for column density estimation are derived in the regions enclosed by blue squares.}
\label{fig_BGPS2945}
\end{center}
\end{figure*}

\begin{figure*}
 \begin{center}
\includegraphics[width=18cm]{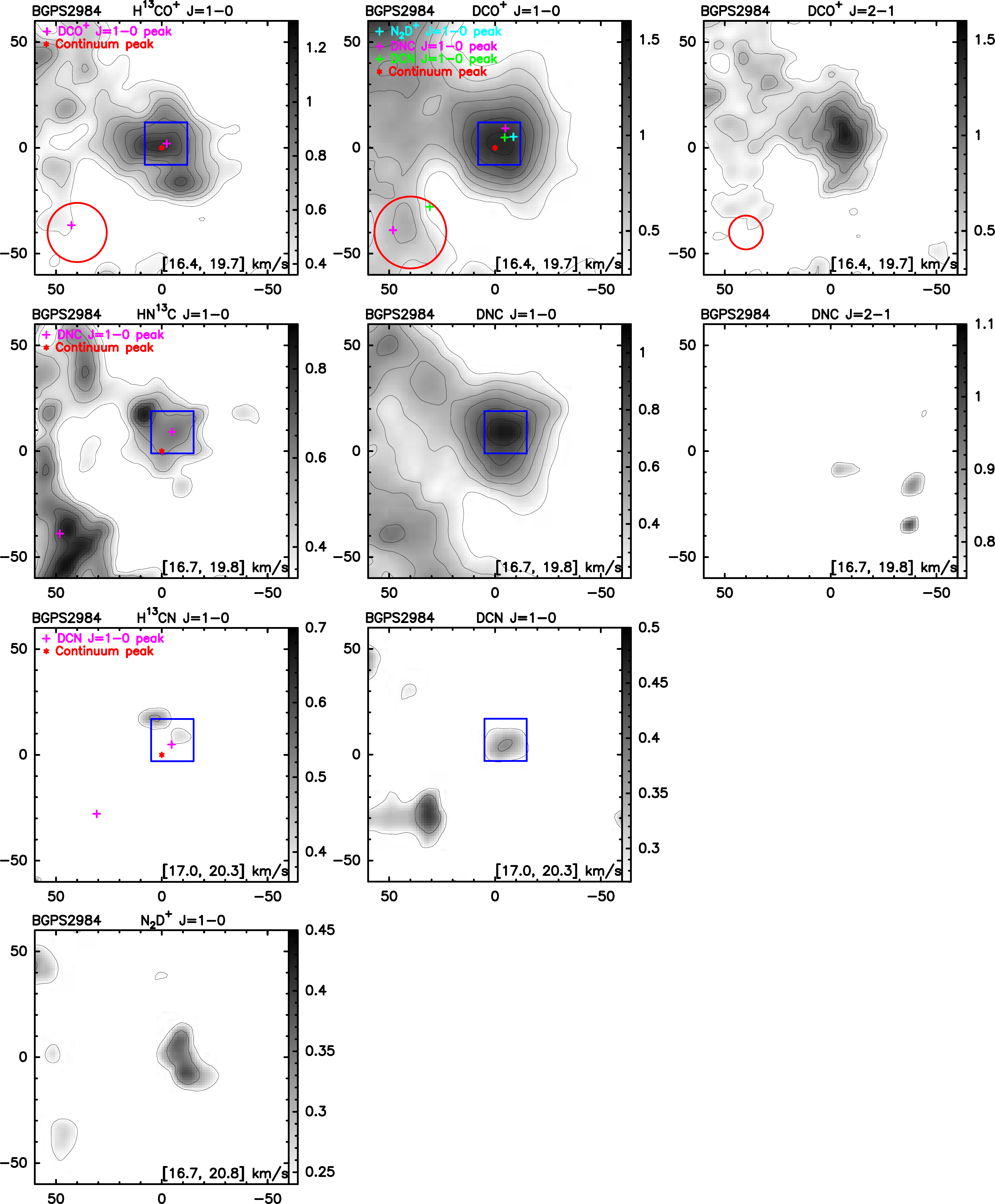}
\caption{The velocity integrated maps of deuterated and $^{13}$C-isotopic molecular lines for BGPS2984. The integrated velocity range, shown at the right-bottom corner, is derived after combining the Gaussian fitting line widths for each pair of deuterated molecular line and $^{13}$C$-$isotopologue.
For H$^{13}$CO$^{+}$, DCO$^{+}$, DNC \textit{J}=1$-$0 and DCO$^{+}$ \textit{J}=2$-$1, the contours start from 3$\sigma$ in steps of 2$\sigma$, with $\sigma$ = 0.12, 0.09, 0.07, and 0.09 K km s$^{-1}$, respectively.
For HN$^{13}$C, H$^{13}$CN, DCN, N$_{2}$D$^{+}$ \textit{J}=1$-$0, and DNC \textit{J}=2$-$1, the contours start from 3$\sigma$ in steps of 1$\sigma$, with $\sigma$ = 0.11, 0.12, 0.09, 0.08, and 0.25 K km s$^{-1}$, respectively.
The crosses represent peaks in the velocity integrated maps of different molecular lines and the red star represents the 1.1mm continuum peak obtained from \citet{2016ApJ...822...59S}. The beam sizes for H$^{13}$CO$^{+}$, DCO$^{+}$ \textit{J}=1$-$0 and DCO$^{+}$ \textit{J}=2$-$1 of $\sim$28$^{\prime\prime}$, $\sim$34$^{\prime\prime}$, and $\sim$16$^{\prime\prime}$ are shown as red circles at the bottom-left corners in their integrated intensity maps.  The averaged spectra for column density estimation are derived in the regions enclosed by blue squares.}
\label{fig_BGPS2984}
\end{center}
\end{figure*}

\begin{figure*}
 \begin{center}
\includegraphics[width=18cm]{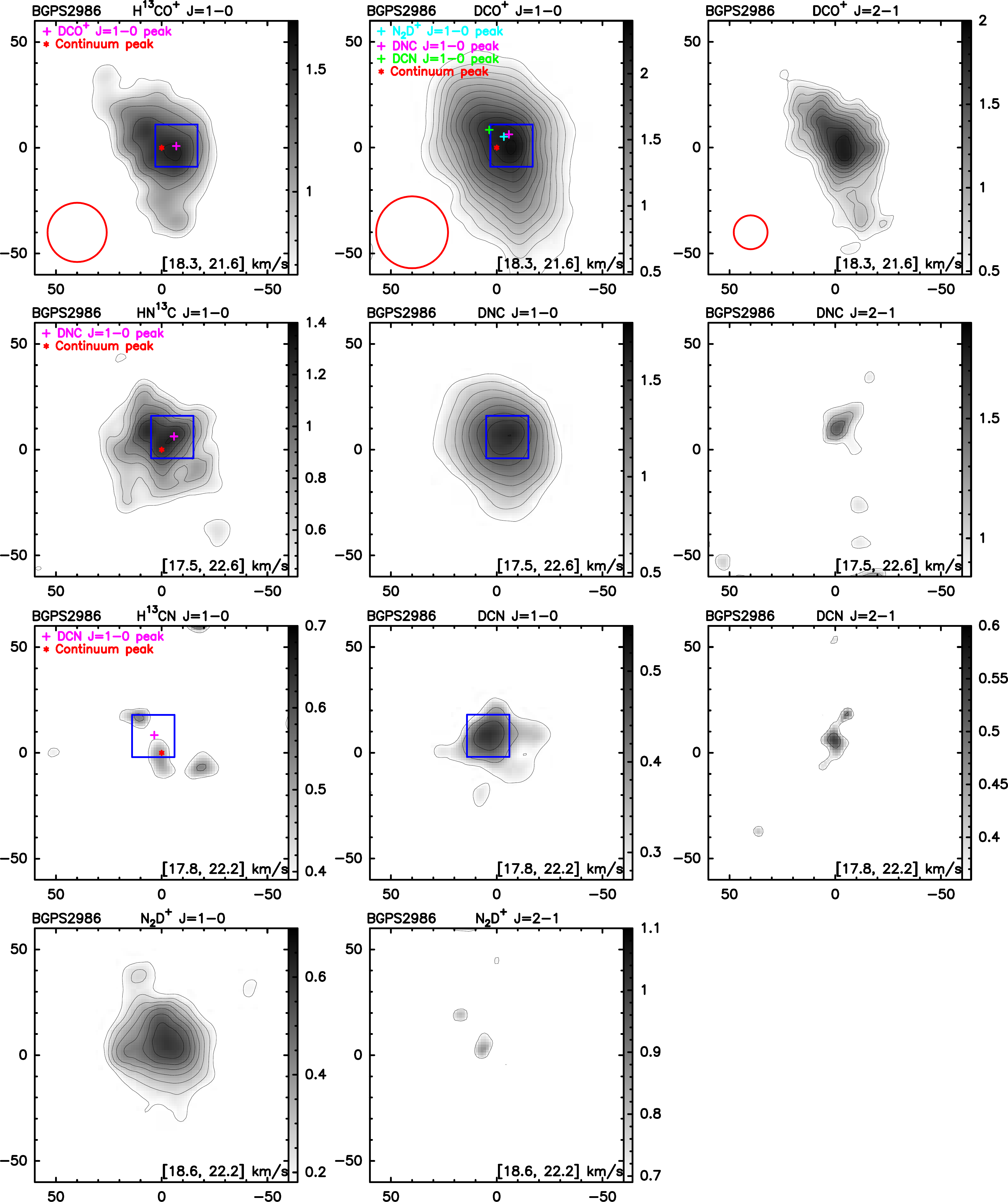}
\caption{The velocity integrated maps of deuterated and $^{13}$C-isotopic molecular lines for BGPS2986. The integrated velocity range, shown at the right-bottom corner, is derived after combining the Gaussian fitting line widths for each pair of deuterated molecular line and $^{13}$C$-$isotopologue.
For H$^{13}$CO$^{+}$, DCO$^{+}$, DNC \textit{J}=1$-$0 and DCO$^{+}$ \textit{J}=2$-$1, the contours start from 6$\sigma$ in steps of 2$\sigma$, with $\sigma$ = 0.11, 0.08, 0.08, and 0.08 K km s$^{-1}$, respectively.
For HN$^{13}$C, H$^{13}$CN, DCN, N$_{2}$D$^{+}$ \textit{J}=1$-$0, and DNC, DCN, N$_{2}$D$^{+}$ \textit{J}=2$-$1, the contours start from 3$\sigma$ in steps of 1$\sigma$, with $\sigma$ = 0.14, 0.13, 0.09, 0.06, 0.28, 0.09, and 0.23 K km s$^{-1}$, respectively.
The crosses represent peaks in the velocity integrated maps of different molecular lines and the red star represents the 1.1mm continuum peak obtained from \citet{2016ApJ...822...59S}. The beam sizes for H$^{13}$CO$^{+}$, DCO$^{+}$ \textit{J}=1$-$0 and DCO$^{+}$ \textit{J}=2$-$1 of $\sim$28$^{\prime\prime}$, $\sim$34$^{\prime\prime}$, and $\sim$16$^{\prime\prime}$ are shown as red circles at the bottom-left corners in their integrated intensity maps.  The averaged spectra for column density estimation are derived in the regions enclosed by blue squares.}
\label{fig_BGPS2986}
\end{center}
\end{figure*}

\begin{figure*}
 \begin{center}
\includegraphics[width=16cm]{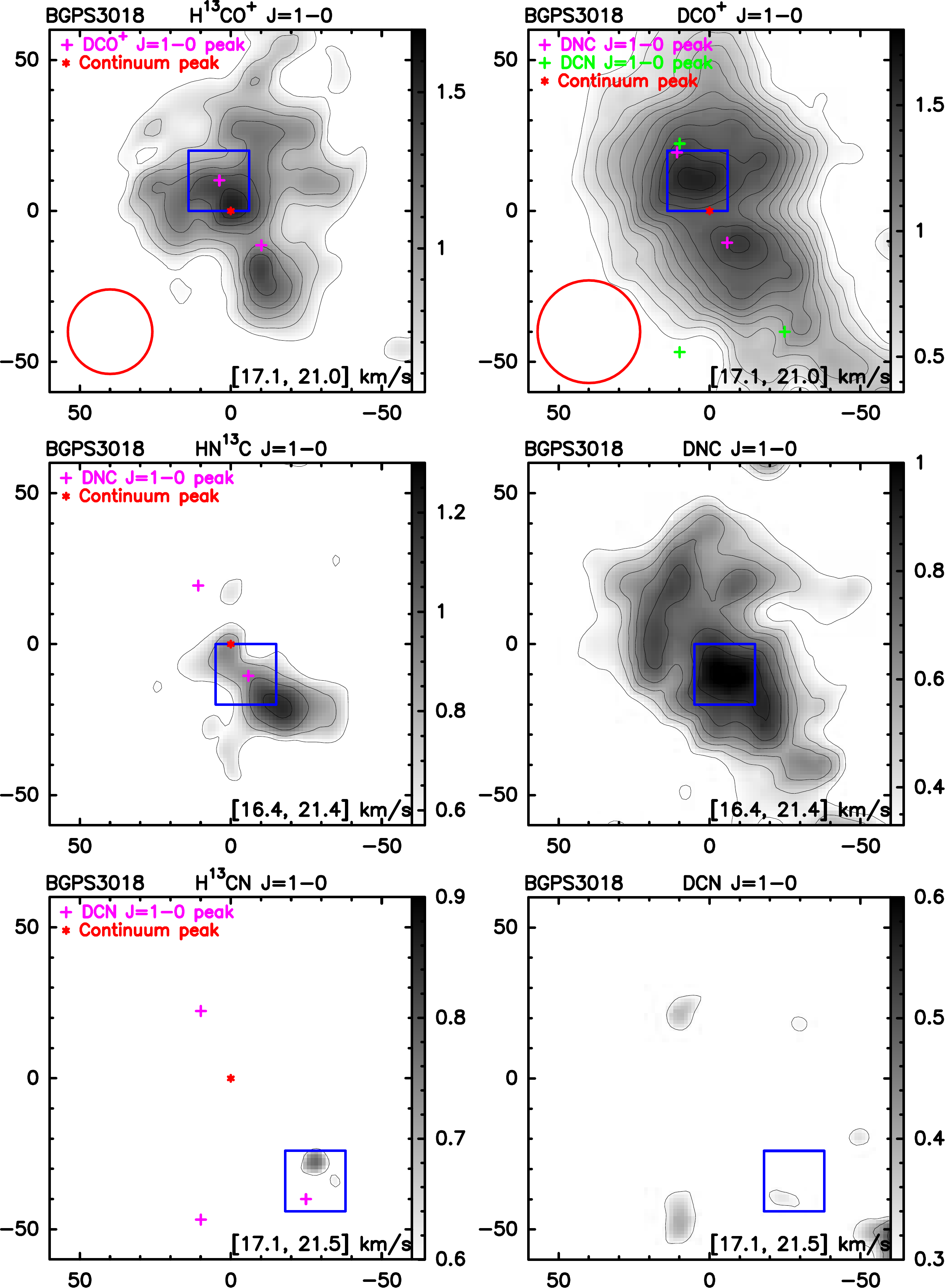}
\caption{The velocity integrated maps of deuterated and $^{13}$C-isotopic molecular lines for BGPS3018. The integrated velocity range, shown at the right-bottom corner, is derived after combining the Gaussian fitting line widths for each pair of deuterated molecular line and $^{13}$C$-$isotopologue.
For H$^{13}$CO$^{+}$, DCO$^{+}$, HN$^{13}$C, DNC,  H$^{13}$CN, and DCN \textit{J}=1$-$0, the contours start from 3$\sigma$ in steps of 1$\sigma$, with $\sigma$ = 0.18, 0.12, 0.19, 0.11, 0.19, and 0.10 K km s$^{-1}$, respectively.
The crosses represent peaks in the velocity integrated maps of different molecular lines and the red star represents the 1.1mm continuum peak obtained from \citet{2016ApJ...822...59S}. The beam sizes for H$^{13}$CO$^{+}$ and DCO$^{+}$ \textit{J}=1$-$0 of $\sim$28$^{\prime\prime}$ and $\sim$34$^{\prime\prime}$ are shown as red circles at the bottom-left corners in their integrated intensity maps. The averaged spectra for column density estimation are derived in the regions enclosed by blue squares.}
\label{fig_BGPS3018}
\end{center}
\end{figure*}

\begin{figure*}
 \begin{center}
\includegraphics[width=18cm]{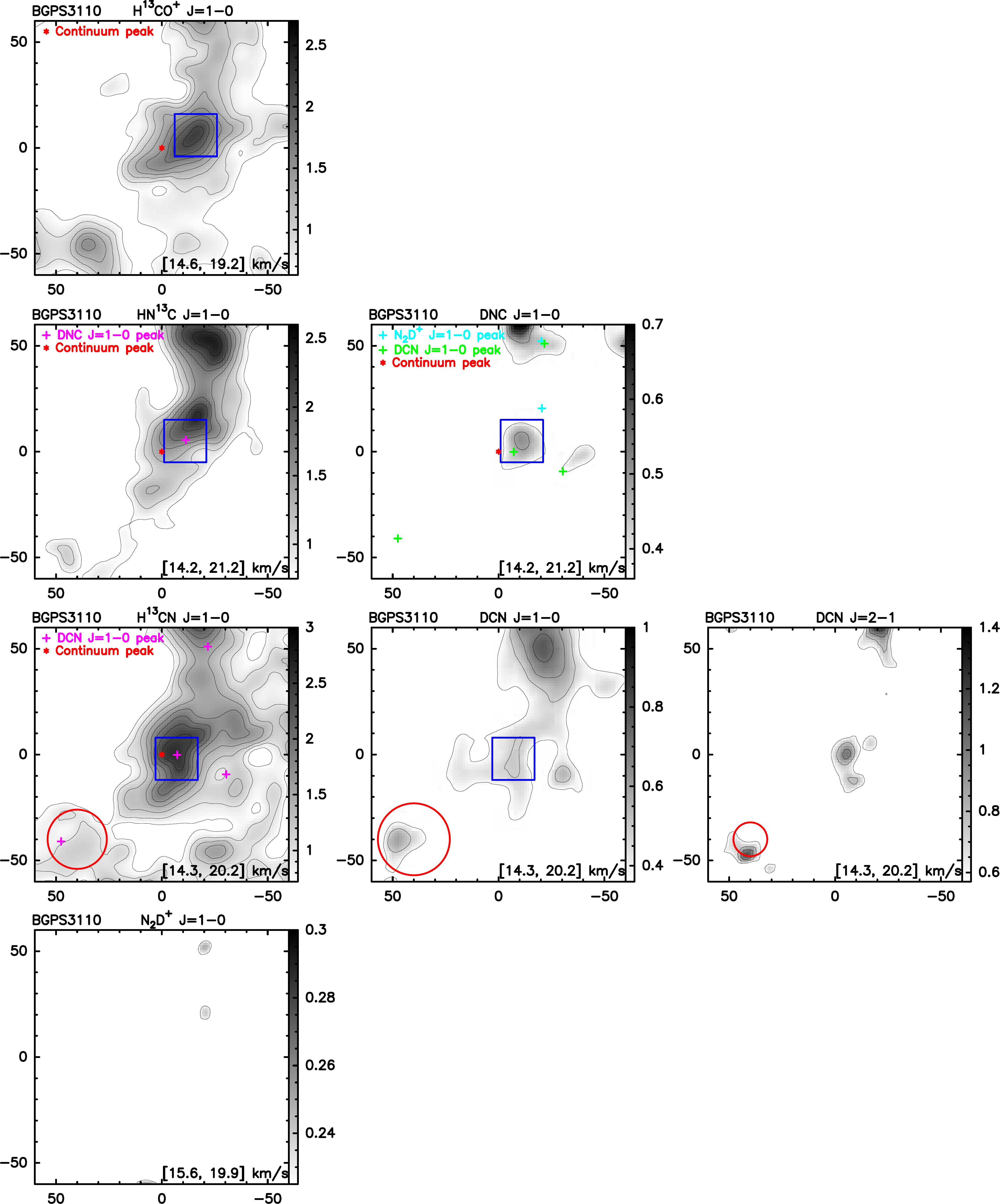}
\caption{The velocity integrated maps of deuterated and $^{13}$C-isotopic molecular lines for BGPS3110. The integrated velocity range, shown at the right-bottom corner, is derived after combining the Gaussian fitting line widths for each pair of deuterated molecular line and $^{13}$C$-$isotopologue.
For H$^{13}$CO$^{+}$, HN$^{13}$C, DNC, H$^{13}$CN, DCN, N$_{2}$D$^{+}$ \textit{J}=1$-$0, and DCN \textit{J}=2$-$1, the contours start from 3$\sigma$ in steps of 1$\sigma$, with $\sigma$ = 0.21, 0.25, 0.12, 0.24, 0.12, 0.08, and 0.19 K km s$^{-1}$, respectively.
The crosses represent peaks in the velocity integrated maps of different molecular lines and the red star represents the 1.1mm continuum peak obtained from \citet{2016ApJ...822...59S}. The beam sizes for H$^{13}$CN, DCN \textit{J}=1$-$0 and DCN \textit{J}=2$-$1 of $\sim$28$^{\prime\prime}$, $\sim$34$^{\prime\prime}$, and $\sim$16$^{\prime\prime}$ are shown as red circles at the bottom-left corners in their integrated intensity maps.  The averaged spectra for column density estimation are derived in the regions enclosed by blue squares.}
\label{fig_BGPS3110}
\end{center}
\end{figure*}

\begin{figure*}
 \begin{center}
\includegraphics[width=18cm]{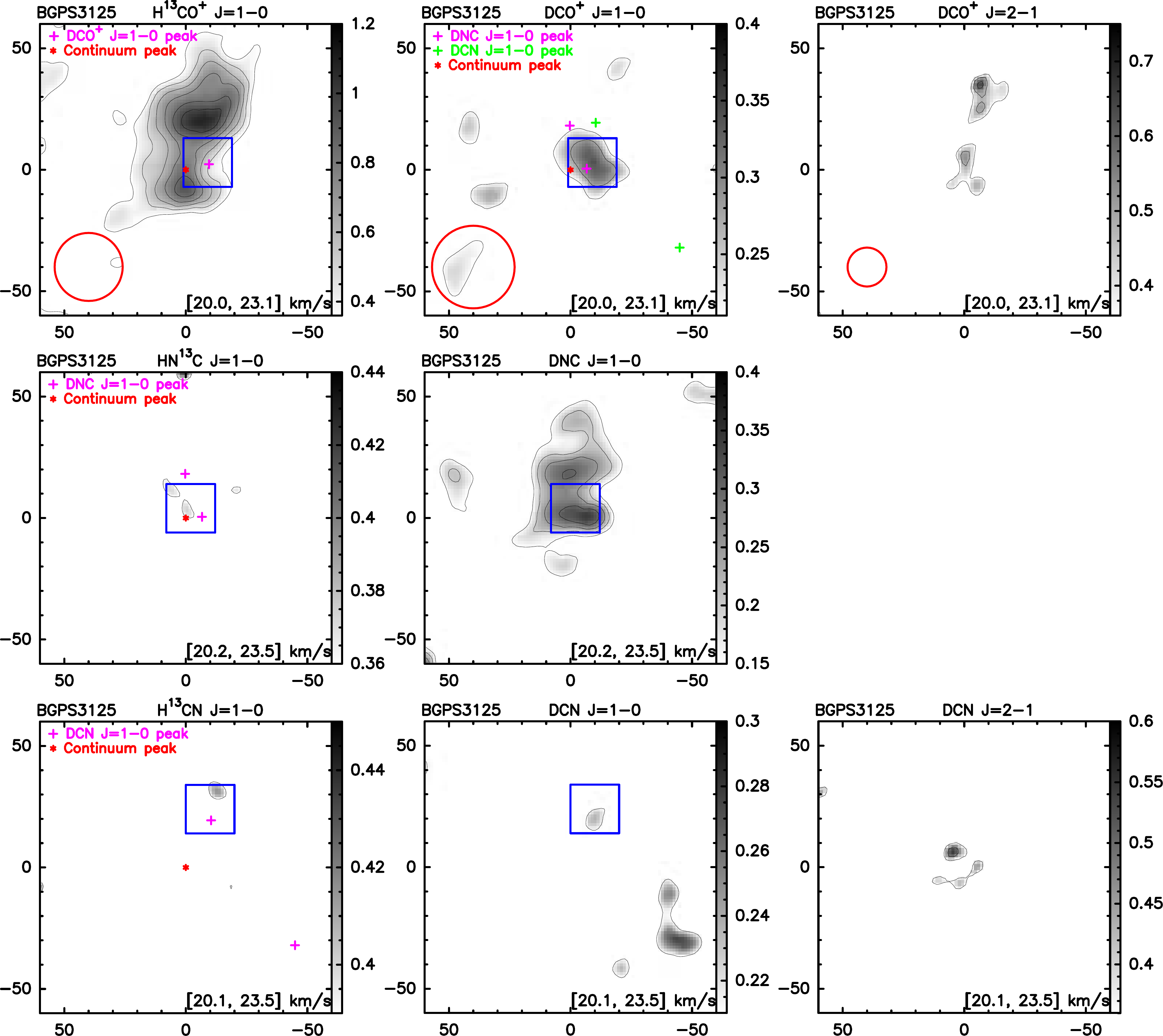}
\caption{The velocity integrated maps of deuterated and $^{13}$C-isotopic molecular lines for BGPS3125. The integrated velocity range, shown at the right-bottom corner, is derived after combining the Gaussian fitting line widths for each pair of deuterated molecular line and $^{13}$C$-$isotopologue.
For H$^{13}$CO$^{+}$, DCO$^{+}$, HN$^{13}$C, DNC, H$^{13}$CN, DCN \textit{J}=1$-$0, and DCO$^{+}$, DCN \textit{J}=2$-$1, the contours start from 3$\sigma$ in steps of 1$\sigma$, with $\sigma$ = 0.12, 0.07, 0.12, 0.05, 0.13, 0.08, 0.11, and 0.12 K km s$^{-1}$, respectively.
The crosses represent peaks in the velocity integrated maps of different molecular lines and the red star represents the 1.1mm continuum peak obtained from \citet{2016ApJ...822...59S}. The beam sizes for H$^{13}$CO$^{+}$, DCO$^{+}$ \textit{J}=1$-$0 and DCO$^{+}$ \textit{J}=2$-$1 of $\sim$28$^{\prime\prime}$, $\sim$34$^{\prime\prime}$, and $\sim$16$^{\prime\prime}$ are shown as red circles at the bottom-left corners in their integrated intensity maps.  The averaged spectra for column density estimation are derived in the regions enclosed by blue squares.}
\label{fig_BGPS3125}
\end{center}
\end{figure*}

\begin{figure*}
 \begin{center}
\includegraphics[width=18cm]{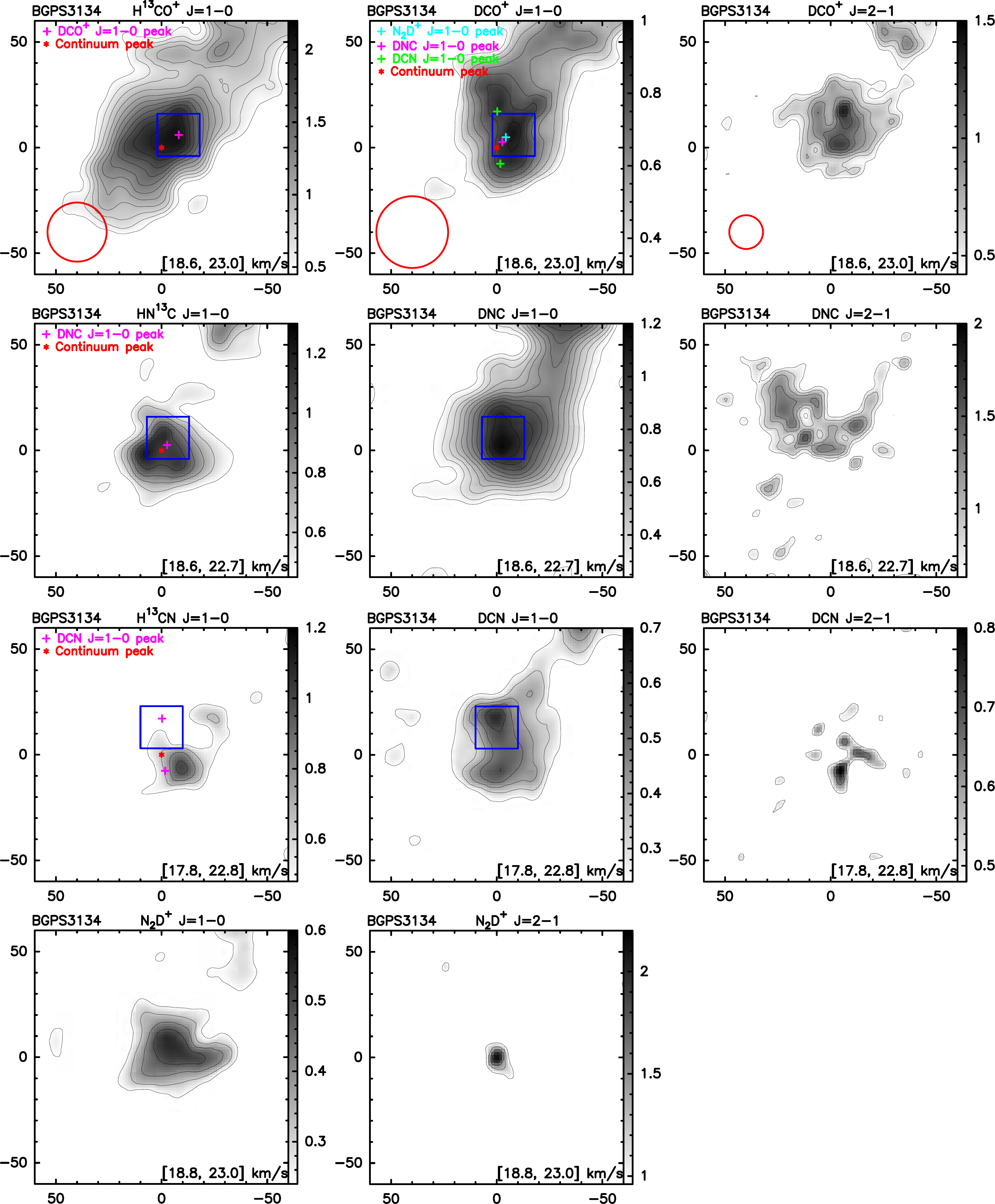}
\caption{The velocity integrated maps of deuterated and $^{13}$C-isotopic molecular lines for BGPS3134. The integrated velocity range, shown at the right-bottom corner, is derived after combining the Gaussian fitting line widths for each pair of deuterated molecular line and $^{13}$C$-$isotopologue.
For H$^{13}$CO$^{+}$, DCO$^{+}$, HN$^{13}$C, DNC, H$^{13}$CN, DCN, N$_{2}$D$^{+}$ \textit{J}=1$-$0, and DCO$^{+}$, DNC, DCN, N$_{2}$D$^{+}$ \textit{J}=2$-$1, the contours start from 3$\sigma$ in steps of 1$\sigma$, with $\sigma$ = 0.15, 0.10, 0.15, 0.08, 0.16, 0.08, 0.14, 0.21, 0.16, and 0.32 K km s$^{-1}$, respectively.
The crosses represent peaks in the velocity integrated maps of different molecular lines and the red star represents the 1.1mm continuum peak obtained from \citet{2016ApJ...822...59S}. The beam sizes for H$^{13}$CO$^{+}$, DCO$^{+}$ \textit{J}=1$-$0 and DCO$^{+}$ \textit{J}=2$-$1 of $\sim$28$^{\prime\prime}$, $\sim$34$^{\prime\prime}$, and $\sim$16$^{\prime\prime}$ are shown as red circles at the bottom-left corners in their integrated intensity maps.  The averaged spectra for column density estimation are derived in the regions enclosed by blue squares.}
\label{fig_BGPS3134}
\end{center}
\end{figure*}

\begin{figure*}
 \begin{center}
\includegraphics[width=18cm]{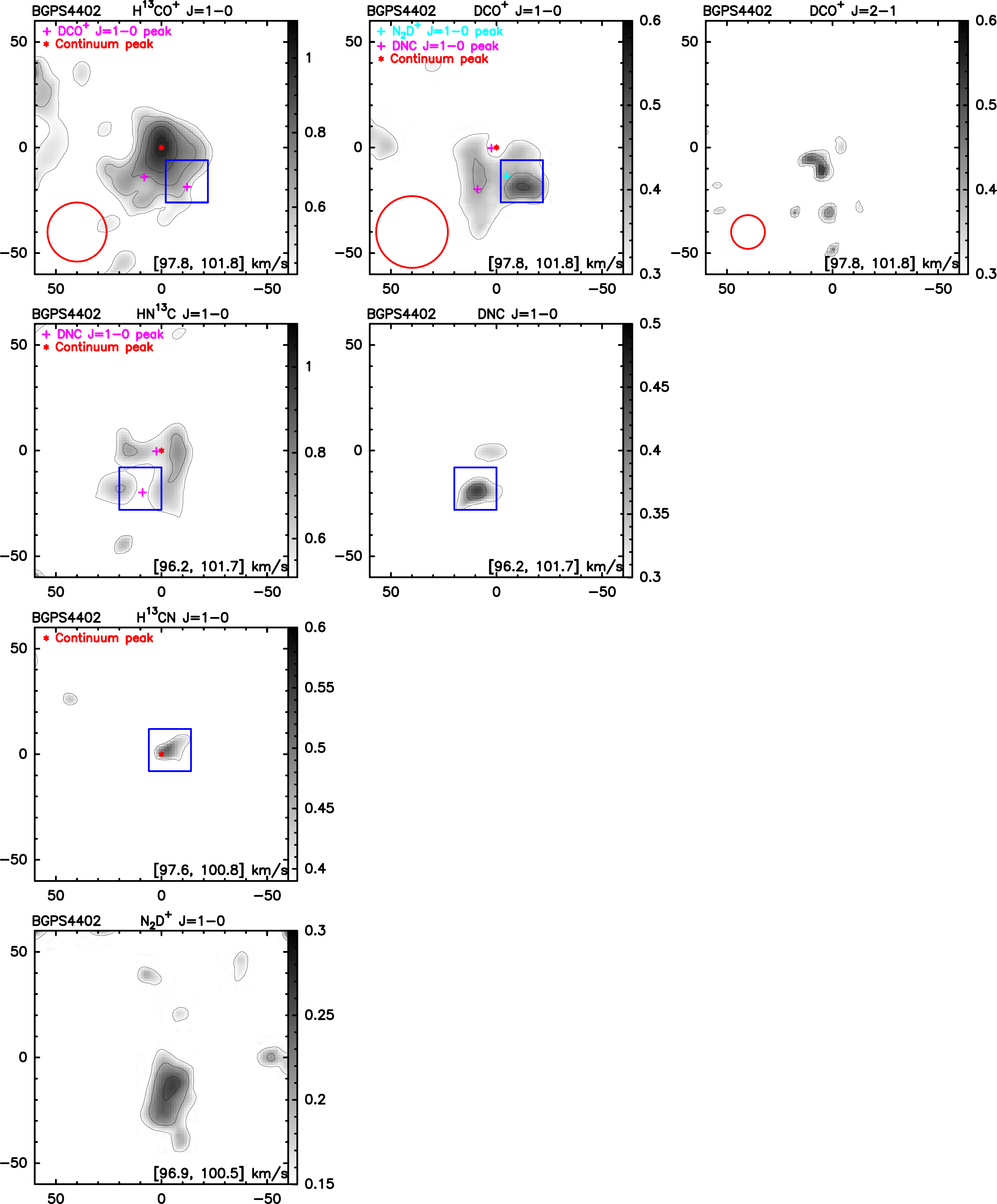}
\caption{The velocity integrated maps of deuterated and $^{13}$C-isotopic molecular lines for BGPS4402. The integrated velocity range, shown at the right-bottom corner, is derived after combining the Gaussian fitting line widths for each pair of deuterated molecular line and $^{13}$C$-$isotopologue.
For H$^{13}$CO$^{+}$, DCO$^{+}$, HN$^{13}$C, DNC, H$^{13}$CN, N$_{2}$D$^{+}$ \textit{J}=1$-$0, and DCO$^{+}$ \textit{J}=2$-$1, the contours start from 3$\sigma$ in steps of 1$\sigma$, with $\sigma$ = 0.14, 0.10, 0.17, 0.10, 0.13, 0.05, and 0.05 K km s$^{-1}$, respectively.
The crosses represent peaks in the velocity integrated maps of different molecular lines and the red star represents the 1.1mm continuum peak obtained from \citet{2016ApJ...822...59S}. The beam sizes for H$^{13}$CO$^{+}$, DCO$^{+}$ \textit{J}=1$-$0 and DCO$^{+}$ \textit{J}=2$-$1 of $\sim$28$^{\prime\prime}$, $\sim$34$^{\prime\prime}$, and $\sim$16$^{\prime\prime}$ are shown as red circles at the bottom-left corners in their integrated intensity maps.  The averaged spectra for column density estimation are derived in the regions enclosed by blue squares.}
\label{fig_BGPS4402}
\end{center}
\end{figure*}

Aiming to study the distribution of deuterated molecules, we made OTF observations towards 11 selected sources. Furthermore, pairs of DCO$^{+}$ and H$^{13}$CO$^{+}$ \textit{J}=1$-$0 (3.2$\times$10$^{4}$ and 6.2$\times$10$^{4}$ cm$^{-3}$), DCN and H$^{13}$CN \textit{J}=1$-$0 (2.6$\times$10$^{5}$ and 4.3$\times$10$^{5}$ cm$^{-3}$), DNC and HN$^{13}$C \textit{J}=1$-$0 (8.2$\times$10$^{4}$ and 9.6$\times$10$^{4}$ cm$^{-3}$) have similar critical densities at a temperature of 10 K, as reported in \citet{2020ApJ...901..145F}. Therefore, the observations of these lines will help us figure out whether the critical densities influence the detectability of the deuterated molecules. Consequently, we focus on the spatial distributions of deuterated molecules, namely, DCO$^{+}$, DCN, DNC, and N$_{2}$D$^{+}$, along with their $^{13}$C$-$isotopologues, H$^{13}$CN, H$^{13}$CO$^{+}$, and HN$^{13}$C, which are listed in Table \ref{obs_para}. Figure \ref{fig_BGPS2693} $-$ \ref{fig_BGPS4402} displays the maps of deuterated molecules and $^{13}$C$-$isotopologues from the OTF observation toward the 11 sources, the centre point of each map is the 1.1~mm continuum peak for each source obtained from \citet{2016ApJ...822...59S}. The regions enclosed by blue squares represent the peak regions of the deuterated molecular lines. The spectra averaged from these regions are presented in Figure \ref{spec_column}. Their parameters are shown in Table \ref{res_para}. We also show the velocity field and line width maps in Figure \ref{2986_velo}$-$\ref{3134_width} and Appendix \ref{app2}, these maps have at least five successive pixels with available values. The integrated maps of HCO$^{+}$, HCN, and HNC \textit{J}=1$-$0 are presented in the Appendix \ref{app1}, but we do not pay much attention to them because they are optically thick molecular lines.

\subsubsection{Distributions among deuterated molecules}

We have successfully mapped the \textit{J}=1$-$0 line emissions of DCO$^{+}$, DNC, DCN, and N$_{2}$D$^{+}$ towards 10, 11, 9, and 8 sources, respectively. These observations yield rms noise levels of approximately 0.05 K, as derived from line-free channels. In the following, we introduce the spatial distributions for each source, in conjunction with the 1.1mm continuum peak positions obtained from \citet{2016ApJ...822...59S}.
\\

\emph{BGPS2693}

The DCO$^{+}$ and DNC \textit{J}=1$-$0 emissions exhibit distinct morphologies, however, both peaking at the 1.1mm continuum peak position.

\emph{BGPS2931}

The positions of highest intensity for deuterated molecules are detected to the southeast of the continuum peak, while DCO$^{+}$ \textit{J}=1$-$0 exhibits more peaks.

\emph{BGPS2940}

DCO$^{+}$, DNC, DCN, and N$_{2}$D$^{+}$ \textit{J}=1$-$0 exhibit significantly different morphologies, but their peaks are located near the continuum peak.

\emph{BGPS2945}

The emissions of DCO$^{+}$ and DNC \textit{J}=1$-$0 extend in the NE-SW direction with a signal-to-noise level exceeding 10. On the contrary, the clumpy DCN and N$_{2}$D$^{+}$ \textit{J}=1$-$0 structures, at approximately 5 times the noise level ($\sim$5$\sigma{\rm area}$), are detected with peaks immediately next to the continuum peak.

\emph{BGPS2984}

All four lines exhibit peaks near the continuum peak, with an additional common peak to the northeast of the central position. Apart from DCO$^{+}$ and DNC \textit{J}=1$-$0, DCN and N$_{2}$D$^{+}$ \textit{J}=1$-$0 exhibit emissions with clumpy morphology, primarily due to limited signal-to-noise levels.

\emph{BGPS2986}

The integrated maps of these four deuterated molecular lines show rounded morphologies, with peaks located at the 1.1mm continuum peak position.

\emph{BGPS3018}

Both DCO$^{+}$ and DNC \textit{J}=1$-$0 emissions exhibit two peaks on opposite sides of the 1.1mm continuum peak position, while several $\sim$3$\sigma_{\rm area}$ DCN clumpy structures are detected.

\emph{BGPS3110}

DNC and DCN \textit{J}=1$-$0 emissions both have peaks near the continuum peak, with DCN \textit{J}=1$-$0 revealing more peaks. N$_{2}$D$^{+}$ \textit{J}=1$-$0 is marginally detected.

\emph{BGPS3125}

The emissions of DCO$^{+}$, DNC, and DCN \textit{J}=1$-$0 display different morphologies. DCO$^{+}$ and DNC \textit{J}=1$-$0 have peaks at the 1.1~mm continuum peak position, while there is no clear correlation among other peaks.

\emph{BGPS3134}

The emissions of DCO$^{+}$, DNC, DCN, and N$_{2}$D$^{+}$ \textit{J}=1$-$0 all extend from the mapping centre to the northwest, with peaks detected surrounding the continuum peak.

\emph{BGPS4402}

The emissions of DCO$^{+}$, DNC, and N$_{2}$D$^{+}$ \textit{J}=1$-$0 display completely different morphologies. DNC \textit{J}=1$-$0 has a peak at the continuum peak position, while DCO$^{+}$ and N$_{2}$D$^{+}$ \textit{J}=1$-$0 do not.
\\

\begin{figure*}
\centering
\begin{tabular}{cc}
\includegraphics[width=82mm]{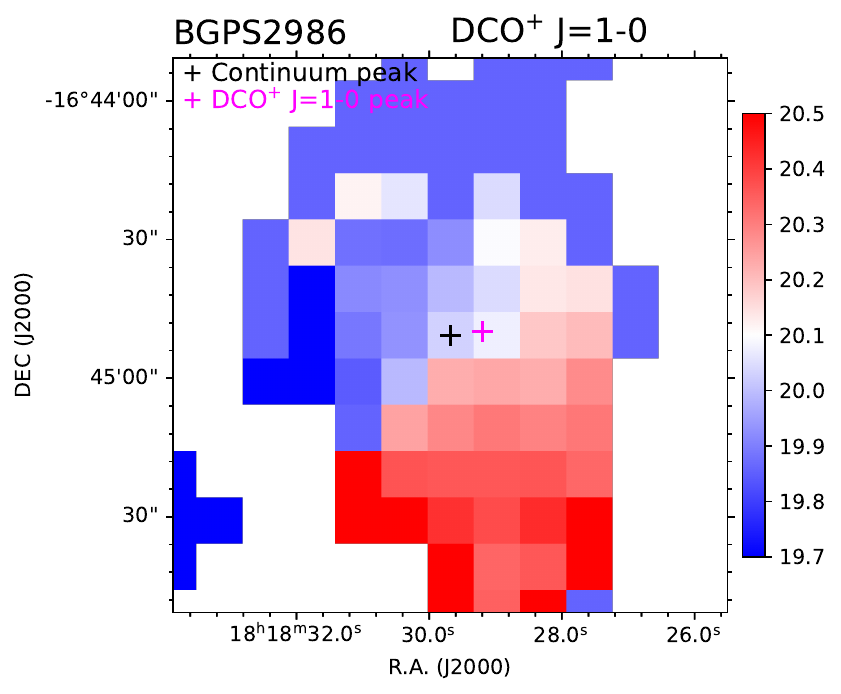}&
\includegraphics[width=82mm]{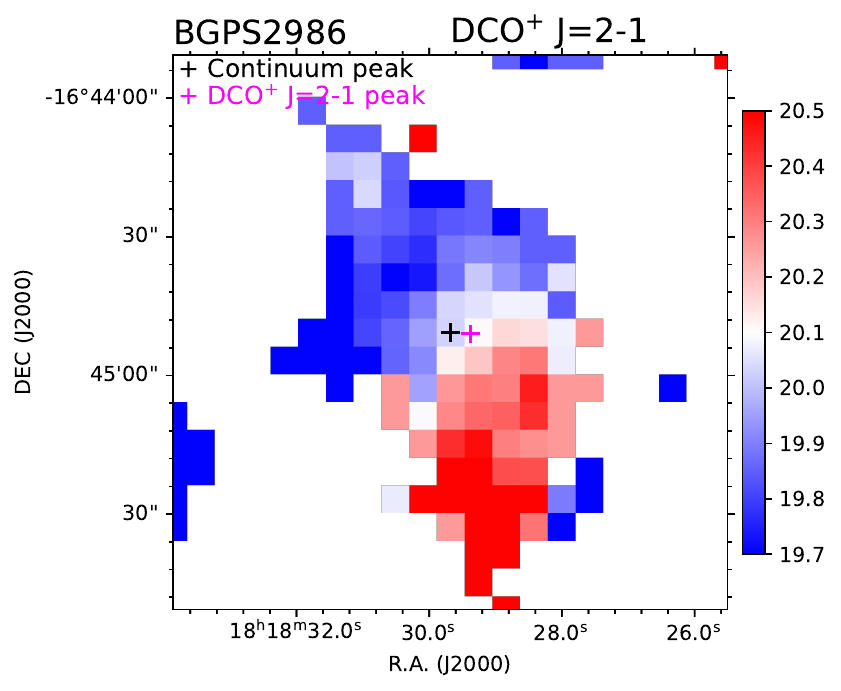}\\
\includegraphics[width=82mm]{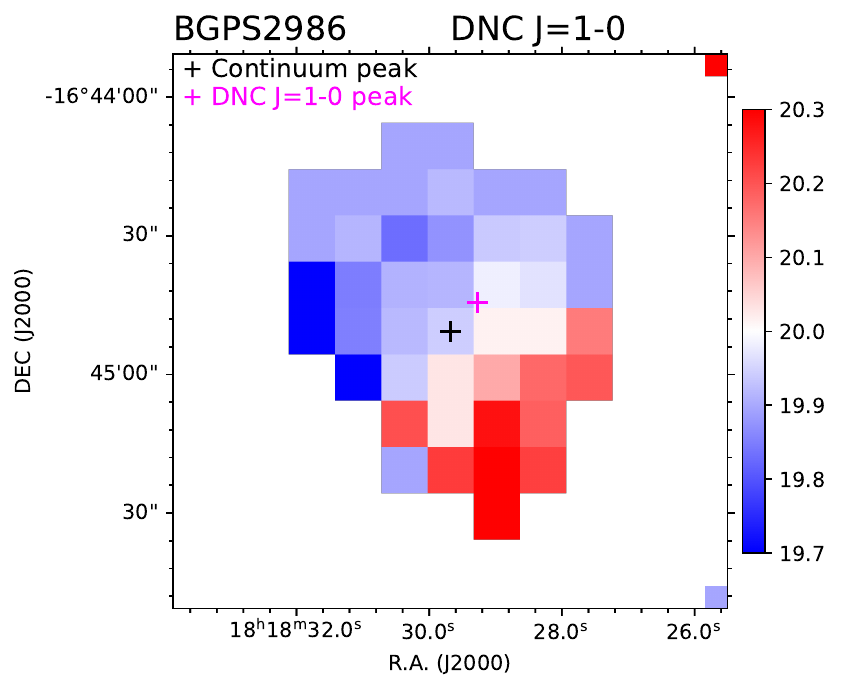}&
\includegraphics[width=82mm]{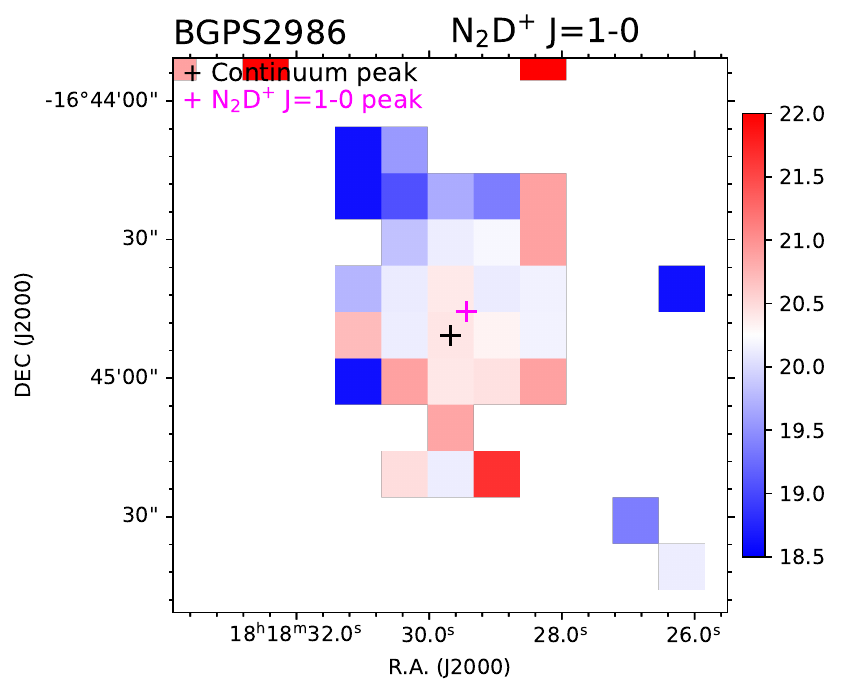}
\end{tabular}
\caption{Velocity field of deuterated molecules for BGPS2986. The crosses represent integrated intensity peaks of different lines and 1.1mm continuum peak.}
\label{2986_velo}
\end{figure*}

\begin{figure*}
\centering
\begin{tabular}{cc}
\includegraphics[width=82mm]{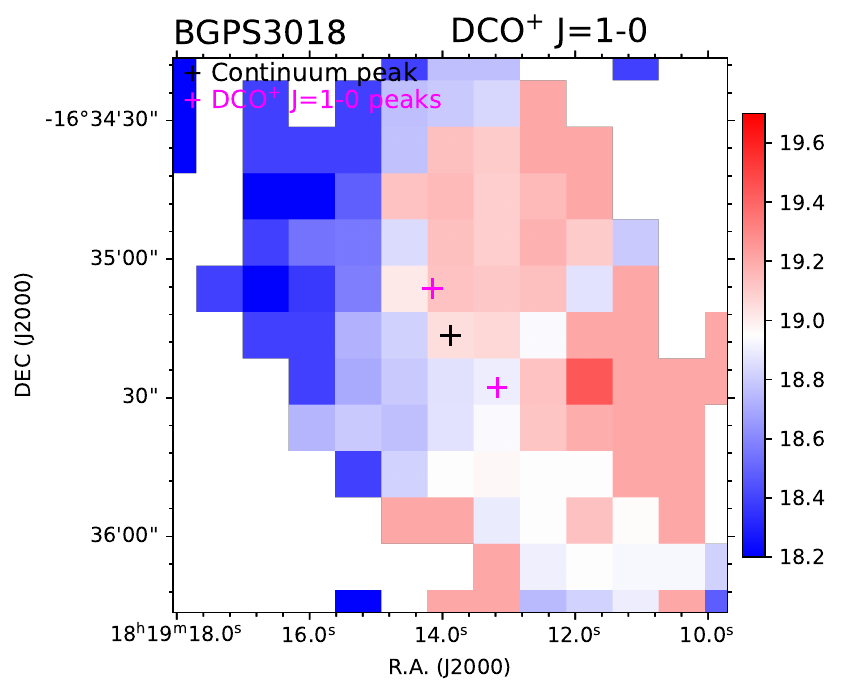}&
\includegraphics[width=82mm]{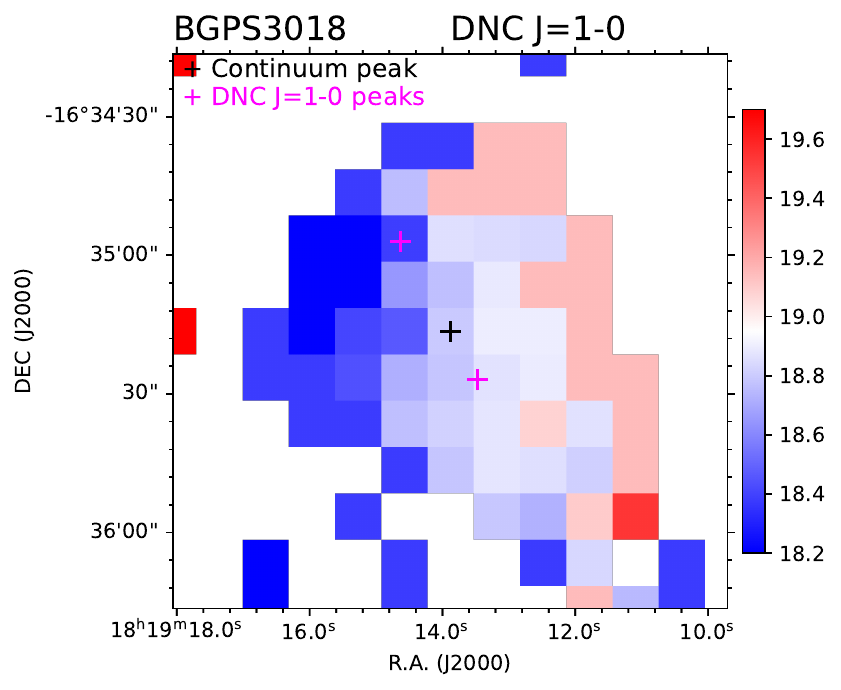}
\end{tabular}
\caption{Line width distribution of deuterated molecules for BGPS3018.}
\label{3018_velo}
\end{figure*}

\begin{figure*}
\centering
\begin{tabular}{cc}
\includegraphics[width=82mm]{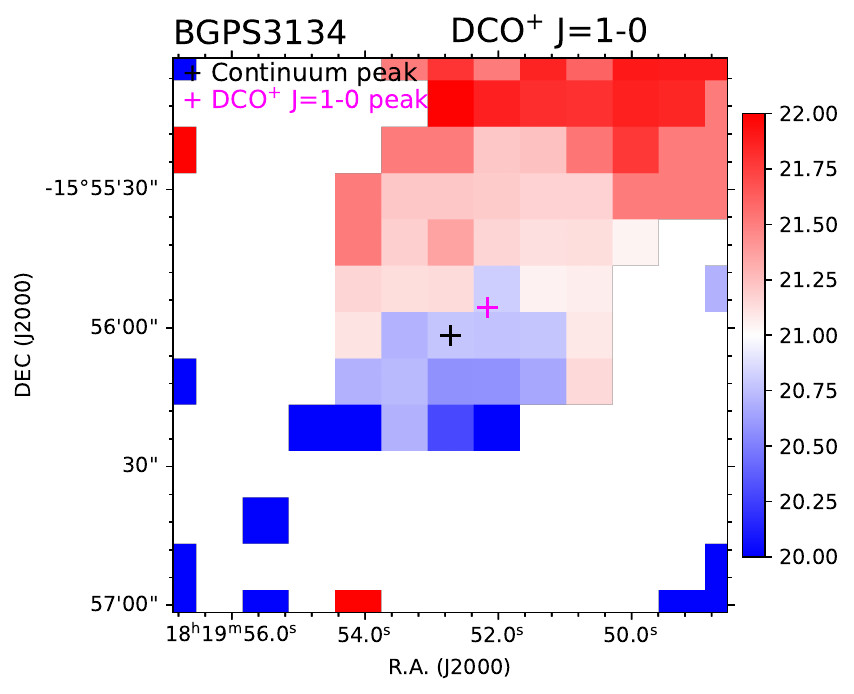}&
\includegraphics[width=82mm]{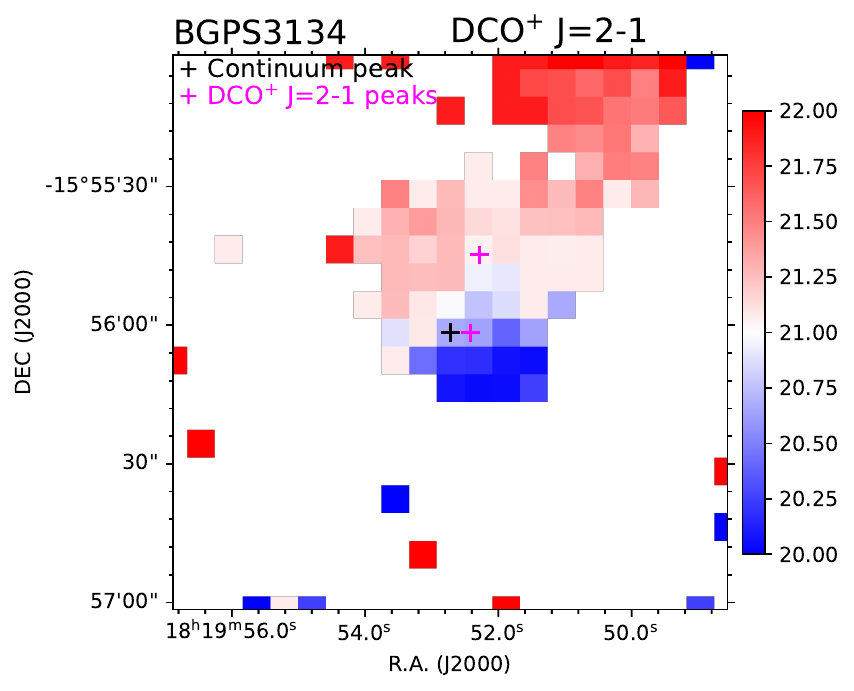}\\
\includegraphics[width=82mm]{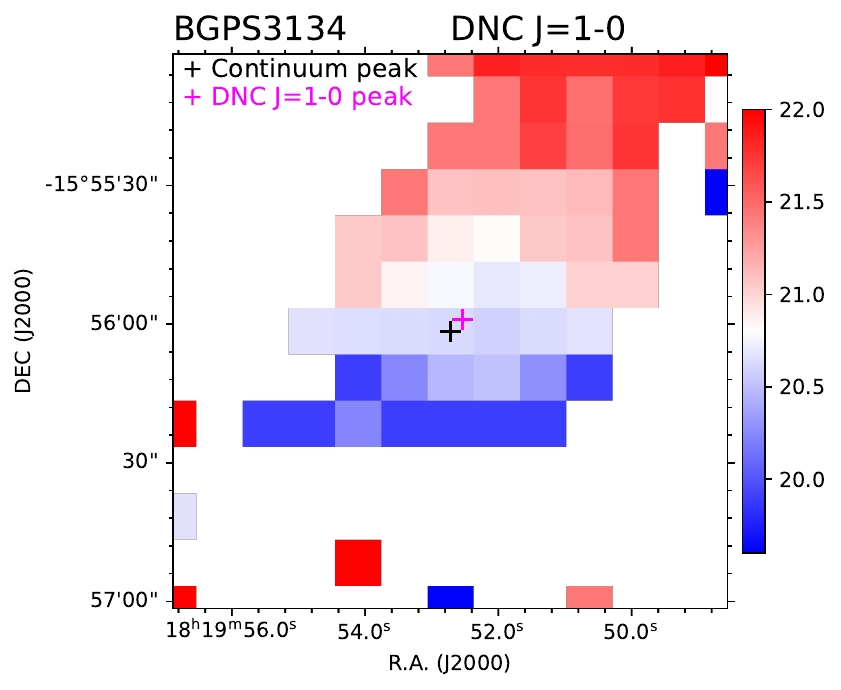}&
\includegraphics[width=82mm]{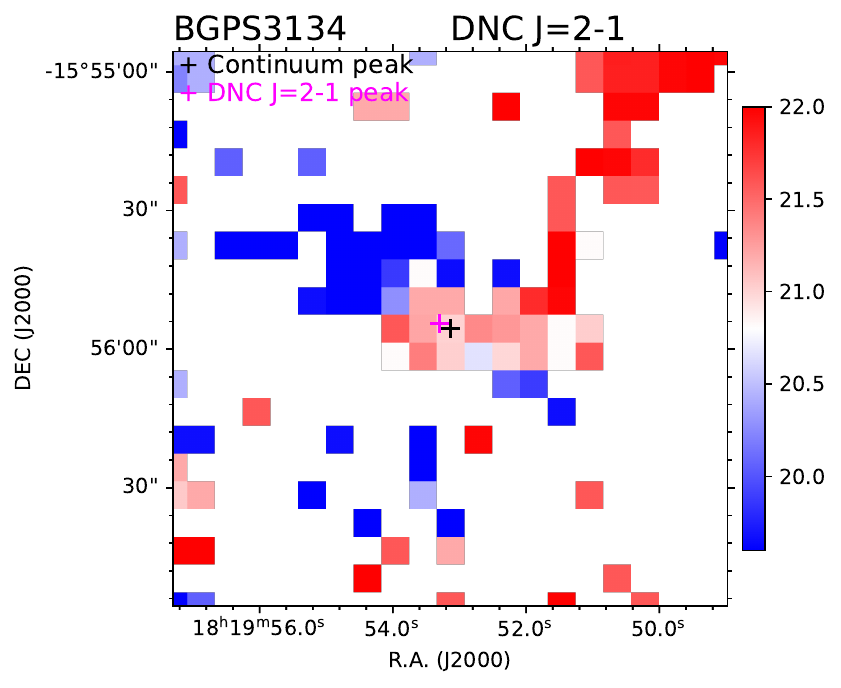}\\
\includegraphics[width=82mm]{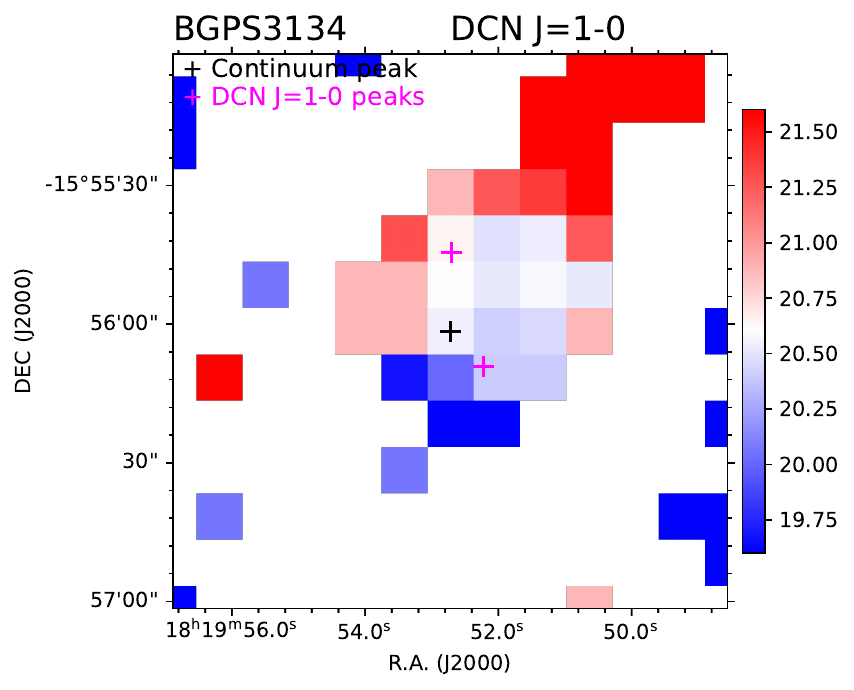}&
\includegraphics[width=82mm]{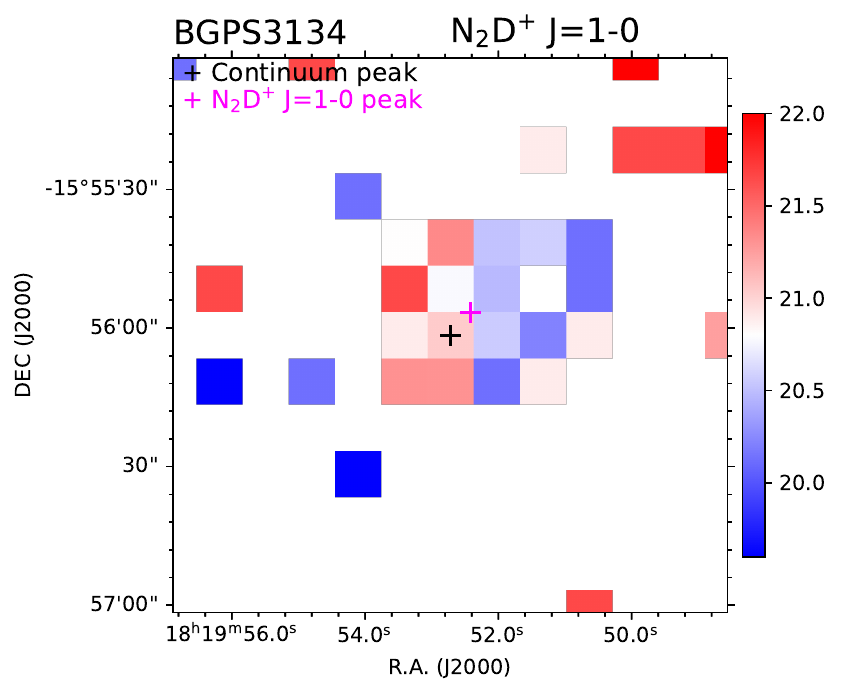}
\end{tabular}
\caption{Line width distribution of deuterated molecules for BGPS3134.}
\label{3134_velo}
\end{figure*}

\begin{figure*}
\centering
\begin{tabular}{cc}
\includegraphics[width=82mm]{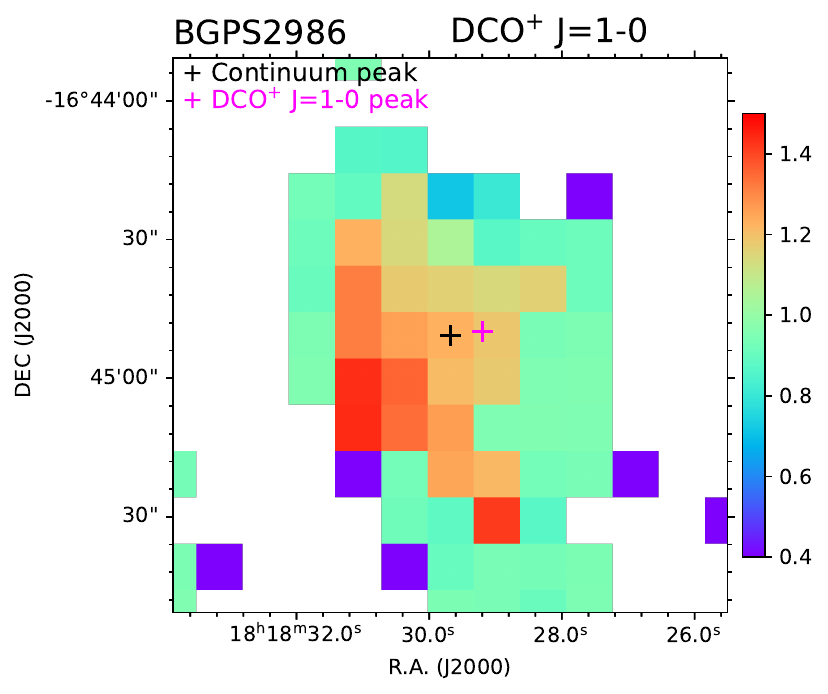}&
\includegraphics[width=82mm]{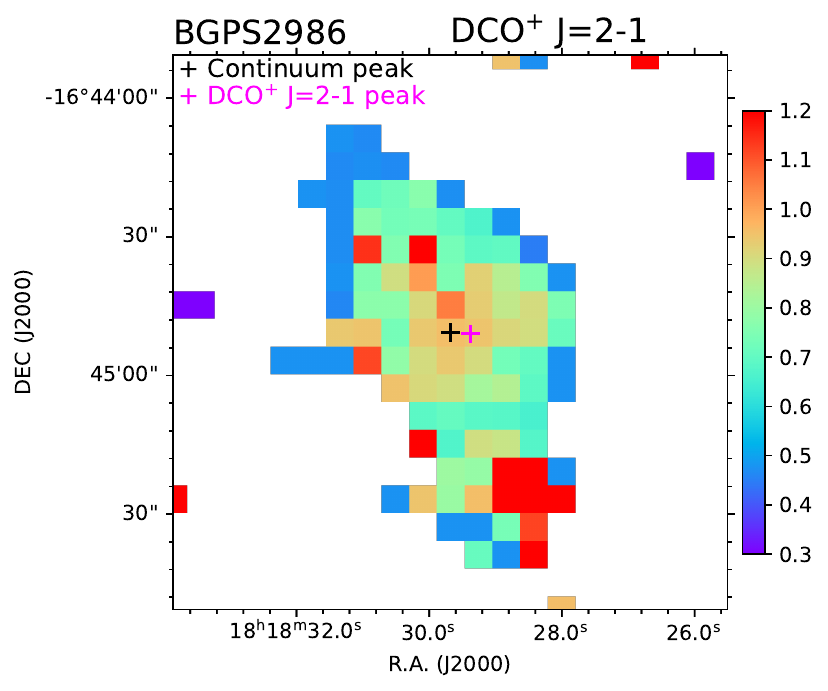}\\
\includegraphics[width=82mm]{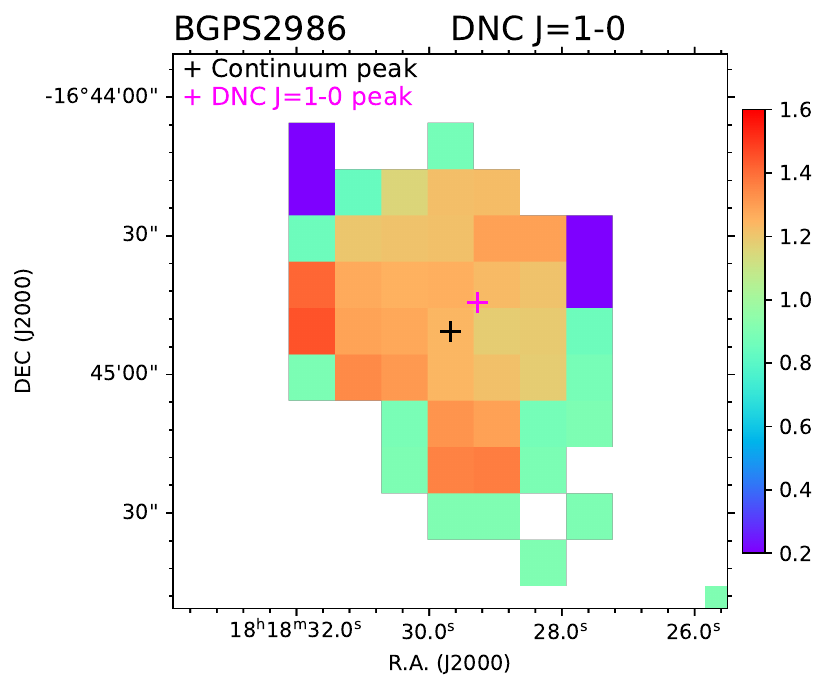}&
\includegraphics[width=82mm]{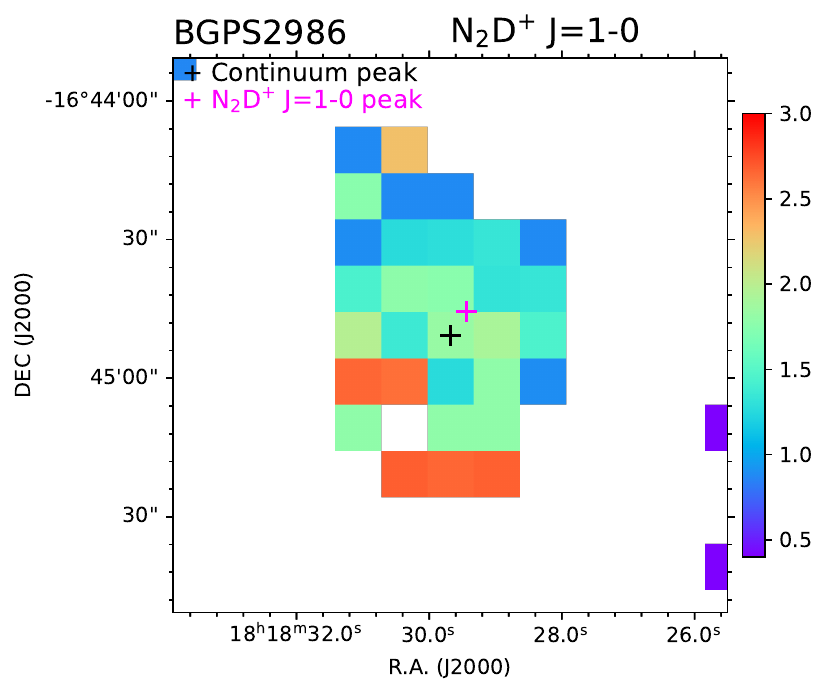}
\end{tabular}
\caption{Line width distribution of deuterated molecules for BGPS2986. The crosses represent integrated intensity peaks of different lines and 1.1mm continuum peak.}
\label{2986_width}
\end{figure*}

\begin{figure*}
\centering
\begin{tabular}{cc}
\includegraphics[width=82mm]{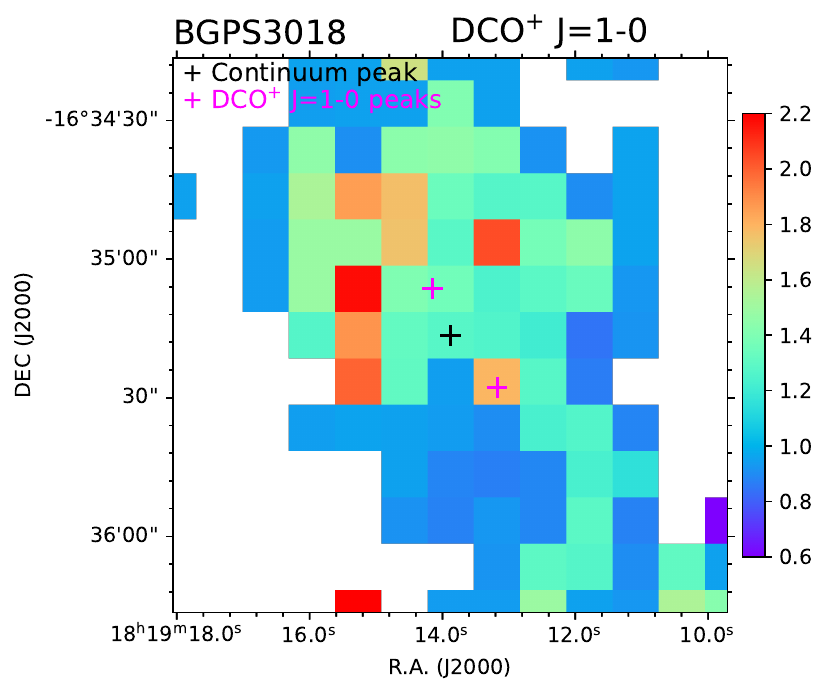}&
\includegraphics[width=82mm]{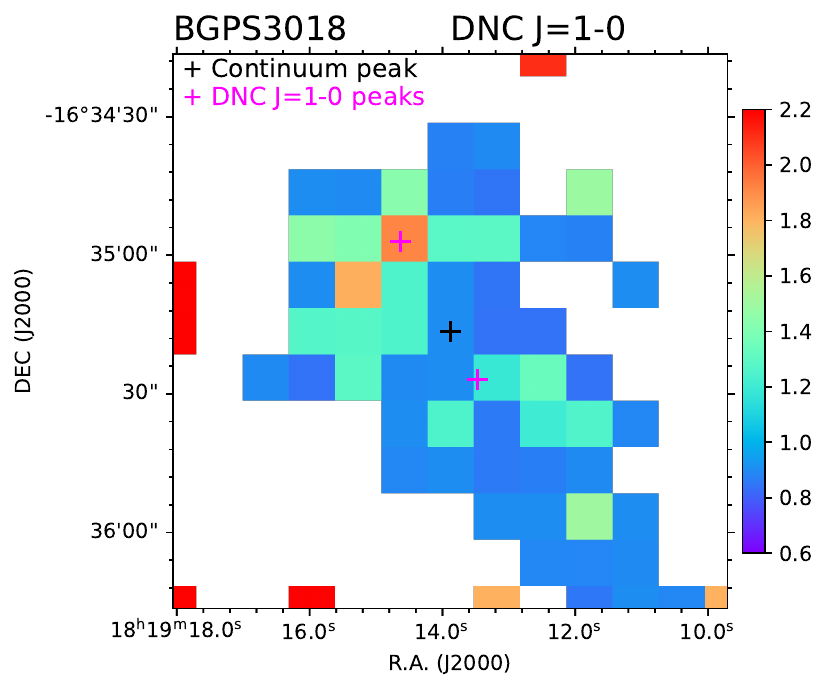}
\end{tabular}
\caption{Line width distribution of deuterated molecules for BGPS3018.}
\label{3018_width}
\end{figure*}

\begin{figure*}
\centering
\begin{tabular}{cc}
\includegraphics[width=82mm]{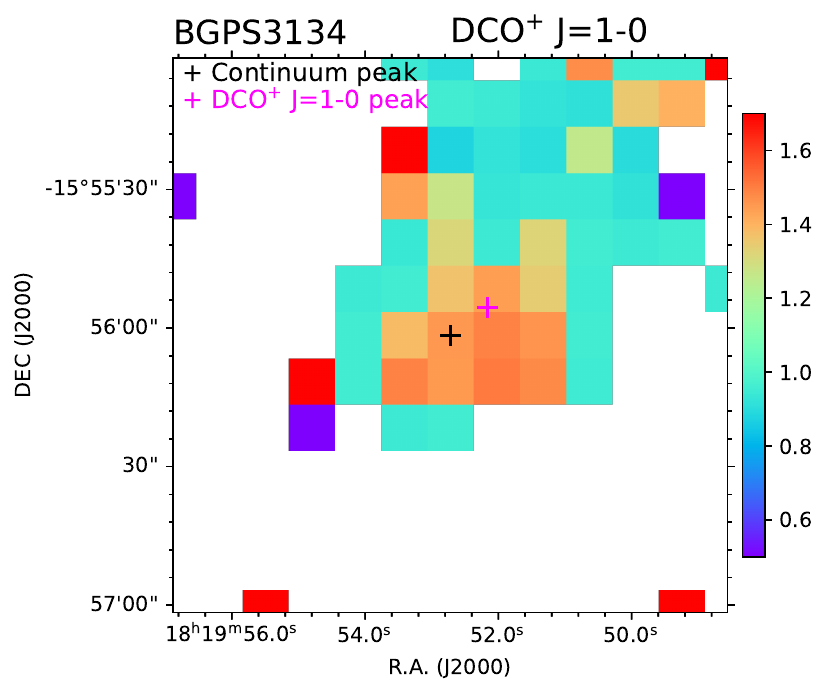}&
\includegraphics[width=82mm]{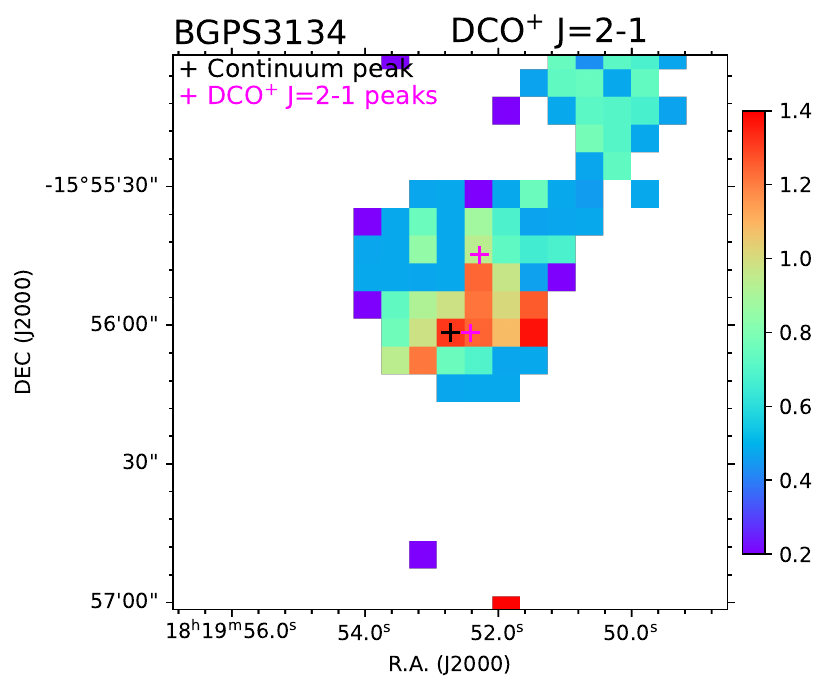}\\
\includegraphics[width=82mm]{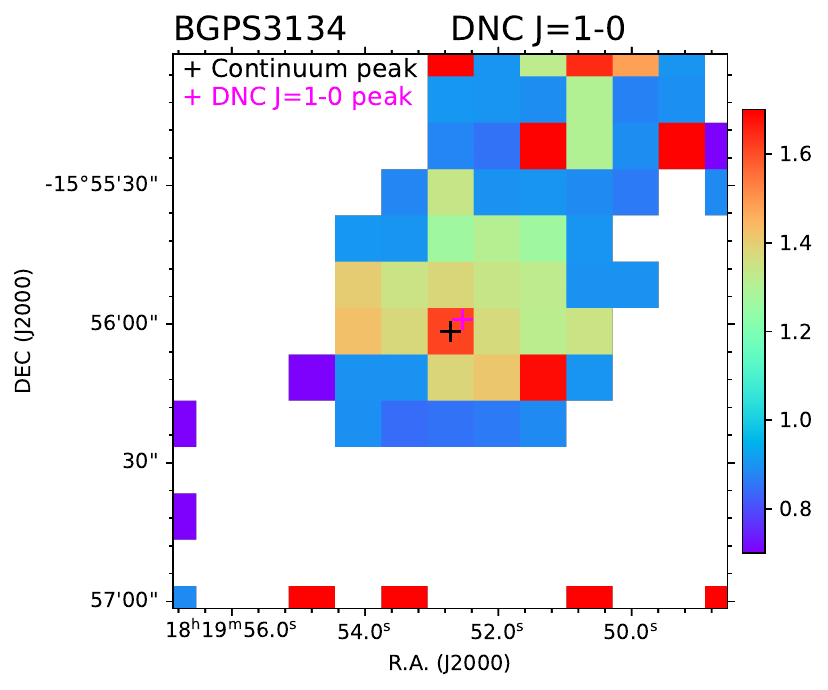}&
\includegraphics[width=82mm]{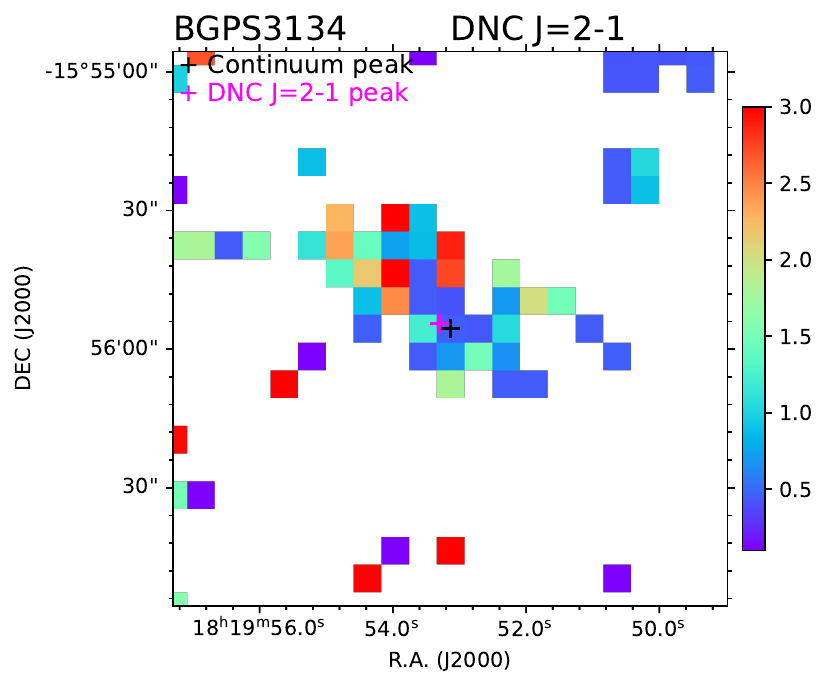}\\
\includegraphics[width=82mm]{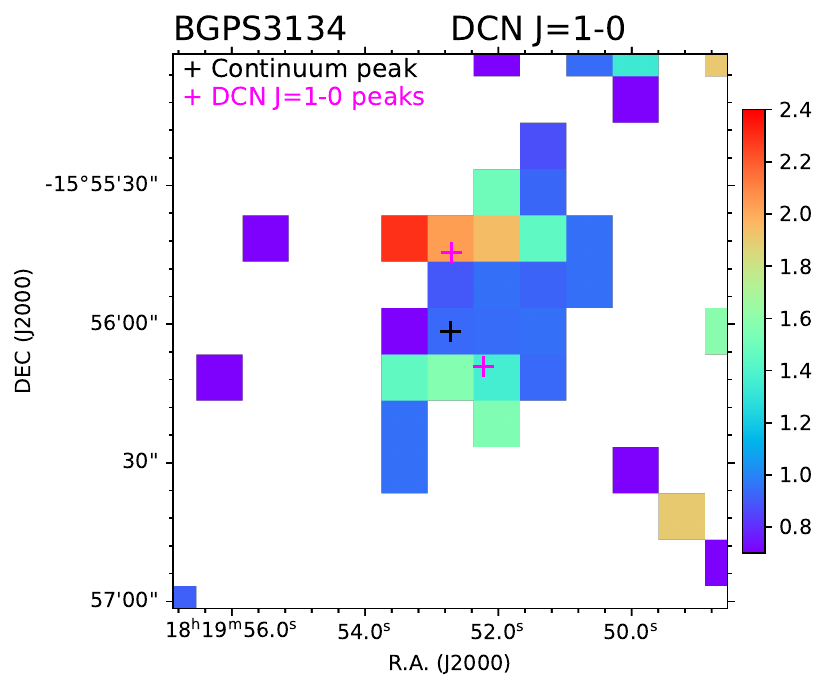}&
\includegraphics[width=82mm]{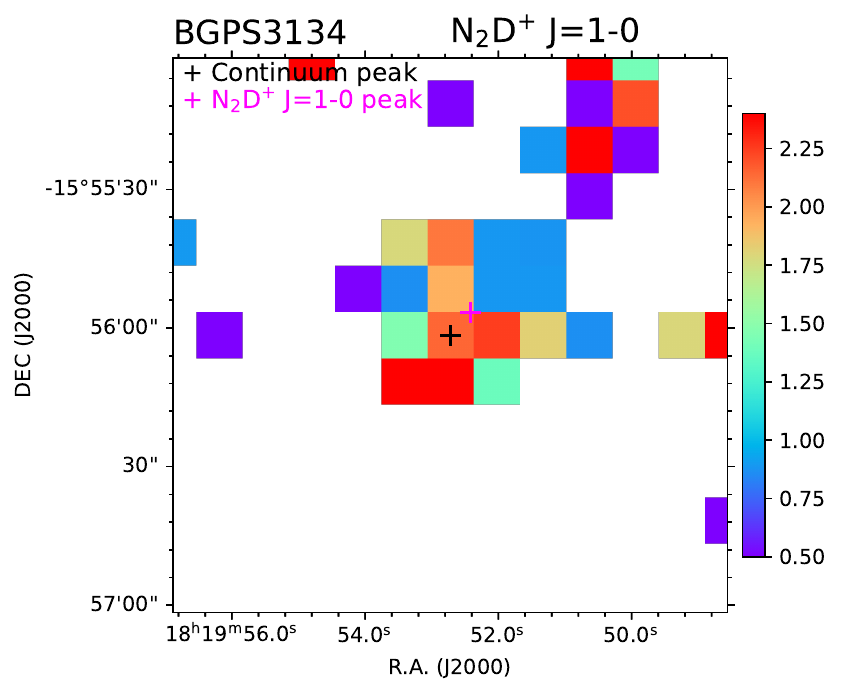}
\end{tabular}
\caption{Line width distribution of deuterated molecules for BGPS3134.}
\label{3134_width}
\end{figure*}

For most of the sources (9 out of 11), considering one beam size, we have not found clear offsets between the peaks of the deuterated molecular lines and the 1.1mm continuum peaks. Additionally, for spectra averaged from regions with the highest deuterated molecular line intensities, the peak intensities of DCO$^{+}$ \textit{J}=1$-$0 have a larger median value (0.759 K) compared to those of DNC \textit{J}=1$-$0 (0.585 K) and DCN \textit{J}=1$-$0 (0.252 K). In Figures \ref{2986_velo}$-$\ref{3134_velo}, we observe clear NE-SW, E-W, and S-N oriented velocity gradients for BGPS2986, BGPS3018, and BGPS3134, respectively. Furthermore, in their line width maps, depicted in Figures \ref{2986_width}$-$\ref{3134_width}, there is an enhancement towards the line peaks.

\subsubsection{Distributions of deuterated molecules and $^{13}$C$-$isotopologues}

The investigation into deuterated molecules and $^{13}$C$-$isotopologues, which are optically thin, could give clues about the distinctions between deuterated molecules and dense gas tracers. By considering the 1.1mm continuum peak positions, we gain insights into the relationship between these molecules and star formation activities. The $^{13}$C lines exhibit rms noise levels of approximately 0.10 K, derived from line-free channels.
\\

\emph{DCO$^{+}$ and H$^{13}$CO$^{+}$ \textit{J}=1$-$0}

We detect DCO$^{+}$ and H$^{13}$CO$^{+}$ \textit{J}=1$-$0 peaks in proximity to the 1.1mm continuum peak positions for eight sources, with the exception of BGPS3125 and BGPS4402. Two sources, BGPS2693 and BGPS4402, present distinctive morphologies for DCO$^{+}$ and H$^{13}$CO$^{+}$. Moreover, in terms of the DCO$^{+}$/H$^{13}$CO$^{+}$ peak intensity ratio, seven out of ten sources display values greater than 1.

\emph{DNC and HN$^{13}$C \textit{J}=1$-$0}

Among the eleven sources examined, seven exhibit DNC and HN$^{13}$C \textit{J}=1$-$0 peaks close to the continuum peak. For BGPS2945, BGPS2984, BGPS2986, and BGPS3134, there is an absence of noticeable dissimilarity in the morphologies of the DNC and HN$^{13}$C emissions. However, in five sources, more HN$^{13}$C peaks are discerned than DNC peaks. In addition, eight out of eleven sources have a DNC/HN$^{13}$C peak intensity ratio exceeding 1.

\emph{DCN and H$^{13}$CN \textit{J}=1$-$0}

Out of the nine sources in which both DCN and H$^{13}$CN \textit{J}=1$-$0 emissions were detected, five display DCN and H$^{13}$CN peaks near the 1.1mm continuum peak positions. Nevertheless, for each source, the morphologies of DCN and H$^{13}$CN \textit{J}=1$-$0 exhibit striking dissimilarities. In particular, almost half of these sources feature a peak H$^{13}$CN \textit{J}=1$-$0 intensity surpassing that of DCN \textit{J}=1$-$0.
\\

In summary, the majority of detected deuterated molecular line emissions and $^{13}$C isotopologue emissions have peaks near the 1.1mm continuum peak positions. The peak intensities of DCN and H$^{13}$CN \textit{J}=1$-$0 in the deuterated peak regions are comparable, while HCO$^{+}$ and HNC display deuterated molecular lines with stronger peak intensities than their $^{13}$C isotopologues.

\subsection{H44$\alpha$ detection}

\begin{figure*}
\centering
\includegraphics[width=160mm]{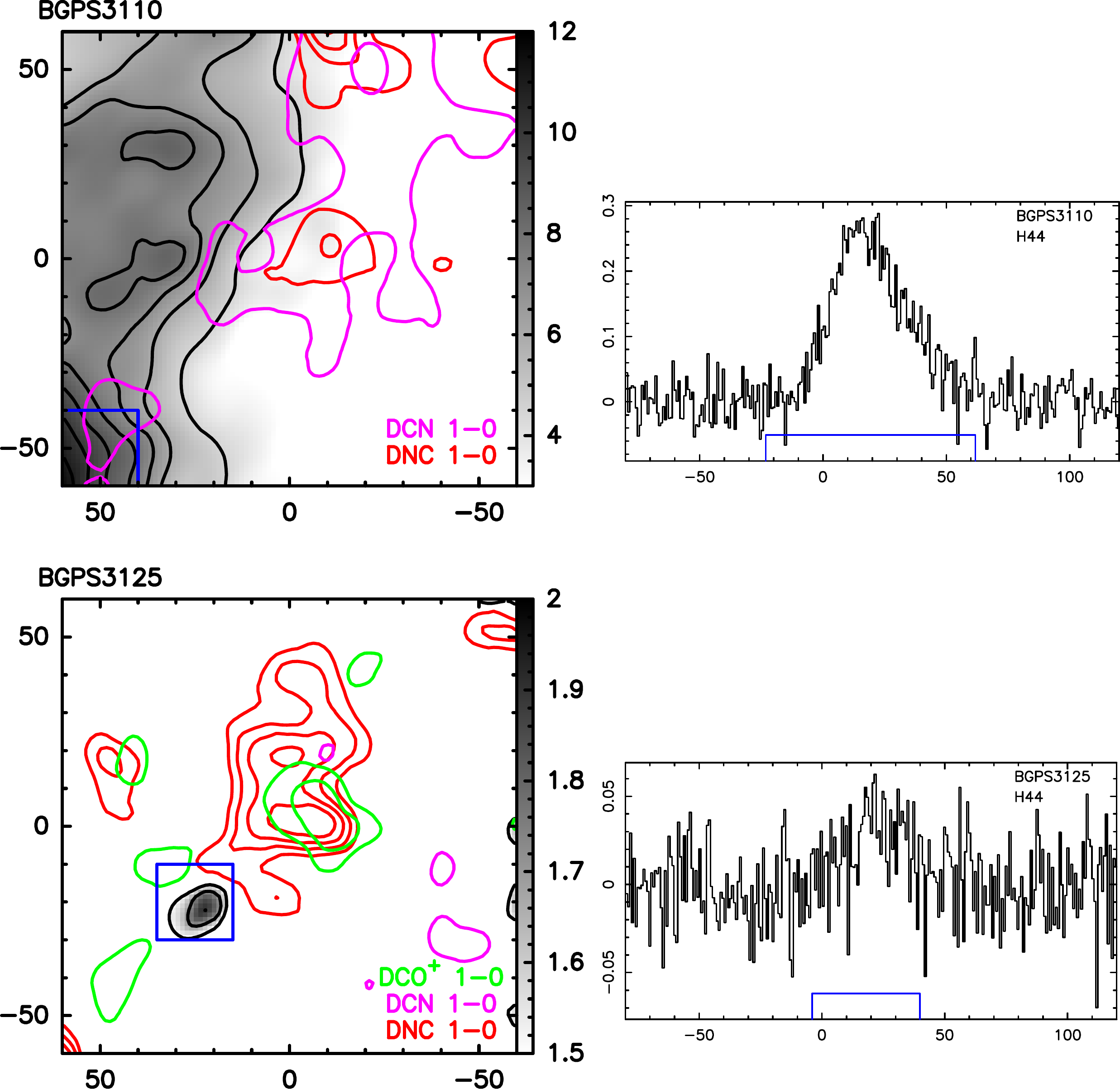}  
\caption{Left: The H44$\alpha$ (black contours and grey-scale maps) versus deuterated lines (color contours) distributions for BGPS3110 and BGPS3125. Right: The H44$\alpha$ spectra for BGPS3110 and BGPS3125. The velocity range that we derived the velocity-integrated intensity maps are shown in the spectra with blue lines. The spectra are averaged from the regions enclosed by blue lines in the velocity-integrated intensity maps. The black contours start from 3$\sigma$ in steps of 1$\sigma$, with $\sigma$ = 1.5 and 0.2 K km s$^{-1}$ for BGPS3110 and BGPS3125. The color contours of the deuterated lines are the same as those in Figure \ref{fig_BGPS3110} and \ref{fig_BGPS3125}.}
\label{H44}
\end{figure*}

Radio recombination lines (RRLs) are the tracers of H{\scriptsize II} regions, and can help us determine whether H {\scriptsize II} regions influence the appearance of deuterated molecules. The H44$\alpha$ line at 76.4 GHz is covered in our OTF observations and is detected in BGPS3110 and BGPS3125. We present H44$\alpha$ versus deuterated line emission distributions and averaged H44$\alpha$ spectra in the central regions (20$^{\prime\prime}$$\times$20$^{\prime\prime}$) in Figure \ref{H44}. For BGPS3110, deuterated line emissions are detected primarily at the edge of H44$\alpha$ emissions. For BGPS3125, the H44$\alpha$ emissions are detected in a tiny area (radius of $\sim$0.09 pc at a distance of $\sim$3.5 kpc), which could be a UCH{\scriptsize II} region, and the deuterated line emissions are not detected towards this region. Thereby the emissions of the deuterated molecules are generated from the massive Starless Clump Candidates other than enhanced by the H{\scriptsize II} regions.

\subsection{Column Density}

In order to investigate the deuterated fraction in the peak regions of the deuterated molecules, we first need to derive their column densities. Assuming local thermodynamic equilibrium (LTE) conditions, and optically thin emission for deuterated molecules and $^{13}$C$-$isotopologues, the column densities of the 1$-$0 molecular lines can be derived following \citet{2015PASP..127..266M}:
\begin{equation}\label{Eq_N}
N_{\rm tot} = \left(\frac{3h}{8 \pi^{3} S \mu^{2} R_{i}}\right) \left(\frac{Q(T_{\rm ex})}{g_{u}}\right) \frac{{\rm exp}\left(\frac{E_{u}}{kT_{ex}}\right)}{{\rm exp}\left(\frac{h\nu}{kT_{ex}}\right)-1} \times \frac{1}{(J_{\nu}(T_{ex})-J_{\nu}(T_{bg}))} \int \frac{T_{R}dv}{f} , \\
\end{equation}
where $h$ is the Planck constant, $S$ is the line strength for linear molecules, $\mu$ is the molecular electric dipole moment, the sum of relative intensities $R_{i}$ = 1 for $\Delta J$ = 1 transitions, $Q(T_{\rm ex})$ is the partition function, $g_{u}$ is the degeneracy of the upper state, $E_{u}$ is the energy of the upper level energy, $k$ is the Boltzmann constant, $T_{\rm ex}$ is the excitation temperature, $\nu$ is the rest frequency of the transition, $\int T_{R}dv$ is the integrated intensity, $f$ is the filling factor assumed to be 1, $J_{\nu}(T)$ is the planck function, and $T_{bg}$ = 2.73 K is the cosmic microwave background.
We adopt the kinetic temperature derived from NH$_{3}$ observations in \citet{2016ApJ...822...59S} as the excitation temperature. For most sources, the uncertainty of the gas kinetic temperatures is lower than 3\%. The spectroscopic parameters are obtained from the CDMS database \citep{2001A&A...370L..49M,2005JMoSt.742..215M,2016JMoSp.327...95E} and JPL catalogues \citep{1998JQSRT..60..883P}.

The estimated column densities are shown in Table \ref{Tab_Column}, ranging from 10$^{11}$ to 10$^{13}$ cm$^{-2}$. For H$^{13}$CN and DCN, the ranges of their column densities are 6.0$\times$10$^{11}$ $-$ 1.2$\times$10$^{13}$ cm$^{-2}$ and 9.8$\times$10$^{11}$ $-$ 3.0$\times$10$^{12}$ cm$^{-2}$, respectively. And the medians are 1.1$\times$10$^{12}$ and 1.5$\times$10$^{12}$ cm$^{-2}$. $N$(H$^{13}$CO$^{+}$) and $N$(DCO$^{+}$) have ranges of 1.4$\times$10$^{12}$ $-$ 9.3$\times$10$^{12}$ cm$^{-2}$ and 1.2$\times$10$^{12}$ $-$ 7.6$\times$10$^{12}$ cm$^{-2}$, respectively. Medians of them are 2.6$\times$10$^{12}$ and 3.8$\times$10$^{12}$ cm$^{-2}$. $N$(HN$^{13}$C) ranges from 8.6$\times$10$^{11}$ to 6.6$\times$10$^{12}$ cm$^{-2}$ and $N$(DNC) ranges from 5.5$\times$10$^{11}$ to 3.6$\times$10$^{12}$ cm$^{-2}$. Their medians are 1.6$\times$10$^{12}$ and 2.2$\times$10$^{12}$ cm$^{-2}$. Most of the estimated column densities have errors lower than 15\%. Although $N$(H$^{13}$CN) of BGPS3110 is the strongest, most of the column densities of H$^{13}$CN are comparable with the column density of HN$^{13}$C, and H$^{13}$CO$^{+}$ has the strongest column density. A similar relationship occurs in $N$(DCN), $N$(DNC), and $N$(DCO$^{+}$). The column densities of DNC and HN$^{13}$C are consistent with the results of studies of massive clumps in the early evolutionary stages of high-mass star formation \citep[e.g.,][]{2012ApJ...747..140S,2015ApJ...803...70S} and cold dark cloud cores \citep[e.g.,][]{2003ApJ...594..859H,2016A&A...591L...2V}. There are also many works \citep[e.g.,][]{2002A&A...381.1026R,2007A&A...464..245R,2012ApJ...749..162O,2012MNRAS.422.1098R,2014A&A...563A..97G,2015A&A...579A..80G,2014A&A...569A..19T,2017ApJ...835..231H} involving studies of $N$(DCN), $N$(H$^{13}$CN), $N$(DCO$^{+}$) and $N$(H$^{13}$CO$^{+}$), our results of column densities conform to those in early phases of massive star formation. In addition, $N$(HN$^{13}$C)/$N$(H$^{13}$CN) has a median value of $~$1.4, which means that HN$^{13}$C \textit{J}=1$-$0 has a higher column density than H$^{13}$CN \textit{J}=1$-$0 for most sources.

According to Equation \ref{Eq_N}, the column density ratios between deuterated molecules and their $^{13}$C$-$isotopologue counterparts can be expressed by integrated intensity ratios: N(DCO$^{+}$)/N(H$^{13}$CO$^{+}$) = $\sim$1.06 $\times$ I(DCO$^{+}$)/I(H$^{13}$CO$^{+}$), N(DCN)/N(H$^{13}$CN) = $\sim$1.31 $\times$ I(DCN)/I(H$^{13}$CN), and N(DNC)/N(H$^{13}$NC) = $\sim$0.97 $\times$ I(DNC)/I(H$^{13}$NC). We have listed these integrated intensity ratios and the corresponding column density ratios in Table \ref{D_C}. The integrated intensity ratios for DCO$^{+}$/H$^{13}$CO$^{+}$ and DNC/HN$^{13}$C closely mirror the column density ratios.

\subsection{Deuterated fraction}\label{frac}

The deuterated fraction (D$_{\rm frac}$) is the column density ratio between a deuterated molecular line and its hydrogenated counterpart. With the derived $^{12}$C/$^{13}$C ratio, we can estimate the deuterated fraction for the 11 selected sources in the OTF observations from the column density ratio between the deuterated molecular line and its $^{13}$C-isotopologue. With the OTF observations, we can obtain the spatial distribution information of the lines and derive a more accurate deuterated fraction. The $^{12}$C/$^{13}$C ratio is a useful tool for studying the stellar nucleosynthesis and chemical evolution of the Milky Way \citep{1994ARA&A..32..191W}, and many observations indicate a gradient of $^{12}$C/$^{13}$C ratios across the Galaxy \citep[e.g.,][]{1976A&A....51..303W,1979MNRAS.188..445W,1980A&A....82...41H,2005ApJ...634.1126M,2023A&A...670A..98Y}. We adopt the $^{12}$C/$^{13}$C ratios by following the relationship in \cite{2005ApJ...634.1126M}:
\begin{equation}\label{Eq_12C/13C}
\rm ^{12}C/^{13}C = 6.21 D_{GC} + 18.71 \\
\end{equation}
The distance from the Galactic center (D$_{gc}$) for each source can be estimated from the kinematic distance from \citet{2016ApJ...822...59S} and the distance from the sun to the Galactic center of $R_{0}$ = 8.15 kpc \citep{2019ApJ...885..131R}. The derived $^{12}$C/$^{13}$C ratios are listed in Table \ref{D_C}. The $^{12}$C/$^{13}$C ratios for the OTF sources have a range of 34.0 to 62.3.

\begin{figure*}
\centering
\includegraphics[width=57mm]{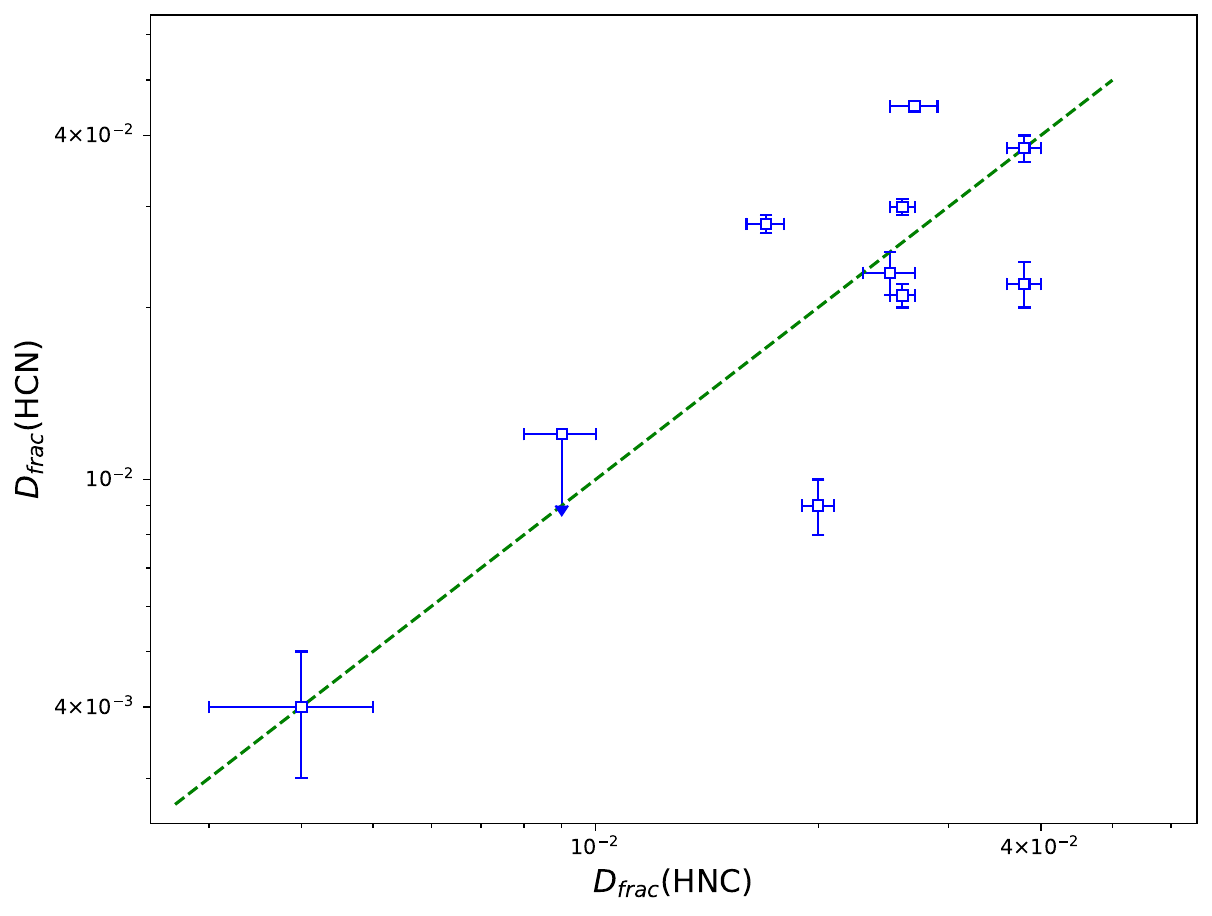}  
\includegraphics[width=57mm]{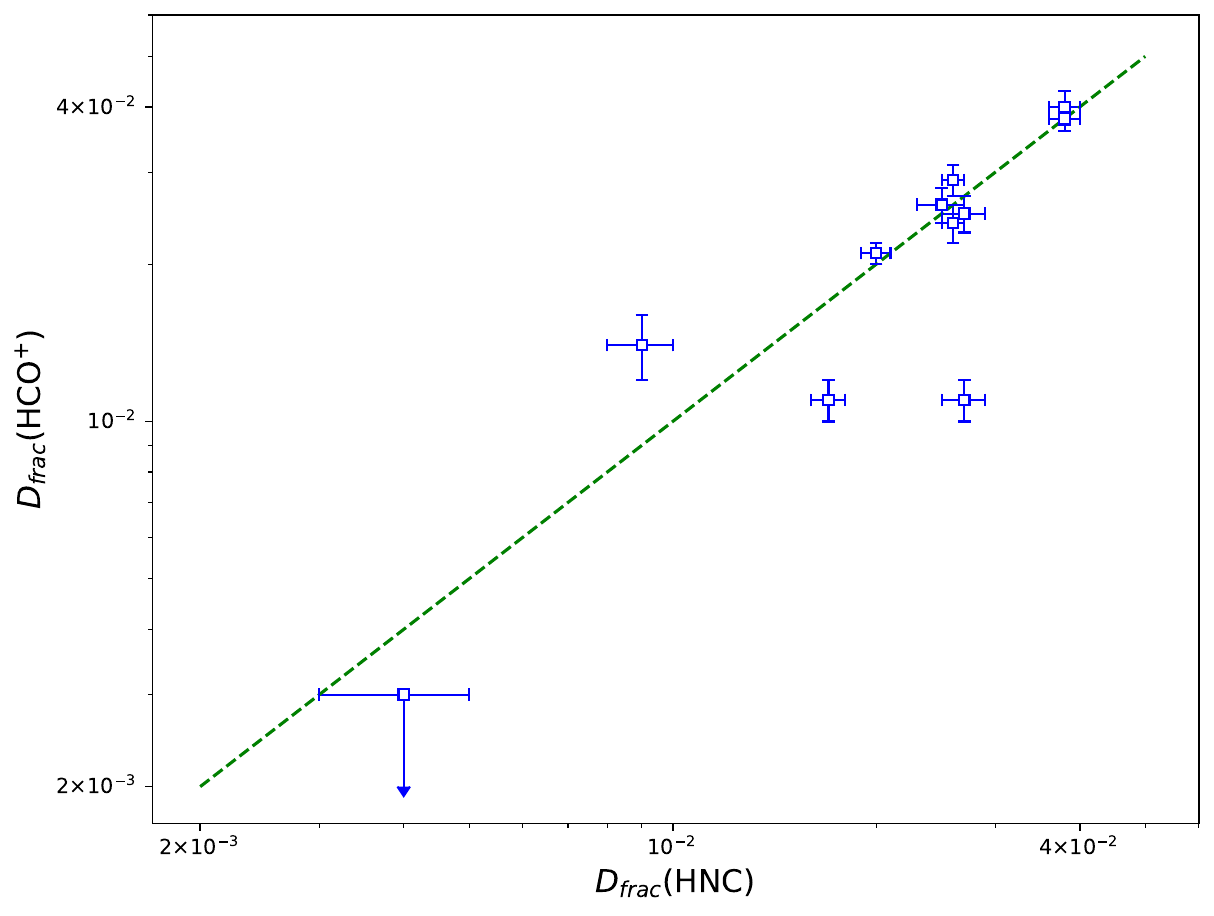} 
\includegraphics[width=57mm]{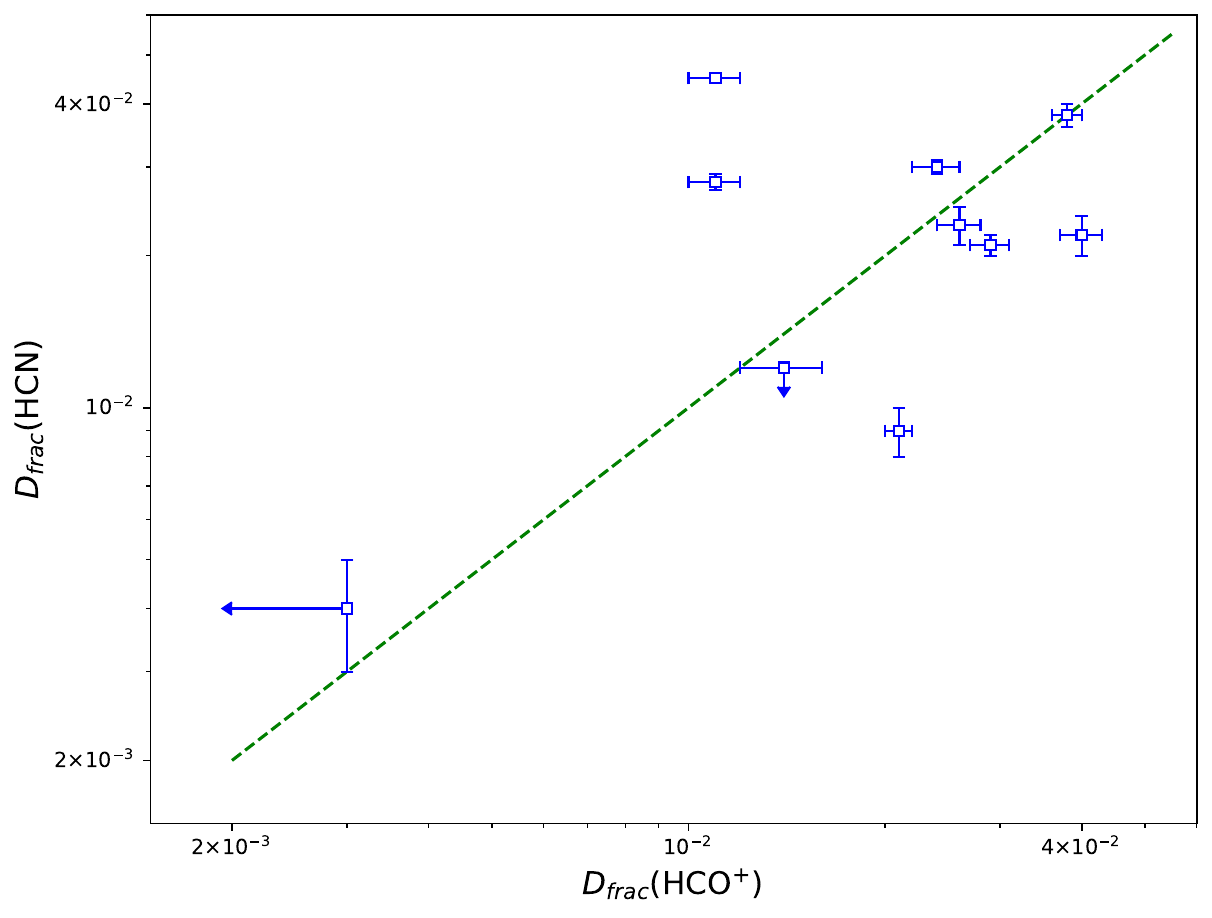} 
\caption{Left: D$_{\rm frac}$(HNC) versus D$_{\rm frac}$(HCN). Middle: D$_{\rm frac}$(HNC) versus D$_{\rm frac}$(HCO$^{+}$). Right: D$_{\rm frac}$(HCN) versus D$_{\rm frac}$(HCO$^{+}$). The arrows show the upper limits of the estimated deuteration fractions. The green dashed lines represent x = y. }
\label{DNC}
\end{figure*}

Based on the $^{12}$C/$^{13}$C ratios and the $N$(DCN)/$N$(H$^{13}$CN), $N$(DCO$^{+}$)/$N$(H$^{13}$CO$^{+}$), and $N$(DNC)/$N$(HN$^{13}$C) values, we can derive the deuterated fractions (shown in Table \ref{D_C}). D$_{\rm frac}$(HCN), D$_{\rm frac}$(HCO$^{+}$) and D$_{\rm frac}$(HNC) have similar values of 0.004$-$0.045, 0.011$-$0.040, and 0.004$-$0.038, respectively. We show the deuterated fractions D$_{\rm frac}$(HNC) versus D$_{\rm frac}$(HCN), D$_{\rm frac}$(HNC) versus D$_{\rm frac}$(HCO$^{+}$), and D$_{\rm frac}$(HCO$^{+}$) versus D$_{\rm frac}$(HCN) in Figure \ref{DNC}. Both D$_{\rm frac}$(HCN) and D$_{\rm frac}$(HCO$^{+}$) show increasing trends with D$_{\rm frac}$(HNC) grows larger. Overall, although DNC and DCO$^{+}$ \textit{J}=1$-$0 have a higher detection rate than DCN \textit{J}=1$-$0, the estimated deuterated fractions are comparable.


\section{Discussion} \label{Dis}

\subsection{Different detection rates for DCO$^{+}$, DCN, and DNC}\label{phy_reason}

In our observations, the emissions of the deuterated molecular lines, DCO$^{+}$, DCN, DNC, and N$_{2}$D$^{+}$ \textit{J}=1$-$0, have peaks at similar positions. But the DCO$^{+}$ \textit{J}=1$-$0 detection rate is higher than DCN \textit{J}=1$-$0 in the single-point observations, with detection rates of DCO$^{+}$, DCN, DNC, and N$_{2}$D$^{+}$ \textit{J}=1$-$0 are 45.5\%, 11.9\%, 50.5\%, and 6.9\%, respectively. In addition, the DCO$^{+}$ emissions tend to have stronger intensities towards the central regions in the OTF observation, while almost half of the DCN \textit{J}=1$-$0 emissions show clumpy structures with an intensity of $\sim$3$\sigma_{\rm area}$. The deuterated molecules have been treated as chemical clocks by several observations \citep[e.g.,][]{2005A&A...433..535F,2006A&A...454L..51B,2014MNRAS.440..448F,2022ApJ...925..144S}, and their detection difference is believed to be caused by deuterium chemistry \citep[e.g.,][]{2015A&A...579A..80G}. In our study, different critical densities could be the main reason for the different detection rates.

The critical density of optically thin deuterated species in the absence of a background continuum can be understood through the equation $n_{\rm crit}$ $\sim$ $A_{jk}/\gamma_{jk}$ \citep{2015PASP..127..299S}, where $A_{jk}$ represents the Einstein A coefficient, and $\gamma_{jk}$ denotes the collision rate. At a temperature of 10 K, the critical densities for DCO$^{+}$, DCN, and DNC in their \textit{J}=1$-$0 transitions are 3.2$\times$10$^{4}$, 2.6$\times$10$^{5}$, and 8.2$\times$10$^{4}$ cm$^{-3}$, respectively, as reported by \citet{2020ApJ...901..145F}. DCO$^{+}$ and DCN have comparable $A_{jk}$ values, but the collisional cross section for collisions with H$_{2}$ of the ionic molecule DCO$^{+}$ is larger than those of neutral species. As a result, DCO$^{+}$ has a critical density of about 10 times lower than that of DCN. Furthermore, \citet{2010MNRAS.404..518S} made a molecular emission model and found that the HNC rate coefficients appear to be larger than the HCN rate coefficients. This also works for DNC and DCN, which makes the critical density of DCN higher than that of DNC. This difference in critical densities would explain these two situations in our OTF observations: (1) The distribution of the DCO$^{+}$ \textit{J}=1$-$0 emission is more extended than that of the DCN emission at the same rms level, which is suitable for mapping BGPS2931, BGPS2940, BGPS2945, BGPS2984, BGPS2984, BGPS3018, BGPS3125, and BGPS3134. (2) There are detections of DCO$^{+}$ and DNC \textit{J}=1$-$0, but no detection of DCN, as occurs in BGPS2693 and BGPS4402. For the remaining massive starless clump candidate, BGPS3110, there is no DCO$^{+}$ \textit{J}=1$-$0 emission detected and the DNC \textit{J}=1$-$0 emissions are found at the edge of the H44$\alpha$ emissions. The DCO$^{+}$ \textit{J}=1$-$0 emission could be dissipated by the nearby H{\scriptsize II} region.

The reactions of these deuterated molecules could be another possibility to explain the differences between DCO$^{+}$ and DCN. Two main pathways lead to the formation of DCO$^{+}$ and DCN: at a temperature below $\sim$ 30K, the primary pathway is via H$_{3}^{+}$ isotopologues; at a temperature of $\sim$30$-$80 K, the dominant pathway is via light hydrocarbons (CH$_{2}$D$^{+}$ and C$_{2}$HD$^{+}$) \citep{2013ApJS..207...27A}. DCO$^{+}$ is considered to be formed mainly from the former pathway and would therefore be expected to be abundant below $\sim$ 30K and be sensitive to freeze-out. DCN is primarily produced from the latter pathway. Therefore, for our sample with a kinetic temperature below 30 K, DCO$^{+}$ is expected to be formed more efficiently than DCN, which then leads to a stronger intensity of DCO$^{+}$ than that of DCN, which is consistent with the results of our OTF observation. For regions with a temperature higher than 30 K, DCN would be formed increasingly and has an intensity higher than the intensity of DCO$^{+}$, which is suitable for the DCN peaks in BPGS2945 and the north-eastern DCN peak in BGPS2984.

Furthermore, the DCN \textit{J}=1$-$0 line exhibits a tripartite hyperfine structure, consisting of F=1$-$1, 2$-$1, and 0$-$1 transitions. Under the assumption of LTE, the intensity of the central F=2$-$1 line significantly accounts for approximately 55.6\% of the total DCN \textit{J}=1$-$0 intensity, which would affect the detection of DCN.

Although the chemistry of HCN and HNC is not clear, the [HCN]/[HNC] ratio has been proposed as a potential tool to probe the temperature in massive star-forming regions \citep[e.g.,][]{2014ApJ...787...74G,2020A&A...635A...4H}, and the ratio decreases with increasing temperature. Observations found the [HCN]/[HNC] ratio $\le$1 in starless cores \citep{2006A&A...456.1037T,2010A&A...513A..41H} and infrared dark clouds \citep{2013MNRAS.431...27L}. The [HCN]/[HNC] ratios of our sources may also have values lower than 1 based on our estimated $N$(HN$^{13}$C)/$N$(H$^{13}$CN) ratios. Considering the estimated D$_{\rm frac}$(HCN) values are found to be similar to the D$_{\rm frac}$(HNC) values in Section \ref{frac}, it is anticipated that the DCN emissions will exhibit a lower intensity compared to the DNC emissions, leading to a higher DNC detection rate.

In single-point observation, we also find difference in detection rates between \textit{J}=1$-$0 and \textit{J}=2$-$1 rotational lines, with detection rates for DCO$^{+}$ and DCN \textit{J}=2$-$1 of 19.8\% and 9.9\%. Our deuterated line survey is performed towards the continuum peak positions from \citet{2016ApJ...822...59S}, which may not be the peak positions of deuterated lines (see Section \ref{res_OTF}). In addition, the beam sizes of the deuterated 2$-$1 rotational lines ($\sim$17$^{\prime\prime}$) are smaller than those of the 1$-$0 rotational lines ($\sim$34$^{\prime\prime}$), which may lead to the sources with DCO$^{+}$ and DCN \textit{J}=1$-$0 lines detection but no \textit{J}=2$-$1 lines detection.

\subsection{Deuterated fraction in SCCs}

Our estimated values of deuterated fractions are similar to the values towards low-mass starless cores \citep[e.g.,][]{2006A&A...455..577T,2011A&A...534A.134M} or massive clumps \citep[e.g.,][]{2015A&A...579A..80G,2019ApJ...883..202F} in infrared dark clouds, indicating that our sources are at an early evolutionary phase. \citet{2012ApJ...747..140S} found DNC/HNC ratio decreases when the star-forming region evolves, and our D$_{\rm frac}$(HNC) have a comparable average value (0.020) with that of their MSX and \textit{Spitzer}-dark sources (0.016), which also reveals that our sources are at similar evolutionary stages to theirs. In addition, the detection rate of DCN in \citet{2015A&A...579A..80G} is much lower than those of DNC and DCO$^{+}$, but the maximum value of D$_{\rm frac}$(HCN) is similar to those of D$_{\rm frac}$(HCO$^{+}$) and D$_{\rm frac}$(HNC). A similar situation can be found in our OTF mapping, which could be explained by different excitation conditions for deuterated molecular lines (see details in Section \ref{phy_reason}). Recently, there have been several studies of deuteration at high angular resolution towards protoplanetary disks (e.g., \citealt{2017ApJ...835..231H}; \citealt{2017A&A...606A.125S}; \citealt{2021ApJS..257...10C}). Although the angular resolution ($\sim$0.3$^{\prime\prime}$$-$0.9$^{\prime\prime}$) is much higher than ours, we have similar deuterated fraction values. In our study, we derive similar deuterated fractions for HCN, HCO$^{+}$, and HNC, more studies on deuterated molecules are needed to prove that the deuterated fractions are chemical clocks.


\section{Summary} \label{Sum}

We first performed single-point observations towards 101 starless clump candidates using the IRAM 30-m telescope, aiming to search for deuterated molecular lines. And then, OTF observations are conducted targeting 11 selected sources. Our results are summarized below.

1. Our single-point observations unveil notable detections, including 46 DCO$^{+}$ \textit{J}=1$-$0, 12 DCN \textit{J}=1$-$0, 51 DNC \textit{J}=1$-$0, 7 N$_{2}$D$^{+}$ \textit{J}=1$-$0, 20 DCO$^{+}$ \textit{J}=2$-$1, and 10 DCN \textit{J}=2$-$1 detections. The SCCs with and without detections of deuterated molecules exhibit similar median values for 1.1~mm intensity, mass, and distance. However, SCCs with deuterated molecule detections display lower median kinetic temperatures and narrower median H$_{2}$CO (1$_{(0,1)}$$-$0$_{(0,0)}$) FWHM values compared to those without such detections.

2. In the OTF observation, nine out of 11 sources are detected with peaks of deuterated molecules near the 1.1mm continuum peak positions. Within the regions containing deuterated molecule peaks, the intensities of DCO$^{+}$ \textit{J}=1$-$0 are stronger than those of DCN and DNC \textit{J}=1$-$0. Additionally, we find clear velocity gradients in three sources. Moreover, most of the emissions of deuterated molecules and $^{13}$C-isotopologues also exhibit peak positions close to those of the 1.1mm continuum peaks. DCN and H$^{13}$CN \textit{J}=1$-$0 display similar peak intensities in the peak regions, while DCO$^{+}$ and DNC \textit{J}=1$-$0 have higher peak intensities than H$^{13}$CO$^{+}$ and HN$^{13}$C \textit{J}=1$-$0.

3. From the OTF observations, we estimate the column density of the deuterated species and $^{13}$C$-$isotopologues for the 11 selected sources within the deuterated peak regions. The estimated column density ranges are 9.8$\times$10$^{11}$ $-$ 3.0$\times$10$^{12}$ cm$^{-2}$ for DCN, 1.2$\times$10$^{12}$ $-$ 7.6$\times$10$^{12}$ cm$^{-2}$ for DCO$^{+}$, and 5.5$\times$10$^{11}$ $-$ 3.6$\times$10$^{12}$ cm$^{-2}$ for DNC. We derive similar deuterated abundances of 0.004$-$0.045, 0.011$-$0.040, and 0.004$-$0.038 for D$_{\rm frac}$(HCN), D$_{\rm frac}$(HCO$^{+}$), and D$_{\rm frac}$(HNC), respectively, despite the different detection rates for DCN, DCO$^{+}$, and DNC \textit{J}=1$-$0.

4. The detection differences among deuterated species of dense gas tracers are likely attributed to their different critical densities and formation pathways.


\begin{acknowledgments}
We thank the anonymous referees for their comments and suggestions, which have improved this work. This work is supported by National Key R\&D Program of China No. 2022YFA1603100, No. 2017YFA0402604, and the National Natural Science Foundation of China (NSFC) grants U1731237 and 11590781. K.Q. acknowledges the science research grant from the China Manned Space Project. K.Y. acknowledges the supports from China Postdoctoral Science Foundation No. 2021M701669. This work is based on observations carried out under project number 131-18 with the IRAM 30m telescope. IRAM is supported by INSU/CNRS (France), MPG (Germany) and IGN (Spain).

\end{acknowledgments}


\clearpage


\startlongtable
\begin{deluxetable*}{c rr r cc rr c cccc c cc}
\tablecaption{Physical parameters of the MSSCs and detection of the deuterated lines.\label{Tab_clumps}}
\tabletypesize{\tiny}
\tablehead{
\colhead{Source} & \colhead{R.A.} & \colhead{Decl.} & \colhead{S$_{\rm 1.1mm}$}   &  \colhead{Mass}  &  \colhead{Distance}  &  \multicolumn {2} {c} {NH$_{3}$}  &  \colhead{}  &  \multicolumn {7} {c} {Detection}  \\
\cline{7-8} \cline{10-16} 
\colhead{} & \colhead{(J2000)} & \colhead{(J2000)} & \colhead{}  &  \colhead{}  &  \colhead{}  &  \colhead{$\nu_{\rm LSR}$}  &  \colhead{$T_{\rm Kin}$}  &  \colhead{}  &  \multicolumn {4} {c} {\textit{J}=1$-$0}  &  \colhead{}  &  \multicolumn {2} {c} {\textit{J}=2$-$1}  \\
\cline{10-13} \cline{15-16} 
\colhead{} & \colhead{(hh:mm:ss)} & \colhead{(dd:mm:ss)} & \colhead{(Jy)}  &  \colhead{(M$_{\odot}$)}  &  \colhead{(kpc)}  &  \colhead{(km s$^{-1}$)}  &  \colhead{(K)}  &  \colhead{}  &  \colhead{DCO$^{+}$}  &  \colhead{DCN}  &  \colhead{DNC}  &  \colhead{N$_{2}$D$^{+}$}  &  \colhead{}  &  \colhead{DCO$^{+}$}  &  \colhead{DCN}  \\
\colhead{(1)} & \colhead{(2)} & \colhead{(3)} & \colhead{(4)}  &  \colhead{(5)}  &  \colhead{(6)}  &  \colhead{(7)}  &  \colhead{(8)}  &  \colhead{}  & \colhead{(9)}  &  \colhead{(10)} &  \colhead{(11)}  &  \colhead{(12)} &  \colhead{}  &  \colhead{(13)}  &  \colhead{(14)} 
}
\startdata
BGPS2427 & 18:09:33.88 & $-$20:47:00.76 &  1.458  &  360.3  &  4.670  &  30.656   &  12.556  &  &    &    &    &     &  &    &    \\
BGPS2430 & 18:08:49.41 & $-$20:40:23.82 &  2.700  &  336.7  &  5.013  &  21.269   &  12.264  &  &    &    &    &    &  &    &    \\
BGPS2432 & 18:09:44.59 & $-$20:47:10.21 &  1.042  &  239.6  &  4.369  &  31.075   &  11.957  &  &    &    &    &    &  &    &    \\
BGPS2437 & 18:10:19.41 & $-$20:50:27.45 &  2.465  &  1081   &  4.437  &    1.868   &  17.324  &  &    &    &    &    &  &    &    \\
BGPS2533 & 18:10:30.29 & $-$20:14:44.20 &  0.755  &  212.2  &  4.975  &  31.835   &  12.267  &  &    &    &  $\surd$  &    &  &    &    \\
BGPS2564 & 18:10:06.08 & $-$18:46:05.64 &  0.227  &  43.03  &  3.013  &  29.572   &  11.758  &  &  $\surd$  &    &  $\surd$  &    &  &  $\surd$  &    \\
BGPS2693 & 18:11:13.56 & $-$17:44:54.85 &  0.387  &  20.71  &  2.103  &  19.134   &  12.440  &  &  $\surd$  &    &  $\surd$  &    &  &  $\surd$  &    \\
BGPS2710 & 18:13:49.04 & $-$17:59:33.25 &  2.948  &  573.8  &  1.200  &  34.594   &  13.976  &  &  $\surd$  &    &  $\surd$  &    &  &  $\surd$  &    \\
BGPS2724 & 18:14:13.61 & $-$17:59:52.02 &  1.010  &  112.5  &  1.185  &  36.251   &  19.897  &  &    &    &    &    &  &    &    \\
BGPS2732 & 18:14:26.85 & $-$17:58:50.93 &  0.318  &  18.97  &  1.191  &  37.474   &  26.836  &  &    &    &    &    &  &    &    \\
BGPS2742 & 18:14:29.10 & $-$17:57:21.83 &  0.510  &  53.37  &  1.183  &  36.011   &  21.974  &  &    &    &    &    &  &    &    \\
BGPS2762 & 18:11:39.52 & $-$17:32:09.40 &  1.376  &  51.88  &  3.304  &  17.849   &  21.549  &  &  $\surd$  &    &  $\surd$  &    &  &  $\surd$  &    \\
BGPS2931 & 18:17:27.51 & $-$17:06:08.42 &  0.539  &  16.25  &  3.285  &  22.771   &  11.711  &  &  $\surd$  &  $\surd$  &  $\surd$  &    &  &  $\surd$  &  $\surd$  \\
BGPS2940 & 18:17:17.15 & $-$17:01:07.47 &  3.492  &  69.98  &  3.366  &  20.035   &  17.795  &  &  $\surd$  &  $\surd$  &  $\surd$  &  $\surd$  &  &  $\surd$  &  $\surd$  \\
BGPS2945 & 18:17:27.35 & $-$17:00:23.66 &  1.316  &  47.59  &  1.178  &  22.683   &  11.666  &  &  $\surd$  &  $\surd$  &  $\surd$  &  $\surd$  &  &  $\surd$  &  $\surd$  \\
BGPS2949 & 18:17:33.74 & $-$16:59:34.94 &  0.901  &  27.35  &  1.279  &  22.520   &  11.548  &  &  $\surd$  &    &  $\surd$  &    &  &  $\surd$  &    \\
BGPS2970 & 18:17:05.08 & $-$16:43:28.66 &  2.076  &  483.6  &  3.568  &  40.010   &  13.606  &  &  $\surd$  &    &  $\surd$  &    &  &  $\surd$  &    \\
BGPS2971 & 18:16:48.12 & $-$16:41:08.91 &  1.054  &  217.7  &  1.815  &  36.545   &  15.689  &  &    &    &    &    &  &    &    \\
BGPS2976 & 18:17:07.84 & $-$16:41:14.59 &  0.493  &  98.07  &  1.830  &  39.630   &  13.792  &  &  $\surd$  &    &  $\surd$  &    &  &  $\surd$  &    \\
BGPS2984 & 18:18:18.23 & $-$16:44:52.26 &  0.787  &  36.16  &  1.855  &  18.422   &  11.486  &  &  $\surd$  &  $\surd$  &  $\surd$  &    &  &  $\surd$  &    \\
BGPS2986 & 18:18:29.68 & $-$16:44:50.69 &  1.092  &  37.08  &  2.014  &  19.966   &  11.634  &  &  $\surd$  &  $\surd$  &  $\surd$  &  $\surd$  &  &  $\surd$  &  $\surd$  \\
BGPS3018 & 18:19:13.88 & $-$16:35:16.47 &  1.671  &  12.60  &  6.522  &  18.958   &  13.472  &  &  $\surd$  &  $\surd$  &  $\surd$  &  $\surd$  &  &  $\surd$  &    \\
BGPS3030 & 18:19:19.68 & $-$16:31:39.82 &  1.929  &  97.79  &  1.784  &  19.006   &  17.197  &  &  $\surd$  &  $\surd$  &  $\surd$  &    &  &  $\surd$  &  $\surd$  \\
BGPS3110 & 18:20:16.27 & $-$16:08:51.13 &  9.667  &  107.2  &  2.005  &  17.687   &  25.166  &  &    &  $\surd$  &  $\surd$  &    &  &    &  $\surd$  \\
BGPS3114 & 18:20:31.50 & $-$16:08:37.80 &64.756  &  4820   &  1.848  &  23.525   &    &  &    &    &    &    &  &    &    \\
BGPS3117 & 18:20:06.68 & $-$16:04:45.75 &  3.166  &  84.77  &  2.001  &  18.574   &  33.383  &  &    &    &    &    &  &    &    \\
BGPS3118 & 18:20:16.17 & $-$16:05:50.72 &  8.842  &  171.3  &  2.001  &  17.210   &  23.681  &  &    &  $\surd$  &    &    &  &    &  $\surd$  \\
BGPS3125 & 18:20:06.11 & $-$16:01:58.02 &  2.180  &  95.55  &  3.466  &  21.531   &  21.251  &  &  $\surd$  &  $\surd$  &  $\surd$  &    &  &  $\surd$  &  $\surd$  \\
BGPS3128 & 18:20:35.27 & $-$16:04:53.81 &  3.765  &  135.1  &  4.170  &  19.751   &  23.170  &  &    &    &    &    &  &    &    \\
BGPS3129 & 18:20:12.99 & $-$16:00:24.13 &  0.463  &  23.49  &  4.289  &  19.716   &  17.865  &  &    &    &  $\surd$  &    &  &    &    \\
BGPS3134 & 18:19:52.72 & $-$15:56:01.56 &  3.168  &  217.9  &  4.083  &  20.465   &  15.919  &  &  $\surd$  &  $\surd$  &  $\surd$  &  $\surd$  &  &  $\surd$  &  $\surd$  \\
BGPS3139 & 18:20:34.24 & $-$15:58:14.00 &  3.046  &  184.9  &  5.408  &  21.798   &  17.232  &  &    &  $\surd$  &  $\surd$  &    &  &    &  $\surd$  \\
BGPS3151 & 18:20:23.19 & $-$15:39:31.96 &  2.323  &  631.7  &  3.379  &  39.401   &  12.337  &  &  $\surd$  &    &  $\surd$  &    &  &    &    \\
BGPS3220 & 18:24:57.03 & $-$13:20:32.39 &  1.927  &  298.1  &  3.874  &  46.130   &  18.524  &  &    &    &    &    &  &    &    \\
BGPS3243 & 18:25:32.74 & $-$13:01:31.05 &  0.554  &  274.8  &  4.597  &  68.468   &  11.712  &  &  $\surd$  &    &    &    &  &  $\surd$  &    \\
BGPS3247 & 18:25:14.45 & $-$12:54:16.74 &  0.796  &  185.2  &  4.447  &  45.168   &  13.037  &  &  $\surd$  &    &  $\surd$  &    &  &  $\surd$  &    \\
BGPS3276 & 18:26:24.92 & $-$12:49:30.07 &  1.054  &  266.9  &  3.379  &  19.647   &  17.192  &  &    &    &    &    &  &    &    \\
BGPS3300 & 18:26:28.42 & $-$12:37:03.98 &  0.601  &  174.7  & 11.668 &  64.147   &  15.401  &  &    &    &  $\surd$  &    &  &    &    \\
BGPS3302 & 18:27:15.23 & $-$12:42:56.45 &  5.207  &  1251   & 11.785 &  66.372   &  18.642  &  &  $\surd$  &    &  $\surd$  &    &  &  $\surd$  &    \\
BGPS3306 & 18:23:34.02 & $-$12:13:52.79 &  0.202  &  64.24  &  4.777  &  57.108   &  12.078  &  &  $\surd$  &    &  $\surd$  &    &  &    &    \\
BGPS3312 & 18:25:44.52 & $-$12:28:34.11 &  0.232  &  32.50  &  5.271  &  47.302   &  17.876  &  &    &    &    &    &  &    &    \\
BGPS3315 & 18:25:33.24 & $-$12:26:50.63 &  8.630  &  1194   &  4.793  &  44.313   &  19.335  &  &    &    &    &    &  &    &    \\
BGPS3344 & 18:26:40.00 & $-$12:25:15.81 &  5.458  &  1631   &  4.314  &  65.694   &  16.138  &  &    &    &  $\surd$  &    &  &    &    \\
BGPS3442 & 18:28:13.51 & $-$11:40:44.94 &  0.835  &  244.4  &  3.442  &  65.752   &  12.545  &  &    &    &    &    &  &    &    \\
BGPS3444 & 18:28:27.26 & $-$11:41:33.99 &  0.291  &  113.9  &  3.297  &  69.686   &  13.059  &  &    &    &    &    &  &    &    \\
BGPS3475 & 18:28:28.28 & $-$11:06:44.16 &  0.718  &  1367   &  3.426  &  75.858   &  17.594  &  &    &    &    &    &  &    &    \\
BGPS3484 & 18:29:15.74 & $-$10:58:28.73 &  0.549  &  958.7  &  3.484  &  56.352   &  18.470  &  &    &    &    &    &  &    &    \\
BGPS3487 & 18:29:22.77 & $-$10:58:01.69 &  1.009  &  1017   &  3.490  &  54.597   &  29.321  &  &    &    &    &    &  &    &    \\
BGPS3534 & 18:30:33.45 & $-$10:24:19.00 &  0.242  &  48.64  &  3.225  &  65.057   &  22.391  &  &    &    &    &    &  &    &    \\
BGPS3604 & 18:30:43.92 & $-$09:34:42.15 &  0.780  &  228.6  & 11.010 &  51.515   &  11.828  &  &  $\surd$  &    &  $\surd$  &    &  &    &    \\
BGPS3606 & 18:29:41.95 & $-$09:24:49.10 &  0.286  &  83.93  &  4.214  &  49.555   &  10.312  &  &  $\surd$  &    &    &    &  &    &    \\
BGPS3608 & 18:31:54.82 & $-$09:39:05.03 &  0.396  &  125.5  &  4.081  &  63.774   &  13.703  &  &    &    &    &    &  &    &    \\
BGPS3627 & 18:31:42.32 & $-$09:24:29.17 &  1.702  &  827.0  &  4.169  &  81.302   &  12.486  &  &    &    &    &    &  &    &    \\
BGPS3656 & 18:32:49.54 & $-$09:21:29.26 &  0.383  &  185.1  &  3.906  &  77.254   &  12.005  &  &    &    &    &    &  &    &    \\
BGPS3686 & 18:34:14.58 & $-$09:18:35.84 &  2.332  &  830.5  &  2.939  &  77.284   &  14.695  &  &  $\surd$  &    &    &    &  &    &    \\
BGPS3705 & 18:34:32.69 & $-$09:14:09.40 &  1.302  &  408.5  &  3.116  &  61.578   &  12.945  &  &  $\surd$  &    &    &    &  &    &    \\
BGPS3710 & 18:34:20.55 & $-$09:10:01.94 &  1.033  &  355.4  &  2.505  &  74.682   &  14.465  &  &    &    &    &    &  &    &    \\
BGPS3716 & 18:34:24.15 & $-$09:08:03.60 &  1.492  &  396.4  &  3.146  &  75.869   &  16.811  &  &    &    &    &    &  &    &    \\
BGPS3736 & 18:33:28.22 & $-$08:55:04.36 &  0.296  &  122.3  &  5.031  &  65.388   &  13.057  &  &    &    &  $\surd$  &    &  &    &    \\
BGPS3822 & 18:33:32.06 & $-$08:32:26.27 &  1.205  &  417.4  &  3.370  &  54.549   &  11.728  &  &  $\surd$  &    &  $\surd$  &    &  &    &    \\
BGPS3833 & 18:33:36.50 & $-$08:30:50.70 &  0.576  &  166.4  &  4.493  &  55.589   &  11.965  &  &    &    &  $\surd$  &    &  &    &    \\
BGPS3892 & 18:35:59.74 & $-$08:38:56.48 &  1.738  &  751.2  &  5.300  &  64.464   &  11.286  &  &    &    &  $\surd$  &    &  &    &    \\
BGPS3922 & 18:33:40.98 & $-$08:14:55.30 &  0.431  &  218.2  &  9.876  &  89.219   &  12.324  &  &    &    &    &    &  &    &    \\
BGPS3924 & 18:34:51.17 & $-$08:23:40.02 &  0.174  &  102.1  &  5.782  &  81.292   &  11.797  &  &    &    &    &  $\surd$  &  &    &    \\
BGPS3982 & 18:34:30.79 & $-$08:02:07.36 &  0.668  &  176.2  & 11.582 &  53.937   &  12.997  &  &    &    &  $\surd$  &    &  &    &    \\
BGPS4029 & 18:35:54.40 & $-$07:59:44.60 &  1.344  &  707.4  &  3.539  &  81.519   &  11.874  &  &  $\surd$  &    &  $\surd$  &    &  &    &    \\
BGPS4082 & 18:35:10.07 & $-$07:39:43.55 &  0.582  &  355.6  &  5.084  &  99.517   &  13.707  &  &  $\surd$  &    &  $\surd$  &    &  &  $\surd$  &    \\
BGPS4085 & 18:33:57.05 & $-$07:29:31.43 &  0.613  &  1100   &  5.087  &  96.624   &  14.983  &  &    &    &    &    &  &    &    \\
BGPS4095 & 18:35:04.00 & $-$07:36:06.46 &  1.502  &  1078   &  5.353  &  112.989 &  13.631  &  &    &    &    &    &  &    &    \\
BGPS4119 & 18:36:29.65 & $-$07:42:06.09 &  0.771  &  1802   &  5.577  &  55.331   &  15.357  &  &  $\surd$  &    &    &    &  &    &    \\
BGPS4135 & 18:37:44.06 & $-$07:48:15.35 &  0.683  &  166.6  &  3.572  &  59.627   &  14.581  &  &  $\surd$  &    &    &    &  &    &    \\
BGPS4140 & 18:36:49.66 & $-$07:40:36.83 &  0.559  &  279.7  &  3.617  &  95.999   &  15.282  &  &    &    &  $\surd$  &    &  &    &    \\
BGPS4145 & 18:36:52.95 & $-$07:39:49.20 &  0.815  &  391.9  &  4.985  &  96.532   &  14.236  &  &    &    &    &    &  &    &    \\
BGPS4191 & 18:37:04.58 & $-$07:33:12.26 &  0.738  &  421.1  &  5.123  &  97.507   &  13.219  &  &  $\surd$  &    &  $\surd$  &    &  &    &    \\
BGPS4230 & 18:35:50.85 & $-$07:12:23.58 &  1.463  &  774.2  &  5.034  &  107.435 &  16.582  &  &    &    &  $\surd$  &    &  &    &    \\
BGPS4294 & 18:38:51.58 & $-$06:55:36.52 &  0.999  &  217.0  &  5.688  &  56.174   &  15.271  &  &    &    &  $\surd$  &    &  &    &    \\
BGPS4297 & 18:38:56.37 & $-$06:55:08.44 &  0.445  &  67.98  &  4.973  &  58.647   &  19.937  &  &    &    &    &    &  &    &    \\
BGPS4346 & 18:38:49.58 & $-$06:31:27.06 &  0.308  &  56.42  &  5.823  &  92.586   &  27.280  &  &    &    &    &    &  &    &    \\
BGPS4347 & 18:38:42.93 & $-$06:30:27.83 &  0.901  &  402.6  &  5.295  &  93.469   &  15.717  &  &  $\surd$  &    &  $\surd$  &    &  &    &    \\
BGPS4354 & 18:38:51.42 & $-$06:29:15.38 &  0.761  &  421.5  &  5.840  &  93.974   &  12.947  &  &  $\surd$  &    &  $\surd$  &    &  &    &    \\
BGPS4356 & 18:37:29.48 & $-$06:18:12.13 &  4.821  &  2676   &  4.453  &  109.935 &  15.780  &  &    &    &  $\surd$  &    &  &    &    \\
BGPS4375 & 18:39:10.19 & $-$06:21:15.90 &  0.287  &  170.4  &  3.793  &  93.072   &  11.790  &  &    &    &    &    &  &    &    \\
BGPS4396 & 18:38:34.74 & $-$05:56:43.97 &  1.687  &  1537   &  4.266  &  112.729 &  12.362  &  &    &    &  $\surd$  &    &  &    &    \\
BGPS4402 & 18:39:28.64 & $-$05:57:58.57 &  0.576  &  330.5  &  4.285  &  99.214   &  13.125  &  &  $\surd$  &    &  $\surd$  &    &  &    &    \\
BGPS4422 & 18:38:47.88 & $-$05:36:16.38 &  0.406  &  240.4  &  3.917  &  110.707 &  13.177  &  &    &    &  $\surd$  &    &  &    &    \\
BGPS4472 & 18:41:17.32 & $-$05:09:56.83 &  0.601  &  105.9  &  3.216  &  46.876   &  14.288  &  &  $\surd$  &    &  $\surd$  &    &  &    &    \\
BGPS4732 & 18:44:23.40 & $-$04:02:01.21 &  6.010  &  3668   &  3.782  &  88.323   &  12.378  &  &  $\surd$  &    &  $\surd$  &    &  &    &    \\
BGPS4827 & 18:44:42.45 & $-$03:44:21.63 &  3.758  &  1326   &  4.928  &  86.095   &  15.926  &  &  $\surd$  &    &  $\surd$  &    &  &    &    \\
BGPS4841 & 18:42:15.65 & $-$03:22:26.19 &  1.050  &  603.0  &  4.266  &  83.981   &  12.030  &  &  $\surd$  &    &  $\surd$  &    &  &    &    \\
BGPS4902 & 18:46:11.36 & $-$03:42:55.73 &  0.834  &  389.2  &  4.656  &  84.140   &  13.527  &  &  $\surd$  &    &    &    &  &    &    \\
BGPS4953 & 18:45:51.82 & $-$03:26:24.16 &  0.489  &  293.8  &  5.502  &  90.783   &  12.690  &  &  $\surd$  &    &  $\surd$  &    &  &    &    \\
BGPS4962 & 18:45:59.61 & $-$03:25:14.53 &  0.791  &  497.7  &  6.092  &  88.308   &  12.537  &  &  $\surd$  &    &    &    &  &    &    \\
BGPS4967 & 18:43:27.80 & $-$03:05:14.94 &  0.515  &  252.0  &  3.681  &  80.309   &  12.161  &  &  $\surd$  &    &  $\surd$  &    &  &    &    \\
BGPS5021 & 18:44:37.07 & $-$02:55:04.40 &  1.138  &  672.7  &  5.181  &  80.052   &  12.107  &  &  $\surd$  &    &  $\surd$  &  $\surd$  &  &    &    \\
BGPS5064 & 18:45:48.44 & $-$02:44:31.65 &  4.001  &  1940   &  5.210  &  100.721 &  15.803  &  &  $\surd$  &    &  $\surd$  &    &  &    &    \\
BGPS5089 & 18:48:49.88 & $-$02:59:47.86 &  1.164  &  402.2  &  6.534  &  85.200   &  15.078  &  &  $\surd$  &    &    &    &  &    &    \\
BGPS5090 & 18:46:35.81 & $-$02:42:30.19 &  0.195  &  142.3  &  5.181  &  96.295   &  11.350  &  &    &    &    &    &  &    &    \\
BGPS5114 & 18:50:23.54 & $-$03:01:31.58 &  2.841  &  651.4  &  3.681  &  65.815   &  14.120  &  &    &    &    &    &  &    &    \\
BGPS5166 & 18:47:54.26 & $-$02:26:07.11 &  0.943  &  525.7  &  6.092  &  102.730 &  15.422  &  &    &    &    &    &  &    &    \\
BGPS5183 & 18:47:00.29 & $-$02:16:38.63 &  0.401  &  429.7  &  6.534  &  113.776 &  12.273  &  &    &    &    &    &  &    &    \\
BGPS5243 & 18:47:54.70 & $-$02:11:10.72 &  0.616  &  425.9  &  5.210  &  95.890   &  12.693  &  &  $\surd$  &    &    &    &  &    &    \\
\enddata
\tablecomments{Columns list the (1) source ID; (2)$-$(3) the coordinates of the SCC sample; (4)$-$(6) 1.1 mm intensity, mass and distance for each SCC obtained from \citet{2016ApJ...822...59S}; (7)$-$(8) LSR velocity and kinetic temperature derived from NH$_{3}$ observations in \citet{2016ApJ...822...59S}, we obtain the peak velocity of HCO$^{+}$ \textit{J}=1$-$0 in \cite{2018ApJ...862...63C} for BGPS 3114 with no NH$_{3}$ detection; (9)$-$(14) detection of the deuterated lines.}
\end{deluxetable*}


\begin{table} 
\caption{Observational parameters.}  \label{obs_para}
\normalsize
\begin{tabular}{lrrccccc}
\hline\hline
\noalign{\smallskip}
Molecules & $\nu$ & E$_{u}$ & $n_{\rm crit}$ & $\Delta$V & F$_{\rm eff}$ & B$_{\rm eff}$ & Observation \\
  & (GHz) & (K) & (cm$^{-3}$) & (km s$^{-1}$) &  &  & \\
   (1) & (2) & (3) & (4) & (5) & (6) & (7) & (8) \\
\noalign{\smallskip}
 \hline
DCO$^{+}$ \textit{J}=1$-$0 & 72.039312 & 3.5 & 3.2E+4 & 0.813 & 98$\%$ & 79$\%$ & 1,2 \\
DCN \textit{J}=1$-$0 & 72.414927 & 3.5 & 2.6E+5 & 0.807 & 98$\%$ & 79$\%$ & 1,2 \\
H$_{2}$CO 1$_{(0,1)}$$-$1$_{(0,0)}$ & 72.837951 & 3.5 & 4.5E+4 & 0.804 & 98$\%$ & 79$\%$ & 1 \\
DNC 1$-$0 & 76.305697 & 3.7 & 8.2E+4 & 0.766 & 98$\%$ & 79$\%$ & 1,2 \\
N$_{2}$D$^{+}$ \textit{J}=1$-$0 & 77.109632 & 3.7 & 5.9E+4 & 0.758 & 98$\%$ & 79$\%$ & 1,2 \\
H$^{13}$CN \textit{J}=1$-$0 & 86.340176 & 4.1 & 4.3E+5 & 0.678 & 98$\%$ & 79$\%$ & 2 \\
H$^{13}$CO$^{+}$ \textit{J}=1$-$0 & 86.754288 & 4.2 & 6.2E+4 & 0.674 & 98$\%$ & 79$\%$ & 2 \\
HN$^{13}$C 1$-$0 & 87.090735 & 4.2 & 9.6E+4 & 0.672 & 98$\%$ & 79$\%$ & 2 \\
HCN \textit{J}=1$-$0 & 88.631847 & 4.3 & 4.7E+5 & 0.660 & 98$\%$ & 79$\%$ & 2 \\
HCO$^{+}$ 1$-$0 & 89.188526 & 4.3 & 7.0E+4 & 0.656 & 98$\%$ & 79$\%$ & 2 \\
HNC 1$-$0 & 90.663564 & 4.4 & 1.4E+5 & 0.645 & 98$\%$ & 79$\%$ & 2 \\
N$_{2}$H$^{+}$ \textit{J}=1$-$0 & 93.173777 & 4.5 & 6.1E+4 & 0.628 & 98$\%$ & 79$\%$ & 2 \\
DCO$^{+}$ \textit{J}=2$-$1 & 144.077285 & 10.4 & 5.3E+5 & 0.406 & 98$\%$ & 78$\%$ & 1,2 \\
DCN \textit{J}=2$-$1 & 144.828002 & 10.4 & 3.9E+6 & 0.404 & 98$\%$ & 78$\%$ & 1,2 \\
DNC 2$-$1 & 152.609739 & 11.0 & 1.2E+6 & 0.383 & 98$\%$ & 78$\%$ & 2 \\
N$_{2}$D$^{+}$ \textit{J}=2$-$1 & 154.217084 & 11.1 & 5.2E+5 & 0.379 & 98$\%$ & 78$\%$ & 2 \\
\hline\hline
\end{tabular}\\
  Note -- Column 1: molecular lines; Column 2: the rest frequency obtained from the CDMS database \citep{2001A&A...370L..49M,2005JMoSt.742..215M,2016JMoSp.327...95E}; Column 3: the upper level energy; Column 4: critical densities at $T_{K}$ = 10 K. For DCO$^{+}$, DCN, DNC, and N$_{2}$D$^{+}$ \textit{J}=2$-$1, values are calculated from the Einstein $A$- and $C$-coefficients from the LAMDA database \citep{2005A&A...432..369S} and the CDMS database, and we assume that the deuterated lines have the same $C$-coefficients as their hydrogenated counterparts; for the rest molecular lines, values are obtained from \citet{2020ApJ...901..145F}; Column 5: the velocity resolution corresponding to 0.195 MHz frequency resolution; Column 6: the forward efficiency; Column 7: the beam efficiency; Column 8: observational references: 1 (single-point observation), 2 (OTF mode observation).
\end{table} 


\begin{table*}
\begin{center}
\caption{Single-point observational parameters of DCO$^{+}$, DCN, and DNC \textit{J}=1$-$0 detections.}  \label{Res_Single1}
\tiny
\begin{tabular}{cccr rccc crrc ccrr}
\hline\hline
\noalign{\smallskip}
Source & & \multicolumn {4} {c} {DCO$^{+}$ \textit{J}=1$-$0} & & \multicolumn {4} {c} {DCN \textit{J}=1$-$0} & & \multicolumn {4} {c} {DNC \textit{J}=1$-$0} \\
\cline{3-6} \cline{8-11} \cline{13-16} 
&  & T$_{\rm mb}$ & $\int$T$_{\rm mb}$dv & V$_{\rm LSR}$ & FWHM &  & T$_{\rm mb}$ & $\int$T$_{\rm mb}$dv & V$_{\rm LSR}$ & FWHM &  & T$_{\rm mb}$ & $\int$T$_{\rm mb}$dv & V$_{\rm LSR}$ & FWHM \\
& & (K) & (K km s$^{-1}$) & (km s$^{-1}$) & (km s$^{-1}$) &  & (K) & (K km s$^{-1}$) & (km s$^{-1}$) & (km s$^{-1}$) &  & (K) & (K km s$^{-1}$) & (km s$^{-1}$) & (km s$^{-1}$) \\
\noalign{\smallskip}
 \hline
BGPS2533 & &  &  &  &  & &  &  &  &  & & 0.193 & 0.38$\pm$0.06 & 31.2$\pm$0.2 & 1.9$\pm$0.2 \\
BGPS2564 & & 0.409 & 0.35$\pm$0.05 & 29.3$\pm$0.1 & 0.8$\pm$0.2 & &  &  &  &  & & 0.248 & 0.22$\pm$0.07 & 29.3$\pm$0.2 & 0.8$\pm$0.2 \\
BGPS2693 & & 1.003 & 0.99$\pm$0.06 & 19.1$\pm$0.1 & 0.9$\pm$0.1 & &  &  &  &  & & 0.697 & 0.90$\pm$0.05 & 19.1$\pm$0.1 & 1.2$\pm$0.1 \\
BGPS2710 & & 0.280 & 0.30$\pm$0.06 & 34.5$\pm$0.1 & 1.0$\pm$0.2 & &  &  &  &  & & 0.274 & 0.44$\pm$0.05 & 34.5$\pm$0.1 & 1.5$\pm$0.2 \\
BGPS2762 & & 0.552 & 0.61$\pm$0.07 & 17.6$\pm$0.1 & 1.1$\pm$0.2 & &  &  &  &  & & 0.193 & 0.33$\pm$0.05 & 17.4$\pm$0.2 & 1.6$\pm$0.2 \\
BGPS2931 & & 1.050 & 1.21$\pm$0.06 & 22.7$\pm$0.1 & 1.1$\pm$0.1 & & 0.160 & 0.26$\pm$0.06 & 22.5$\pm$0.1 & 1.5$\pm$0.3 & & 0.654 & 0.90$\pm$0.06 & 22.7$\pm$0.2 & 1.3$\pm$0.1 \\
BGPS2940 & & 1.236 & 2.21$\pm$0.07 & 20.0$\pm$0.1 & 1.7$\pm$0.1 & & 0.241 & 0.48$\pm$0.11 & 20.3$\pm$0.2 & 1.9$\pm$0.3 & & 0.685 & 1.39$\pm$0.06 & 19.9$\pm$0.1 & 1.9$\pm$0.1 \\
BGPS2945 & & 1.466 & 1.65$\pm$0.08 & 22.6$\pm$0.1 & 1.1$\pm$0.1 & & 0.227 & 0.29$\pm$0.07 & 22.7$\pm$0.2 & 1.1$\pm$0.3 & & 0.766 & 1.07$\pm$0.06 & 22.7$\pm$0.1 & 1.3$\pm$0.1 \\
BGPS2949 & & 0.552 & 0.67$\pm$0.09 & 22.5$\pm$0.1 & 1.1$\pm$0.2 & &  &  &  &  & & 0.202 & 0.24$\pm$0.04 & 22.5$\pm$0.1 & 1.1$\pm$0.2 \\
BGPS2970 & & 0.234 & 0.26$\pm$0.05 & 40.0$\pm$0.2 & 1.1$\pm$0.2 & &  &  &  &  & & 0.140 & 0.34$\pm$0.06 & 39.5$\pm$0.2 & 2.2$\pm$0.2 \\
BGPS2976 & & 0.383 & 0.45$\pm$0.06 & 39.6$\pm$0.1 & 1.1$\pm$0.2 & &  &  &  &  & & 0.267 & 0.33$\pm$0.05 & 39.6$\pm$0.1 & 1.2$\pm$0.2 \\
BGPS2984 & & 1.423 & 1.23$\pm$0.05 & 18.5$\pm$0.1 & 0.8$\pm$0.2 & & 0.281 & 0.24$\pm$0.06 & 18.6$\pm$0.1 & 0.8$\pm$0.1 & & 0.717 & 0.93$\pm$0.04 & 18.4$\pm$0.1 & 1.2$\pm$0.1 \\
BGPS2986 & & 1.684 & 2.32$\pm$0.07 & 20.1$\pm$0.1 & 1.3$\pm$0.1 & & 0.436 & 0.54$\pm$0.10 & 19.8$\pm$0.1 & 1.2$\pm$0.3 & & 1.105 & 1.84$\pm$0.06 & 20.0$\pm$0.1 & 1.6$\pm$0.1 \\
BGPS3018 & & 1.077 & 1.49$\pm$0.07 & 19.0$\pm$0.1 & 1.3$\pm$0.1 & & 0.114 & 0.23$\pm$0.08 & 19.0$\pm$0.3 & 1.9$\pm$0.3 & & 0.458 & 0.75$\pm$0.06 & 18.9$\pm$0.1 & 1.6$\pm$0.2 \\
BGPS3030 & & 0.300 & 0.32$\pm$0.05 & 19.3$\pm$0.2 & 1.0$\pm$0.3 & & 0.263 & 0.22$\pm$0.07 & 19.0$\pm$0.1 & 0.8$\pm$0.1 & & 0.331 & 0.46$\pm$0.04 & 18.9$\pm$0.1 & 1.3$\pm$0.2 \\
BGPS3110 & &  &  &  &  & & 0.150 & 0.43$\pm$0.09 & 17.6$\pm$0.2 & 3.0$\pm$0.5 & & 0.149 & 0.46$\pm$0.07 & 17.3$\pm$0.2 & 2.9$\pm$0.2 \\
BGPS3118 & &  &  &  &  & & 0.184 & 0.20$\pm$0.07 & 17.5$\pm$0.2 & 1.1$\pm$0.3 & &  &  &  &  \\
BGPS3125 & & 0.094 & 0.15$\pm$0.03 & 21.7$\pm$0.2 & 1.6$\pm$0.3 & & 0.220 & 0.18$\pm$0.03 & 21.6$\pm$0.2 & 0.8$\pm$0.2 & & 0.241 & 0.40$\pm$0.05 & 21.3$\pm$0.1 & 1.6$\pm$0.2 \\
BGPS3129 & &  &  &  &  & &  &  &  &  & & 0.164 & 0.28$\pm$0.08 & 18.9$\pm$0.2 & 1.6$\pm$0.2 \\
BGPS3134 & & 0.708 & 1.03$\pm$0.15 & 21.1$\pm$0.1 & 1.4$\pm$0.2 & & 0.199 & 0.32$\pm$0.12 & 20.1$\pm$0.3 & 1.5$\pm$0.3 & & 0.705 & 1.30$\pm$0.11 & 20.9$\pm$0.1 & 1.7$\pm$0.2 \\
BGPS3139 & &  &  &  &  & & 0.358 & 0.31$\pm$0.11 & 20.1$\pm$0.3 & 0.8$\pm$0.1 & & 0.509 & 0.69$\pm$0.05 & 20.1$\pm$0.1 & 1.3$\pm$0.1 \\
BGPS3151 & & 0.480 & 0.68$\pm$0.10 & 39.2$\pm$0.1 & 1.3$\pm$0.2 & &  &  &  &  & & 0.347 & 0.49$\pm$0.06 & 39.3$\pm$0.1 & 1.3$\pm$0.2 \\
BGPS3243 & & 0.350 & 0.41$\pm$0.09 & 68.0$\pm$0.1 & 1.1$\pm$0.3  & &  &  &  &  & &  &  &  &  \\
BGPS3247 & & 0.398 & 0.58$\pm$0.12 & 44.9$\pm$0.1 & 1.3$\pm$0.2 & &  &  &  &  & & 0.367 & 0.56$\pm$0.06 & 44.9$\pm$0.1 & 1.5$\pm$0.2 \\
BGPS3300 & &  &  &  &  & &  &  &  &  & & 0.146 & 0.23$\pm$0.07 & 64.2$\pm$0.2 & 1.5$\pm$0.3 \\
BGPS3302 & & 0.214 & 0.38$\pm$0.07 & 66.6$\pm$0.2 & 1.7$\pm$0.3 & &  &  &  &  & & 0.168 & 0.35$\pm$0.07 & 66.4$\pm$0.2 & 2.0$\pm$0.3 \\
BGPS3306 & & 0.260 & 0.22$\pm$0.06 & 56.9$\pm$0.1 & 0.8$\pm$0.2 & &  &  &  &  & & 0.169 & 0.25$\pm$0.05 & 57.3$\pm$0.2 & 1.5$\pm$0.3 \\
BGPS3344 & &  &  &  &  & &  &  &  &  & & 0.126 & 0.23$\pm$0.06 & 65.8$\pm$0.2 & 1.6$\pm$0.4 \\
BGPS3604 & & 0.689 & 0.90$\pm$0.08 & 51.5$\pm$0.1 & 1.2$\pm$0.1 & &  &  &  &  & & 0.253 & 0.41$\pm$0.05 & 51.3$\pm$0.1 & 1.5$\pm$0.2 \\
BGPS3606 & & 0.287 & 0.56$\pm$0.09 & 49.5$\pm$0.2 & 1.9$\pm$0.3 & &  &  &  &  & &  &  &  &  \\
BGPS3686 & & 0.271 & 0.23$\pm$0.06 & 77.6$\pm$0.2 & 0.8$\pm$0.2  & &  &  &  &  & &  &  &  &  \\
BGPS3710 & & 0.132 & 0.33$\pm$0.10 & 74.9$\pm$0.3 & 2.4$\pm$0.4  & &  &  &  &  & &  &  &  &  \\
BGPS3736 & &  &  &  &  & &  &  &  &  & & 0.112 & 0.18$\pm$0.04 & 65.0$\pm$0.2 & 1.6$\pm$0.2 \\
BGPS3822 & & 0.293 & 0.42$\pm$0.19 & 54.8$\pm$0.3 & 1.4$\pm$0.3 & &  &  &  &  & & 0.280 & 0.46$\pm$0.08 & 54.4$\pm$0.2 & 1.6$\pm$0.3 \\
BGPS3833 & &  &  &  &  & &  &  &  &  & & 0.176 & 0.34$\pm$0.06 & 55.4$\pm$0.2 & 1.8$\pm$0.2 \\
BGPS3892 & &  &  &  &  & &  &  &  &  & & 0.199 & 0.57$\pm$0.09 & 64.2$\pm$0.2 & 2.7$\pm$0.4 \\
BGPS3982 & &  &  &  &  & &  &  &  &  & & 0.137 & 0.22$\pm$0.06 & 54.1$\pm$0.2 & 1.6$\pm$0.3 \\
BGPS4029 & & 0.321 & 0.68$\pm$0.10 & 81.4$\pm$0.1 & 2.0$\pm$0.3 & &  &  &  &  & & 0.200 & 0.46$\pm$0.07 & 81.5$\pm$0.2 & 2.4$\pm$0.3 \\
BGPS4082 & & 0.297 & 0.25$\pm$0.06 & 99.7$\pm$0.1 & 0.8$\pm$0.1 & &  &  &  &  & & 0.231 & 0.29$\pm$0.07 & 99.4$\pm$0.2 & 1.2$\pm$0.2\\
BGPS4119 & & 0.147 & 0.38$\pm$0.08 & 55.7$\pm$0.2 & 2.5$\pm$0.4  & &  &  &  &  & &  &  &  &  \\
BGPS4135 & & 0.399 & 0.50$\pm$0.09 & 58.3$\pm$0.1 & 1.2$\pm$0.3 & &  &  &  &  & &  &  &  &  \\
BGPS4140 & &  &  &  &  & &  &  &  &  & & 0.195 & 0.18$\pm$0.04 & 95.7$\pm$0.2 & 0.9$\pm$0.2 \\
BGPS4191 & & 0.332 & 0.28$\pm$0.05 & 97.4$\pm$0.1 & 0.8$\pm$0.2 & &  &  &  &  & & 0.124 & 0.34$\pm$0.12 & 97.6$\pm$0.4 & 2.6$\pm$0.5 \\
BGPS4230 & &  &  &  &  & &  &  &  &  & & 0.224 & 0.41$\pm$0.05 & 107.6$\pm$0.1 & 1.7$\pm$0.2 \\
BGPS4347 & & 0.294 & 0.34$\pm$0.07 & 93.5$\pm$0.1 & 1.1$\pm$0.2 & &  &  &  &  & & 0.223 & 0.37$\pm$0.05 & 93.4$\pm$0.1 & 1.6$\pm$0.2 \\
BGPS4354 & & 0.100 & 0.34$\pm$0.09 & 94.2$\pm$0.4 & 3.2$\pm$0.4 & &  &  &  &  & & 0.135 & 0.35$\pm$0.09 & 93.7$\pm$0.3 & 2.5$\pm$0.4 \\
BGPS4356 & &  &  &  &  & &  &  &  &  & & 0.132 & 0.21$\pm$0.09 & 110.2$\pm$0.1 & 1.5$\pm$0.3 \\
BGPS4396 & &  &  &  &  & &  &  &  &  & & 0.217 & 0.30$\pm$0.06 & 112.5$\pm$0.1 & 1.3$\pm$0.3 \\
BGPS4402 & & 0.222 & 0.42$\pm$0.05 & 99.5$\pm$0.1 & 1.8$\pm$0.2 & &  &  &  &  & & 0.200 & 0.29$\pm$0.06 & 99.2$\pm$0.1 & 1.4$\pm$0.2 \\
BGPS4422 & &  &  &  &  & &  &  &  &  & & 0.114 & 0.27$\pm$0.06 & 111.4$\pm$0.2 & 2.2$\pm$0.4 \\
BGPS4472 & & 0.378 & 0.32$\pm$0.03 & 46.9$\pm$0.1 & 0.8$\pm$0.4 & &  &  &  &  & & 0.206 & 0.31$\pm$0.04 & 46.7$\pm$0.1 & 1.4$\pm$0.2 \\
BGPS4732 & & 0.399 & 0.71$\pm$0.06 & 88.4$\pm$0.1 & 1.7$\pm$0.2 & &  &  &  &  & & 0.283 & 0.44$\pm$0.04 & 88.2$\pm$0.1 & 1.5$\pm$0.1 \\
BGPS4827 & & 0.150 & 0.30$\pm$0.07 & 86.1$\pm$0.2 & 1.9$\pm$0.4 & &  &  &  &  & & 0.169 & 0.26$\pm$0.05 & 85.9$\pm$0.2 & 1.5$\pm$0.3 \\
BGPS4841 & & 0.180 & 0.33$\pm$0.06 & 84.0$\pm$0.2 & 1.7$\pm$0.3 & &  &  &  &  & & 0.107 & 0.20$\pm$0.06 & 83.9$\pm$0.2 & 1.8$\pm$0.4 \\
BGPS4902 & & 0.227 & 0.32$\pm$0.08 & 84.2$\pm$0.2 & 1.4$\pm$0.2  & &  &  &  &  & &  &  &  &  \\
BGPS4953 & & 0.218 & 0.40$\pm$0.06 & 90.6$\pm$0.1 & 1.7$\pm$0.2 & &  &  &  &  & & 0.135 & 0.28$\pm$0.06 & 91.0$\pm$0.2 & 2.0$\pm$0.4 \\
BGPS4962 & & 0.203 & 0.17$\pm$0.04 & 88.2$\pm$0.1 & 0.8$\pm$0.2  & &  &  &  &  & &  &  &  &  \\
BGPS4967 & & 0.298 & 0.42$\pm$0.06 & 80.2$\pm$0.1 & 1.3$\pm$0.3 & &  &  &  &  & & 0.203 & 0.47$\pm$0.06 & 80.3$\pm$0.1 & 2.2$\pm$0.3 \\
BGPS5021 & & 0.187 & 0.27$\pm$0.06 & 80.0$\pm$0.1 & 1.4$\pm$0.3 & &  &  &  &  & & 0.216 & 0.38$\pm$0.05 & 80.2$\pm$0.1 & 1.7$\pm$0.2 \\
BGPS5064 & & 0.170 & 0.14$\pm$0.04 & 100.6$\pm$0.1 & 0.8$\pm$0.1 & &  &  &  &  & & 0.123 & 0.24$\pm$0.05 & 100.8$\pm$0.2 & 1.9$\pm$0.3 \\
BGPS5089 & & 0.237 & 0.20$\pm$0.03 & 85.1$\pm$0.1 & 0.8$\pm$0.1 & &  &  &  &  & &  &  &  &  \\
BGPS5243 & & 0.307 & 0.53$\pm$0.13 & 95.8$\pm$0.2 & 1.6$\pm$0.4  & &  &  &  &  & &  &  &  &  \\
\hline\hline
\end{tabular}
\end{center}
\end{table*}


\begin{table*}
\begin{center}
\caption{Single-point observational parameters of DCO$^{+}$, DCN \textit{J}=2$-$1 and N$_{2}$D$^{+}$ \textit{J}=1$-$0 detections.}  \label{Res_Single2}
\tiny
\begin{tabular}{cccrrccccrrcccrr}
\hline\hline
\noalign{\smallskip}
Source & & \multicolumn {4} {c} {N$_{2}$D$^{+}$ \textit{J}=1$-$0} & & \multicolumn {4} {c} {DCO$^{+}$ \textit{J}=2$-$1} & & \multicolumn {4} {c} {DCN \textit{J}=2$-$1} \\
\cline{3-6} \cline{8-11} \cline{13-16} 
&  & T$_{\rm mb}$ & $\int$T$_{\rm mb}$dv & V$_{\rm LSR}$ & FWHM &  & T$_{\rm mb}$ & $\int$T$_{\rm mb}$dv & V$_{\rm LSR}$ & FWHM &  & T$_{\rm mb}$ & $\int$T$_{\rm mb}$dv & V$_{\rm LSR}$ & FWHM \\
& & (K) & (K km s$^{-1}$) & (km s$^{-1}$) & (km s$^{-1}$) &  & (K) & (K km s$^{-1}$) & (km s$^{-1}$) & (km s$^{-1}$) &  & (K) & (K km s$^{-1}$) & (km s$^{-1}$) & (km s$^{-1}$) \\
\noalign{\smallskip}
 \hline
BGPS2564 & &  &  &  &  & & 0.223 & 0.23$\pm$0.05 & 29.4$\pm$0.1 & 1.0$\pm$0.3 & &  &  &  &  \\
BGPS2693 & &  &  &  &  & & 1.054 & 0.76$\pm$0.05 & 19.1$\pm$0.1 & 0.7$\pm$0.1 & &  &  &  &  \\
BGPS2710 & &  &  &  &  & & 0.215 & 0.34$\pm$0.05 & 34.5$\pm$0.1 & 1.5$\pm$0.2 & &  &  &  &  \\
BGPS2762 & &  &  &  &  & & 0.698 & 0.77$\pm$0.03 & 17.5$\pm$0.1 & 1.0$\pm$0.1 & &  &  &  &  \\
BGPS2931 & &  &  &  &  & & 1.209 & 0.91$\pm$0.03 & 22.7$\pm$0.1 & 0.7$\pm$0.1 & & 0.244 & 0.14$\pm$0.02 & 22.6$\pm$0.1 & 0.6$\pm$0.2 \\
BGPS2940 & & 0.151 & 0.31$\pm$0.07 & 20.0$\pm$0.2 & 2.0$\pm$0.3 & & 1.816 & 2.64$\pm$0.05 & 19.9$\pm$0.1 & 1.4$\pm$0.1 & & 0.281 & 0.51$\pm$0.06 & 19.9$\pm$0.1 & 1.7$\pm$0.2 \\
BGPS2945 & & 0.092 & 0.30$\pm$0.04 & 23.2$\pm$0.1 & 3.1$\pm$0.3 & & 1.308 & 1.26$\pm$0.03 & 22.7$\pm$0.1 & 0.9$\pm$0.1 & & 0.245 & 0.19$\pm$0.04 & 22.6$\pm$0.1 & 0.7$\pm$0.1 \\
BGPS2949 & &  &  &  &  & & 0.376 & 0.39$\pm$0.03 & 22.5$\pm$0.1 & 1.0$\pm$0.1 & &  &  &  &  \\
BGPS2970 & &  &  &  &  & & 0.108 & 0.21$\pm$0.03 & 40.0$\pm$0.1 & 1.9$\pm$0.2 & &  &  &  &  \\
BGPS2976 & &  &  &  &  & & 0.320 & 0.45$\pm$0.04 & 39.7$\pm$0.1 & 1.3$\pm$0.1 & &  &  &  &  \\
BGPS2984 & &  &  &  &  & & 1.006 & 0.70$\pm$0.03 & 18.4$\pm$0.1 & 0.6$\pm$0.1 & &  &  &  &  \\
BGPS2986 & & 0.323 & 0.68$\pm$0.12 & 20.3$\pm$0.2 & 2.0$\pm$0.4 & & 1.397 & 1.76$\pm$0.03 & 20.1$\pm$0.1 & 1.2$\pm$0.1 & & 0.238 & 0.25$\pm$0.05 & 19.9$\pm$0.1 & 1.0$\pm$0.2 \\
BGPS3018 & & 0.108 & 0.22$\pm$0.06 & 19.1$\pm$0.3 & 1.9$\pm$0.1 & & 0.864 & 0.89$\pm$0.04 & 18.9$\pm$0.1 & 1.0$\pm$0.1 & &  &  &  &  \\
BGPS3030 & &  &  &  &  & & 0.410 & 0.35$\pm$0.05 & 19.1$\pm$0.1 & 0.8$\pm$0.2 & & 0.232 & 0.15$\pm$0.03 & 18.9$\pm$0.1 & 0.6$\pm$0.1 \\
BGPS3110 & &  &  &  &  & &  &  &  &  & & 0.248 & 0.39$\pm$0.08 & 17.7$\pm$0.1 & 1.5$\pm$0.2 \\
BGPS3118 & &  &  &  &  & &  &  &  &  & & 0.228 & 0.23$\pm$0.04 & 17.6$\pm$0.1 & 1.0$\pm$0.2 \\
BGPS3125 & &  &  &  &  & & 0.548 & 0.42$\pm$0.04 & 21.5$\pm$0.1 & 0.7$\pm$0.1 & & 0.506 & 0.42$\pm$0.04 & 21.2$\pm$0.1 & 0.8$\pm$0.1 \\
BGPS3134 & & 0.212& 0.70$\pm$0.12 & 21.1$\pm$0.3 & 3.0$\pm$0.3 & & 1.155 & 1.09$\pm$0.08 & 21.1$\pm$0.1 & 0.9$\pm$0.1 & & 0.430 & 0.35$\pm$0.05 & 21.0$\pm$0.1 & 0.8$\pm$0.1 \\
BGPS3139 & &  &  &  &  & &  &  &  &  & & 0.390 & 0.28$\pm$0.05 & 19.8$\pm$0.1 & 0.7$\pm$0.1 \\
BGPS3243 & &  &  &  &  & & 0.277 & 0.37$\pm$0.11 & 68.5$\pm$0.1 & 1.3$\pm$0.2 & &  &  &  &  \\
BGPS3247 & &  &  &  &  & & 0.442 & 0.33$\pm$0.04 & 45.1$\pm$0.1 & 0.7$\pm$0.1 & & & & & \\
BGPS3302 & &  &  &  &  & & 0.180 & 0.17$\pm$0.04 & 66.2$\pm$0.1 & 0.9$\pm$0.2 & &  &  &  &  \\
BGPS3924 & & 0.139 & 0.21$\pm$0.06 & 80.9$\pm$0.2 & 1.5$\pm$0.3 & &  &  &  &  & &  &  &  &  \\
BGPS4082 & &  &  &  &  & & 0.248 & 0.41$\pm$0.08 & 99.6$\pm$0.1 & 1.6$\pm$0.2 & & & & & \\
BGPS5021 & & 0.100 & 0.11$\pm$0.05 & 80.2$\pm$0.3 & 1.4$\pm$0.3 & &  &  &  &  & &  &  &  &  \\
\hline\hline
\end{tabular}
\end{center}
\end{table*}


\begin{table*}
\begin{center}
\caption{Single-point observational parameters of H$_{2}$CO(1$_{(0,1)}$$-$0$_{(0,0)}$) detections.}  \label{Res_Single3}
\tiny
\begin{tabular}{cc rrrrc c cc rrrrc}
\hline\hline
Source & & \multicolumn {5} {c} {H$_{2}$CO(1$_{(0,1)}$$-$0$_{(0,0)}$)} & & Source & & \multicolumn {5} {c} {H$_{2}$CO(1$_{(0,1)}$$-$0$_{(0,0)}$)} \\
\cline{3-7} \cline{11-15}
&  & T$_{\rm mb}$ & $\int$T$_{\rm mb}$dv & V$_{\rm LSR}$ & FWHM & D$-$isotopes & & & & T$_{\rm mb}$ & $\int$T$_{\rm mb}$dv & V$_{\rm LSR}$ & FWHM & D$-$isotopes \\
& & (K) & (K km s$^{-1}$) & (km s$^{-1}$) & (km s$^{-1}$) & detection &  &  &  & (K) & (K km s$^{-1}$) & (km s$^{-1}$) & (km s$^{-1}$) & detection \\
 \hline
BGPS2564 & & 0.499 & 0.42$\pm$0.08 & 29.4$\pm$0.1 & 0.8$\pm$0.1 & Y & & BGPS3656 & & 0.320 & 0.27$\pm$0.05 & 77.2$\pm$0.1 & 0.8$\pm$0.2 &   \\
BGPS2693 & & 0.964 & 1.03$\pm$0.06 & 19.1$\pm$0.1 & 1.0$\pm$0.1 & Y & & BGPS3686 & & 0.390 & 1.22$\pm$0.11 & 77.2$\pm$0.1 & 2.9$\pm$0.3 & Y \\
BGPS2710 & & 0.575 & 1.26$\pm$0.07 & 34.6$\pm$0.1 & 2.2$\pm$0.2 & Y & & BGPS3710 & & 0.435 & 1.89$\pm$0.14 & 74.5$\pm$0.2 & 4.1$\pm$0.4 & Y \\
BGPS2724 & & 0.826 & 3.02$\pm$0.11 & 35.9$\pm$0.1 & 3.5$\pm$0.1 &   & & BGPS3716 & & 0.436 & 1.45$\pm$0.09 & 75.0$\pm$0.1 & 3.1$\pm$0.3 &    \\
BGPS2732 & & 0.461 & 1.79$\pm$0.14 & 36.8$\pm$0.1 & 3.6$\pm$0.3 &   & & 		 & & 0.266 & 0.99$\pm$0.09 & 78.3$\pm$0.1 & 3.5$\pm$0.3 &    \\
		 & & 0.147 & 0.12$\pm$0.07 & 39.4$\pm$0.2 & 0.8$\pm$0.1 &   & & BGPS3736 & & 0.438 & 0.93$\pm$0.07 & 65.5$\pm$0.1 & 2.0$\pm$0.2 & Y \\
BGPS2742 & & 0.421 & 1.81$\pm$0.09 & 35.9$\pm$0.1 & 4.1$\pm$0.3 &   & & 		 & & 0.146 & 0.31$\pm$0.07 & 68.2$\pm$0.2 & 2.0$\pm$0.5 &    \\
BGPS2762 & & 1.366 & 2.96$\pm$0.08 & 18.1$\pm$0.1 & 2.0$\pm$0.1 & Y & & BGPS3822 & & 0.568 & 1.25$\pm$0.12 & 54.5$\pm$0.1 & 2.1$\pm$0.2 & Y\\
BGPS2931 & & 0.716 & 1.43$\pm$0.07 & 22.9$\pm$0.1 & 1.9$\pm$0.2 & Y & & BGPS3833 & & 0.239 & 0.47$\pm$0.13 & 56.3$\pm$0.3 & 1.9$\pm$0.3 & Y \\
BGPS2940 & & 3.308 & 11.4$\pm$0.10 & 20.0$\pm$0.1 & 3.2$\pm$0.1 & Y & & BGPS3892 & & 0.253 & 0.47$\pm$0.10 & 63.3$\pm$0.3 & 1.7$\pm$0.2 & Y \\
BGPS2945 & & 0.382 & 1.20$\pm$0.10 & 19.5$\pm$0.1 & 2.9$\pm$0.3 & Y & & 		 & & 0.367 & 0.58$\pm$0.18 & 65.3$\pm$0.2 & 1.5$\pm$0.3 &  \\
		 & & 1.143 & 1.85$\pm$0.08 & 22.7$\pm$0.1 & 1.5$\pm$0.1 &   & & BGPS3922 & & 0.310 & 0.35$\pm$0.07 & 89.4$\pm$0.1 & 1.1$\pm$0.2 &    \\
BGPS2949 & & 0.581 & 2.39$\pm$0.17 & 20.8$\pm$0.1 & 3.9$\pm$0.3 & Y & & BGPS3982 & & 0.339 & 0.88$\pm$0.12 & 54.0$\pm$0.2 & 2.5$\pm$0.3 & Y \\
		 & & 0.899 & 0.76$\pm$0.10 & 22.7$\pm$0.1 & 0.8$\pm$0.1 &   & & BGPS4029 & & 0.481 & 1.71$\pm$0.13 & 81.5$\pm$0.1 & 3.4$\pm$0.3 & Y \\
BGPS2970 & & 0.665 & 1.69$\pm$0.13 & 40.2$\pm$0.1 & 2.4$\pm$0.2 &   & & BGPS4082 & & 0.610 & 1.59$\pm$0.14 & 99.5$\pm$0.1 & 2.4$\pm$0.3 & Y \\
BGPS2971 & & 0.340 & 1.16$\pm$0.18 & 36.8$\pm$0.2 & 3.2$\pm$0.4 &   & & BGPS4085 & & 0.302 & 0.84$\pm$0.10 & 96.6$\pm$0.2 & 2.6$\pm$0.3 &    \\
		 & & 0.267 & 0.55$\pm$0.17 & 40.4$\pm$0.2 & 1.9$\pm$0.3 &   & & BGPS4095 & & 0.128 & 0.27$\pm$0.17 & 109.5$\pm$0.4 & 1.9$\pm$0.3 &    \\
BGPS2976 & & 0.532 & 1.16$\pm$0.09 & 39.6$\pm$0.1 & 2.1$\pm$0.2 & Y & & 		 & & 0.331 & 0.95$\pm$0.16 & 112.5$\pm$0.2 & 2.6$\pm$0.4 &    \\
BGPS2984 & & 0.846 & 1.12$\pm$0.06 & 18.6$\pm$0.1 & 1.2$\pm$0.1 & Y & & BGPS4135 & & 0.361 & 0.60$\pm$0.07 & 58.3$\pm$0.1 & 1.6$\pm$0.2 & Y \\
BGPS2986 & & 1.332 & 2.19$\pm$0.07 & 20.1$\pm$0.1 & 1.5$\pm$0.1 & Y & &  		 & & 0.159 & 0.26$\pm$0.07 & 61.7$\pm$0.3 & 1.6$\pm$0.2 &  \\
BGPS3018 & & 0.930 & 1.52$\pm$0.07 & 19.0$\pm$0.1 & 1.5$\pm$0.1 & Y & & BGPS4140 & & 0.499 & 0.90$\pm$0.08 & 95.8$\pm$0.1 & 1.7$\pm$0.2 & Y \\
BGPS3030 & & 0.769 & 1.21$\pm$0.11 & 19.1$\pm$0.1 & 1.5$\pm$0.1 & Y & & BGPS4145 & & 0.278 & 0.81$\pm$0.12 & 96.4$\pm$0.2 & 2.8$\pm$0.3 &    \\
BGPS3110 & & 2.552 & 9.54$\pm$0.10 & 17.2$\pm$0.1 & 3.5$\pm$0.1 & Y & & BGPS4191 & & 0.563 & 1.47$\pm$0.10 & 97.2$\pm$0.1 & 2.4$\pm$0.2 & Y \\
BGPS3114 & & 0.405 & 1.62$\pm$0.10 & 22.6$\pm$0.1 & 3.8$\pm$0.2 &   & & BGPS4230 & & 0.786 & 1.99$\pm$0.08 & 107.7$\pm$0.1 & 2.4$\pm$0.1 & Y \\
BGPS3117 & & 1.280 & 3.12$\pm$0.08 & 18.7$\pm$0.1 & 2.3$\pm$0.1 &   & & BGPS4294 & & 0.417 & 1.58$\pm$0.11 & 56.3$\pm$0.1 & 3.6$\pm$0.3 & Y \\
BGPS3118 & & 1.941 & 5.13$\pm$0.15 & 16.8$\pm$0.1 & 2.5$\pm$0.1 & Y & & BGPS4297 & & 0.560 & 1.88$\pm$0.11 & 58.2$\pm$0.1 & 3.2$\pm$0.2 &    \\
BGPS3125 & & 0.275 & 0.60$\pm$0.11 & 18.5$\pm$0.2 & 2.1$\pm$0.3 & Y & & BGPS4346 & & 0.567 & 2.81$\pm$0.16 & 92.5$\pm$0.1 & 4.7$\pm$0.2 &    \\
		 & & 0.793 & 1.65$\pm$0.11 & 21.4$\pm$0.1 & 2.0$\pm$0.1 &   & & 	     & & 0.139 & 0.66$\pm$0.14 & 97.0$\pm$0.3 & 4.5$\pm$0.4 &    \\
BGPS3128 & & 1.190 & 3.78$\pm$0.11 & 18.4$\pm$0.1 & 3.0$\pm$0.1 &   & & BGPS4347 & & 0.905 & 3.36$\pm$0.14 & 93.5$\pm$0.1 & 3.5$\pm$0.1 & Y \\
BGPS3129 & & 1.054 & 1.83$\pm$0.12 & 19.5$\pm$0.1 & 1.6$\pm$0.1 & Y & & BGPS4354 & & 0.410 & 1.88$\pm$0.10 & 93.4$\pm$0.1 & 4.3$\pm$0.3 & Y \\
BGPS3134 & & 1.766 & 3.50$\pm$0.13 & 20.9$\pm$0.1 & 1.9$\pm$0.1 & Y & & BGPS4356 & & 0.691 & 1.81$\pm$0.15 & 109.9$\pm$0.1 & 2.5$\pm$0.3 & Y \\
BGPS3139 & & 2.819 & 4.55$\pm$0.14 & 20.1$\pm$0.1 & 1.5$\pm$0.1 & Y & & 		 & & 0.361 & 0.90$\pm$0.15 & 113.0$\pm$0.2 & 2.3$\pm$0.3 &   \\
BGPS3151 & & 0.659 & 1.71$\pm$0.16 & 39.6$\pm$0.1 & 2.4$\pm$0.2 & Y & & BGPS4375 & & 0.446 & 0.89$\pm$0.09 & 93.1$\pm$0.1 & 1.9$\pm$0.2 &    \\
BGPS3220 & & 0.654 & 3.33$\pm$0.18 & 46.4$\pm$0.1 & 4.8$\pm$0.3 &   & & BGPS4396 & & 0.339 & 1.04$\pm$0.07 & 96.9$\pm$0.1 & 2.9$\pm$0.2 & Y \\
		 & & 0.205 & 0.36$\pm$0.07 & 50.9$\pm$0.2 & 1.7$\pm$0.2 &   & & 		 & & 0.320 & 1.07$\pm$0.08 & 112.8$\pm$0.1 & 3.2$\pm$0.2 &   \\
BGPS3243 & & 0.546 & 1.01$\pm$0.08 & 68.2$\pm$0.1 & 1.7$\pm$0.1& Y  & & BGPS4402 & & 0.339 & 1.10$\pm$0.08 & 99.3$\pm$0.1 & 3.1$\pm$0.3 & Y \\
BGPS3247 & & 0.760 & 1.45$\pm$0.10 & 45.0$\pm$0.1 & 1.8$\pm$0.2 & Y & & 		 & & 0.178 & 0.31$\pm$0.07 & 102.5$\pm$0.2 & 1.7$\pm$0.2 &   \\
BGPS3276 & & 0.598 & 1.70$\pm$0.10 & 66.9$\pm$0.1 & 2.7$\pm$0.2 &   & & BGPS4422 & & 0.397 & 0.92$\pm$0.08 & 110.6$\pm$0.1 & 2.2$\pm$0.2 & Y \\
BGPS3300 & & 0.332 & 0.94$\pm$0.10 & 64.6$\pm$0.3 & 2.7$\pm$0.2 & Y & & BGPS4472 & & 0.537 & 2.37$\pm$0.10 & 46.9$\pm$0.1 & 4.1$\pm$0.3 & Y \\
		 & & 0.231 & 0.58$\pm$0.07 & 66.4$\pm$0.3 & 2.4$\pm$0.3 &   & & BGPS4732 & & 0.460 & 0.98$\pm$0.05 & 88.4$\pm$0.1 & 2.0$\pm$0.1 & Y \\
BGPS3302 & & 0.656 & 3.20$\pm$0.10 & 66.8$\pm$0.1 & 4.6$\pm$0.2 & Y & & BGPS4827 & & 0.753 & 1.67$\pm$0.09 & 86.3$\pm$0.1 & 2.1$\pm$0.1 & Y \\
BGPS3306 & & 0.463 & 0.73$\pm$0.06 & 57.0$\pm$0.1 & 1.5$\pm$0.1 & Y & & BGPS4841 & & 0.467 & 1.13$\pm$0.08 & 83.5$\pm$0.1 & 2.3$\pm$0.2 & Y \\
BGPS3312 & & 0.397 & 0.55$\pm$0.07 & 47.0$\pm$0.1 & 1.3$\pm$0.2 &   & & BGPS4902 & & 0.209 & 0.58$\pm$0.09 & 80.2$\pm$0.2 & 2.6$\pm$0.1 & Y \\
BGPS3315 & & 0.234 & 0.40$\pm$0.11 & 42.8$\pm$0.2 & 1.6$\pm$0.2 &   & & 		 & & 0.893 & 3.16$\pm$0.10 & 84.3$\pm$0.1 & 3.3$\pm$0.1 &  \\
		 & & 0.572 & 1.93$\pm$0.15 & 47.2$\pm$0.1 & 3.2$\pm$0.3 &   & & BGPS4953 & & 0.435 & 1.16$\pm$0.08 & 90.8$\pm$0.1 & 2.5$\pm$0.2 & Y \\
BGPS3344 & & 0.284 & 0.78$\pm$0.12 & 65.6$\pm$0.2 & 2.6$\pm$0.3 & Y & & BGPS4962 & & 0.319 & 0.82$\pm$0.09 & 88.6$\pm$0.1 & 2.4$\pm$0.2 & Y \\
BGPS3442 & & 0.219 & 0.50$\pm$0.08 & 65.7$\pm$0.2 & 2.1$\pm$0.3 &   & & BGPS4967 & & 0.432 & 0.81$\pm$0.06 & 80.5$\pm$0.1 & 1.8$\pm$0.2 & Y \\
BGPS3475 & & 0.237 & 0.39$\pm$0.10 & 75.1$\pm$0.2 & 1.6$\pm$0.2 &   & & BGPS5021 & & 0.464 & 1.74$\pm$0.09 & 80.0$\pm$0.1 & 3.5$\pm$0.2 & Y \\
		 & & 0.241 & 0.73$\pm$0.12 & 77.6$\pm$0.3 & 2.9$\pm$0.3 &   & & BGPS5064 & & 0.684 & 1.75$\pm$0.10 & 100.7$\pm$0.1 & 2.4$\pm$0.2 & Y \\
BGPS3484 & & 0.331 & 0.95$\pm$0.08 & 56.0$\pm$0.1 & 2.7$\pm$0.2 &   & & BGPS5089 & & 0.585 & 1.33$\pm$0.05 & 85.3$\pm$0.1 & 2.1$\pm$0.1 & Y  \\
BGPS3487 & & 0.203 & 0.33$\pm$0.12 & 53.3$\pm$0.3 & 1.5$\pm$0.3 &   & & BGPS5090 & & 0.277 & 0.76$\pm$0.10 & 96.3$\pm$0.2 & 2.6$\pm$0.3 &    \\
		 & & 0.332 & 0.99$\pm$0.16 & 56.0$\pm$0.2 & 2.8$\pm$0.3 &   & & BGPS5114 & & 0.467 & 1.37$\pm$0.13 & 65.9$\pm$0.2 & 2.8$\pm$0.2 &    \\
BGPS3604 & & 0.723 & 1.64$\pm$0.09 & 51.5$\pm$0.1 & 2.1$\pm$0.1 & Y & & BGPS5166 & & 0.649 & 1.27$\pm$0.10 & 102.7$\pm$0.1 & 1.8$\pm$0.2 &   \\
BGPS3606 & & 0.193 & 0.28$\pm$0.08 & 49.5$\pm$0.2 & 1.4$\pm$0.2 & Y & & BGPS5183 & & 0.214 & 0.50$\pm$0.10 & 113.9$\pm$0.2 & 2.2$\pm$0.3 &    \\
BGPS3608 & & 0.245 & 0.90$\pm$0.13 & 64.4$\pm$0.2 & 3.5$\pm$0.3 &   & & BGPS5243 & & 0.478 & 1.15$\pm$0.11 & 95.9$\pm$0.1 & 2.3$\pm$0.3 & Y \\
BGPS3627 & & 0.119 & 0.26$\pm$0.11 & 81.1$\pm$0.3 & 2.1$\pm$0.3 &   & &  & &  &  &  &  &  \\
		 & & 0.151 & 0.41$\pm$0.12 & 84.0$\pm$0.3 & 2.5$\pm$0.3 &   & &  & &  &  &  &  &  \\
\hline\hline
\end{tabular}
\end{center}
\end{table*}


\begin{table}
\movetabledown=60mm
\begin{rotatetable}
\begin{center}
\scriptsize
\caption{OTF observational parameters of DCO$^{+}$, H$^{13}$CO$^{+}$, DNC, HN$^{13}$C, DCN, and  H$^{13}$CN \textit{J}=1$-$0 for the 11 sources.}  \label{res_para}
\begin{tabular}{c lrrrr c clrrrr c}
\hline\hline
\noalign{\smallskip}
Source & Molecular species & T$_{\rm mb}$ & $\int$T$_{\rm mb}$dv & V$_{\rm LSR}$ & FWHM & Centre & Source & Molecular species & T$_{\rm mb}$ & $\int$T$_{\rm mb}$dv & V$_{\rm LSR}$ & FWHM & Centre \\
  &  & (K) & (K km s$^{-1}$) & (km s$^{-1}$) & (km s$^{-1}$) &  &  &  & (K) & (K km s$^{-1}$) & (km s$^{-1}$) & (km s$^{-1}$)  &  \\
\noalign{\smallskip}
 \hline
BGPS2693 & H$^{13}$CO$^{+}$ \textit{J}=1$-$0 & 0.532 & 0.43$\pm$0.06 & 19.1$\pm$0.1 & 0.8$\pm$0.2 & [2$^{\prime\prime}$, 2$^{\prime\prime}$] & BGPS2931 & H$^{13}$CO$^{+}$ \textit{J}=1$-$0 & 0.768 & 0.79$\pm$0.05 & 22.7$\pm$0.1 & 1.0$\pm$0.1  & [10$^{\prime\prime}$, $-$35$^{\prime\prime}$] \\
 & DCO$^{+}$ \textit{J}=1$-$0 & 0.671 & 0.58$\pm$0.06 & 19.0$\pm$0.1 & 0.8$\pm$0.2 &  &  & DCO$^{+}$ \textit{J}=1$-$0 & 1.135 & 1.42$\pm$0.06 & 22.7$\pm$0.1 & 1.2$\pm$0.1  &  \\
 & &  &  &  &  &  & &  &  &  &   &  &  \\
 &  &  &  &  &  &  &  & H$^{13}$CN \textit{J}=1$-$0 & 0.206 & 0.17$\pm$0.05 & 22.8$\pm$0.1 & 0.7$\pm$0.1 & [8$^{\prime\prime}$, $-$17$^{\prime\prime}$] \\
 &  &  &  &  &  &  &  & DCN \textit{J}=1$-$0 & 0.253 & 0.25$\pm$0.08 & 22.7$\pm$0.2 & 1.1$\pm$0.2  &  \\
 & &  &  &  &  &  & &  &  &  &   &  &  \\
 & HN$^{13}$C \textit{J}=1$-$0 & 0.456 & 0.41$\pm$0.06 & 19.0$\pm$0.1 & 0.9$\pm$0.1 & [0$^{\prime\prime}$, 3$^{\prime\prime}$] &  & HN$^{13}$C \textit{J}=1$-$0 & 0.505 & 0.64$\pm$0.05 & 22.6$\pm$0.1 & 1.2$\pm$0.1  & [17$^{\prime\prime}$, $-$33$^{\prime\prime}$] \\
 & DNC \textit{J}=1$-$0 & 0.685 & 0.65$\pm$0.07 & 19.0$\pm$0.1 & 0.9$\pm$0.1 &  &  & DNC \textit{J}=1$-$0 & 0.759 & 1.26$\pm$0.06 & 22.8$\pm$0.1 & 1.6$\pm$0.1  &  \\
	\hline
BGPS2940 & H$^{13}$CO$^{+}$ \textit{J}=1$-$0 & 0.821 & 2.00$\pm$0.08 & 19.9$\pm$0.1 & 2.3$\pm$0.1 & [$-$10$^{\prime\prime}$, 2$^{\prime\prime}$] & BGPS2945 & H$^{13}$CO$^{+}$ \textit{J}=1$-$0 & 0.596 & 0.61$\pm$0.04 & 22.6$\pm$0.1 & 1.0$\pm$0.1  & [$-$12$^{\prime\prime}$, 0$^{\prime\prime}$]  \\
 & DCO$^{+}$ \textit{J}=1$-$0 & 0.985 & 1.92$\pm$0.05 & 20.0$\pm$0.1 & 1.8$\pm$0.1 &  &  & DCO$^{+}$ \textit{J}=1$-$0 & 0.847 & 0.86$\pm$0.03 & 22.6$\pm$0.1 & 1.0$\pm$0.1  &  \\
 & &  &  &  &  &  & &  &  &  &   &  &  \\
 & H$^{13}$CN \textit{J}=1$-$0 & 0.387 & 1.47$\pm$0.22 & 20.2$\pm$0.2 & 3.6$\pm$0.3 & [$-$9$^{\prime\prime}$, 6$^{\prime\prime}$] &  & H$^{13}$CN \textit{J}=1$-$0 & 0.240 & 0.22$\pm$0.06 & 22.9$\pm$0.2 & 0.9$\pm$0.2  & [$-$14$^{\prime\prime}$, $-$12$^{\prime\prime}$]  \\
 & DCN \textit{J}=1$-$0 & 0.255 & 0.51$\pm$0.07 & 20.1$\pm$0.1 & 1.9$\pm$0.3 &  &  & DCN \textit{J}=1$-$0 & 0.290 & 0.32$\pm$0.06 & 22.7$\pm$0.1 & 1.0$\pm$0.1  &  \\
 & &  &  &  &  &  & &  &  &  &   &  &  \\
 & HN$^{13}$C \textit{J}=1$-$0 & 0.457 & 1.02$\pm$0.08 & 19.7$\pm$0.1 & 2.1$\pm$0.2 & [$-$11$^{\prime\prime}$, 4$^{\prime\prime}$] &  & HN$^{13}$C \textit{J}=1$-$0 & 0.653 & 0.63$\pm$0.05 & 22.5$\pm$0.1 & 0.9$\pm$0.1  & [$-$9$^{\prime\prime}$, $-$3$^{\prime\prime}$]  \\
 & DNC \textit{J}=1$-$0 & 0.475 & 1.00$\pm$0.04 & 20.1$\pm$0.1 & 2.0$\pm$0.1 &  &  & DNC \textit{J}=1$-$0 & 0.739 & 1.05$\pm$0.04 & 22.7$\pm$0.1 & 1.3$\pm$0.1  &  \\
	\hline
BGPS2984 & H$^{13}$CO$^{+}$ \textit{J}=1$-$0 & 0.909 & 0.81$\pm$0.04 & 18.5$\pm$0.1 & 0.8$\pm$0.1 & [$-$2$^{\prime\prime}$, 2$^{\prime\prime}$] & BGPS2986 & H$^{13}$CO$^{+}$ \textit{J}=1$-$0 & 0.994 & 1.33$\pm$0.06 & 19.9$\pm$0.1 & 1.3$\pm$0.1  & [$-$7$^{\prime\prime}$, 1$^{\prime\prime}$] \\
 & DCO$^{+}$ \textit{J}=1$-$0 & 1.176 & 1.27$\pm$0.05 & 18.4$\pm$0.1 & 1.0$\pm$0.1 &  &  & DCO$^{+}$ \textit{J}=1$-$0 & 1.409 & 1.86$\pm$0.05 & 20.0$\pm$0.1 & 1.2$\pm$0.1  &  \\%
 & &  &  &  &  &  & &  &  &  &   &  &  \\
 & H$^{13}$CN \textit{J}=1$-$0 & 0.195 & 0.29$\pm$0.05 & 18.6$\pm$0.1 & 1.4$\pm$0.1 & [$-$5$^{\prime\prime}$, 7$^{\prime\prime}$] &  & H$^{13}$CN \textit{J}=1$-$0 & 0.262 & 0.40$\pm$0.08 & 20.1$\pm$0.1 & 1.5$\pm$0.2  & [4$^{\prime\prime}$, 8$^{\prime\prime}$] \\
 & DCN \textit{J}=1$-$0 & 0.252 & 0.28$\pm$0.07 & 18.4$\pm$0.1 & 1.1$\pm$0.2 &  &  & DCN \textit{J}=1$-$0 & 0.314 & 0.41$\pm$0.06 & 20.1$\pm$0.1 & 1.2$\pm$0.2  &  \\
 & &  &  &  &  &  & &  &  &  &   &  &  \\
 & HN$^{13}$C \textit{J}=1$-$0 & 0.570 & 0.58$\pm$0.06 & 18.1$\pm$0.1 & 1.0$\pm$0.1 & [$-$5$^{\prime\prime}$, 9$^{\prime\prime}$] &  & HN$^{13}$C \textit{J}=1$-$0 & 0.742 & 1.01$\pm$0.06 & 19.7$\pm$0.1 & 1.3$\pm$0.1  & [$-$5$^{\prime\prime}$, 6$^{\prime\prime}$] \\
 & DNC \textit{J}=1$-$0 & 0.752 & 0.92$\pm$0.04 & 18.4$\pm$0.1 & 1.2$\pm$0.1 &  &  & DNC \textit{J}=1$-$0 & 1.000 & 1.48$\pm$0.03 & 20.0$\pm$0.1 & 1.4$\pm$0.1  &  \\%
 	\hline
BGPS3018 & H$^{13}$CO$^{+}$ \textit{J}=1$-$0 & 0.464 & 0.67$\pm$0.08 & 19.1$\pm$0.1 & 1.4$\pm$0.1 & [4$^{\prime\prime}$, 10$^{\prime\prime}$] & BGPS3110 & H$^{13}$CO$^{+}$ \textit{J}=1$-$0 & 0.535 & 1.93$\pm$0.11 & 17.0$\pm$0.1 & 3.4$\pm$0.2  & [$-$16$^{\prime\prime}$, 6$^{\prime\prime}$] \\
 & DCO$^{+}$ \textit{J}=1$-$0 & 0.651 & 0.86$\pm$0.06 & 19.1$\pm$0.1 & 1.3$\pm$0.1 &  &  &  &  &  &  &   &  \\
 & &  &  &  &  &  & &  &  &  &   &  &  \\
 & H$^{13}$CN \textit{J}=1$-$0 & 0.156 & 0.34$\pm$0.07 & 19.6$\pm$0.2 & 2.1$\pm$0.3 & [$-$28$^{\prime\prime}$, $-$34$^{\prime\prime}$] &  & H$^{13}$CN \textit{J}=1$-$0 & 0.629 & 2.07$\pm$0.16 & 17.4$\pm$0.1 & 3.1$\pm$0.2  & [$-$7$^{\prime\prime}$, $-$2$^{\prime\prime}$] \\
 & DCN \textit{J}=1$-$0 & 0.237 & 0.20$\pm$0.06 & 19.1$\pm$0.1 & 0.8$\pm$0.1 &  &  & DCN \textit{J}=1$-$0 & 0.172 & 0.40$\pm$0.07 & 17.4$\pm$0.2 & 2.2$\pm$0.3  &  \\
 & &  &  &  &  &  & &  &  &  &   &  &  \\
 & HN$^{13}$C \textit{J}=1$-$0 & 0.227 & 0.41$\pm$0.09 & 18.7$\pm$0.2 & 1.7$\pm$0.3 & [$-$5$^{\prime\prime}$, $-$10$^{\prime\prime}$] &  & HN$^{13}$C \textit{J}=1$-$0 & 0.590 & 1.63$\pm$0.13 & 16.8$\pm$0.1 & 2.6$\pm$0.2  & [$-$11$^{\prime\prime}$, 5$^{\prime\prime}$] \\
 & DNC \textit{J}=1$-$0 & 0.271 & 0.54$\pm$0.05 & 19.0$\pm$0.1 & 1.9$\pm$0.2 &  &  & DNC \textit{J}=1$-$0 & 0.147 & 0.36$\pm$0.06 & 17.7$\pm$0.1 & 2.3$\pm$0.2  &  \\
  	\hline
BGPS3125 & H$^{13}$CO$^{+}$ \textit{J}=1$-$0 & 0.901 & 0.64$\pm$0.05 & 21.5$\pm$0.1 & 0.7$\pm$0.2 & [$-$9$^{\prime\prime}$, 3$^{\prime\prime}$] & BGPS3134 & H$^{13}$CO$^{+}$ \textit{J}=1$-$0 & 1.097 & 1.68$\pm$0.06 & 21.1$\pm$0.1 & 1.4$\pm$0.1  & [$-$8$^{\prime\prime}$, 6$^{\prime\prime}$] \\
 & DCO$^{+}$ \textit{J}=1$-$0 & 0.120 & 0.31$\pm$0.08 & 21.8$\pm$0.2 & 2.4$\pm$0.4 &  &  & DCO$^{+}$ \textit{J}=1$-$0 & 0.431 & 0.83$\pm$0.05 & 20.9$\pm$0.1 & 1.8$\pm$0.1  &  \\
 & &  &  &  &  &  & &  &  &  &   &  &  \\
 & H$^{13}$CN \textit{J}=1$-$0 & 0.174 & 0.19$\pm$0.06 & 21.8$\pm$0.1 & 1.0$\pm$0.3 & [$-$10$^{\prime\prime}$, 24$^{\prime\prime}$] &  & H$^{13}$CN \textit{J}=1$-$0 & 0.301 & 0.26$\pm$0.06 & 20.6$\pm$0.1 & 0.7$\pm$0.1  & [0$^{\prime\prime}$, 13$^{\prime\prime}$] \\
 & DCN \textit{J}=1$-$0 & 0.085 & 0.20$\pm$0.04 & 21.6$\pm$0.3 & 2.2$\pm$0.3 &  &  & DCN \textit{J}=1$-$0 & 0.229 & 0.41$\pm$0.08 & 20.7$\pm$0.2 & 2.0$\pm$0.3 &  \\
 & &  &  &  &  &  & &  &  &  &   &  &  \\
 & HN$^{13}$C \textit{J}=1$-$0 & 0.344 & 0.24$\pm$0.04 & 21.2$\pm$0.1 & 0.7$\pm$0.2 & [$-$2$^{\prime\prime}$, 4$^{\prime\prime}$] &  & HN$^{13}$C \textit{J}=1$-$0 & 0.425 & 0.89$\pm$0.06 & 20.7$\pm$0.1 & 2.0$\pm$0.1  & [$-$3$^{\prime\prime}$, 6$^{\prime\prime}$] \\
 & DNC \textit{J}=1$-$0 & 0.226 & 0.20$\pm$0.03 & 21.6$\pm$0.1 & 0.8$\pm$0.2 &  &  & DNC \textit{J}=1$-$0 & 0.585 & 1.12$\pm$0.05 & 20.7$\pm$0.1 & 1.8$\pm$0.1  &  \\
   	\hline
BGPS4402 & H$^{13}$CO$^{+}$ \textit{J}=1$-$0 & 0.304 & 0.62$\pm$0.07 & 99.3$\pm$0.1 & 1.9$\pm$0.2 & [$-$12$^{\prime\prime}$, $-$16$^{\prime\prime}$] &  & &  &  &  &   &  \\
 & DCO$^{+}$ \textit{J}=1$-$0 & 0.249 & 0.36$\pm$0.05 & 99.3$\pm$0.1 & 1.4$\pm$0.2 &  &  & &  &  &  &   &  \\
 & &  &  &  &  &  & &  &  &  &   &  &  \\
 & H$^{13}$CN \textit{J}=1$-$0 & 0.153 & 0.26$\pm$0.06 & 99.7$\pm$0.2 & 1.6$\pm$0.2 & [$-$4$^{\prime\prime}$, 2$^{\prime\prime}$] &  & &  &  &  &   &  \\
 & &  &  &  &  &  & &  &  &  &   &  &  \\
 & HN$^{13}$C \textit{J}=1$-$0 & 0.282 & 0.46$\pm$0.07 & 99.1$\pm$0.1 & 1.6$\pm$0.2 & [10$^{\prime\prime}$, $-$18$^{\prime\prime}$] &  & &  &  &  &   &  \\
 & DNC \textit{J}=1$-$0 & 0.129 & 0.21$\pm$0.02 & 99.2$\pm$0.2 & 1.5$\pm$0.2 &  &  & &  &  &  &   &  \\
\hline\hline
\end{tabular}
\end{center}
\end{rotatetable}
\end{table}


\begin{table*}
\centering
\caption{Estimated column densities of optically thin molecular \textit{J}=1$-$0 lines in the 11 sources from the OTF observations.}  \label{Tab_Column}
\normalsize
\begin{tabular}{cccccccc}
\hline\hline
\noalign{\smallskip}
Source & T$_{\rm Kin}$ & $N$(H$^{13}$CN) & $N$(DCN) & $N$(H$^{13}$CO$^{+}$) & $N$(DCO$^{+}$) & $N$(HN$^{13}$C) & $N$(DNC) \\
  & (K) & ($\times$10$^{12}$ cm$^{-2}$) & ($\times$10$^{12}$ cm$^{-2}$) & ($\times$10$^{12}$ cm$^{-2}$) & ($\times$10$^{12}$ cm$^{-2}$) & ($\times$10$^{12}$ cm$^{-2}$) & ($\times$10$^{12}$ cm$^{-2}$) \\
 \noalign{\smallskip}
 \hline
\noalign{\smallskip}
BGPS2693 & 12.440 & ...       & ...      & 1.4$\pm$0.2 & 1.9$\pm$0.2 & 1.1$\pm$0.2 & 1.7$\pm$0.2 \\
BGPS2931 & 11.711 & 0.60$\pm$0.15 & 1.1$\pm$0.3 & 2.6$\pm$0.2 & 4.9$\pm$0.2 & 1.6$\pm$0.1 & 3.1$\pm$0.2 \\
BGPS2940 & 17.795 &  6.5$\pm$0.8 & 2.9$\pm$0.4 & 7.5$\pm$0.3 & 7.6$\pm$0.2 & 3.3$\pm$0.2 & 3.2$\pm$0.1 \\
BGPS2945 & 11.666 & 0.77$\pm$0.18 & 1.5$\pm$0.2 & 3.0$\pm$0.2 & 4.9$\pm$0.2 & 1.6$\pm$0.1 & 2.6$\pm$0.1 \\
BGPS2984 & 11.486 &  1.0$\pm$0.2 & 1.3$\pm$0.3 & 2.6$\pm$0.1 & 4.4$\pm$0.2 & 1.5$\pm$0.2 & 2.2$\pm$0.1 \\
BGPS2986 & 11.634 &  1.5$\pm$0.2 & 1.9$\pm$0.3 & 4.1$\pm$0.2 & 6.2$\pm$0.2 & 2.6$\pm$0.2 & 3.6$\pm$0.1 \\
BGPS3018 & 13.472 &  1.3$\pm$0.3 & 0.98$\pm$0.21 & 2.1$\pm$0.3 & 2.8$\pm$0.2 & 1.1$\pm$0.3 & 1.4$\pm$0.1 \\
BGPS3110 & 25.166 &   12$\pm$0.9 & 3.0$\pm$0.5 & 9.3$\pm$0.6 & ...      & 6.6$\pm$0.5 & 1.5$\pm$0.3 \\
BGPS3125 & 21.251 &  0.95$\pm$0.22 & 1.3$\pm$0.3 & 2.5$\pm$0.2 & 1.3$\pm$0.3 &0.86$\pm$0.13 &0.71$\pm$0.11 \\
BGPS3134 & 15.919 &  1.1$\pm$0.2 & 2.2$\pm$0.4 & 5.9$\pm$0.2 & 3.1$\pm$0.2 & 2.6$\pm$0.2 & 3.3$\pm$0.1 \\
BGPS4402 & 13.125 &  0.97$\pm$0.13 & ...      & 1.9$\pm$0.2 & 1.2$\pm$0.2 & 1.2$\pm$0.2 &0.55$\pm$0.06 \\
\hline\hline
\end{tabular}
\end{table*}


\begin{table}
\movetabledown=65mm
\begin{rotatetable}
\begin{center}
\normalsize
\caption{Estimated integrated intensity ratio, column density ratio and deuterated fraction from the OTF observations.} \label{D_C}
\begin{tabular}{ccccccccccccc}
\hline\hline
Source & Distance & D$_{\rm gc}$ & $^{12}$C/$^{13}$C & $\frac{I(\rm DCN)}{I(\rm H^{13}CN)}$ & $\frac{N(\rm DCN)}{N(\rm H^{13}CN)}$ & $\frac{I(\rm DCO^{+})}{I(\rm H^{13}CO^{+})}$ & $\frac{N(\rm DCO^{+})}{N(\rm H^{13}CO^{+})}$ & $\frac{I(\rm DNC)}{I(\rm HN^{13}C)}$ & $\frac{N(\rm DNC)}{N(\rm HN^{13}C)}$ & D$_{\rm frac}$(\rm HCN) & D$_{\rm frac}$(\rm HCO$^{+}$) & D$_{\rm frac}$(\rm HNC) \\
  & (kpc) & (kpc) &  &  &  &  &  &  &  &  &  & \\
 \hline
BGPS2693 & 2.103 & 6.115 & 56.684 & ... 	              & ...                      & 1.33$\pm$0.09 & 1.39$\pm$0.09 & 1.58$\pm$0.11 & 1.54$\pm$0.10 & ...                          & 0.025$\pm$0.002 & 0.027$\pm$0.002 \\
BGPS2931 & 3.285 & 5.024 & 49.909 & 2.07$\pm$0.10 & 1.89$\pm$0.20 & 1.80$\pm$0.06 & 1.88$\pm$0.07 & 1.95$\pm$0.06 & 1.91$\pm$0.05 & 0.038$\pm$0.006 & 0.038$\pm$0.002 & 0.038$\pm$0.002 \\
BGPS2940 & 3.366 & 4.950 & 49.450 & 0.34$\pm$0.03 & 0.45$\pm$0.04 & 0.96$\pm$0.05 & 1.02$\pm$0.05 & 0.98$\pm$0.05 & 0.98$\pm$0.05 & 0.009$\pm$0.001 & 0.021$\pm$0.001 & 0.020$\pm$0.001 \\
BGPS2945 & 1.178 & 7.013 & 62.261 & 1.45$\pm$0.03 & 1.87$\pm$0.06 & 1.53$\pm$0.12 & 1.58$\pm$0.13 & 1.66$\pm$0.06 & 1.61$\pm$0.06 & 0.030$\pm$0.001 & 0.024$\pm$0.002 & 0.026$\pm$0.001 \\
BGPS2984 & 1.855 & 6.369 & 58.261 & 0.97$\pm$0.08 & 1.24$\pm$0.12 & 1.57$\pm$0.12 & 1.65$\pm$0.14 & 1.58$\pm$0.04 & 1.54$\pm$0.04 & 0.021$\pm$0.001 & 0.029$\pm$0.002 & 0.026$\pm$0.001 \\
BGPS2986 & 2.014 & 6.219 & 57.330 & 1.02$\pm$0.07 & 1.32$\pm$0.09 & 1.40$\pm$0.12 & 1.46$\pm$0.12 & 1.46$\pm$0.12 & 1.42$\pm$0.12 & 0.023$\pm$0.002 & 0.026$\pm$0.002 & 0.025$\pm$0.002 \\
BGPS3018 & 6.522 & 2.462 & 33.999 & 0.60$\pm$0.07 & 0.76$\pm$0.09 & 1.28$\pm$0.10 & 1.34$\pm$0.10 & 1.33$\pm$0.11 & 1.28$\pm$0.11 & 0.022$\pm$0.002 & 0.040$\pm$0.003 & 0.038$\pm$0.002 \\
BGPS3110 & 2.005 & 6.236 & 57.436 & 0.20$\pm$0.04 & 0.25$\pm$0.05 & ...                      & ...                      & 0.22$\pm$0.03 & 0.22$\pm$0.03 & 0.004$\pm$0.001 & ...                          & 0.004$\pm$0.001 \\
BGPS3125 & 3.466 & 4.889 & 49.071 & 1.05$\pm$0.10 & 1.38$\pm$0.11 & 0.48$\pm$0.05 & 0.53$\pm$0.05 & 0.95$\pm$0.08 & 0.83$\pm$0.07 & 0.028$\pm$0.002 & 0.011$\pm$0.001 & 0.017$\pm$0.001 \\
BGPS3134 & 4.083 & 4.344 & 45.686 & 2.28$\pm$0.10 & 2.05$\pm$0.11 & 0.49$\pm$0.03 & 0.52$\pm$0.04 & 1.26$\pm$0.11 & 1.25$\pm$0.11 & 0.045$\pm$0.005 & 0.011$\pm$0.001 & 0.027$\pm$0.002 \\
BGPS4402 & 4.285 & 4.706 & 47.934 & ...                      & ...                      & 0.59$\pm$0.06 & 0.61$\pm$0.07 & 0.45$\pm$0.07 & 0.44$\pm$0.06 & ...                          & 0.014$\pm$0.002 & 0.009$\pm$0.001 \\
\hline\hline
\end{tabular}
\end{center}
\end{rotatetable}
\end{table}


\begin{figure*}
 \begin{center}
\includegraphics[width=16.5cm]{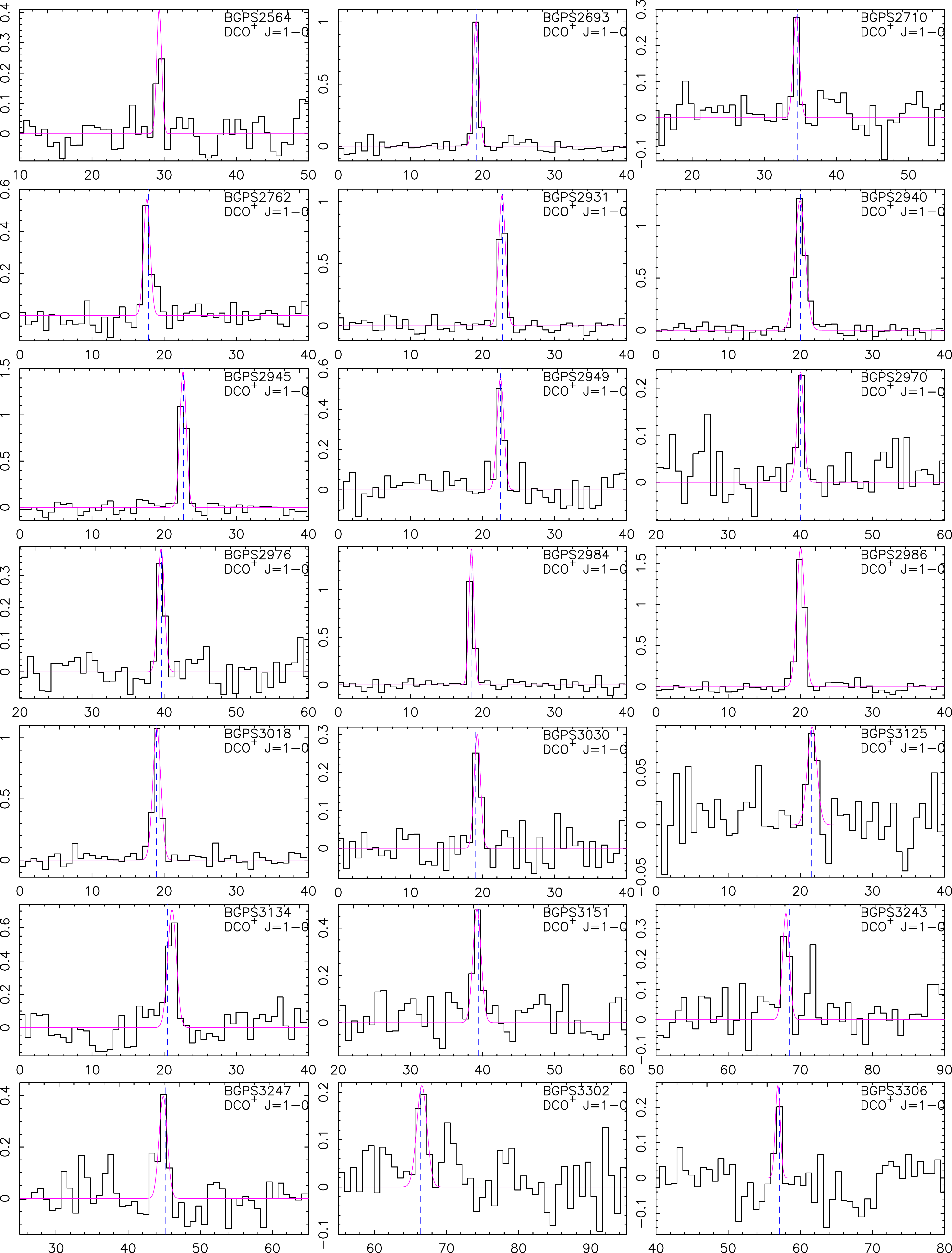}
\caption{Line profiles of the 46 DCO$^{+}$ \textit{J}=1$-$0 detected clumps. The Gaussian fitting result for each spectrum is shown in magenta. The blue dashed vertical line represents the LSR velocity of NH$_{3}$ for each SCC derived from \citet{2016ApJ...822...59S}. The x-axis is velocity in km s$^{-1}$, and the y-axis is T$_{\rm mb}$ in kelvin.}
\label{DCO+1-0}
\end{center}
\end{figure*}

\begin{figure*}
\addtocounter{figure}{-1}
 \begin{center}
\includegraphics[width=16.5cm]{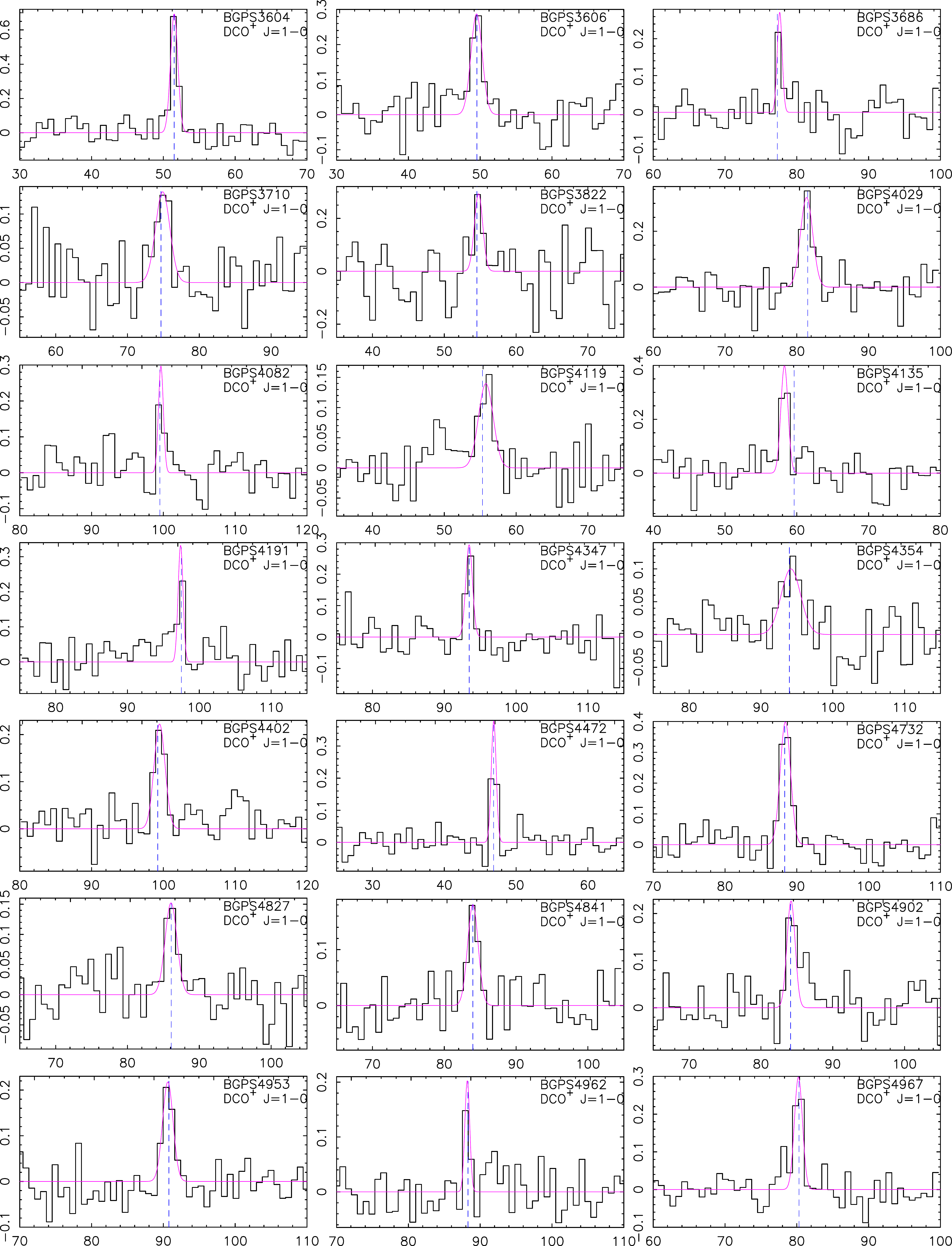}
\caption{Continued.}
\end{center}
\end{figure*}

\begin{figure*}
\addtocounter{figure}{-1}
 \begin{center}
\includegraphics[width=16.5cm]{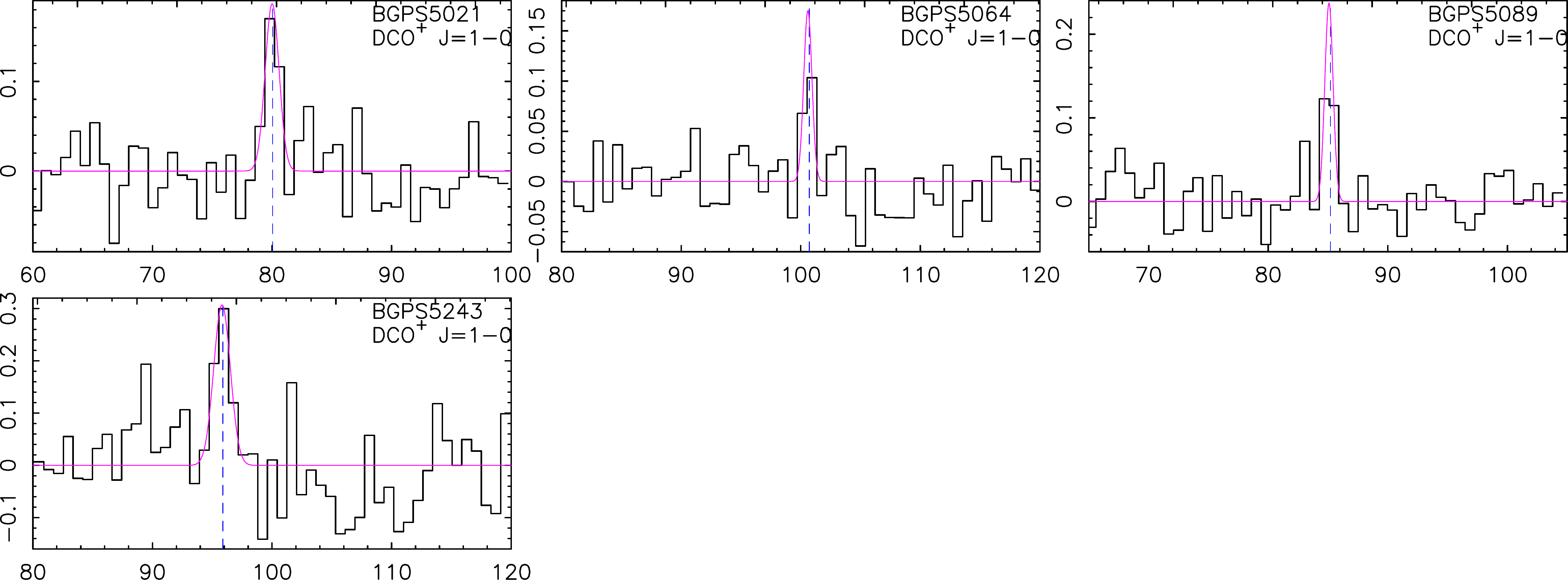}
\caption{Continued.}
\end{center}
\end{figure*}


\begin{figure*}
 \begin{center}
\includegraphics[width=16.5cm]{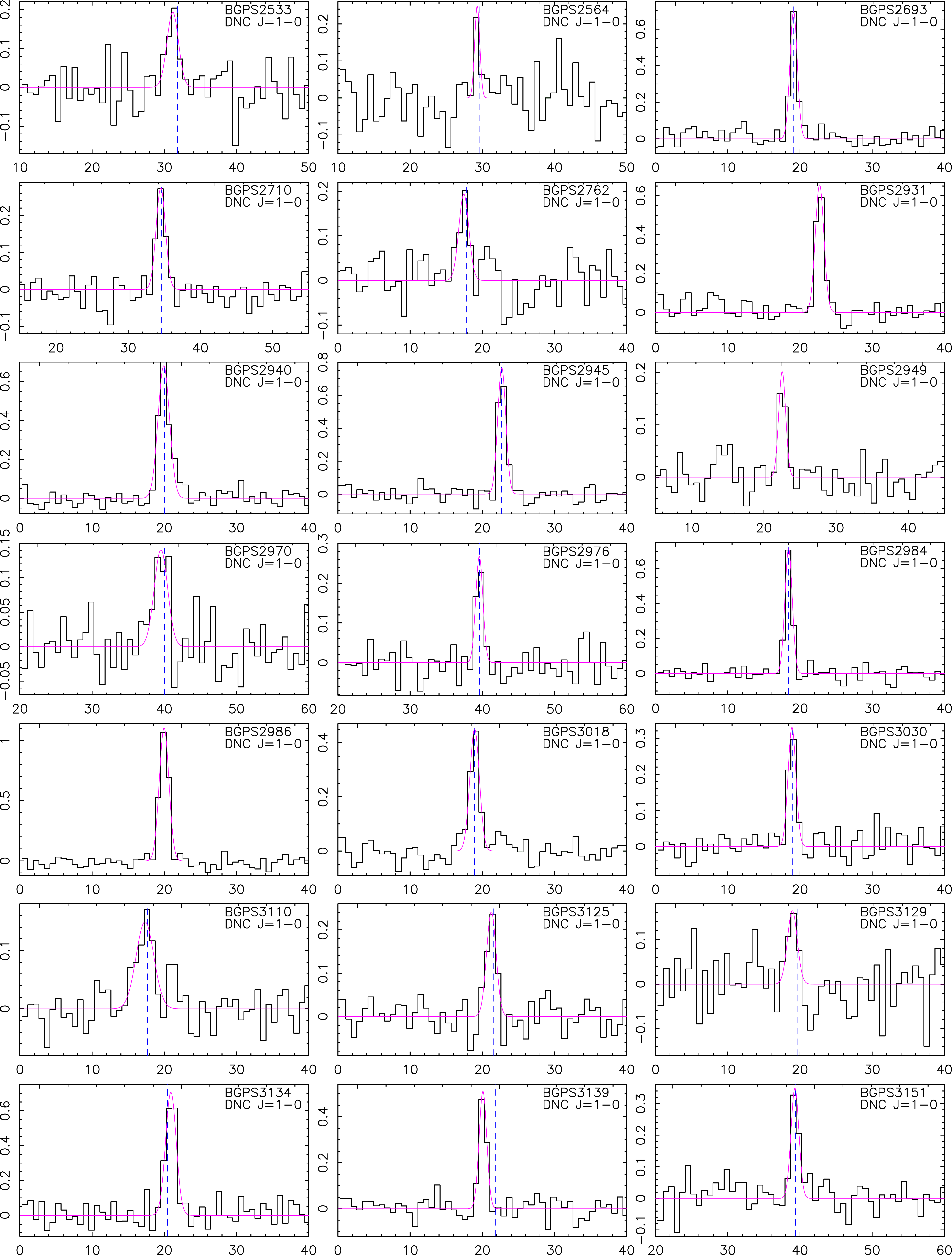}
\caption{Line profiles of the 51 DNC \textit{J}=1$-$0 detected clumps. The Gaussian fitting result for each spectrum is shown in magenta. The blue dashed vertical line represents the LSR velocity of NH$_{3}$ for each SCC derived from \citet{2016ApJ...822...59S}. The x-axis is velocity in km s$^{-1}$, and the y-axis is T$_{\rm mb}$ in kelvin.}
\label{DNC1-0}
\end{center}
\end{figure*}

\begin{figure*}
\addtocounter{figure}{-1}
 \begin{center}
\includegraphics[width=16.5cm]{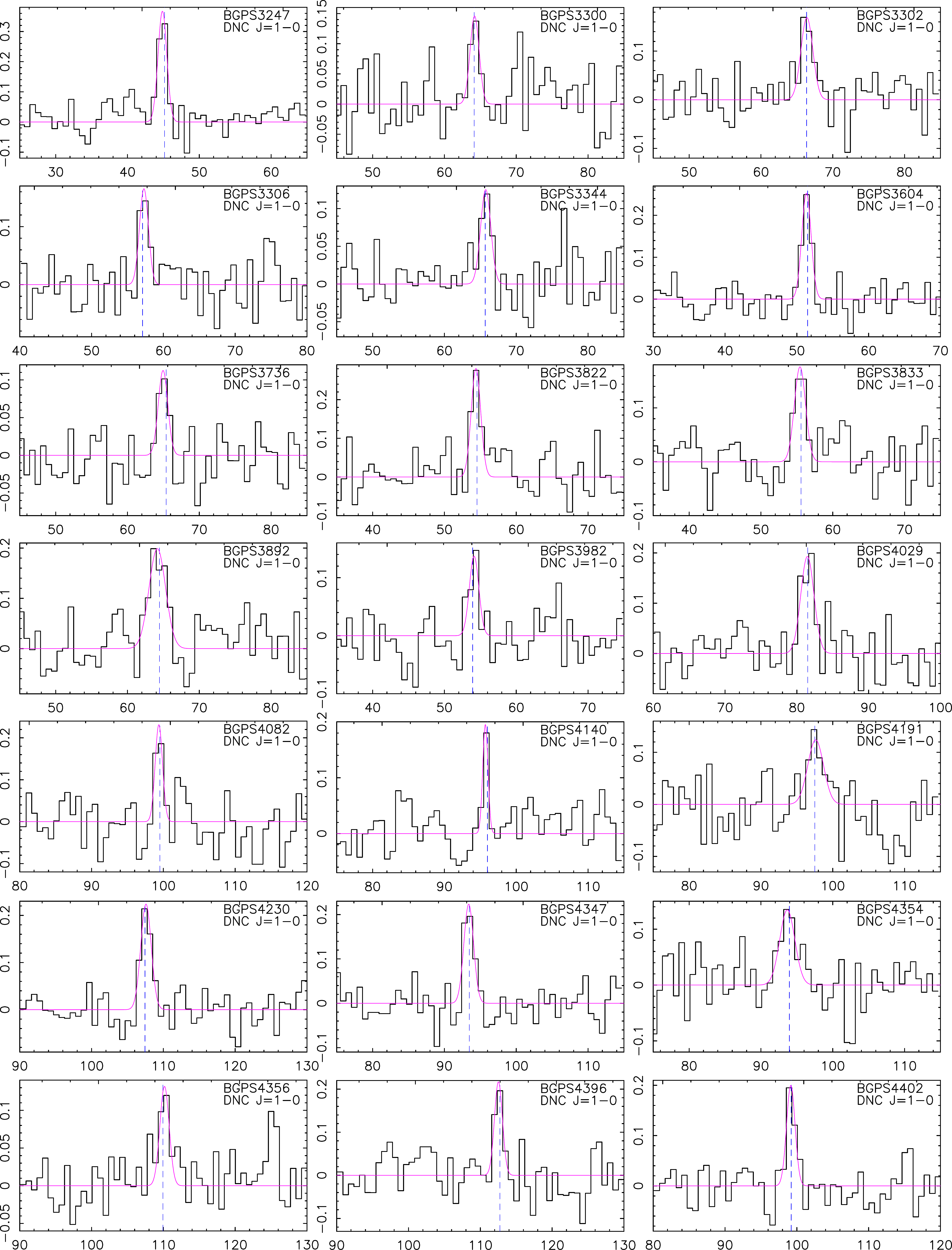}
\caption{Continued.}
\end{center}
\end{figure*}

\begin{figure*}
\addtocounter{figure}{-1}
 \begin{center}
\includegraphics[width=16.5cm]{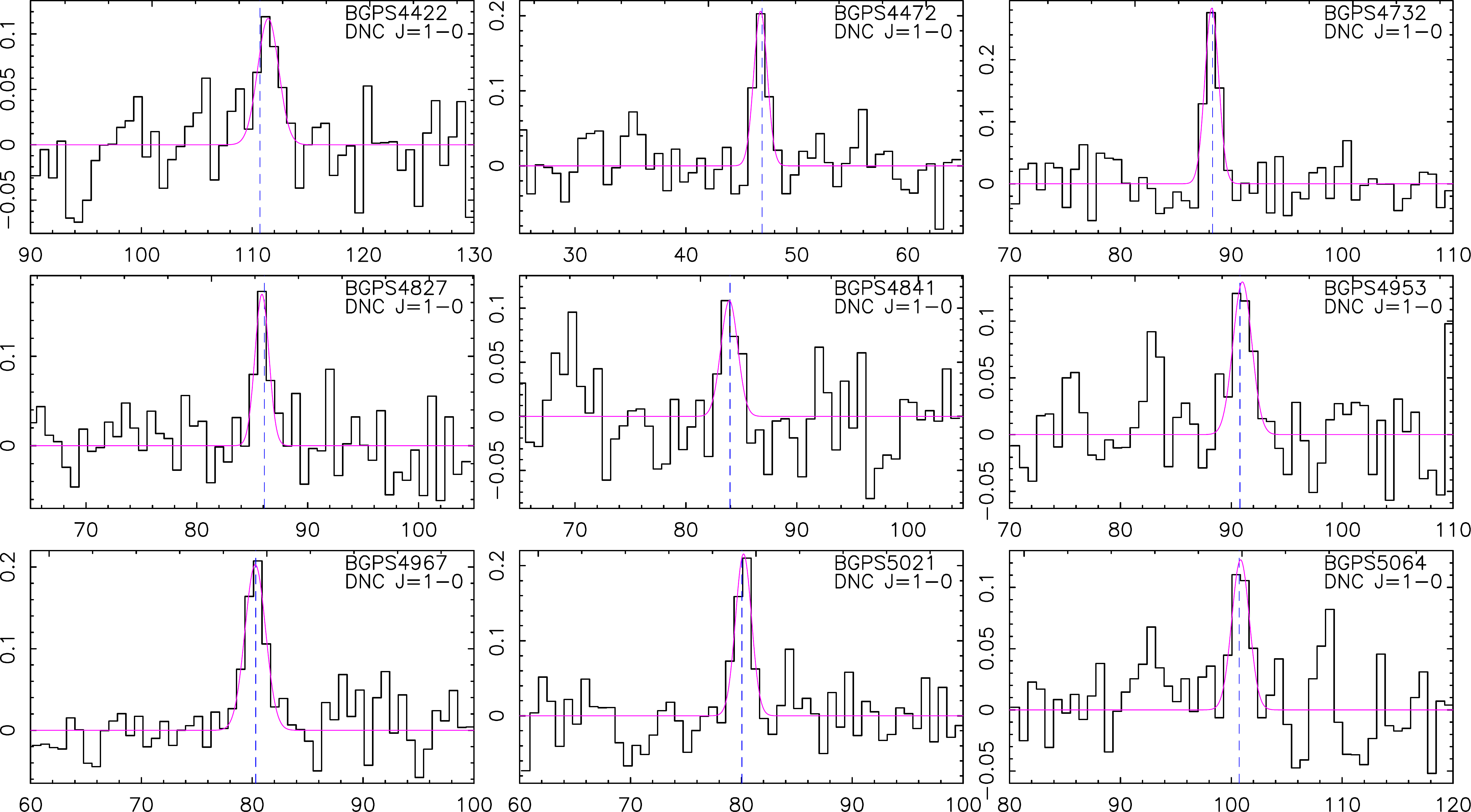}
\caption{Continued.}
\end{center}
\end{figure*}


\begin{figure*}
 \begin{center}
\includegraphics[width=16.5cm]{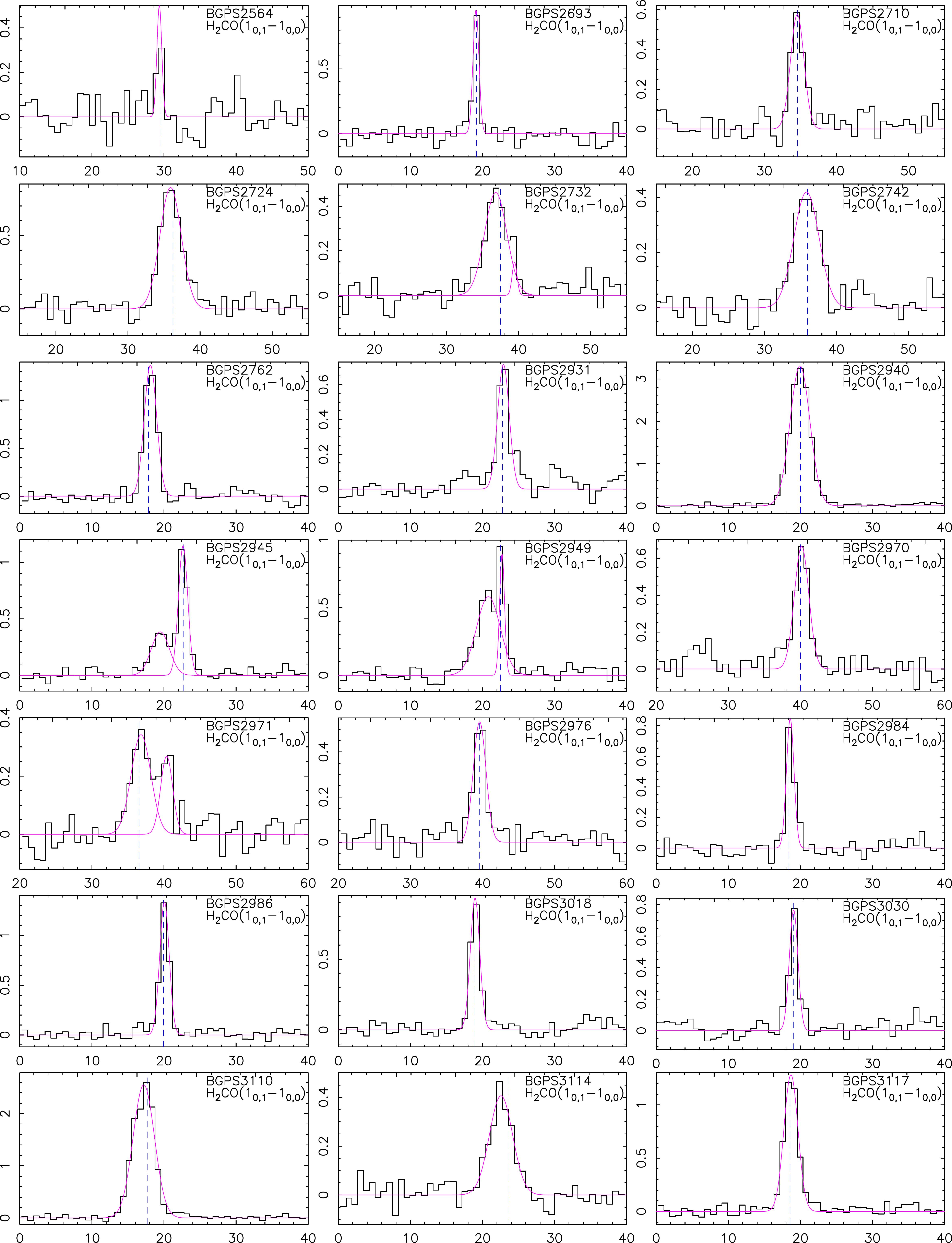}
\caption{Line profiles of the 91 H$_{2}$CO(1$_{(0,1)}$$-$0$_{(0,0)}$) detected clumps. The Gaussian fitting result for each spectrum is shown in magenta. The blue dashed vertical line represents the LSR velocity of NH$_{3}$ for each MSCC derived from \citet{2016ApJ...822...59S}. The x-axis is velocity in km s$^{-1}$, and the y-axis is T$_{\rm mb}$ in kelvin.}
\label{H2CO}
\end{center}
\end{figure*}

\begin{figure*}
\addtocounter{figure}{-1}
 \begin{center}
\includegraphics[width=16.5cm]{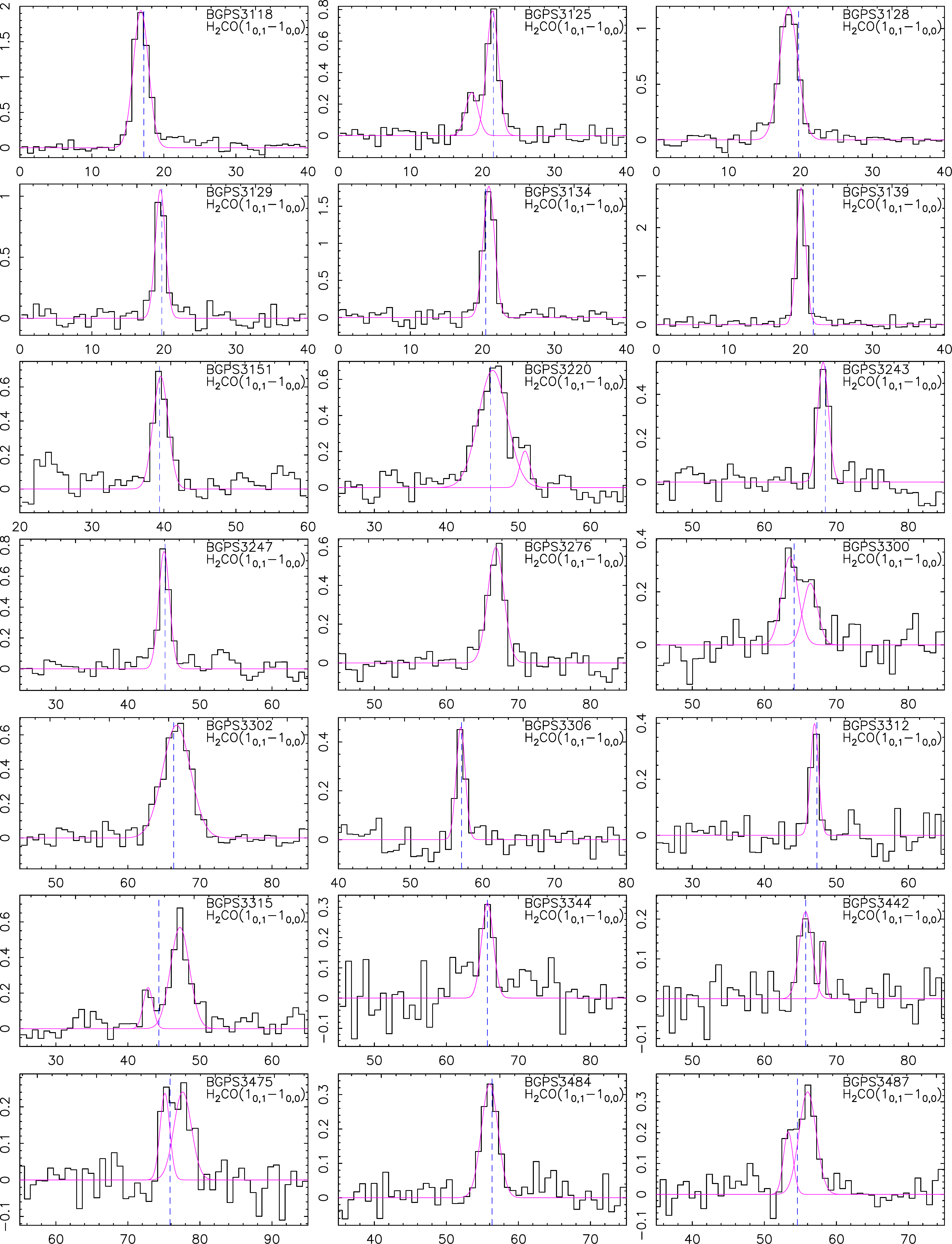}
\caption{Continued.}
\end{center}
\end{figure*}

\begin{figure*}
\addtocounter{figure}{-1}
 \begin{center}
\includegraphics[width=16.5cm]{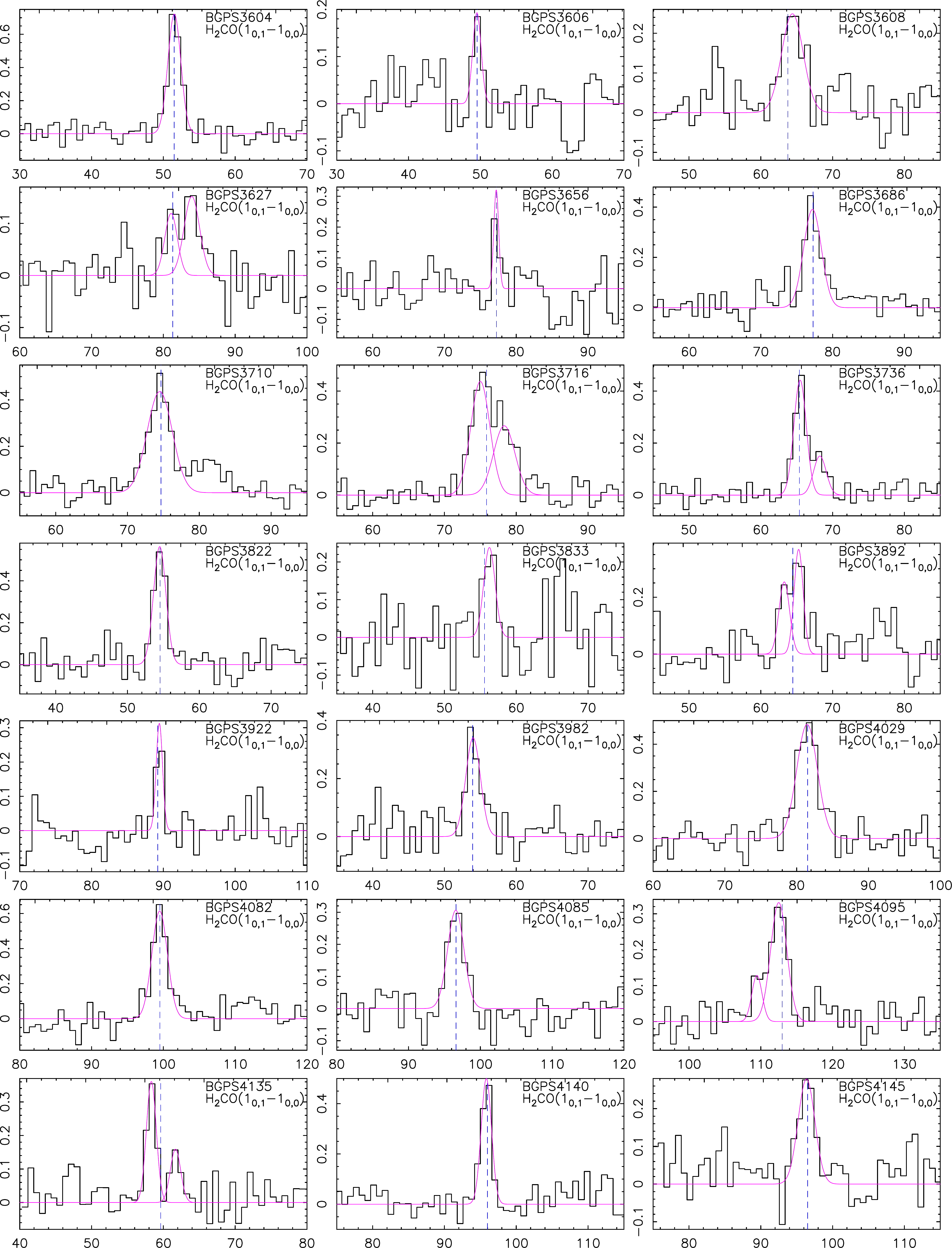}
\caption{Continued.}
\end{center}
\end{figure*}

\begin{figure*}
\addtocounter{figure}{-1}
 \begin{center}
\includegraphics[width=16.5cm]{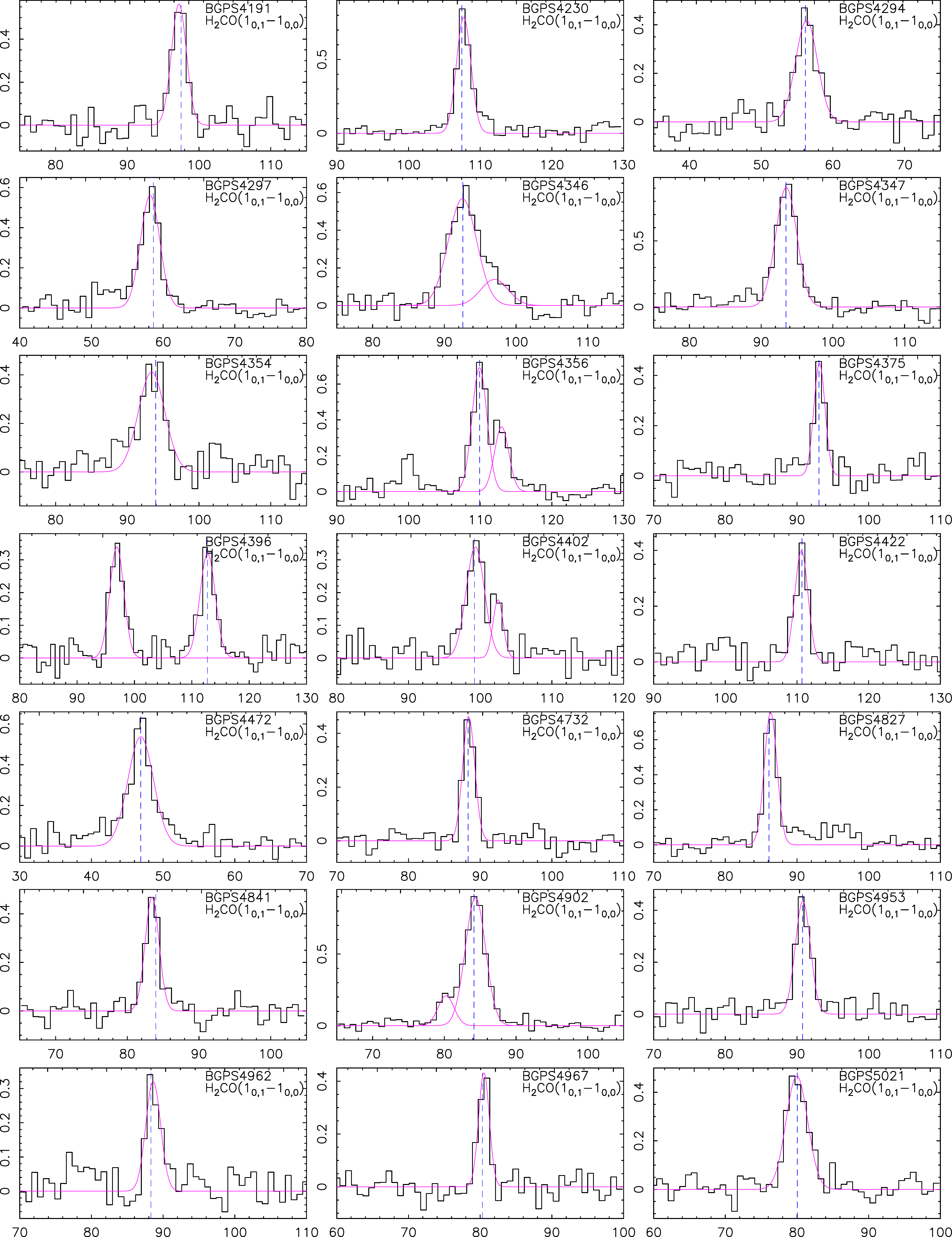}
\caption{Continued.}
\end{center}
\end{figure*}

\begin{figure*}
\addtocounter{figure}{-1}
 \begin{center}
\includegraphics[width=16.5cm]{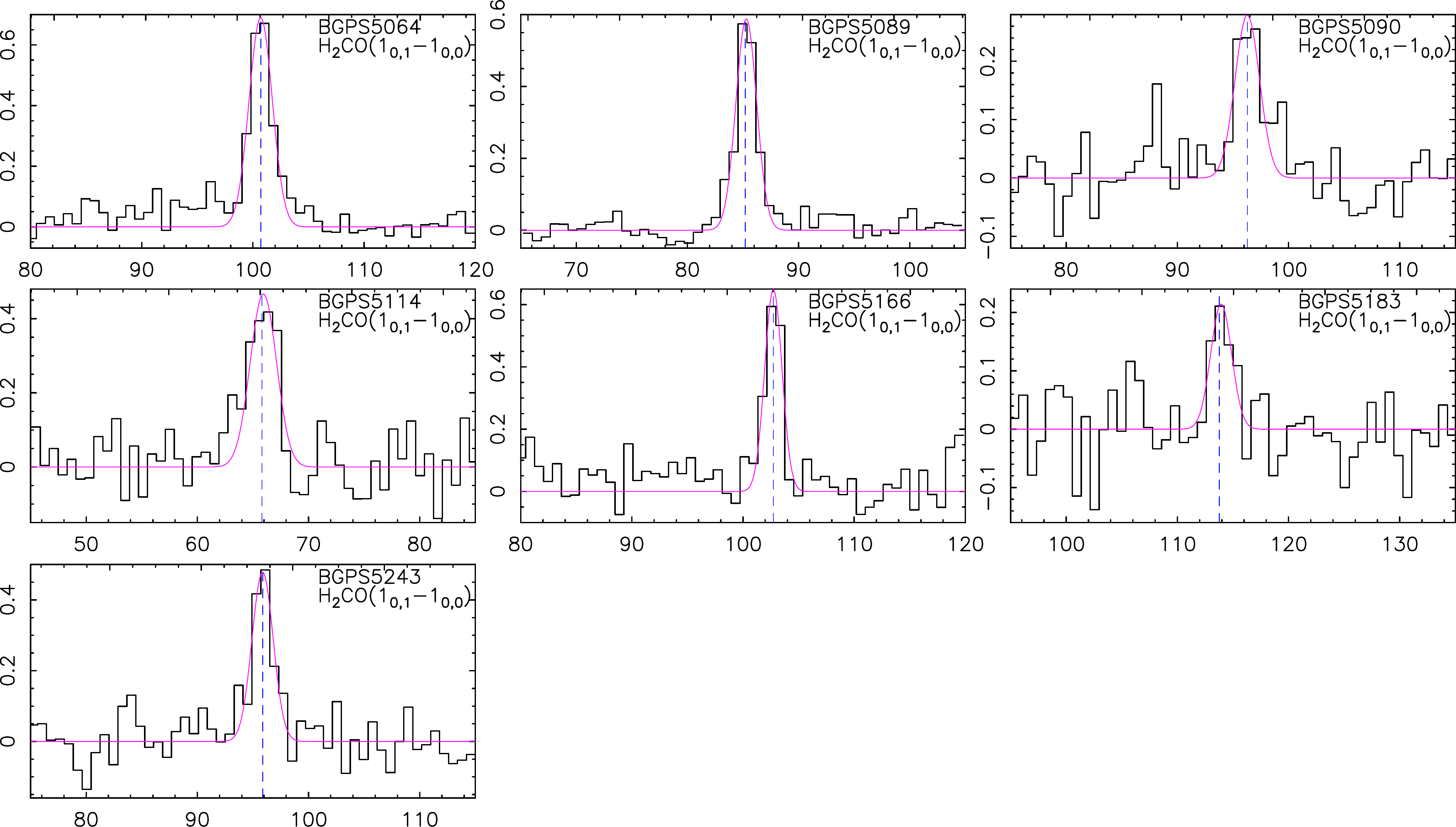}
\caption{Continued.}
\end{center}
\end{figure*}


\maxdeadcycles=1000


\begin{figure*}
 \begin{center}
\includegraphics[width=16.5cm]{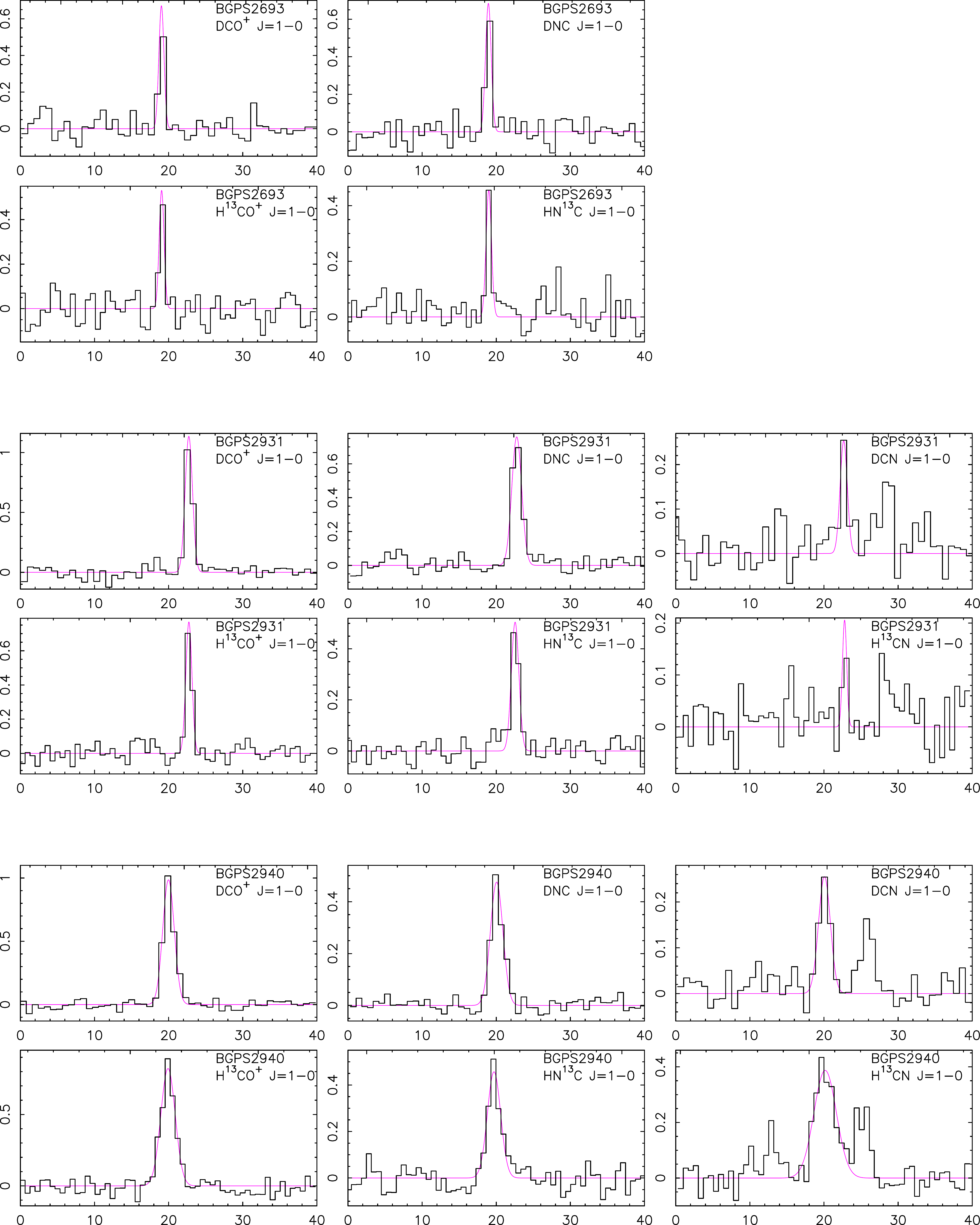}
\caption{Averaged DCO$^{+}$, H$^{13}$CO$^{+}$, DNC, HN$^{13}$C, DCN, and  H$^{13}$CN \textit{J}=1$-$0 spectra of the 11 OTF sources. These spectrum are averaged from the region enclosed by blue squares in Figures \ref{fig_BGPS2693}$-$\ref{fig_BGPS4402}. These regions represent the peak regions of the deuterated molecular lines. The Gaussian fitting result for each spectrum is shown in magenta. The x-axis is velocity in km s$^{-1}$, and the y-axis is T$_{\rm mb}$ in kelvin.}
\label{spec_column}
\end{center}
\end{figure*}

\begin{figure*}
\addtocounter{figure}{-1}
 \begin{center}
\includegraphics[width=16.5cm]{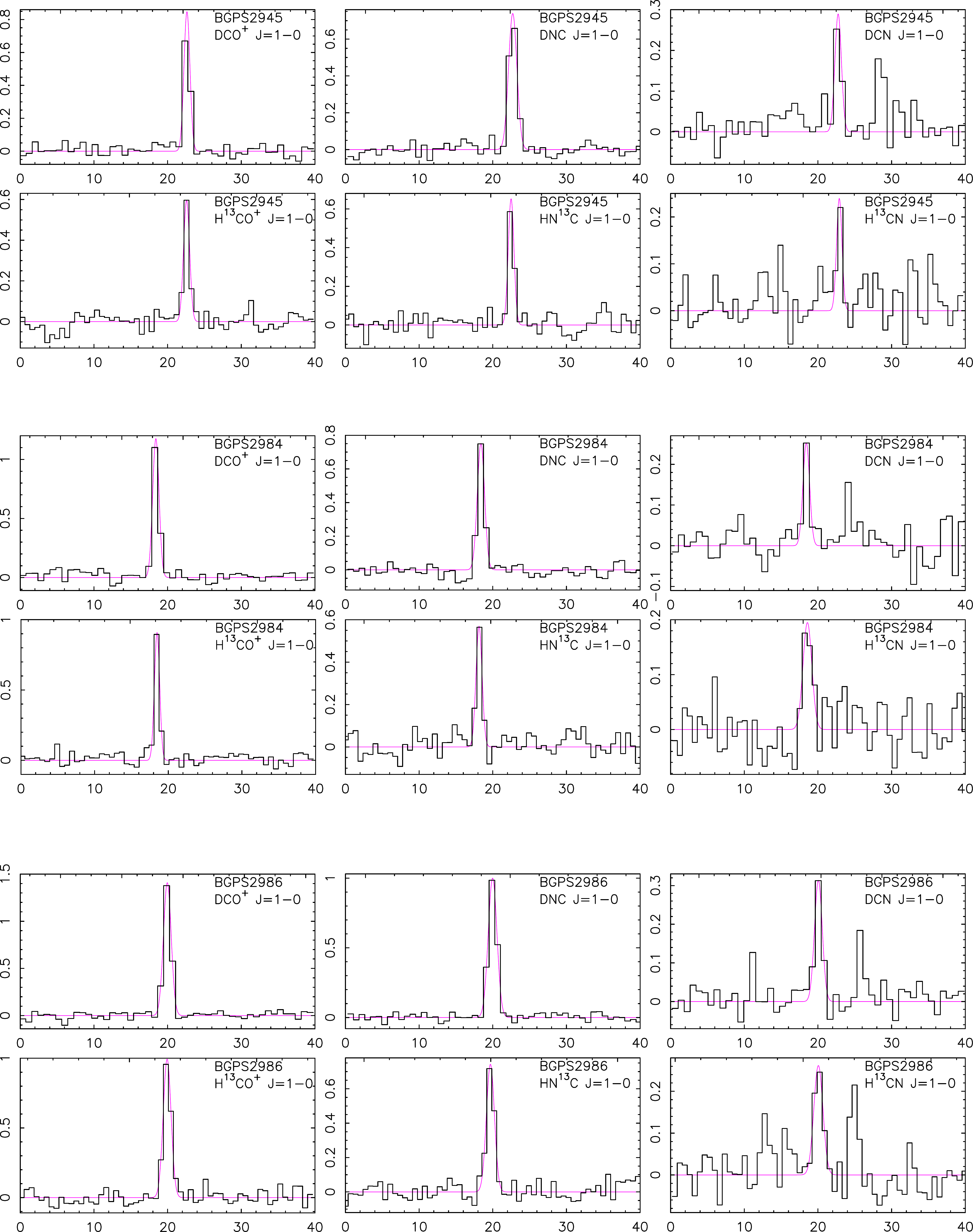}
\caption{Continued.}
\end{center}
\end{figure*}

\begin{figure*}
\addtocounter{figure}{-1}
 \begin{center}
\includegraphics[width=16.5cm]{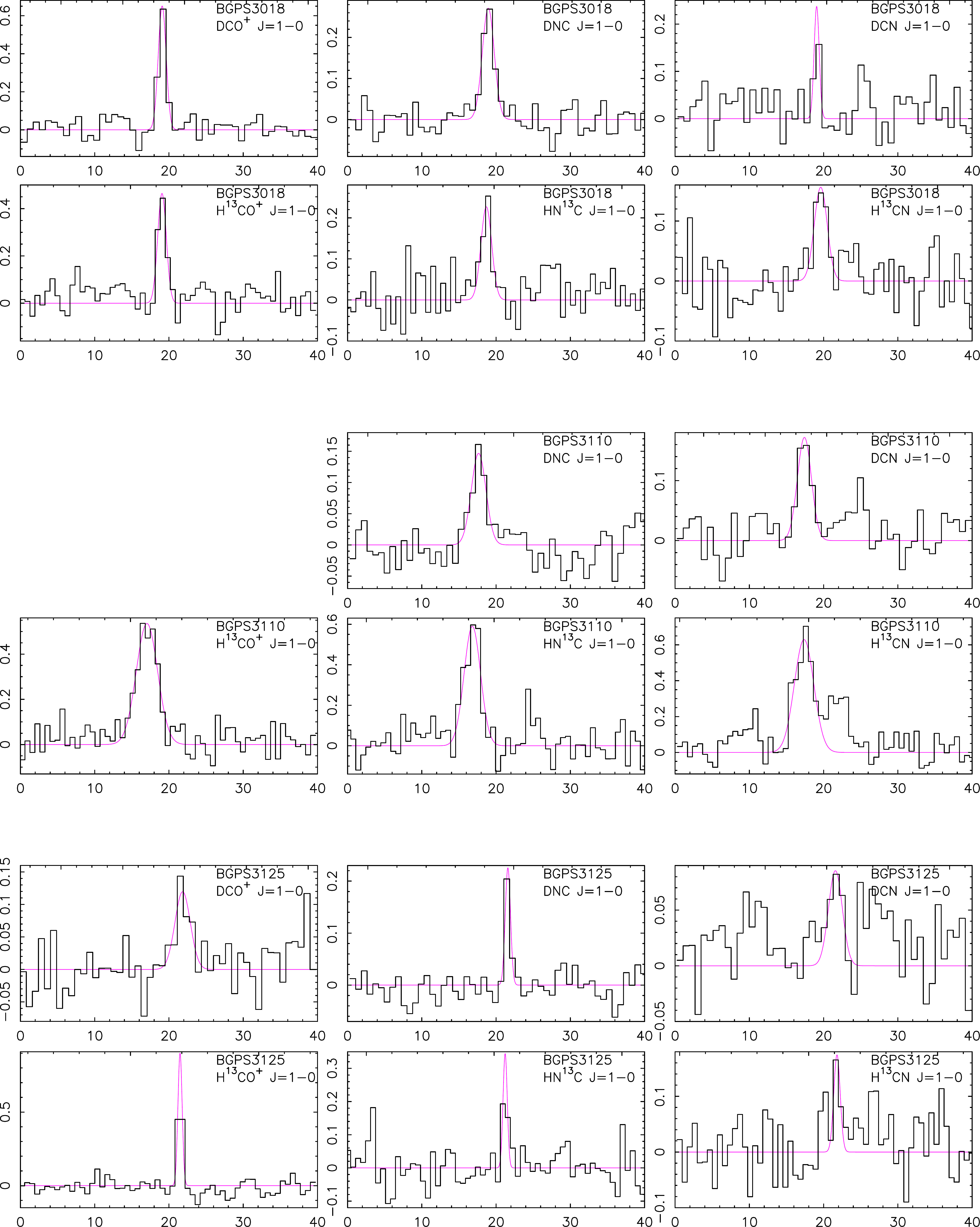}
\caption{Continued.}
\end{center}
\end{figure*}

\begin{figure*}
\addtocounter{figure}{-1}
 \begin{center}
\includegraphics[width=16.5cm]{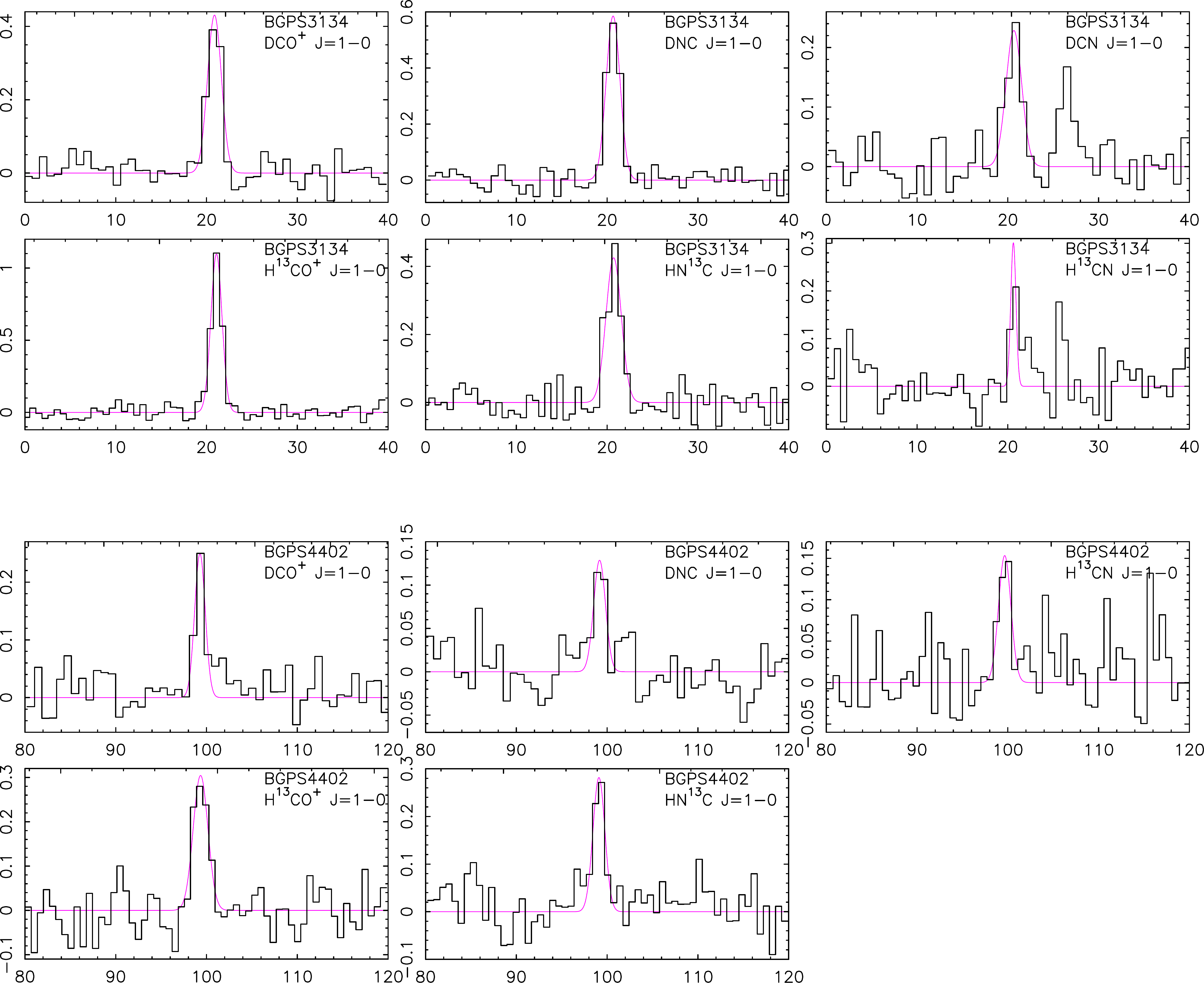}
\caption{Continued.}
\end{center}
\end{figure*}


\clearpage



\clearpage

\begin{appendix}

\section{Integrated maps of HCO$^{+}$, HCN, HNC, and N$_{2}$H$^{+}$ \textit{J}=1$-$0} \label{app1}

\begin{figure*}
\centering
\begin{tabular}{c}
\includegraphics[width=160mm]{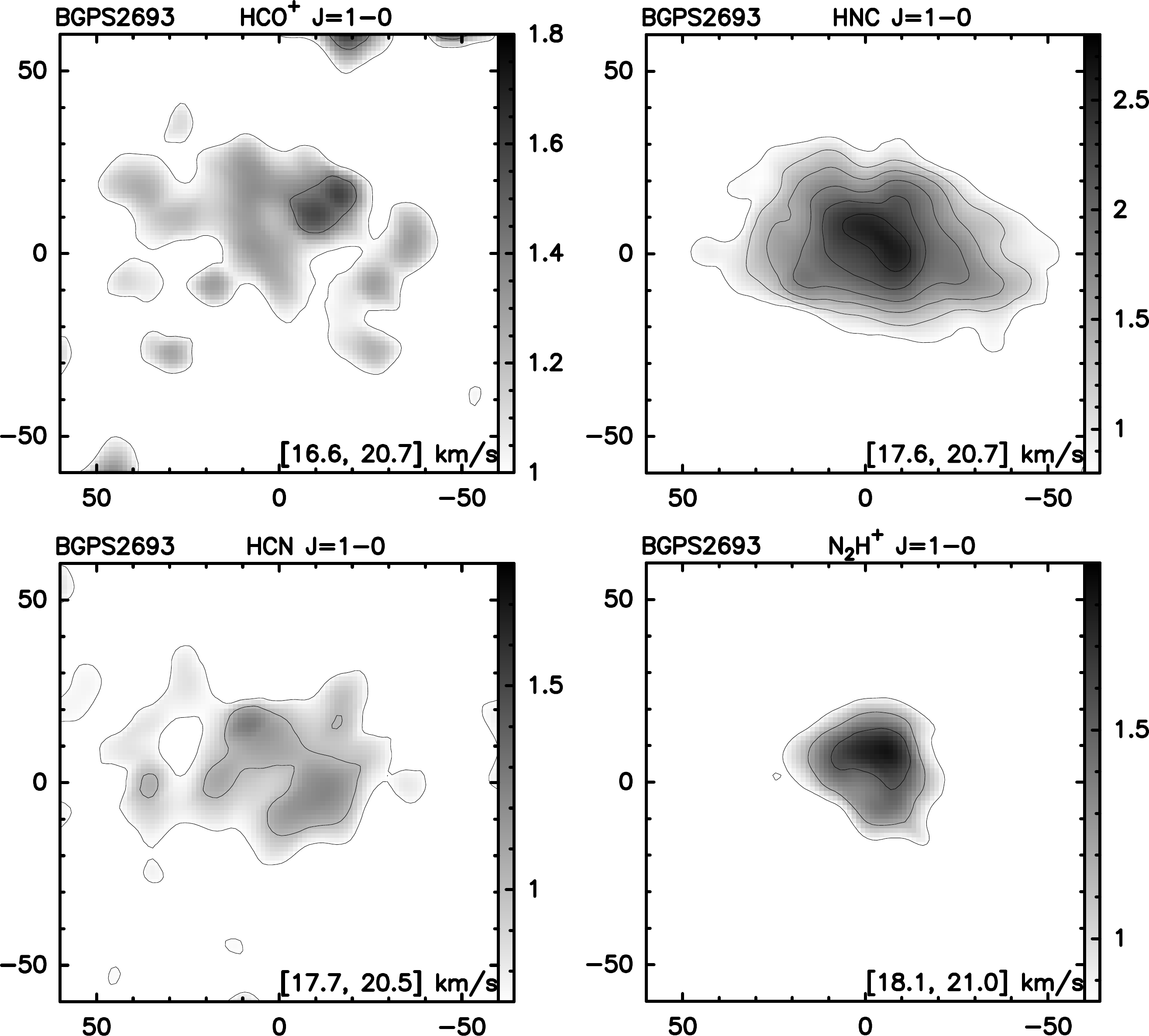}
\end{tabular}
\caption{HCO$^{+}$, HCN, HNC, and N$_{2}$H$^{+}$ \textit{J}=1$-$0 integrated maps of BGPS2693. The integrated velocity range is shown at the right-bottom corner. For HCO$^{+}$, HCN, HNC, and N$_{2}$H$^{+}$ \textit{J}=1$-$0, the contours start from 5$\sigma$ in steps of 2$\sigma$, with $\sigma$ = 0.20, 0.15, 0.16, and 0.17 K km s$^{-1}$.}
	\label{2693}
\end{figure*}

\begin{figure*}
\centering
\begin{tabular}{c}
\includegraphics[width=160mm]{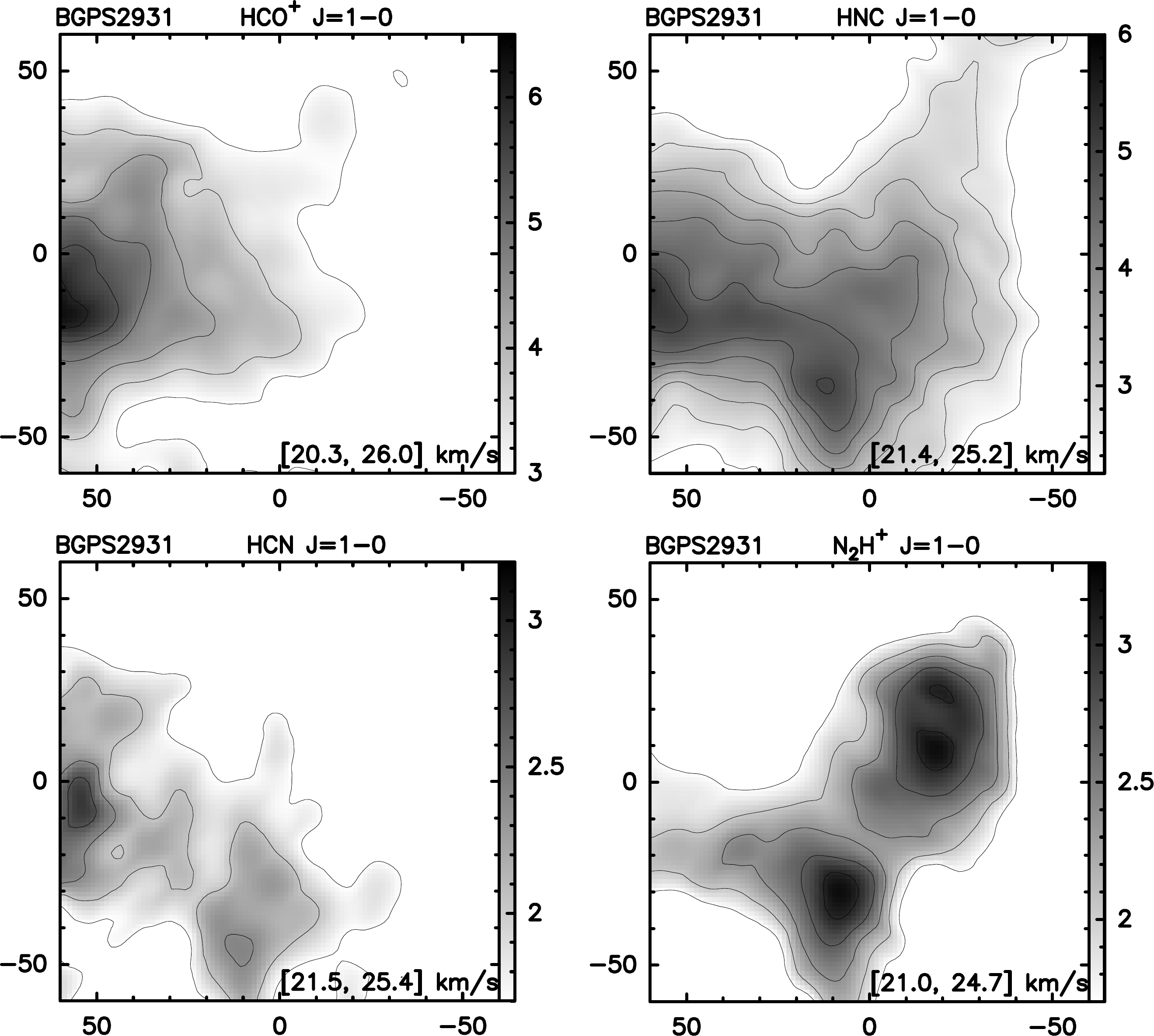}
\end{tabular}
\caption{HCO$^{+}$, HCN, HNC, and N$_{2}$H$^{+}$ \textit{J}=1$-$0 integrated maps of BGPS2931. The integrated velocity range is shown at the right-bottom corner. For HCO$^{+}$ and HNC \textit{J}=1$-$0, the contours start from 15$\sigma$ in steps of 3$\sigma$, with $\sigma$ = 0.20 and 0.15 K km s$^{-1}$. For HCN and N$_{2}$H$^{+}$ \textit{J}=1$-$0, the contours start from 10$\sigma$ in steps of 2$\sigma$, with $\sigma$ = 0.17 and 0.17 K km s$^{-1}$.}
	\label{2931}
\end{figure*}

\begin{figure*}
\centering
\begin{tabular}{c}
\includegraphics[width=160mm]{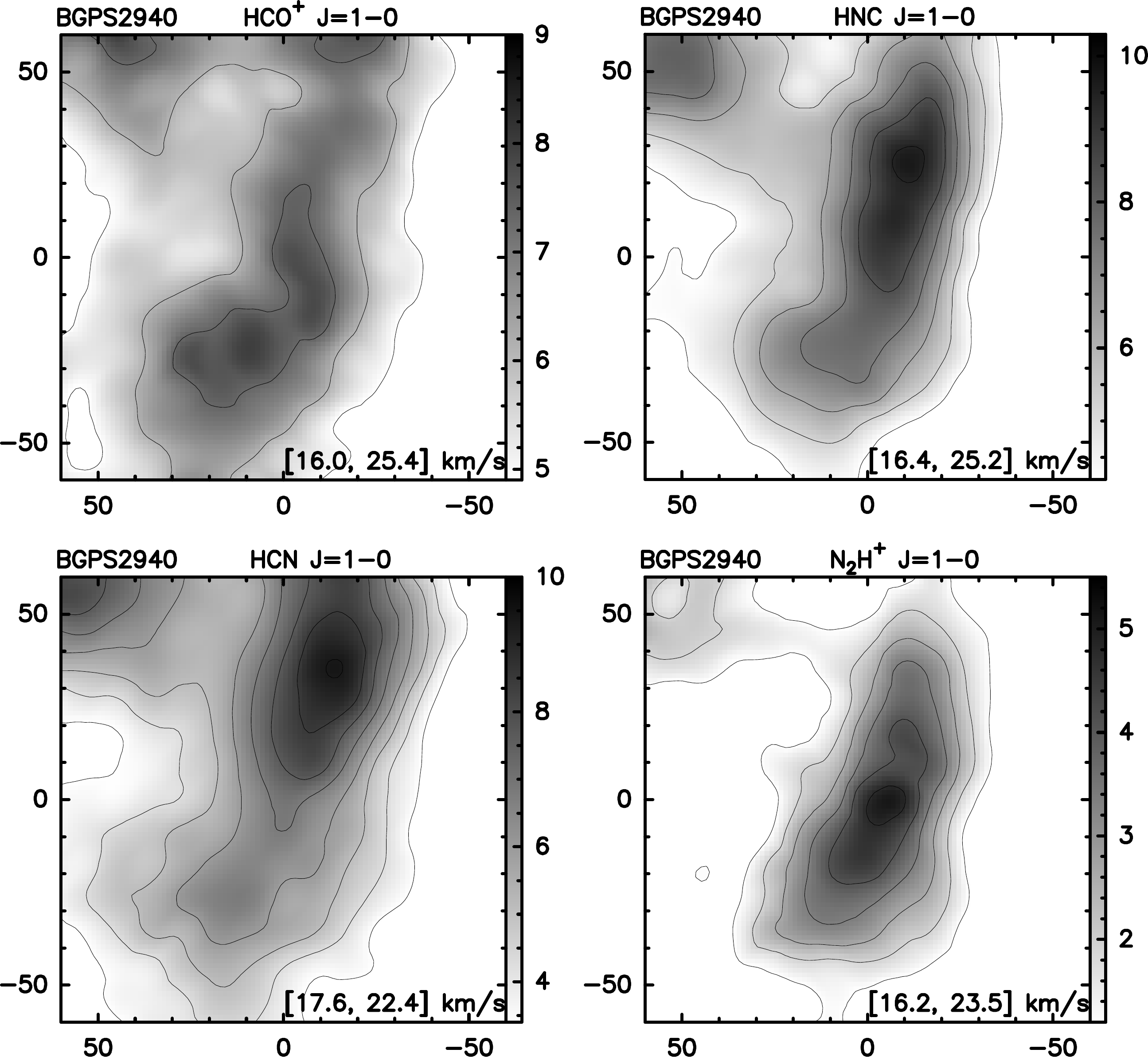}
\end{tabular}
\caption{HCO$^{+}$, HCN, HNC, and N$_{2}$H$^{+}$ \textit{J}=1$-$0 integrated maps of BGPS2940. The integrated velocity range is shown at the right-bottom corner. For HCO$^{+}$, HCN, and HNC \textit{J}=1$-$0, the contours start from 20$\sigma$ in steps of 5$\sigma$, with $\sigma$ = 0.25, 0.17, and 0.21 K km s$^{-1}$. For N$_{2}$H$^{+}$ \textit{J}=1$-$0, the contours start from 5$\sigma$ in steps of 3$\sigma$, with $\sigma$ = 0.24 K km s$^{-1}$.}
	\label{2940}
\end{figure*}

\begin{figure*}
\centering
\begin{tabular}{c}
\includegraphics[width=160mm]{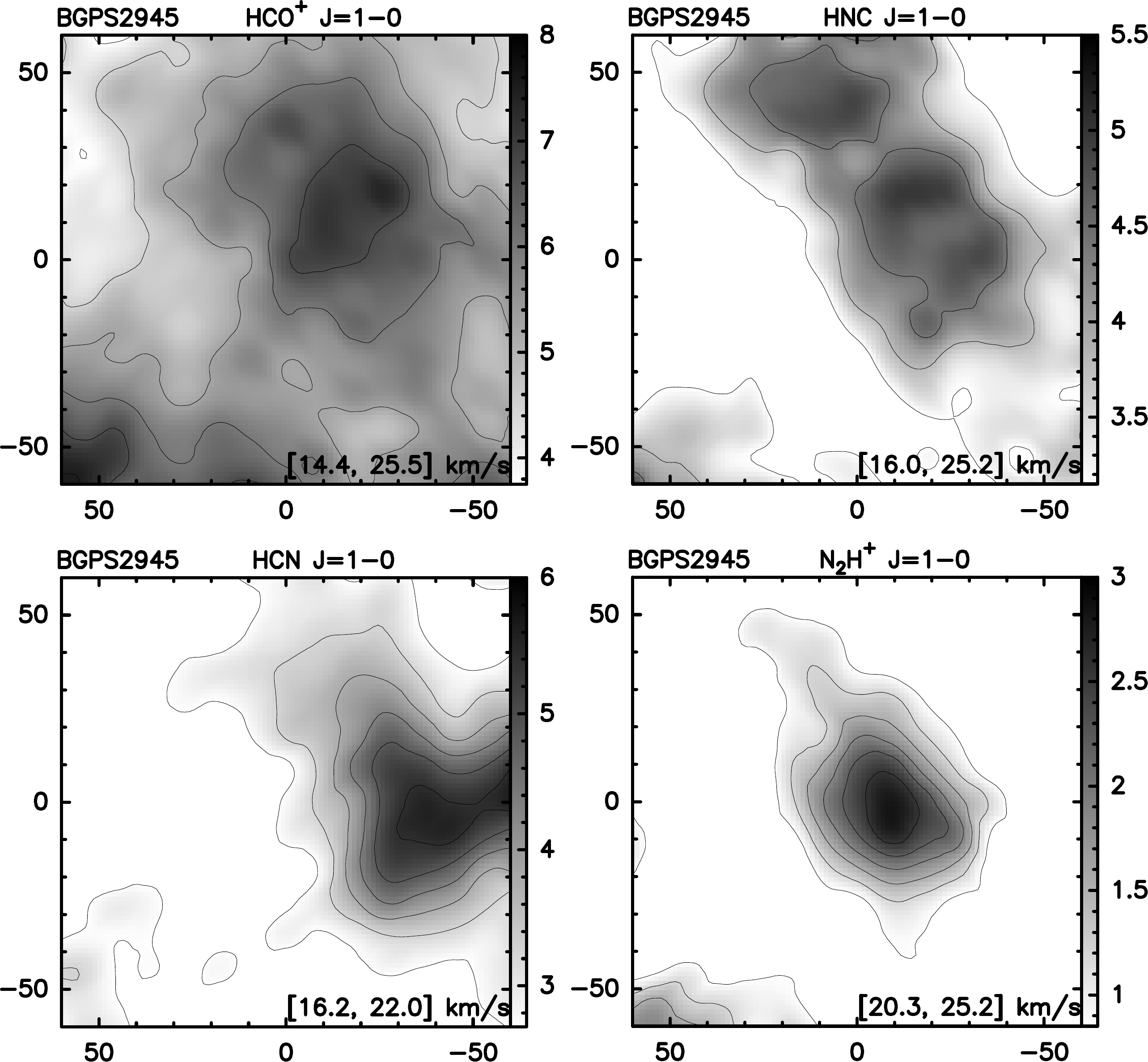}
\end{tabular}
\caption{HCO$^{+}$, HCN, HNC, and N$_{2}$H$^{+}$ \textit{J}=1$-$0 integrated maps of BGPS2945. The integrated velocity range is shown at the right-bottom corner. For HCO$^{+}$, HCN, and HNC \textit{J}=1$-$0, the contours start from 15$\sigma$ in steps of 3$\sigma$, with $\sigma$ = 0.25, 0.18, and 0.21 K km s$^{-1}$. For N$_{2}$H$^{+}$ \textit{J}=1$-$0, the contours start from 5$\sigma$ in steps of 2$\sigma$, with $\sigma$ = 0.17 K km s$^{-1}$.}
	\label{2945}
\end{figure*}

\begin{figure*}
\centering
\begin{tabular}{c}
\includegraphics[width=160mm]{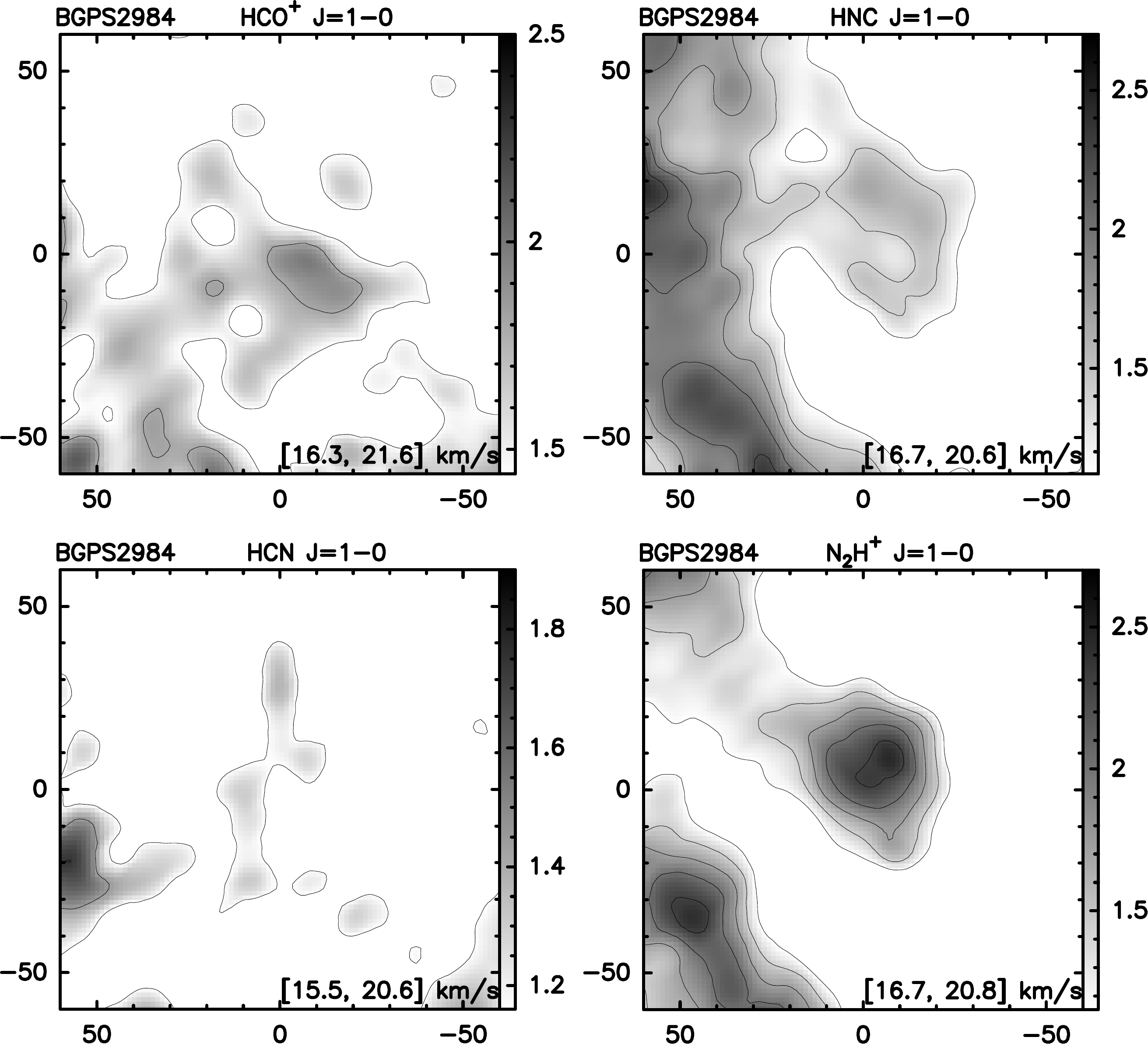}
\end{tabular}
\caption{HCO$^{+}$, HCN, HNC, and N$_{2}$H$^{+}$ \textit{J}=1$-$0 integrated maps of BGPS2984. The integrated velocity range is shown at the right-bottom corner. For HCO$^{+}$, HCN, HNC, and N$_{2}$H$^{+}$ \textit{J}=1$-$0, the contours start from 8$\sigma$ in steps of 2$\sigma$, with $\sigma$ = 0.18, 0.17, 0.14, and 0.15 K km s$^{-1}$.}
	\label{2984}
\end{figure*}

\begin{figure*}
\centering
\begin{tabular}{c}
\includegraphics[width=160mm]{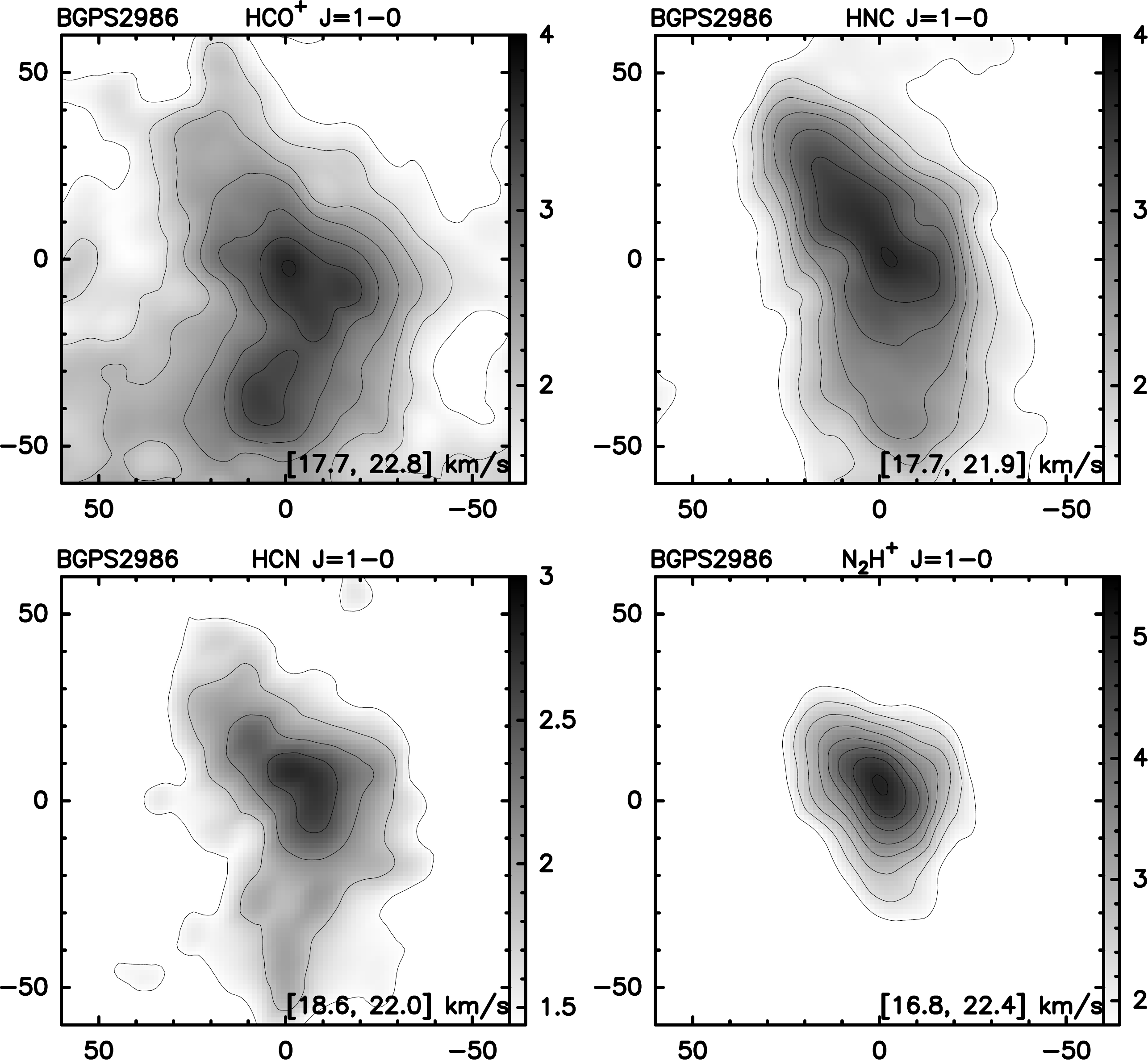}
\end{tabular}
\caption{HCO$^{+}$, HCN, HNC, and N$_{2}$H$^{+}$ \textit{J}=1$-$0 integrated maps of BGPS2986. The integrated velocity range is shown at the right-bottom corner. For HCO$^{+}$, HCN, HNC, and N$_{2}$H$^{+}$ \textit{J}=1$-$0, the contours start from 12$\sigma$ in steps of 3$\sigma$, with $\sigma$ = 0.12, 0.12, 0.12, and 0.15 K km s$^{-1}$.}
	\label{2986}
\end{figure*}

\begin{figure*}
\centering
\begin{tabular}{c}
\includegraphics[width=160mm]{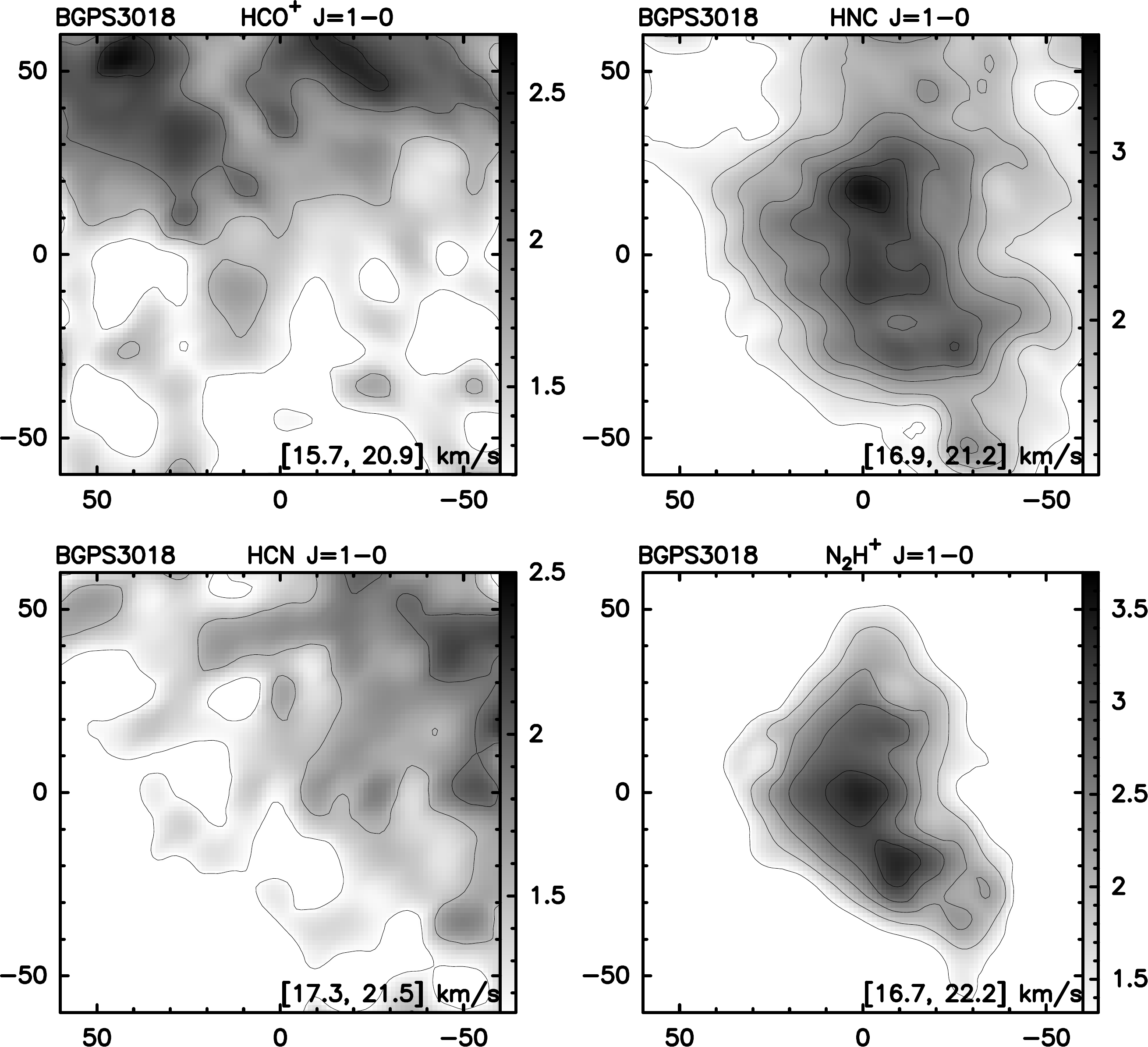}
\end{tabular}
\caption{HCO$^{+}$, HCN, HNC, and N$_{2}$H$^{+}$ \textit{J}=1$-$0 integrated maps of BGPS3018. The integrated velocity range is shown at the right-bottom corner. For HCO$^{+}$, HCN, HNC, and N$_{2}$H$^{+}$ \textit{J}=1$-$0, the contours start from 6$\sigma$ in steps of 2$\sigma$, with $\sigma$ = 0.20, 0.19, 0.18, and 0.22 K km s$^{-1}$.}
	\label{3018}
\end{figure*}

\begin{figure*}
\centering
\begin{tabular}{c}
\includegraphics[width=160mm]{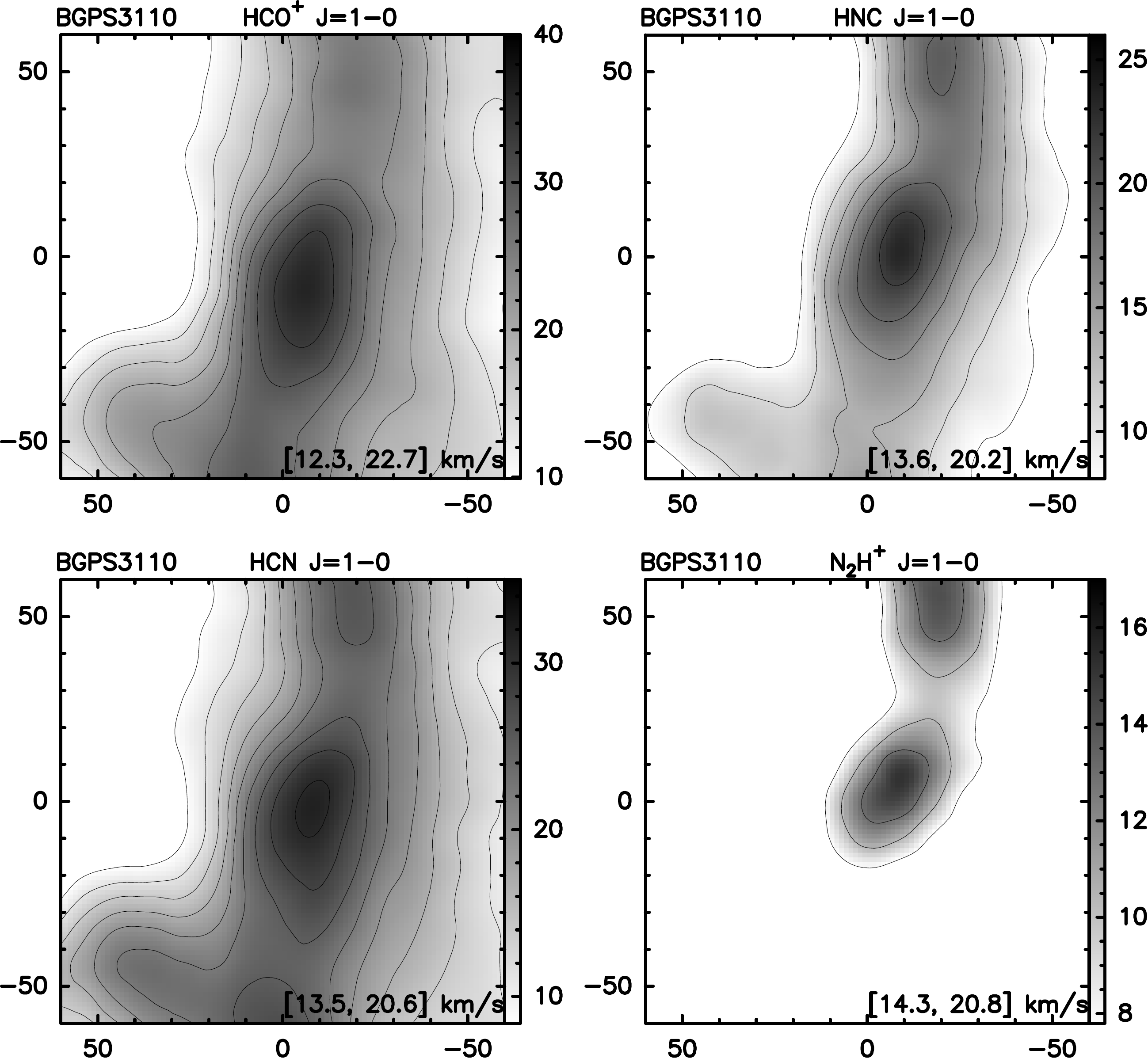}
\end{tabular}
\caption{HCO$^{+}$, HCN, HNC, and N$_{2}$H$^{+}$ \textit{J}=1$-$0 integrated maps of BGPS3110. The integrated velocity range is shown at the right-bottom corner. For HCO$^{+}$, HCN, HNC, and N$_{2}$H$^{+}$ \textit{J}=1$-$0, the contours start from 30$\sigma$ in steps of 10$\sigma$, with $\sigma$ = 0.33, 0.28, 0.27, and 0.26 K km s$^{-1}$.}
	\label{3110}
\end{figure*}

\begin{figure*}
\centering
\begin{tabular}{c}
\includegraphics[width=160mm]{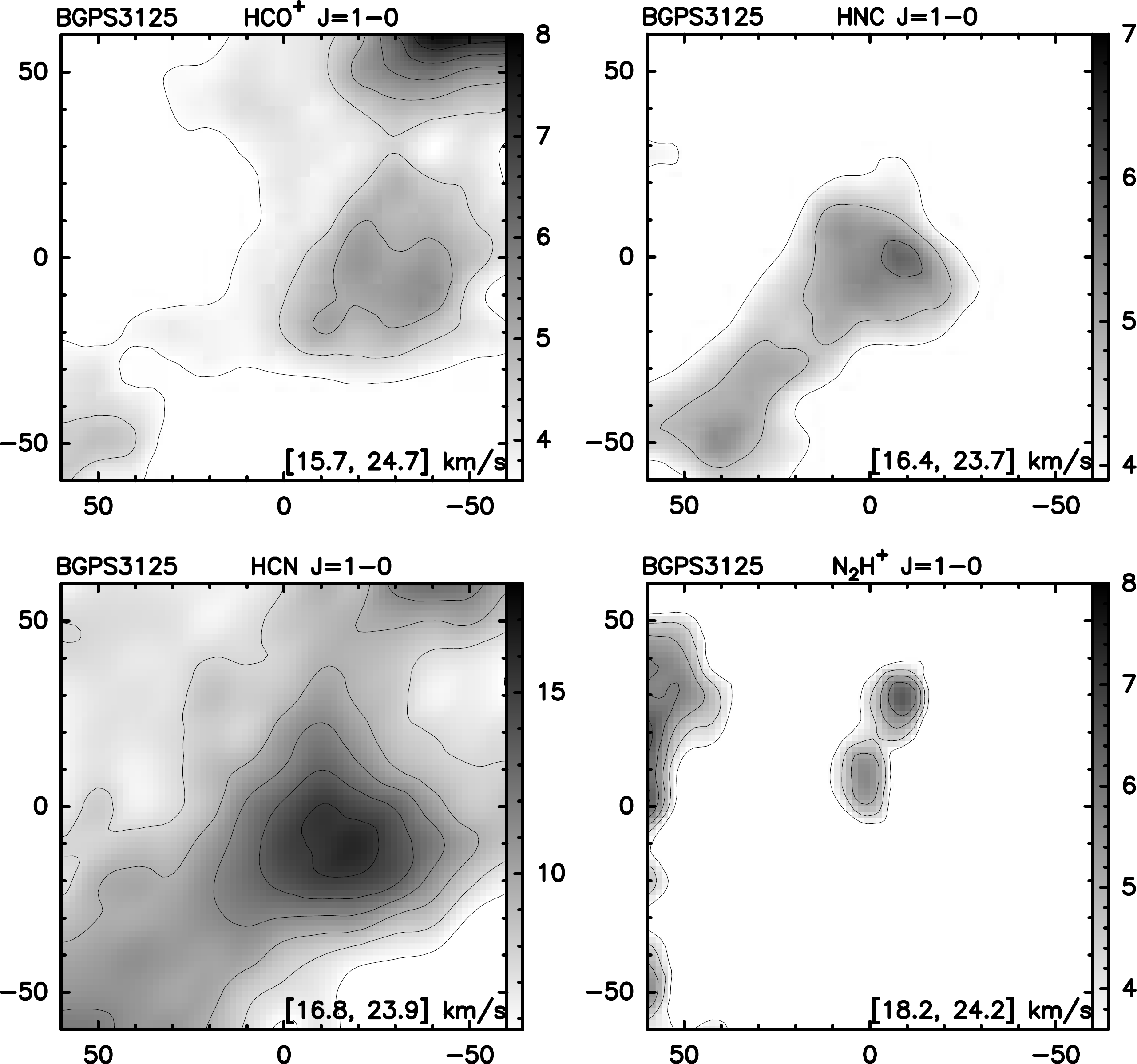}
\end{tabular}
\caption{HCO$^{+}$, HCN, HNC, and N$_{2}$H$^{+}$ \textit{J}=1$-$0 integrated maps of BGPS3125. The integrated velocity range is shown at the right-bottom corner. For HCO$^{+}$, HNC, and N$_{2}$H$^{+}$ \textit{J}=1$-$0, the contours start from 15$\sigma$ in steps of 3$\sigma$, with $\sigma$ = 0.24, 0.26, and 0.24 K km s$^{-1}$. For HCN \textit{J}=1$-$0, the contours start from 30$\sigma$ in steps of 10$\sigma$, with $\sigma$ = 0.19 K km s$^{-1}$.}
	\label{3125}
\end{figure*}

\begin{figure*}
\centering
\begin{tabular}{c}
\includegraphics[width=160mm]{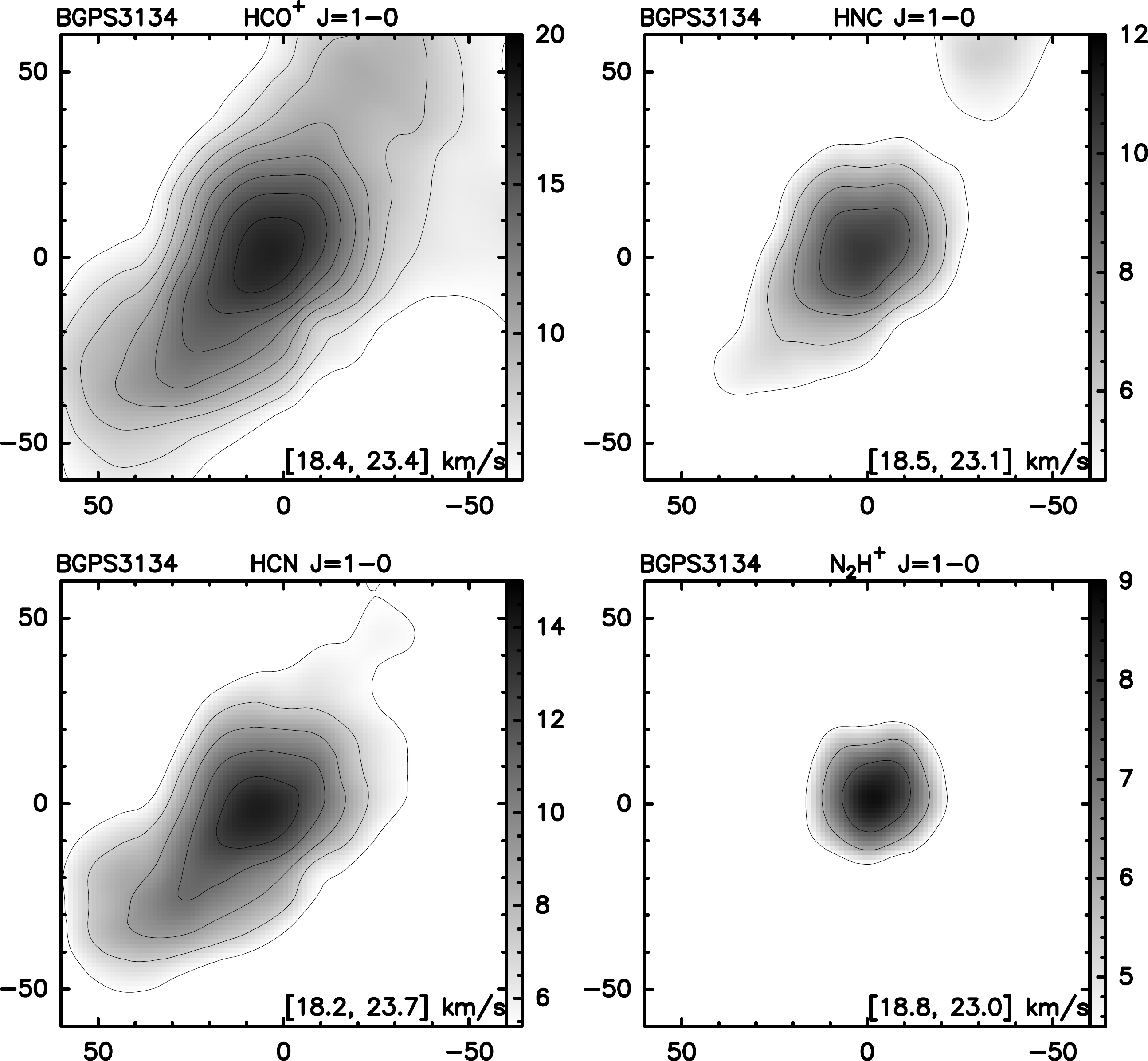}
\end{tabular}
\caption{HCO$^{+}$, HCN, HNC, and N$_{2}$H$^{+}$ \textit{J}=1$-$0 integrated maps of BGPS3134. The integrated velocity range is shown at the right-bottom corner. For HCO$^{+}$, HCN, HNC, and N$_{2}$H$^{+}$ \textit{J}=1$-$0, the contours start from 30$\sigma$ in steps of 10$\sigma$, with $\sigma$ = 0.17, 0.18, 0.15, and 0.15 K km s$^{-1}$.}
	\label{3134}
\end{figure*}

\begin{figure*}
\centering
\begin{tabular}{c}
\includegraphics[width=160mm]{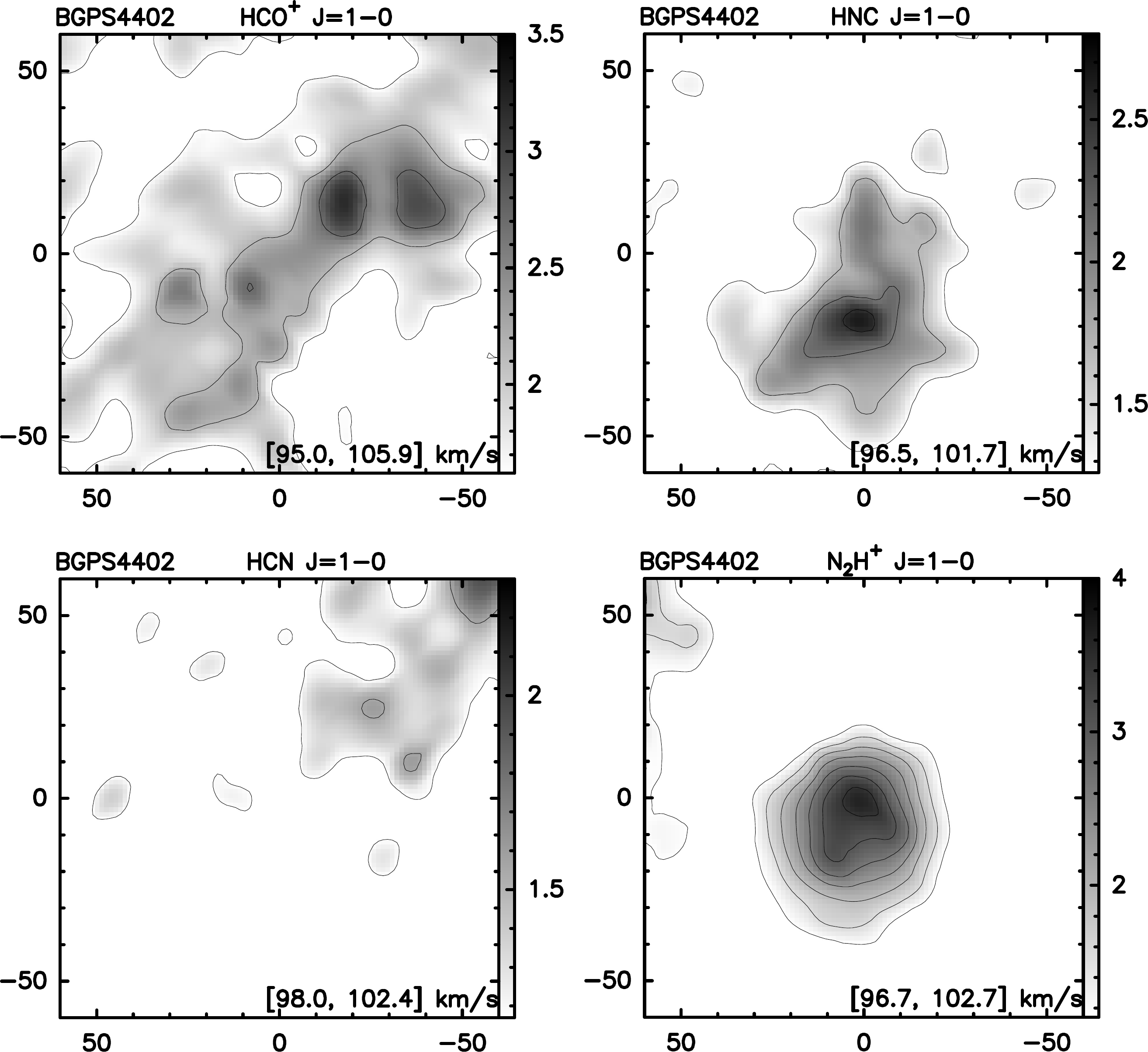}
\end{tabular}
\caption{HCO$^{+}$, HCN, HNC, and N$_{2}$H$^{+}$ \textit{J}=1$-$0 integrated maps of BGPS4402. The integrated velocity range is shown at the right-bottom corner. For HCO$^{+}$, HCN, HNC, and N$_{2}$H$^{+}$ \textit{J}=1$-$0, the contours start from 6$\sigma$ in steps of 2$\sigma$, with $\sigma$ = 0.27, 0.19, 0.21, and 0.19 K km s$^{-1}$.}
	\label{4402}
\end{figure*}

In this section, we present the HCO$^{+}$, HCN, HNC, and N$_{2}$H$^{+}$ \textit{J}=1$-$0 integrated maps for each source in the OTF observations in Figures \ref{2693}$-$\ref{4402}.

\clearpage


\clearpage

\section{Velocity field and line width distribution of sources in the OTF observation} \label{app2}

\begin{figure*}
\centering
\begin{tabular}{cc}
\includegraphics[width=82mm]{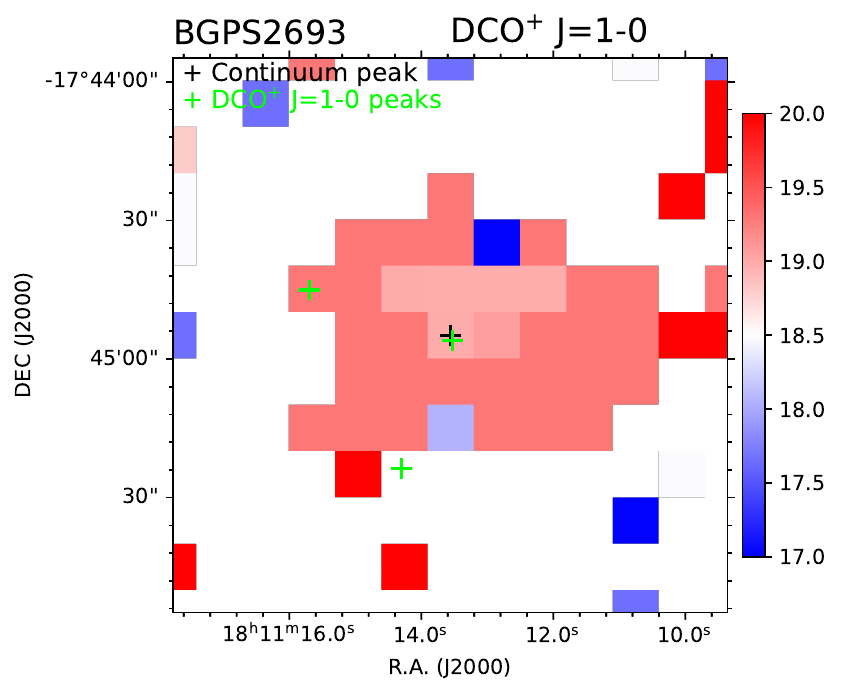}&
\includegraphics[width=82mm]{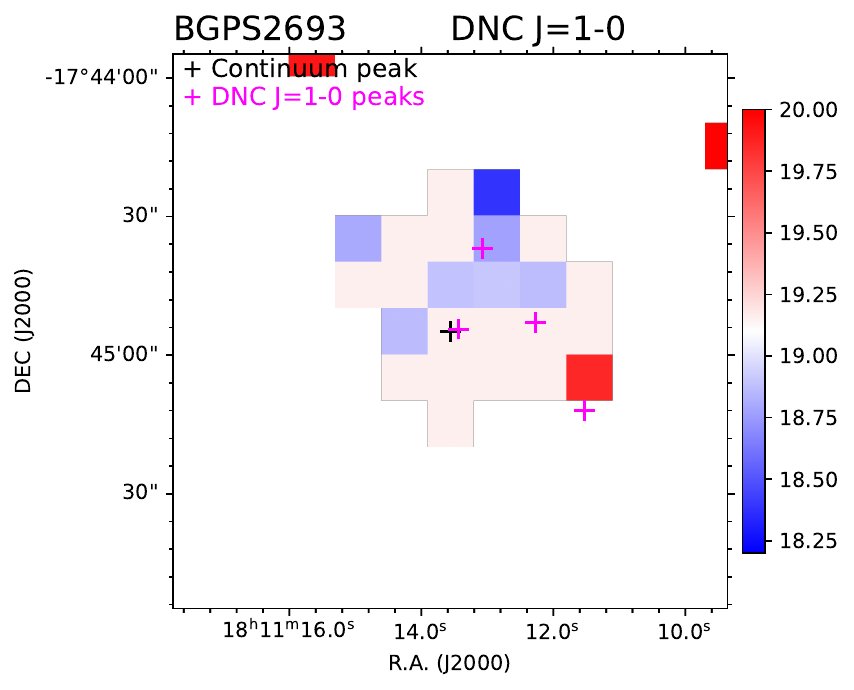}
\end{tabular}
\caption{Line width distribution of deuterated molecules for BGPS2693.}
\label{2693_velo}
\end{figure*}

\begin{figure*}
\centering
\begin{tabular}{cc}
\includegraphics[width=82mm]{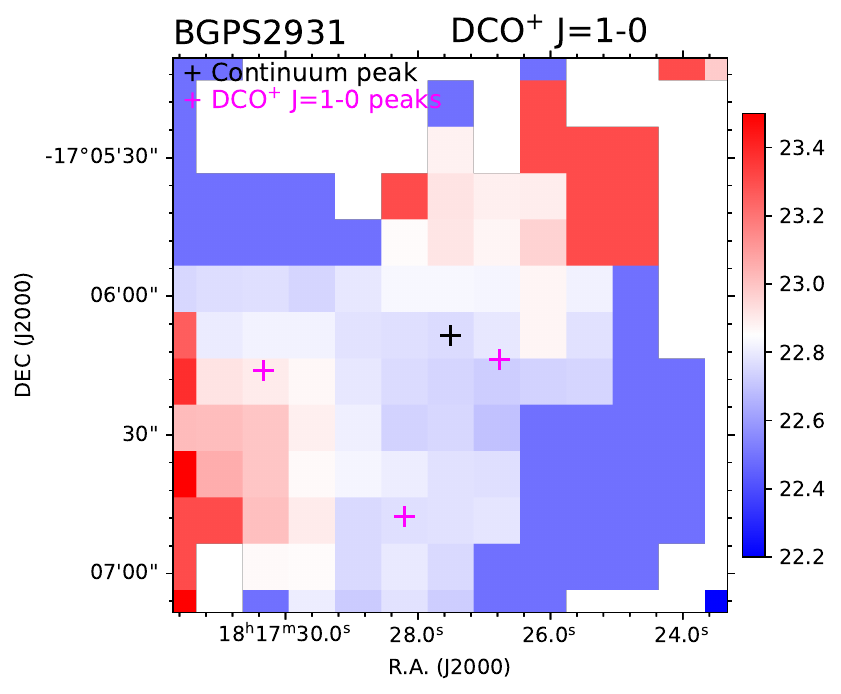}&
\includegraphics[width=82mm]{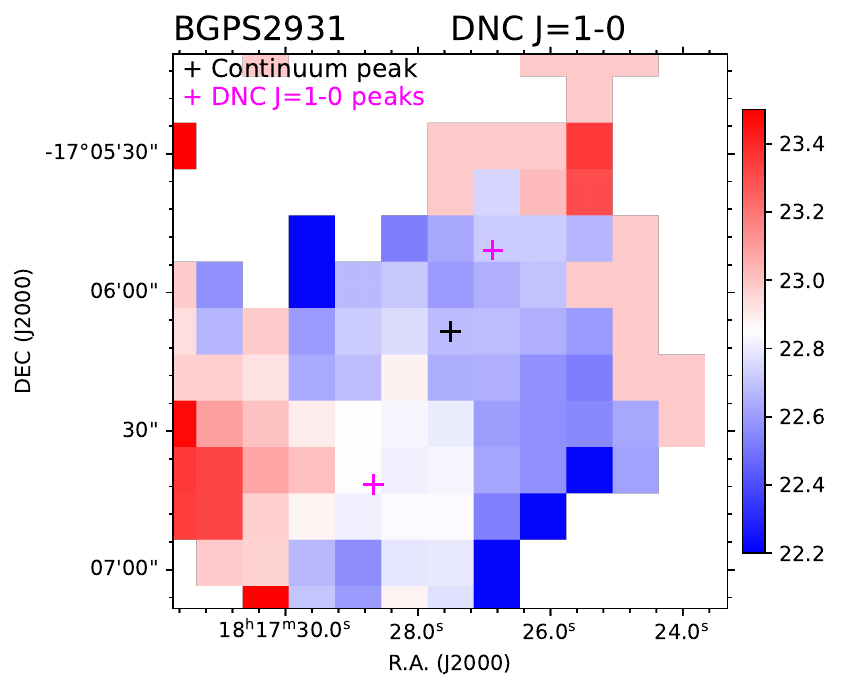}
\end{tabular}
\caption{Line width distribution of deuterated molecules for BGPS2931.}
\label{2931_velo}
\end{figure*}

\begin{figure*}
\centering
\begin{tabular}{cc}
\includegraphics[width=82mm]{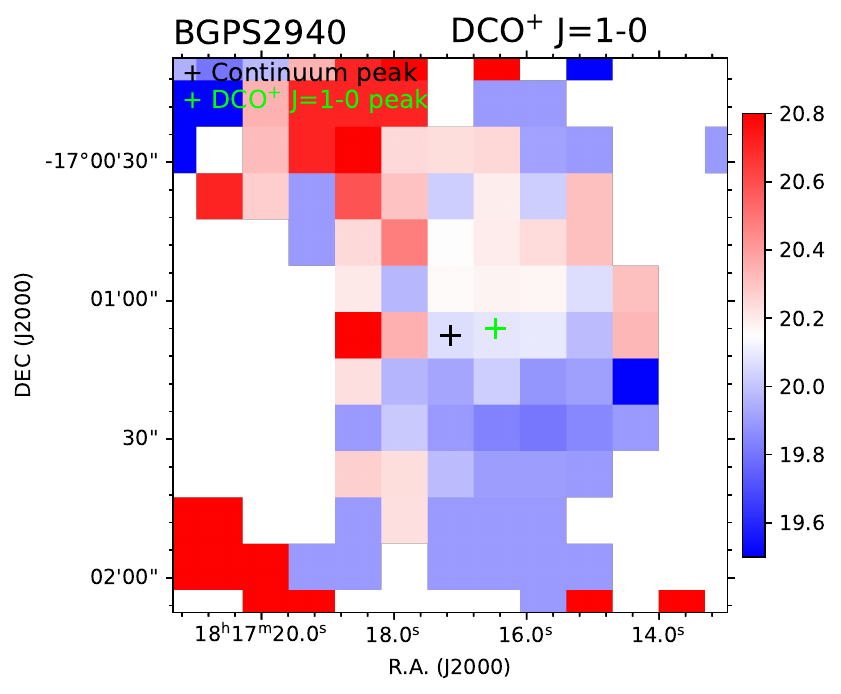}&
\includegraphics[width=82mm]{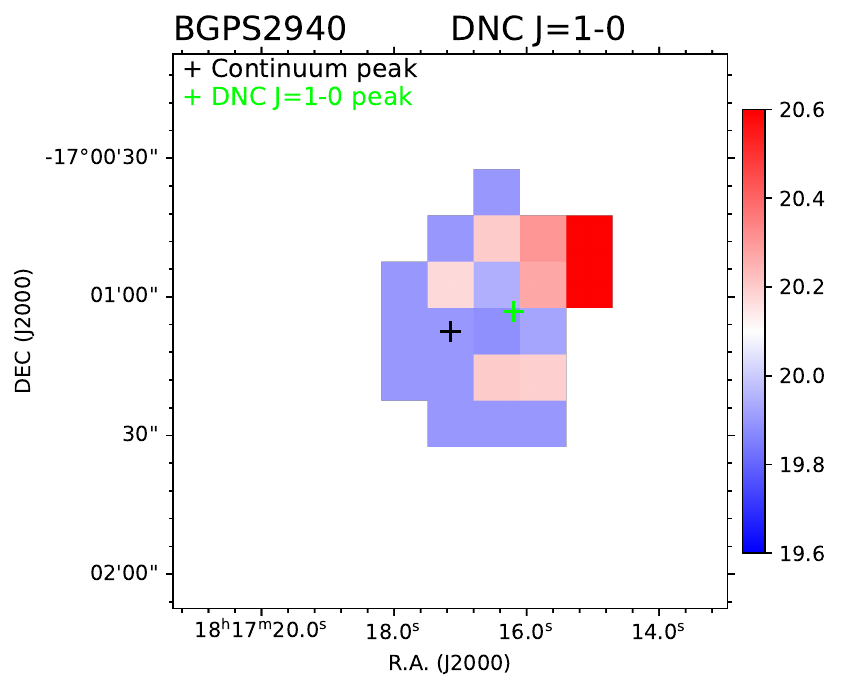}\\
\includegraphics[width=82mm]{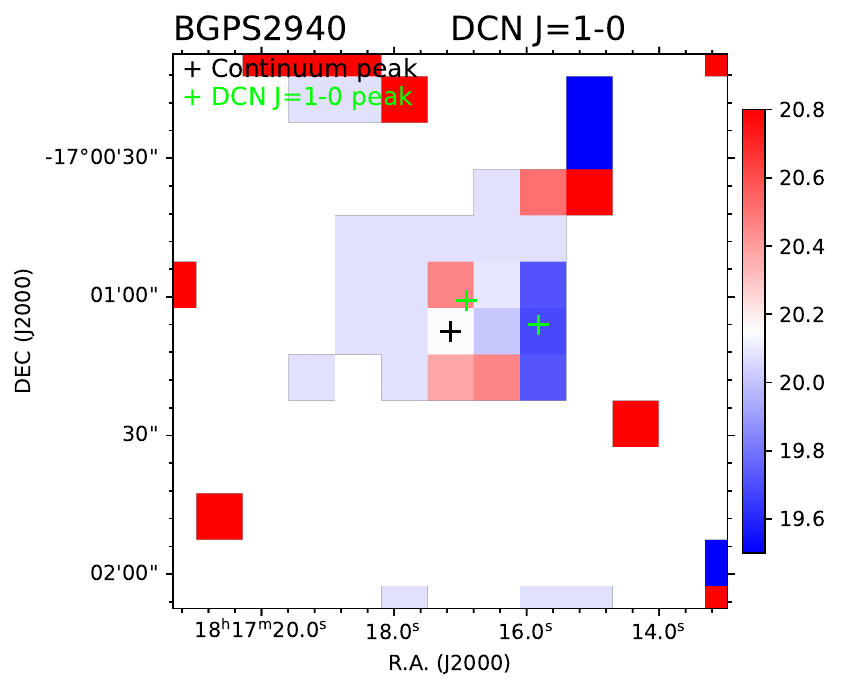}&
\includegraphics[width=82mm]{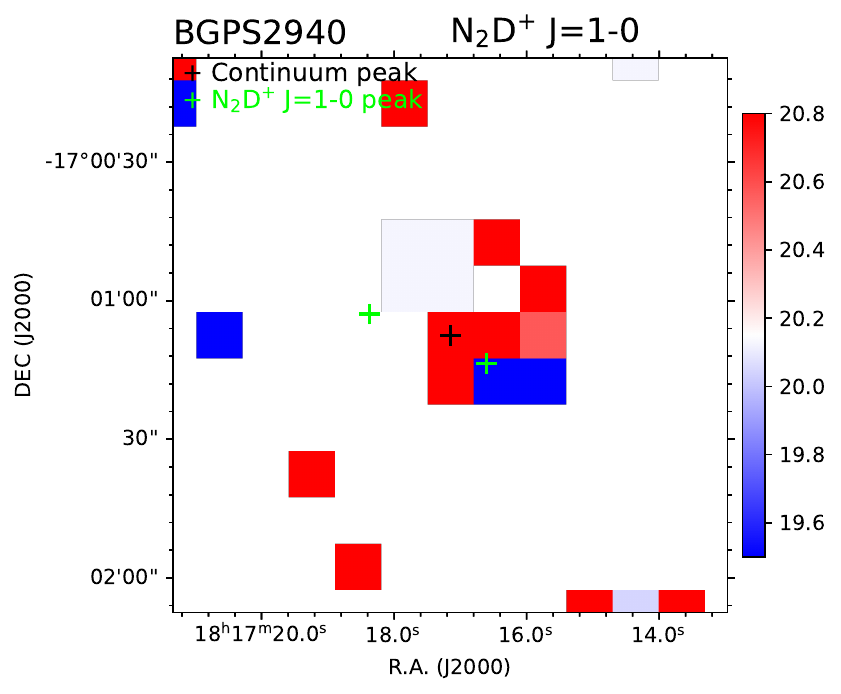}
\end{tabular}
\caption{Line width distribution of deuterated molecules for BGPS2940.}
\label{2940_velo}
\end{figure*}

\begin{figure*}
\centering
\begin{tabular}{cc}
\includegraphics[width=82mm]{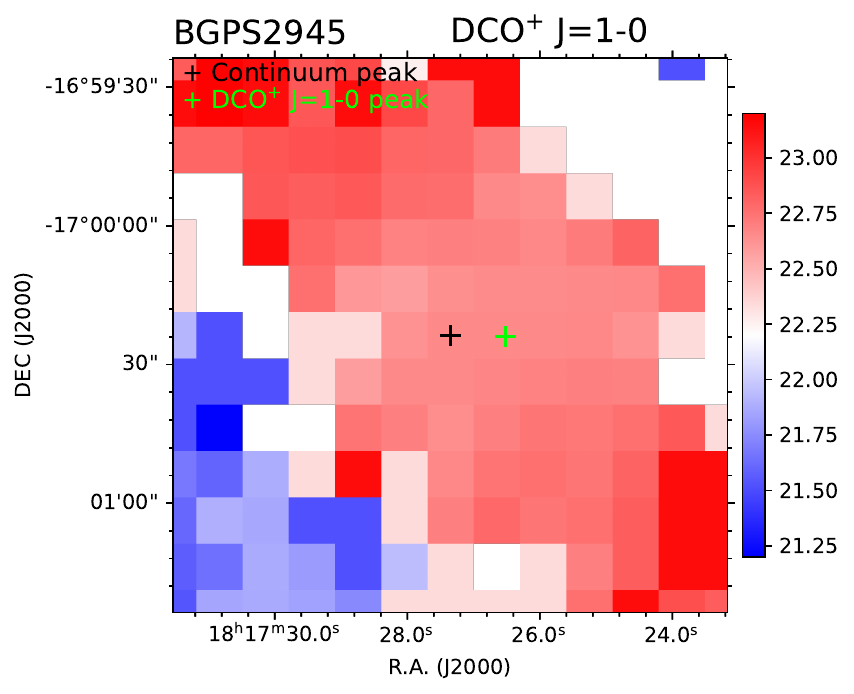}&
\includegraphics[width=82mm]{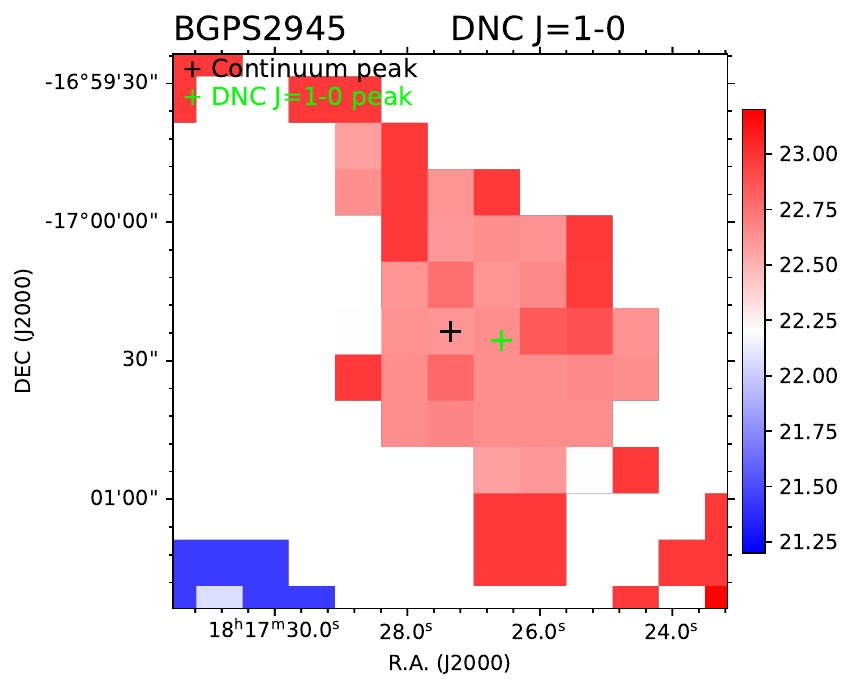}
\end{tabular}
\caption{Line width distribution of deuterated molecules for BGPS2945.}
\label{2945_velo}
\end{figure*}

\begin{figure*}
\centering
\begin{tabular}{cc}
\includegraphics[width=82mm]{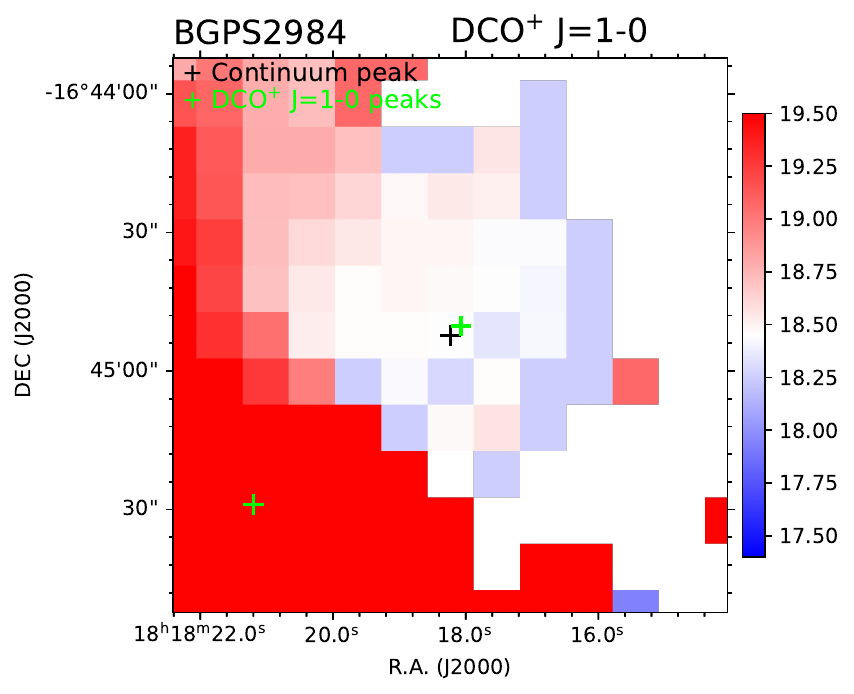}&
\includegraphics[width=82mm]{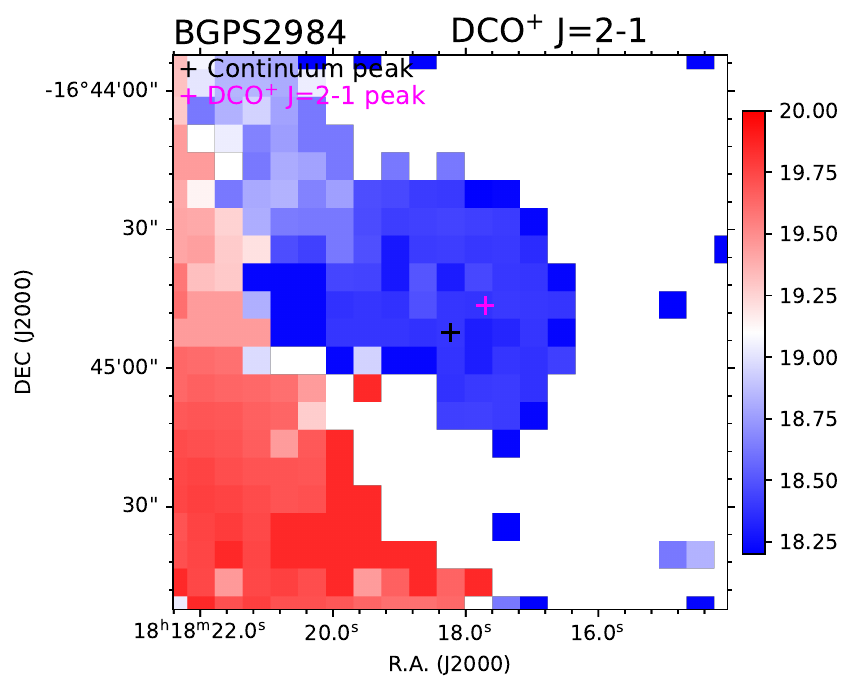}\\
\includegraphics[width=82mm]{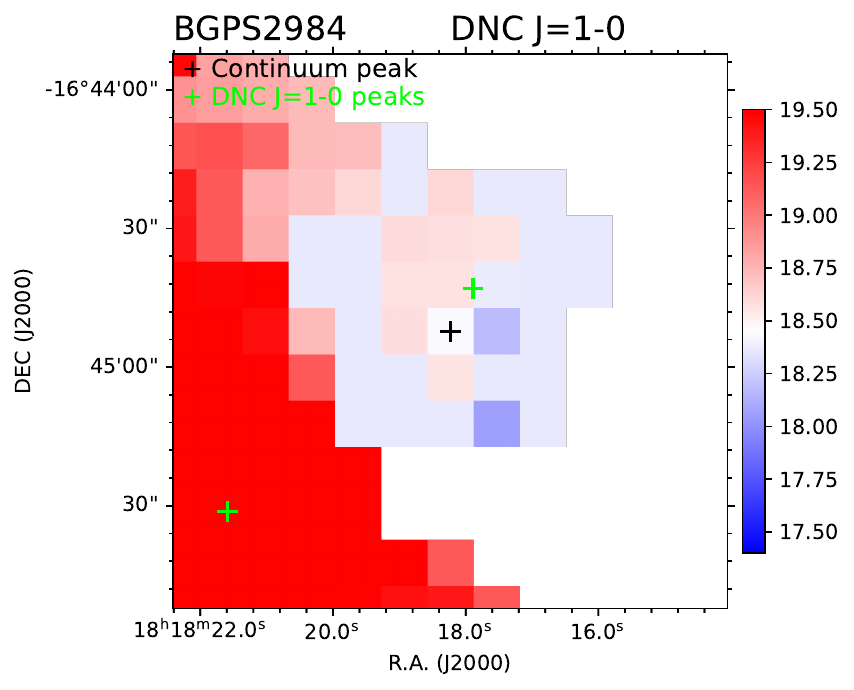}&
\end{tabular}
\caption{Line width distribution of deuterated molecules for BGPS2984.}
\label{2984_velo}
\end{figure*}

\begin{figure*}
\centering
\begin{tabular}{cc}
\includegraphics[width=82mm]{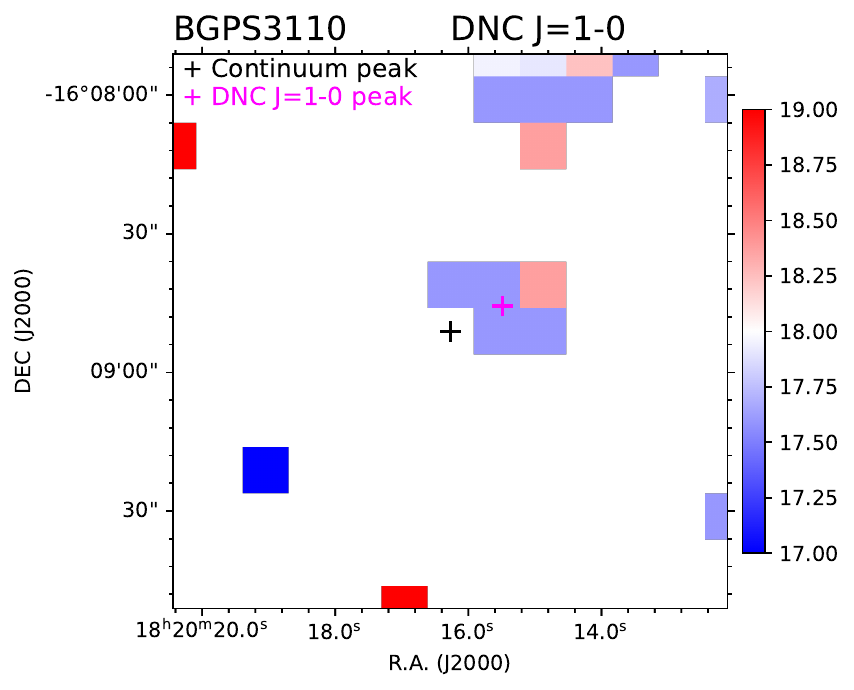}&
\includegraphics[width=82mm]{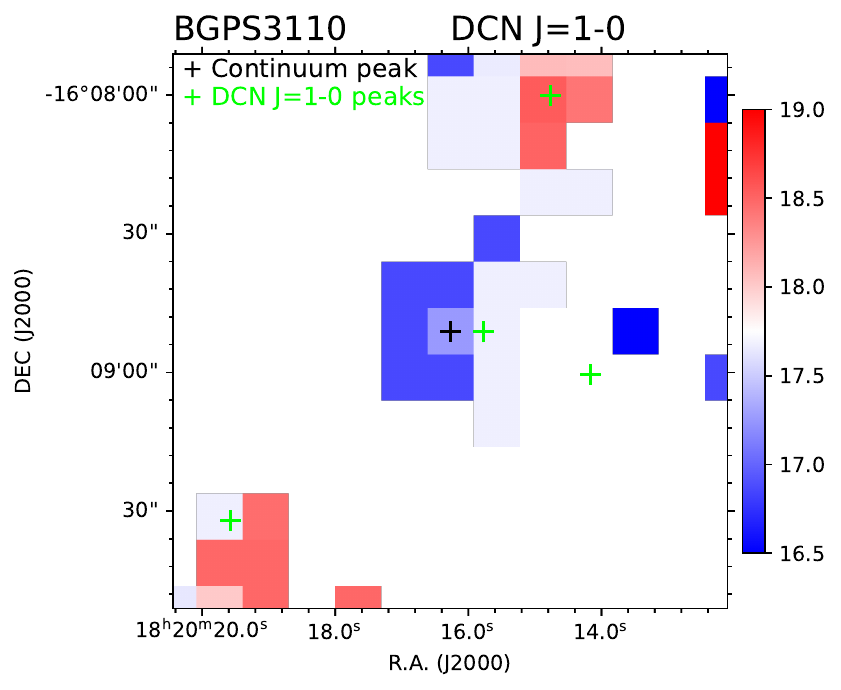}
\end{tabular}
\caption{Line width distribution of deuterated molecules for BGPS3110.}
\label{3110_velo}
\end{figure*}

\begin{figure*}
\centering
\begin{tabular}{cc}
\includegraphics[width=82mm]{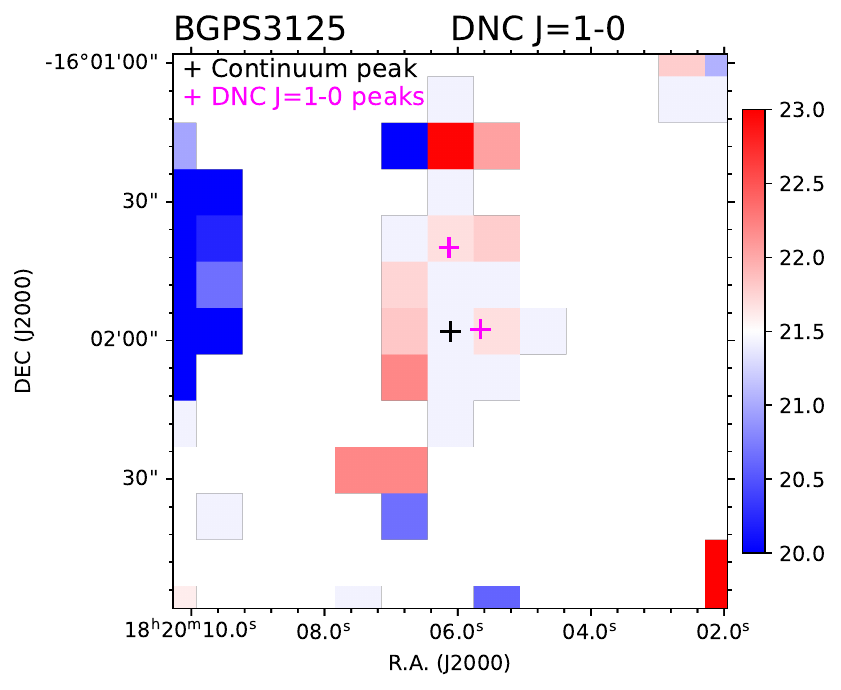}
\end{tabular}
\caption{Line width distribution of deuterated molecules for BGPS3125.}
\label{3125_velo}
\end{figure*}

\begin{figure*}
\centering
\begin{tabular}{cc}
\includegraphics[width=82mm]{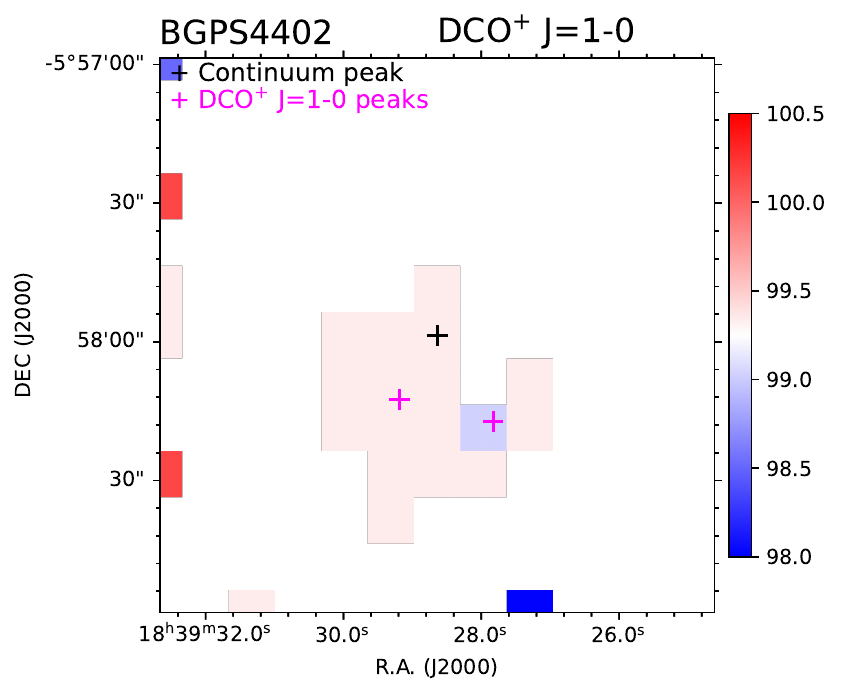}
\end{tabular}
\caption{Line width distribution of deuterated molecules for BGPS4402.}
\label{4402_velo}
\end{figure*}

\begin{figure*}
\centering
\begin{tabular}{cc}
\includegraphics[width=82mm]{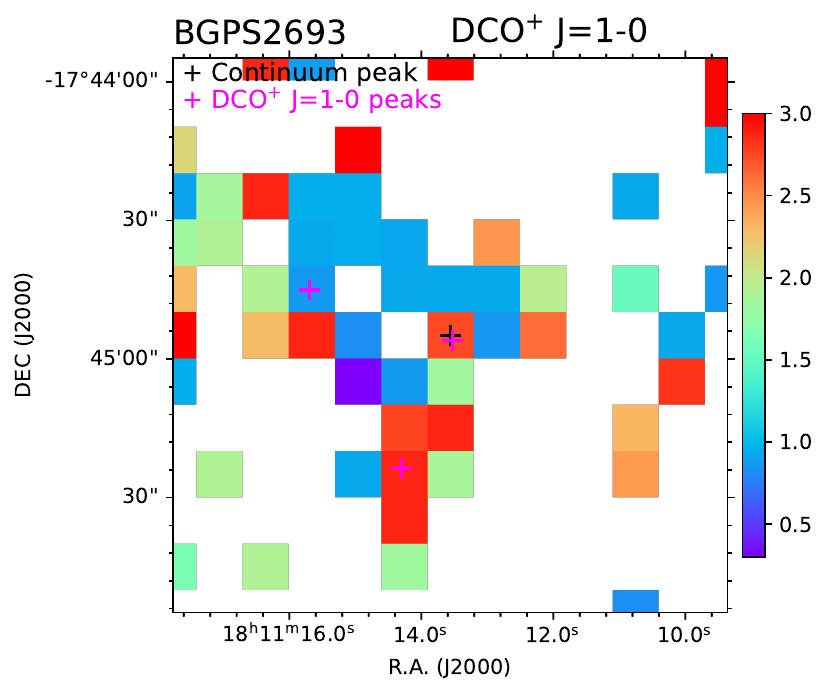}&
\includegraphics[width=82mm]{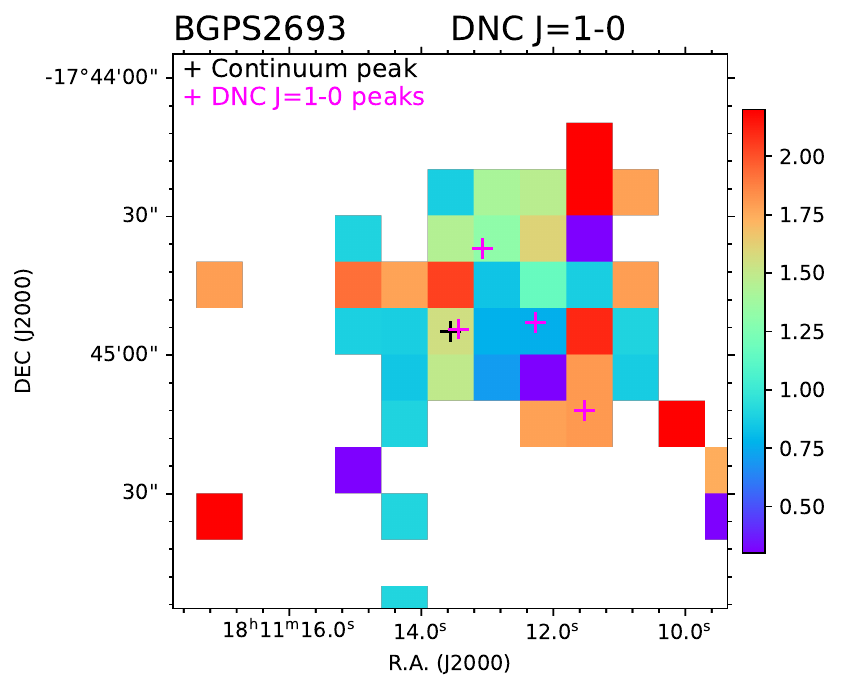}
\end{tabular}
\caption{Line width distribution of deuterated molecules for BGPS2693.}
\label{2693_width}
\end{figure*}

\begin{figure*}
\centering
\begin{tabular}{cc}
\includegraphics[width=82mm]{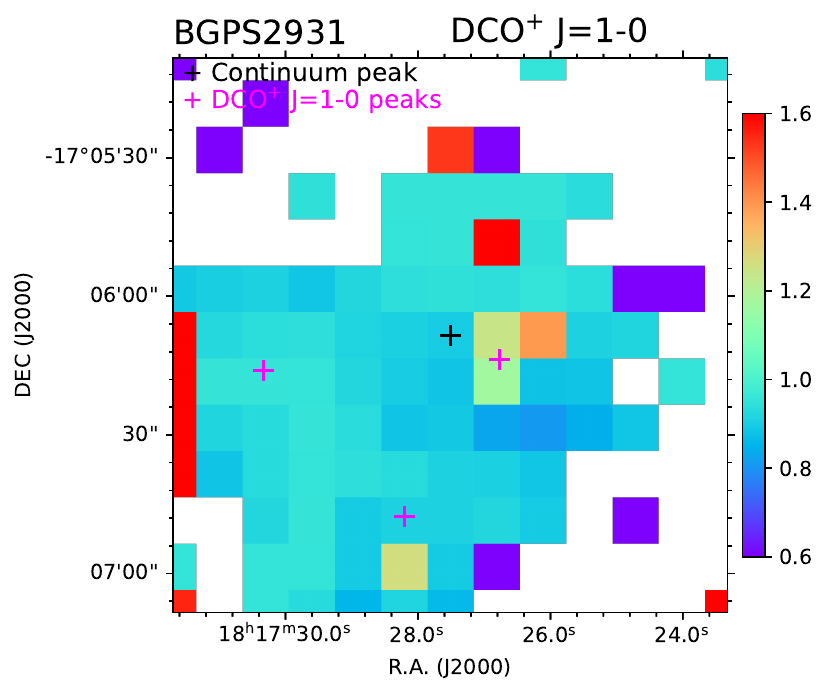}&
\includegraphics[width=82mm]{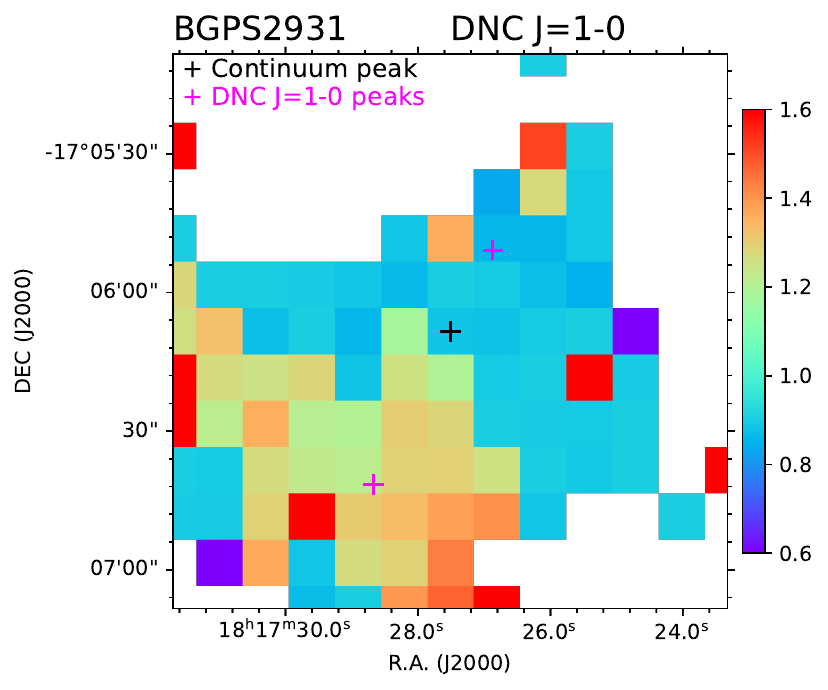}
\end{tabular}
\caption{Line width distribution of deuterated molecules for BGPS2931.}
\label{2931_width}
\end{figure*}

\begin{figure*}
\centering
\begin{tabular}{cc}
\includegraphics[width=82mm]{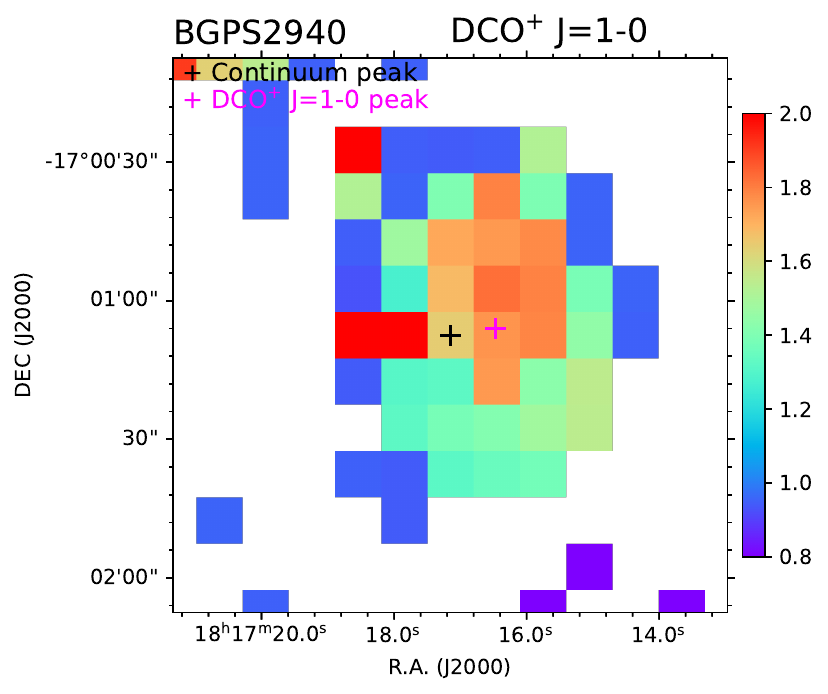}&
\includegraphics[width=82mm]{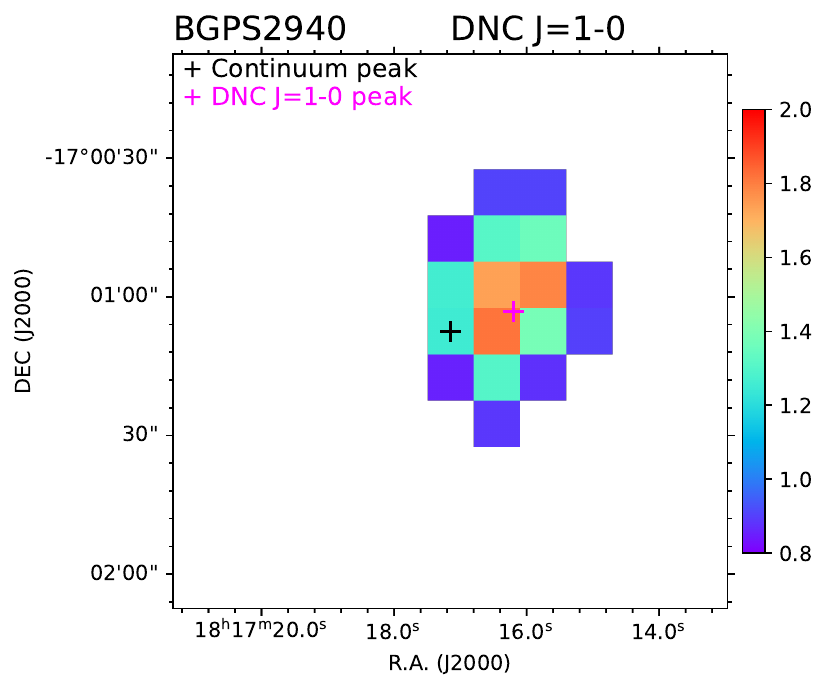}\\
\includegraphics[width=82mm]{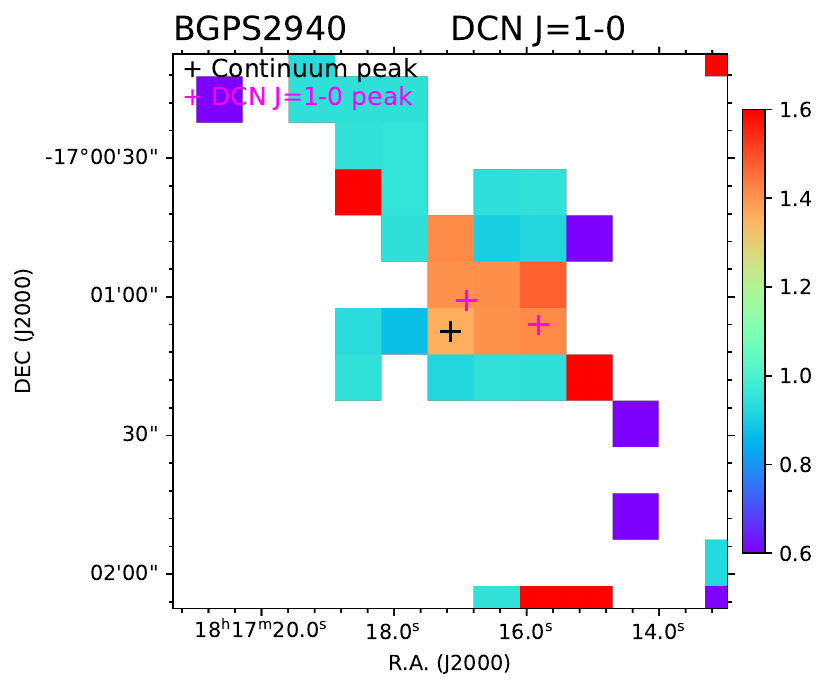}&
\includegraphics[width=82mm]{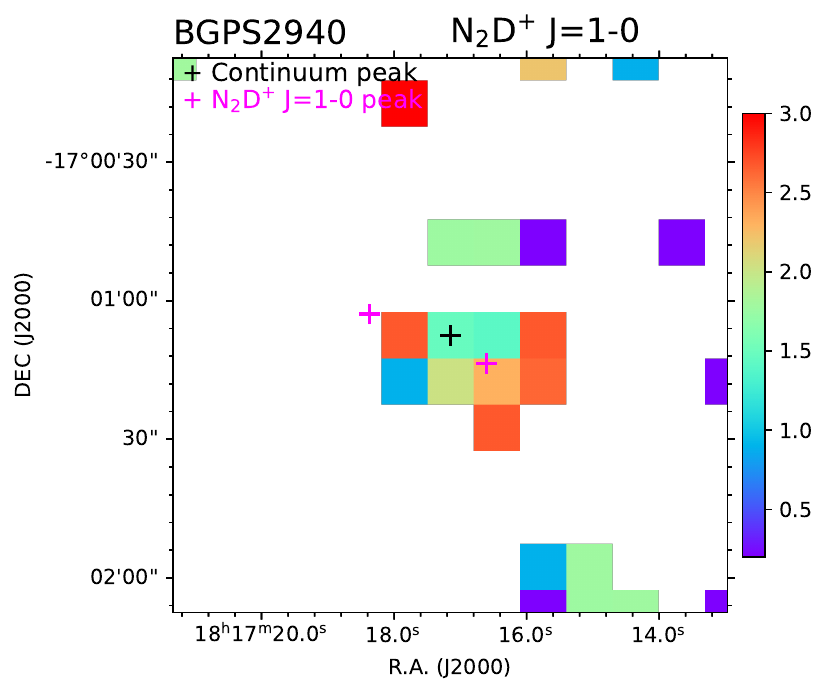}
\end{tabular}
\caption{Line width distribution of deuterated molecules for BGPS2940.}
\label{2940_width}
\end{figure*}

\begin{figure*}
\centering
\begin{tabular}{cc}
\includegraphics[width=82mm]{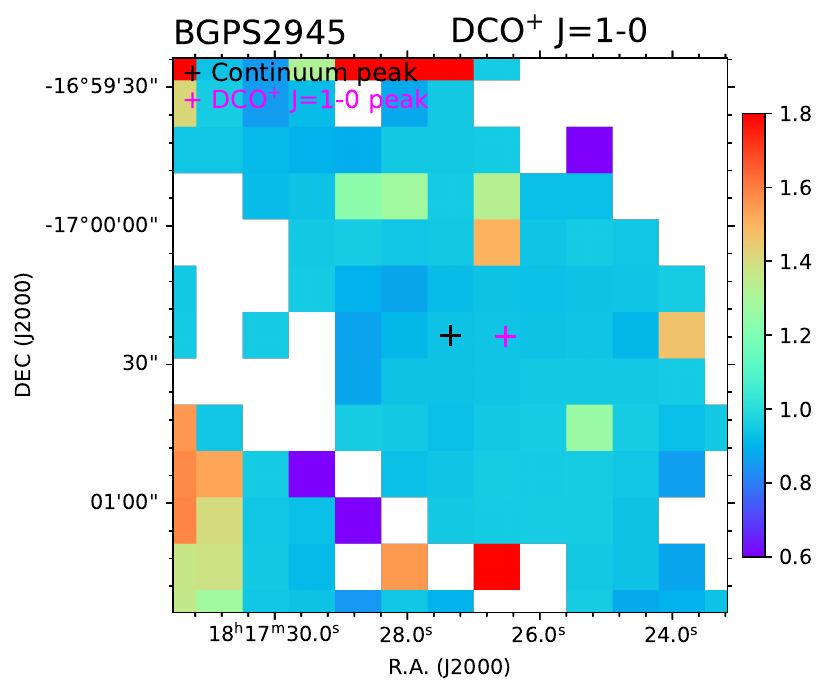}&
\includegraphics[width=82mm]{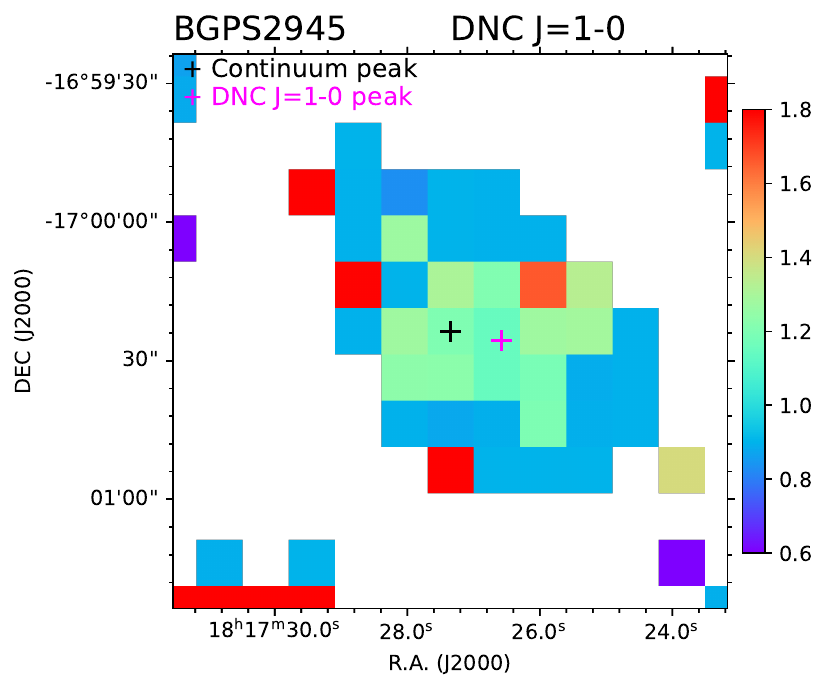}
\end{tabular}
\caption{Line width distribution of deuterated molecules for BGPS2945.}
\label{2945_width}
\end{figure*}

\begin{figure*}
\centering
\begin{tabular}{cc}
\includegraphics[width=82mm]{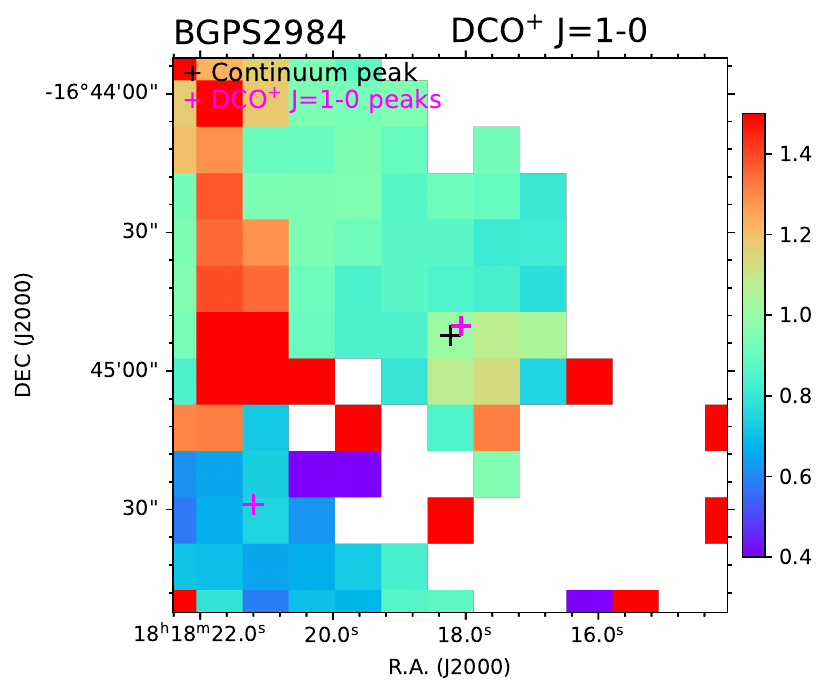}&
\includegraphics[width=82mm]{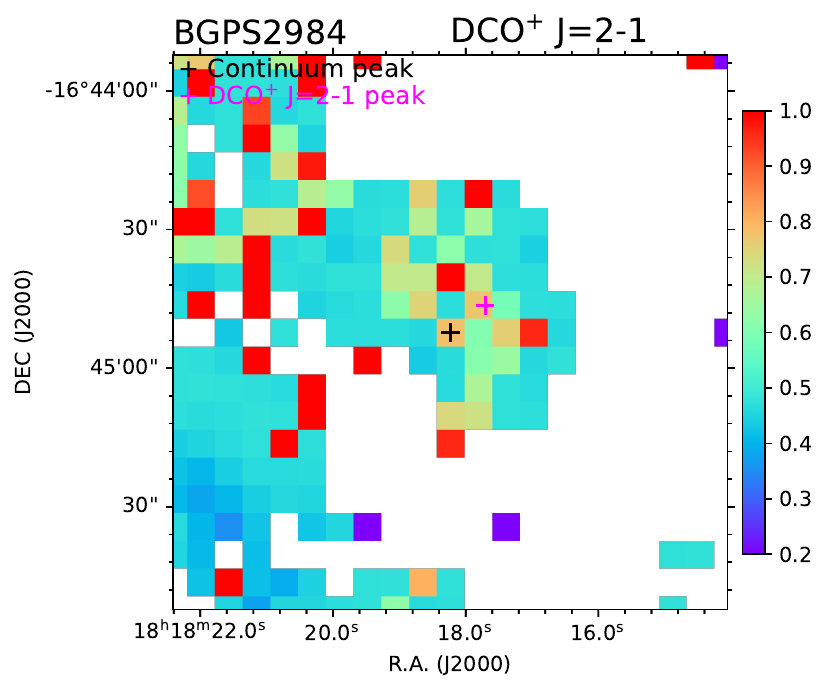}\\
\includegraphics[width=82mm]{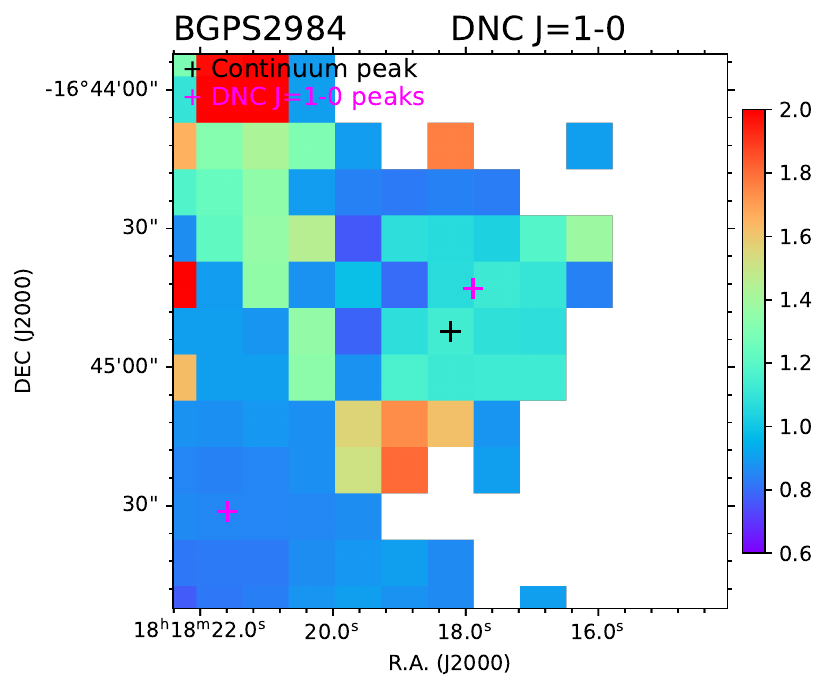}&
\end{tabular}
\caption{Line width distribution of deuterated molecules for BGPS2984.}
\label{2984_width}
\end{figure*}

\begin{figure*}
\centering
\begin{tabular}{cc}
\includegraphics[width=82mm]{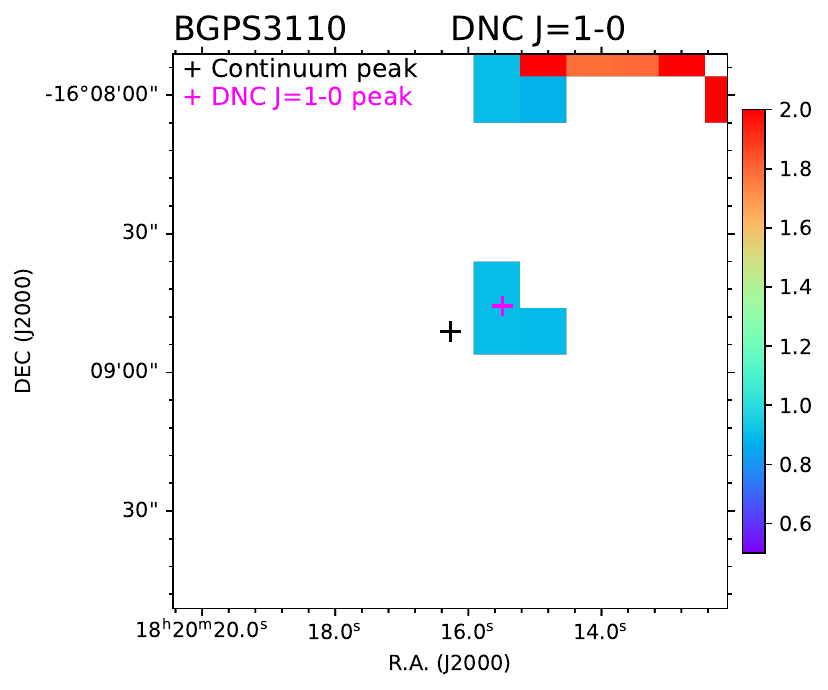}&
\includegraphics[width=82mm]{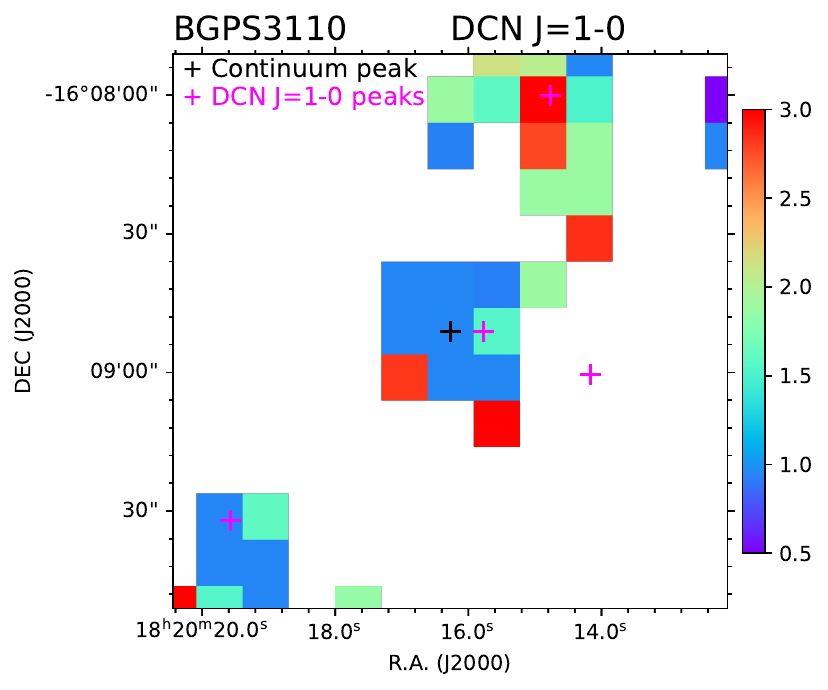}
\end{tabular}
\caption{Line width distribution of deuterated molecules for BGPS3110.}
\label{3110_width}
\end{figure*}

\begin{figure*}
\centering
\begin{tabular}{cc}
\includegraphics[width=82mm]{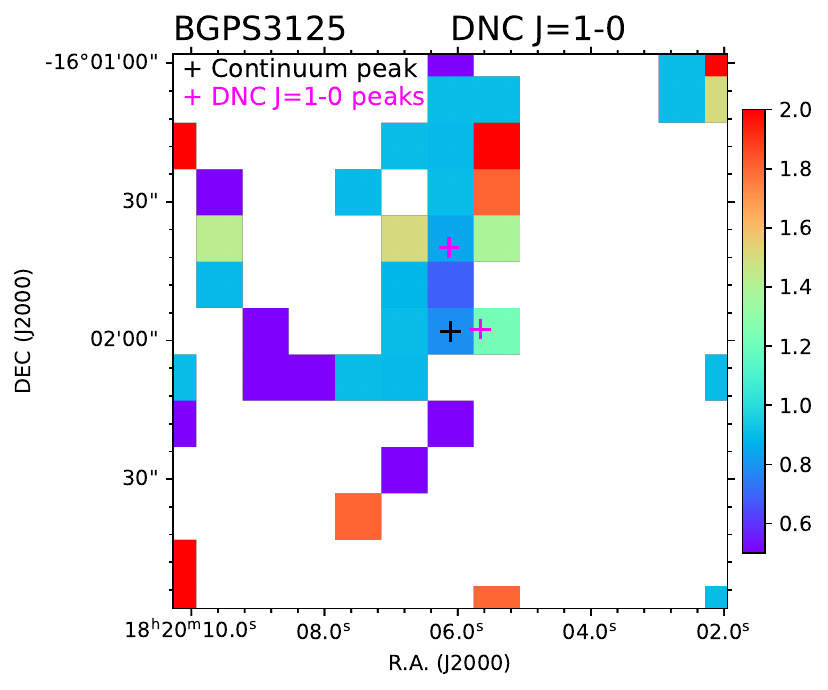}
\end{tabular}
\caption{Line width distribution of deuterated molecules for BGPS3125.}
\label{3125_width}
\end{figure*}

\begin{figure*}
\centering
\begin{tabular}{cc}
\includegraphics[width=82mm]{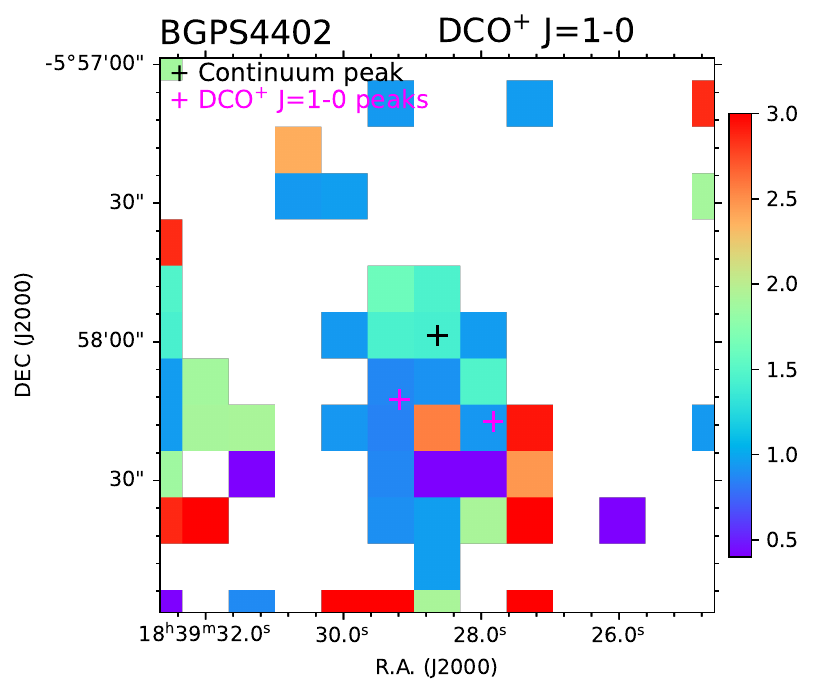}
\end{tabular}
\caption{Line width distribution of deuterated molecules for BGPS4402.}
\label{4402_width}
\end{figure*}

In this section, we present the velocity field and line width distribution of sources in the OTF observation in Figures \ref{2693_velo}$-$\ref{4402_velo} and \ref{2693_width}$-$\ref{4402_width}.

\clearpage


\clearpage


\end{appendix}


\end{document}